\documentclass[a4paper,11pt]{memoir}

\usepackage[bitstream-charter]{mathdesign}

\usepackage[utf8]{inputenc}
\usepackage[english]{babel}
\usepackage{blindtext}
\blindmathtrue

\usepackage{lscape}
\usepackage{bibunits}
\usepackage{subcaption}
\usepackage[inline]{enumitem}
\usepackage[table]{xcolor}
\usepackage{arydshln} % dashed vertical line
\usepackage{multirow}
\usepackage{ebproof}
\usepackage{fancyvrb}
\usepackage{hyperref}
\usepackage[noend]{algpseudocode}
\usepackage{algorithm}
\usepackage{amsmath}
\usepackage{stmaryrd}
\usepackage{amsthm}
\usepackage{thmtools,thm-restate}
\usepackage{graphicx}
\usepackage{tikz}
\usetikzlibrary{automata,shapes,arrows,positioning,calc,trees,fit,decorations.markings,matrix}
\usepackage{mathpazo}
\usepackage{tgpagella}
\usepackage{todonotes}
\newtheorem{lemma}{Lemma}[section]
\newtheorem{theorem}[lemma]{Theorem}
\newtheorem{proposition}[lemma]{Proposition}
\newtheorem{corollary}[lemma]{Corollary}
\theoremstyle{definition}
\newtheorem{definition}{Definition}
\newtheorem{invariant}{Invariant}
\newtheorem{example}{Example}
\newtheorem{inlinealgorithm}{Algorithm}

\usepackage[final]{microtype}

\DeclareMathOperator{\lfp}{lfp}

\newlength\drop
\makeatletter
\newcommand*\titleM{\begingroup
\setlength\drop{0.08\textheight}
\centering
\vspace*{\drop}
{\Huge\bfseries \sffamily Stream Processing Using Grammars and Regular Expressions}\\[\baselineskip]
\vfill
{\large\bfseries Ulrik Terp Rasmussen}\\
  {DIKU, Department of Computer Science\\
  University of Copenhagen, Denmark}\\[3em]
{\scshape \@date}\\
\vfill
{\large\bfseries PhD Thesis}\\
{\large\itshape This thesis has been submitted to the PhD School of the Faculty of Science, \\University of Copenhagen, Denmark}\\
\endgroup}
\makeatother

\addto\captionsenglish{%
}

\begin{document}

\begin{titlingpage}
\calccentering{\unitlength}                         % Calculate center length and stores in unitlength
\begin{adjustwidth*}{\unitlength}{-\unitlength}     % Adjust center
    \begin{adjustwidth}{-1cm}{-1cm}                 % Extra lage front page
      \titleM
    \end{adjustwidth}
\end{adjustwidth*}
\end{titlingpage}

\frontmatter

\renewcommand{\abstractname}{Abstract}
\begin{abstract}
  In this dissertation we study regular expression based parsing and the use of grammatical specifications for the synthesis of fast, streaming string-processing programs.

  In the first part we develop two linear-time algorithms for regular expression based parsing with Perl-style \emph{greedy} disambiguation. The first algorithm operates in two passes in a semi-streaming fashion, using a constant amount of working memory and an auxiliary tape storage which is written in the first pass and consumed by the second. The second algorithm is a single-pass and optimally streaming algorithm which outputs as much of the parse tree as is semantically possible based on the input prefix read so far, and resorts to buffering as many symbols as is required to resolve the next choice. Optimality is obtained by performing a PSPACE-complete pre-analysis on the regular expression.

  In the second part we present \emph{Kleenex}, a language for expressing high-performance streaming string processing programs as regular grammars with embedded semantic actions, and its compilation to streaming string transducers with worst-case linear-time performance. Its underlying theory is based on transducer decomposition into oracle and action machines, and a finite-state specialization of the streaming parsing algorithm presented in the first part. In the second part we also develop a new linear-time streaming parsing algorithm for \emph{parsing expression grammars} (PEG) which generalizes the regular grammars of Kleenex. The algorithm is based on a bottom-up tabulation algorithm reformulated using least fixed points and evaluated using an instance of the \emph{chaotic iteration} scheme by Cousot and Cousot.
\end{abstract}
\clearpage
\thispagestyle{empty}
~
\clearpage

\renewcommand{\abstractname}{Resumé}
\begin{abstract}
I denne afhandling beskæftiger vi os med parsing med regulære udtryk samt anvendelsen af grammatiske specifikationer til syntese af hurtige, strømmende programmer til strengprocessering.

I første del udvikler vi to algoritmer til parsing med regulære udtryk i lineær tid, og med \emph{grådig} afgørelse af flertydigheder i stil med Perl. Den første algoritme består af to faser, der afvikles på en semi-strømmende facon med konstant størrelse arbejdslager, samt et ekstra båndlager der henholdsvis skrives og læses af hver af de to faser. Den anden algoritme består af en enkelt fase og er optimalt strømmende i den forstand, at den udskriver så meget af parse-træet, som det er semantisk muligt ud fra det præfix af inddata, der på det givne tidspunkt er blevet indlæst. Algoritmen falder tilbage til buffering af så mange inputsymboler, som det er nødvendigt for at kunne afgøre næste valg. Optimalitet opnås ved hjælp af en PSPACE-fuldstændig præanalyse af det regulære udtryk.

I anden del præsenterer vi \emph{Kleenex}, et sprog til at udtrykke højtydende, strømmende strengprocesseringsprogrammer som regulære grammatikker med indlejrede semantiske handlinger, samt dets oversættelse til streaming string transducers med worst-case lineær tids ydelse. Den underliggende teori er baseret på dekomponering af transducere i orakel- og handlingsmaskiner, samt en specialisering af den strømmende parsingalgoritme fra den første del som en endelig tilstandsmaskine. I anden del udvikler vi også en ny lineær tids, strømmende parsing algoritme til \emph{parsing expression grammars} (PEG) der generaliserer de regulære grammatikker fra Kleenex. Algoritmen er baseret på en bottom-up tabelopstillingsalgoritme, der reformuleres ved brug af mindste fikspunkter, og som beregnes ved hjælp af en instans af Cousot og Cousots \emph{chaotic iteration}.
\end{abstract}

\renewcommand{\abstractname}{Abstract}

\clearpage

\tableofcontents

\clearpage

\listoffigures

\clearpage
 
\chapter{Preface}

This dissertation has been submitted to the PhD School of Science, Faculty of Science, University of Copenhagen, in partial fulfillment of the degree of PhD at Department of Computer Science (DIKU).

The dissertation is written as a synopsis of four enclosed research papers, including three peer-reviewed conference papers and one, as of yet, unpublished manuscript. Chapter~\ref{chap:introduction} presents a brief introduction to the two topics of this dissertation. Chapters~\ref{chap:regex-parsing} and \ref{chap:streaming-transduction} each give a more comprehensive overview of the respective topic, including an outline of the area of research, the main problems to be solved, and my contribution in relation to existing work in the literature. Each chapter concludes with a brief outline of the perspectives for future work.

I could not have written this dissertation alone, so at this point I would like to take the opportunity to thank the people who have helped me along the way. First of all, the material presented here is the result of close collaboration with my coauthors, to whom I would like to express my sincere gratitude.

To Fritz Henglein, my supervisor, thank you for giving me both enormous freedom in my research and expert guidance when needed. Your passion and never-ending spirit has been a constant source of inspiration.

To Bjørn, thank you for being a great colleague, friend, office mate and travel companion.

To Dexter Kozen, thank you for hosting me for a wonderful five months at Cornell University. To all of my current and past office mates at both Cornell and DIKU, and to my colleagues in the DIKU APL section, thank you for providing a pleasant and stimulating work environment.

To my family, thank you for your support and understanding during my work on this dissertation.

To Lotte, thank you for everything.

\vspace{1em}

{
\hfill {\underline{\hspace{12em}}}

\noindent\hfill {Ulrik Terp Rasmussen}
}

%%% Local Variables:
%%% mode: latex
%%% TeX-master: "thesis"
%%% End:

\mainmatter

\begin{bibunit}[abbrv]
{
% Calligraphy face
\newcommand{\C}[1]{\mathcal{#1}}

% Language denotation
\newcommand{\Lang}[1]{\C{L} \llbracket {#1} \rrbracket}

% Type denotation
\newcommand{\Type}[1]{\C{T} \llbracket {#1} \rrbracket}

% Set of terms
\newcommand{\Term}{\mathsf{Term}}

% List type constructor
\newcommand{\List}{\mathsf{List}}

% Injectors
\newcommand{\inl}{\mathsf{left}~}
\newcommand{\inr}{\mathsf{right}~}

% Coding
\newcommand{\Code}{\mathsf{code}}
\newcommand{\Decode}{\mathsf{decode}}

\tikzstyle{sstate}=[state,inner sep=3pt,minimum size=0pt]

% Closer underline
\newcommand{\ul}[1]{\underline{\smash{#1}}\vphantom{#1}}

% Highlighting
\newcommand{\highlight}[1]{{\setlength{\fboxsep}{0pt}\colorbox{gray!20}{#1}}}

% Semantic action
\newcommand{\sact}[1]{{\setlength{\fboxsep}{0pt}\colorbox{blue!20}{\texttt{#1}}}}

% Problem box
\newcommand{\problembox}[1]{\begin{center}%
    \fbox{\begin{minipage}{0.9\linewidth}%
      {#1}%
    \end{minipage}}%
\end{center}}

%%% Local Variables: 
%%% mode: latex
%%% TeX-master: "thesis"
%%% End: 

\chapter{Introduction}
\label{chap:introduction}

Programmers need to make several trade-offs when writing software. Most important, the software has to be correct while at the same time being able to handle all reasonably expected inputs and usage scenarios. In addition, the underlying implementation should be as simple as possible so that it can be maintained and adjusted without major risks of introducing errors. Furthermore, the software should also be efficient in the sense that requests are answered within a reasonable time frame, and there should be no way to bring the software to use excessive amounts of time and space, either by an adversary or by accident.

In this dissertation, we will focus on programs that process data in its simplest form: strings of symbols. In practice, programs of this kind can be found performing a variety of tasks including processing of user input, decoding of data formats and protocols, automated word processing, and searching over large amounts of sequential data, from log files and source code to sequenced DNA. The one task all programs have in common is the task of deducing the underlying structure of the data in order to be able to process it in a meaningful way. This task can be quite difficult to tackle in itself, and is not made easier by virtue of having to take into account the trade-offs mentioned earlier. There is therefore a need for general solutions and tools that can help overcome these challenges, thereby reducing the time and risk associated with software development.

Any general solution will also have to make trade-offs, so we should not expect a single approach to be able to solve all of our problems once and for all. In this dissertation, we will approach the problem from the perspective of automata theory and formal languages, which already have deep roots in the theoretical and practical aspects of parsing. It is the ultimate goal to provide a new set of methods based on solid foundations which can be used to build string-processing programs with strong performance guarantees, while still being flexible and expressive enough to not increase development costs.

The narrative of this dissertation can roughly be divided into two parts. %The first part is about techniques for parsing using regular expressions, and how this can be achieved in a streaming fashion. The second part is about how we can use this to build a compiler for a domain-specific programming language for high-efficiency streaming text transformations over regular languages, and finally we explore extensions to formalisms that are more expressive than regular expressions.

\subsection{Regular Expression Based Parsing}

The first part will be concerned with \emph{regular expressions}, an algebraic formalism with a well-understood theory that is commonly used to express patterns of strings. Their conciseness and attractive computational properties have made them popular as a language for expressing string search and input validation programs. Since we rarely validate an input string without the intention of using it, most practical regular expression implementations also provide facilities for breaking up the string into parts based on the specified pattern, a process also known as \emph{parsing}. However, the classical theory does not account for the issues that are normally associated with parsing, and as a result these data extraction facilities have been built as ad-hoc extensions on top of implementations of the classical interpretation of regular expressions as pure string patterns. This approach has missed some opportunities for greater expressivity, and has also resulted in the loss of the attractive performance guarantees that were associated with regular expressions in the first place.

We take a different approach, and work from a generalized theory of regular expressions that take parsing into account. From this perspective we find new algorithmic methods for solving the \emph{regular expression parsing problem}: Given a regular expression and an input string, what is its associated \emph{parse tree}?

\subsection{Grammar-Based Stream Processing}

In the second part we focus on formalisms for specifying string processing programs which operate based on the syntactic structure of their inputs. Programs of this kind perform a range of useful tasks, including advanced text substitution, filtering and formatting of logging data, as well as implementations of data exchange formats. As data of this kind is often generated at a high rate, string processing programs have to operate in a \emph{streaming fashion} where they only store a small part of the input string in memory at any time. Writing and maintaining software which keeps track of the technical details of streaming while also dealing with the complexities of a data format is a challenging task.

We propose the use of \emph{syntax-directed translation schemes} as a suitable formalism for expressing such programs. The formalism based on existing formalisms for describing string patterns, such as regular expressions, extended with \emph{embedded semantic actions}---arbitrary program fragments which are executed based on how a given input string is matched by the specified pattern. We study two different formalisms, and methods for efficiently running specifications written in them in a streaming fashion. The first of these have been used in the design and implementation of the high-performance streaming string processing language \emph{Kleenex}.

%%% Local Variables:
%%% mode: latex
%%% TeX-master: "thesis"
%%% End:

\chapter{Regular Expression Based Parsing}
\label{chap:regex-parsing}

This chapter is concerned with the problem of parsing using \emph{regular expressions}, which are mathematical expressions for denoting sets of strings, first introduced by Kleene to describe sets of events in mathematical models of the nervous system~\cite{kleene1956}. After their practical application for text search was pointed out by Thompson~\cite{thompson68}, regular expressions became a popular language for specifying complex text search patterns. They now enjoy applications in many diverse areas, including text editing~\cite{pike1987}, querying of data formats~\cite{clarke1997}, detection of code duplication~\cite{schwarz2014} and searching in sequenced DNA data~\cite{mahalingam2008}. Their popularity primarily stems from their simplicity, conciseness, and attractive theoretical properties. Most important, a computer program only has to spend time proportional to the length of a string in order to decide if it belongs to the set described by a given regular expression, guaranteeing that a search query will return within reasonable time.

Over the years, implementations have moved away from these theoretical foundations, and the nomenclature ``regex'' is now informally used to refer to the implemented versions of the original ``regular expressions'', with which they have little in common apart from syntax. Operators were added in order to increase the number of patterns that could be expressed, notably backreferences and recursion, and mechanisms for limited parsing in the form of \emph{capturing groups} were introduced to accommodate advanced text substitution. Most of these extensions seem to have been added based on what was possible to implement as extensions to the existing search algorithms, and as a result the theoretical properties were lost: Matching a string against a regex can take exponential time in the length of the input, and it is not uncommon to see performance bugs due to seemingly innocent-looking regexes that suddenly trigger this behavior for rare pathological inputs\footnote{\url{http://stackstatus.net/post/147710624694/outage-postmortem-july-20-2016}}\footnote{\url{http://davidvgalbraith.com/how-i-fixed-atom/}}.

We will take a different approach, and work from a more general theory of regular expressions which takes the issues related to parsing into account. By changing our perspective on the problem, we reveal new and efficient algorithms for solving the core problem related to the use of regular expressions for data extraction. Furthermore, we will see that the generalization offers an increase in expressivity, enabling new and interesting applications of regular expressions.

We give a semi-formal exposition of the theory of regular expressions in Section~\ref{sec:regular-expressions-in-theory}, including its relation to finite automata. In Section~\ref{sec:regular-expressions-in-practice}, we show how popular ``regex'' software packages have extended this theory and discuss the trade-offs. We present the main problem of \emph{regular expression based parsing} in Section~\ref{sec:regular-expression-based-parsing}, and relate it to a computational model called \emph{finite transducers} in Section~\ref{sec:parsing:as-transduction}. In Section~\ref{sec:parsing-techniques} we review the current approaches to solving this problem, and in Section~\ref{sec:contributions} we present our contributions. We conclude this chapter in Section~\ref{sec:parsing:conclusions}.

\section{Regular Expressions In Theory}
\label{sec:regular-expressions-in-theory}

A regular expression (RE) is a formal mathematical expression using a limited set of operators. Their purpose is to serve as concise specifications of sets of strings with certain desirable properties.

It is assumed that some finite set of symbols $\Sigma$, also called the \emph{alphabet}, is given. The alphabet specifies the valid symbols that may occur in the strings described by an RE. For example, $\Sigma$ could be the set of the 256 bytes that can be represented by 8-bit words, or the full set of Unicode code points---in the remainder of this chapter we will just assume that $\Sigma$ is the set of lowercase letters $\{\texttt{a},\texttt{b}, ..., \texttt{z}\}$. The \emph{infinite} set of all strings over $\Sigma$ is written $\Sigma^*$, that is
\[ \Sigma^* = \{ \varepsilon, \texttt{a}, \texttt{b}, ..., \texttt{z}, \texttt{aa}, \texttt{ab}, ..., \texttt{az}, ..., \texttt{ba}, \texttt{bb}, ..., \texttt{bz}, ... \} \]
and so on, where $\varepsilon$ stands for the \emph{empty string}. Of course, appending the empty string to another string $u$ results in the same string again: $\varepsilon u = u = u \varepsilon$. We will generally use letters $u,v,w$ to refer to strings, and will avoid using them as symbols.

The syntax of REs can be compactly described by a generative grammar:
\[ E ::= a \mid \epsilon \mid E_1^* \mid E_1 E_2 \mid E_1 + E_2  \]
That is, the simplest REs consist of a single symbol $a$ from $\Sigma$ or the ``unit expression'' $\epsilon$. Smaller REs $E_1, E_2$ can be combined to form larger ones by the ``star operator'' $E_1^*$, the ``sequence operator'' $E_1 E_2$ or the ``sum operator'' $E_1 + E_2$. These are listed in increasing order of precedence, i.e. $\texttt{ab}^* + \texttt{c}$ is parenthesized as $(\texttt{a}(\texttt{b}^*)) + \texttt{c}$. Sequence and sum associate to the right, so $E_1 E_2 E_3$ and $E_1 + E_2 + E_3$ parenthesize as $E_1 (E_2 E_3)$ and $E_1 + (E_2 + E_3)$, respectively.

The usual interpretation of REs is as denotations of \emph{formal languages}, each of which is a subset of $\Sigma^*$. The sets $\{\texttt{cnn}, \texttt{bbc} \}$, $\{ \texttt{a}, \texttt{aa}, \texttt{aaa}, ... \}$ and $\emptyset = \{\}$ are all examples of such, where the last is the degenerate case of the \emph{empty language}. In order to define the meaning of REs, we will first need to introduce some operations on languages. Given two languages $A$ and $B$, we can combine them into a new language $AB$ formed by concatenating every string in $A$ with every string in $B$, or formally:
\[ AB = \{ uv \mid u \in A, v \in B \}. \]
For example, if $A = \{ \texttt{ab}, \texttt{cd} \}$ and $B = \{\texttt{e}, \texttt{f}\}$, then $AB = \{\texttt{abe}, \texttt{abf}, \texttt{cde}, \texttt{cdf}\}$. Concatenation can be iterated any number of times for a single language: For any number $n \geq 0$, define
\[ A^n = \underbrace{A A \cdots A}_{\text{$n$ times}} \]
where $A^0 = \{ \varepsilon \}$ is defined as the language containing just the empty string. For example, if $A = \{\texttt{a},\texttt{b}\}$, then $A^3 = \{\texttt{aaa}, \texttt{aab}, \texttt{aba}, \texttt{abb}, \texttt{baa}, \texttt{bab}, \texttt{bba}, \texttt{bbb} \}$. The last language operation we will need is also the most powerful. For a language $A$, write $A^*$ for the language formed by taking any number of strings from $A$ and concatenating them. Formally, this is the language
\[ A^* = \bigcup_{n=0}^\infty A^n = A^0 \cup A^1 \cup A^2 \cup ... \]
This is a quite powerful operation. For example, if we view the alphabet $\Sigma$ as a language of single-symbol strings, then $\Sigma^*$ is exactly the infinite set of all strings containing symbols from $\Sigma$. For another example:
\[
\{\texttt{ab}, \texttt{c}\}^* = \{ \varepsilon, \texttt{ab}, \texttt{c}, \texttt{abc}, \texttt{cba}, \texttt{abab}, \texttt{cc}, \texttt{abcab}, ... \}.
\]

Every RE $E$ is a description of a language $\Lang{E}$ which is built using the operations we have just defined. The mapping from syntax to language operators should be quite apparent, and is formally defined as follows:
\begin{align*}
  \Lang{a} ={}& \{ a \} & \Lang{\epsilon} ={}& \{\varepsilon\} \\
  \Lang{E_1^*} ={}& \Lang{E_1}^* & \Lang{E_1E_2} ={}& \Lang{E_1}\Lang{E_2} \\
  \Lang{E_1 + E_2} ={}& \Lang{E_1} \cup \Lang{E_2}
\end{align*}

It can be quite instructive to view an RE $E$ as a \emph{pattern} whose meaning as such is the set of strings $\Lang{E}$ matched by it. This view also hints to their practical use for text search. For example, consider the following pattern:
\[ (\texttt{he} + \texttt{she})(\texttt{was} + \texttt{is})((\texttt{very})^* + \texttt{not})(\texttt{happy} + \texttt{hungry} + \texttt{sad}) \]
Ignoring the issue of word spacing, this matches an infinite number of variations of sentences of the following kind:
\begin{center}
  \texttt{he was not hungry}, \quad \texttt{she is very happy}, \quad \texttt{she is very very hungry}, \\
  \texttt{she is not sad}, \quad \texttt{he is very very sad}, $...$
\end{center}

While REs offer a lot of expressive power, there are many languages that they cannot express. For example, there is no way to specify the language of all strings of the form \[ \underbrace{\texttt{a} \texttt{a} \cdots \texttt{a}}_{\text{$n$ times}} \underbrace{\texttt{b} \texttt{b} \cdots \texttt{b}}_{\text{$n$ times}} \]
that is, strings with the same number of occurrences of $a$s and $b$s, but with all $a$s occurring before the $b$s. Patterns of this kind may occur in practice in the form of strings of matching parentheses, so surely it would be useful to be able to express them. However, this restriction of expressive power is deliberate. In order to see why, we have to look at the computational properties of REs.

\subsection{Finite Automata}

At this point we have established the semantics of REs, and we have illustrated their power and limitations as a language for constructing string patterns. We now briefly review a general solution to the \emph{recognition problem}:
\problembox{Given an RE $E$ and a string $u$, is $u$ in the language $\Lang{E}$?}
The limited expressivity of REs turns out to be an advantage when solving this problem, as it allows every RE to be converted into a particularly simple type of program called a \emph{finite automaton}~\cite{kozen1997}.

Automata are usually defined using state diagrams as follows:
\begin{center}
\begin{tikzpicture}[>=stealth',auto]
  \node[initial,state] at (0,0) (S0) {$0$};
  \node[state,accepting] at (3,-1.5) (S1) {$1$};
  \node[state,accepting] at (3,1.5) (S2) {$2$};
  \draw[->,bend left] (S0)  to node {$\texttt{a}$} (S2);
  \draw[->,bend left] (S2)  to node {$\texttt{b}$} (S1);
  \draw[->] (S1) to node[swap] {$\varepsilon$} (S0);
  \draw[->] (S2) to node {$\texttt{a}$} (S0);
  \draw[->,loop above] (S0) to node {$\texttt{c}$} (S0);
  \draw[->,loop above] (S2) to node {$\texttt{a}$} (S2);
  \draw[->,bend left] (S1) to node {$\texttt{b}$} (S0);
\end{tikzpicture}
\end{center}
The circles are called \emph{states}, and the numbers within are names identifying them. The arrows between states are called \emph{transitions}, and are labeled by either a single symbol or the empty string. A single state is a designated \emph{starting state} and is marked as such. Each state is either \emph{accepting} or \emph{not accepting}, with accepting states drawn as double circles.

With every automaton $M$ is associated a set of strings $\Lang{M}$, in the same way that every RE is associated with one. However, where the language of an RE is defined in terms of language operators, the language of an automaton is defined in terms of a process. A specific set of rules specify how to ``run'' an automaton on some string $u$ by keeping track of a number of ``pebbles'' that are placed on the states. The rules are as follows:
\begin{enumerate}
\item Place a pebble on the starting state and on any state reachable via one or more $\varepsilon$-transitions.
\item For each symbol in $u$ from left to right:
  \begin{enumerate}
  \item Pick up all pebbles, remembering what states had pebbles on them.
  \item For every state that had pebbles on it and has a transition matching the next symbol, put a pebble on the destination state.
  \item Put a pebble on all states that can be reached via one or more $\epsilon$-transitions from a state with pebbles on it.
  \end{enumerate}
\end{enumerate}

If at least one accepting state has pebbles on it when all symbols have been processed, then $u$ is \emph{accepted} by the automaton, and we say that $u$ is in $\Lang{M}$; otherwise it is \emph{rejected}. Note that we are not concerned with the number of pebbles on each state, just that it has a non-zero amount. The following demonstrates an accepting run of the automaton on the string $\texttt{aabca}$:
\[
\begin{array}{p{1.25cm}p{1.25cm}p{1.25cm}p{1.25cm}p{1.25cm}p{1.25cm}l}
  start & $\texttt{a}$ & $\texttt{a}$ & $\texttt{b}$ & $\texttt{c}$ & $\texttt{a}$ & \\
  $\{0\}$ & $\{2\}$ & $\{2,0\}$ & $\{1,0\}$ & $\{0\}$ & $\{2\}$ & \text{accept}
\end{array}
\]

Our interest in finite automata comes from the fact that this process is very efficient to implement on a computer, which only has to look at each input symbol once. Since the set of states with pebbles on them can never be larger than the total amount of states, it will always take time proportional to the length of the string to decide whether it is recognized by the automaton or not.

There is a deep connection between REs and automata, namely that for every RE $E$ there is an automaton $M$ such that $\Lang{E} = \Lang{M}$~\cite{kleene1956}. In other words, automata provide the recipe for efficiently ``running'' regular expressions. For an example, consider the RE $(a^*b)^*$ which has the following associated automaton:
\begin{center}
\begin{tikzpicture}[>=stealth',auto,x=1cm,y=0.75cm,scale={0.75}]
  \node[sstate] at (0,0) (loop0) {$1$};
  \node[sstate] at (-2,2) (a0) {$2$};
  \node[sstate] at (2,2) (a1) {$3$};
  \node[sstate] at (4,0) (b0) {$4$};
  \node[sstate] at (8,0) (b1) {$5$};
  \node[initial,sstate] at (4,-2) (loop1) {$0$};
  \node[sstate,accepting] at (8,-2) (accept) {$6$};
  \draw[->] (a0) to node {$\texttt{a}$} (a1);
  \draw[->] (b0) to node {$\texttt{b}$} (b1);
  \draw[->] (loop0) to node {$\varepsilon$} (a0);
  \draw[->] (a1) to node {$\varepsilon$} (loop0);
  \draw[->] (loop0) to node {$\varepsilon$} (b0);
  \draw[->] (b1) to node[swap] {$\varepsilon$} (loop1);
  \draw[->] (loop1) to node {$\varepsilon$} (loop0);
  \draw[->] (loop1) to node {$\varepsilon$} (accept);
\end{tikzpicture}
\end{center}
Another remarkable fact is that this connection also holds in the other direction: for every automaton, there is an RE denoting its language. This equivalence with finite automata explains our hesitance towards adding more expressive power to REs, as they have exactly the amount of power they need while still enabling us to use efficient techniques based on finite automata to implement them.

The finite automata presented in this section are also called \emph{non-deterministic finite automata} (NFA), due to the fact that there can be more than one state with pebbles on it when running it. This is done to distinguish them from the special case where exactly on state can ever have pebbles on it, in which case the automaton is called \emph{deterministic} (DFA). Any NFA can be converted to a DFA~\cite{rabin1959}, although this may cause the number of states to increase exponentially. On the other hand, running a DFA is often even faster than running an NFA, so this optimization can pay off when the input size is significantly larger than the RE itself.

\section{Regular Expressions In Practice}
\label{sec:regular-expressions-in-practice}

In this section we briefly review some of the history of regex implementations and point out the central differences between them and the REs presented in the previous section. For a more comprehensive account of both the history and features of regexes, we refer to the book by Friedl~\cite{friedl97}.

REs were popularized in computing from 1968. Thompson~\cite{thompson68} pointed out the application of text search and implemented them in the editors QED and \texttt{ed}. This later led to the creation of the specialized UNIX search tool \texttt{grep}, whose name comes from the \texttt{ed} command $\texttt{g/}re\texttt{/p}$ for RE based searching~\cite{friedl97}. Also in 1968, Johnson, Porter, Ackley, Ross \cite{johnson1968} applied REs for building lexical analyzers for programming language compilers.

In 1986 Harry Spencer wrote the first general regex library for incorporation in other software. This was later adopted and heavily extended by Larry Wall in his Perl~\cite{perl} programming language, which became popular for practical string manipulation tasks, in part due to regexes being built into its syntax. Its popularity eventually led to the notion of ``Perl compatible regexes'' (PCRE) for referring to implementations that behaved like Perl's. PCRE regexes have now made it into most other popular programming languages, either built into the language syntax or provided via a library.

There are several important differences between regexes and REs. On the surface, the syntax is slightly different from that of REs: the RE $\texttt{a}(\texttt{b}^*+\texttt{c})\texttt{d}^*$ would typically be written as the regex \texttt{a(b*|c)d*}, with variations depending on the exact flavor. In the following we will highlight the notable semantic differences that separate regex implementations from their theoretical counterparts.

\subsection{Capturing Groups}
The recognition semantics are extended to \emph{matching} via the notion of \emph{capturing groups} which are used to extract information about \emph{how} a string is matched by a regex, and not just whether it is in its language or not. For every parenthesis in a regex, the implementation will report the position of the substring that matched the enclosed subexpression. The user can later refer to the substrings matched by capturing groups by using the names \texttt{\textbackslash{}1}, \texttt{\textbackslash{}2}, ..., and so on---the groups are numbered from left to right according to their opening parentheses. The following example shows a how a PCRE implementation will report matches when given the input \texttt{aaaabbbb}:

\[
  \texttt{a*(aa(bb|bbb))b*} \qquad\qquad \texttt{a a }\overbrace{\texttt{a a }\underbrace{\texttt{b b}}_{\texttt{\textbackslash{}2}}}^{\texttt{\textbackslash{}1}}\texttt{ b b }
\]
The example illustrates that even though capturing groups seem benign, they introduce a new problem of \emph{ambiguity}. The second capturing group could also have matched the substring \texttt{bbb}, but the left alternative was chosen by the implementation. This behavior is shared by all PCRE implementations, which prefer left alternatives over right ones. However, a second popular flavor of regex specified by the POSIX standard~\cite{ieee1003.1-2008} would pick the right alternative based on a policy which maximizes the lengths of submatches from left to right. The differences in matching semantics of different regex flavors has resulted in much confusion among users, with the issue further exacerbated by the fact that the POSIX regex specification is unnecessarily obscure, leading to a situation where no two tools claiming to be POSIX actually report the same results for all inputs~\cite{fowlerPOSIX,okui2013,kuklewicz,sulzmann2014}.

Another limitation of capturing groups is that they do not work well when combined with the star operator. For example, when the regex \texttt{(a*b)*} is matched against input \texttt{abaaab}, it is ambiguous whether the implementation should return \texttt{ab} or \texttt{aaab} for the substring \texttt{\textbackslash{}1}. In some usage scenarios, the user might even want the position of both submatches (e.g. for reading a list of data items), but that is not possible under the regex matching paradigm.

\subsection{Backreferences and Backtracking}
Capturing groups enable another regex extension called \emph{backreferences} that allow users to write \texttt{\textbackslash{}1}, \texttt{\textbackslash{}2}, ... to refer back to the substrings matched by capturing groups earlier in the pattern. For example, the regex \texttt{(a*)b\textbackslash{}1} will match any string of the form
\[ \underbrace{\texttt{a} \texttt{a} \cdots \texttt{a}}_{\text{$n$ times}} \texttt{b} \underbrace{\texttt{a} \texttt{a} \cdots \texttt{a}}_{\text{$n$ times}} \]
No RE or finite automaton can express such a language.
Implementations with backreferences typically work by using an extended form of automata where transitions can mark the beginnings and endings of a capturing groups, and other transitions can refer back to them. Consider the regex \texttt{(aa|a)\textbackslash{}1a}, which has the following extended automaton:
\begin{center}
\begin{tikzpicture}[>=stealth',auto,x=0.65cm,y=0.4cm]
  \node[sstate,initial] at (-1,0) (S0) {$0$};
  \node[sstate] at (1,0) (S1) {$1$};
  \node[sstate] at (3,2) (S2) {$2$};
  \node[sstate] at (5,2) (S3) {$3$};
  \node[sstate] at (7,2) (S4) {$4$};
  \node[sstate] at (3,-2) (S5) {$5$};
  \node[sstate] at (7,-2) (S6) {$6$};
  \node[sstate] at (9,0) (S7) {$7$};
  \node[sstate] at (11,0) (S8) {$8$};
  \node[sstate] at (13,0) (S9) {$9$};
  \node[sstate,accepting] at (15,0) (S10) {$10$};
  \draw[->] (S0) to node{$($} (S1);
  \draw[->] (S1) to node{$\varepsilon$} (S2);
  \draw[->] (S2) to node{\texttt{a}} (S3);
  \draw[->] (S3) to node{\texttt{a}} (S4);
  \draw[->] (S4) to node{$\varepsilon$} (S7);
  \draw[->] (S1) to node[swap]{$\varepsilon$} (S5);
  \draw[->] (S5) to node{\texttt{a}} (S6);
  \draw[->] (S6) to node[swap]{$\varepsilon$} (S7);
  \draw[->] (S7) to node{$)$} (S8);
  \draw[->] (S8) to node{\texttt{\textbackslash{}1}} (S9);
  \draw[->] (S9) to node{\texttt{a}} (S10);
\end{tikzpicture}
\end{center}
We can run it using a strategy called \emph{backtracking}, which resembles the pebble strategy from the previous section, but with the restriction that at most one state can have a pebble on it at any time. When the pebble can move to more than one state, we choose one arbitrarily and remember the state we came from in case we need to go back and try another one. If at any point the pebble cannot move to a new state but more symbols are in the string, then we backtrack to the last choice point and try the next alternative. This process is repeated until all input symbols have been consumed and the pebble is on an accepting state, or until all possible pebble movements have been exhausted. We write down the position in the input string whenever the pebble crosses an opening and closing parenthesis. When the pebble tries to transition a backreference, it uses these positions to check whether the remainder of the string matches the substring given by the last recorded positions.

For example, on input \texttt{aaa} in the automaton depicted above, the pebble first moves along the states 0,1,2,3,4,7,8, recording that the group \texttt{\textbackslash{}1} matches \texttt{aa}. But now the pebble cannot move from 8 to 9, since only \texttt{a} is left, and \texttt{aa} is needed. It backtracks to the last choice (state 1) and now continues along states 5,6,7,8, recording that \texttt{\textbackslash{}1} matches \texttt{a}. Since \texttt{aa} remains, the pebble can now move along states 9 and 10, and the automaton accepts.

This strategy works, and is used in PCRE style regex implementations. They agree on disambiguation by systematically always trying left alternatives first. For this reason, the PCRE disambiguation policy is also called \emph{greedy} or \emph{first-match} disambiguation. Although backtracking is usually fast in practice, it has the disadvantage that some regexes and inputs can make it spend an atrocious amount of time---exponential in the length of the input string.

As stated in the introduction of this chapter, the exponential time worst case does occur in practice, and also opens up systems to attacks by adversaries who construct input data to deliberately trigger the worst-case behavior of certain regexes. This type of attack, known as \emph{regular expression denial-of-service} (REDoS)~\cite{owasp2015}, has motivated a lot of recent research in methods for identifying vulnerable regexes~\cite{kirrage2013,sumi2014,berglund2014,DBLP:journals/corr/RathnayakeT14,weideman2016}.

\section{Regular Expression Based Parsing}
\label{sec:regular-expression-based-parsing}

After reading the previous sections it should be clear that regex implementations make trade-offs between expressivity and predictable performance. These trade-offs are perfectly acceptable for situations like ad-hoc text manipulation and automation of non-critical tasks, but as we pointed out there are also scenarios where we need greater matching expressivity (capturing groups under the star operator) and/or hard performance guarantees (exponential time performance bugs in mission-critical systems). In this section we address the first issue by presenting a generalization of the language semantics for REs into one of \emph{parsing}. The goal of formulating a new semantics for REs is to separate the specification of what we want to solve from its actual implementation, giving us freedom to try out different approaches. The second issue will then be addressed by finding new ways to implement the parsing semantics efficiently, which is the main topic of our contributions.

\subsection{Terms and Types}
Before we can define the parsing semantics for REs, we need to introduce the notions of \emph{terms} and \emph{types}.

Terms act as generalizations of strings with extra information about how they are matched. They can be compactly described by the following generative grammar:
\[
  V ::= a \mid \varepsilon \mid (V_1, V_2) \mid \inl V_1 \mid \inr V_2 \mid [V_1, V_2, ..., V_n]
\]
In other words, symbols $a$ and the empty string $\varepsilon$ are the smallest terms. Larger terms can be build as follows: if $V_1, V_2$ are terms, then $(V_1,V_2)$ is a term; $\inl V_1$ is a term; and $\inr V_2$ is a term. Finally, if $V_1, V_2, ..., V_n$ are terms, then the list $[V_1, V_2, ..., V_n]$ is a term. In particular, the empty list $[]$ is a term.

Terms have structure, and can easily be decomposed into smaller parts by a program. It is instructive to view them as upside-down trees, with the smallest terms at the bottom. For example, the term $V=((\texttt{a},\varepsilon), \inl [\inl \texttt{a}, \inr \texttt{b}])$ has the following tree structure:
\begin{center}
\begin{tikzpicture}[scale=0.8]
  \tikzstyle{level 1}=[sibling distance=2.5cm,level distance=1.25cm]
  \tikzstyle{level 2}=[sibling distance=1.25cm]
  \node {$(\cdot,\cdot)$}
  child {node {$(\cdot,\cdot)$}
    child {child {child {node {$\texttt{a}$}}}}
    child {child {child {node {$\varepsilon$}}}}
  }
  child {node {$\inr \cdot$}
    child {node {$[\cdots]$}
      child {node {$\inl \cdot$}
        child {node {$\texttt{a}$}}
      }
      child {node {$\inr \cdot$}
        child {node {$\texttt{b}$}}
      }
    }
  };
\end{tikzpicture}
\end{center}
This illustrates how terms are merely strings with more structure: by removing anything above the symbols at the bottom, we are left with the string $\texttt{a}\varepsilon{}\texttt{ab} = \texttt{aab}$. This is also called the \emph{flattening} of $V$, and we write $|V| = \texttt{aab}$.

We write $\Term_\Sigma$ to refer to the infinite set of all terms over the alphabet $\Sigma$. We will call subsets $T$ of terms for \emph{types}, analogously to the way we called subsets of $\Sigma^*$ languages in the previous sections. We also need to define analogous operations for concatenation, union and star.

Given two types $T$ and $U$, write $T \times U$ for their \emph{product} which is the set of all pairs whose first and second components are from $T$ and $U$, respectively. Formally, $T \times U = \{ (V_1, V_2) \mid V_1 \in T, V_2 \in U \}$.
For example,
\begin{align*}
  &\{\inl \texttt{a}, \inr \texttt{b}\} \times \{[], [\texttt{a}], [\texttt{a},\texttt{a}], ... \} \\
  &\qquad {}= \{ (\inl \texttt{a}, []), (\inr \texttt{b}, []), (\inl \texttt{a}, [\texttt{a}]), (\inr \texttt{b}, [\texttt{a}]), ... \}
\end{align*}
The product operation is analogous to the concatenation operation $AB$ for languages, but with the difference that it reveals whenever there are multiple ways of obtaining the same string. Consider the following example, which shows language concatenation in the top and type product in the bottom:
\begin{align*}
  \{\texttt{a}, \texttt{ab}\} \{\texttt{bc}, \texttt{c}\} ={}& \{ \texttt{abc}, \texttt{ac}, \texttt{abbc} \} \\
  \{\texttt{a},~ (\texttt{a},\texttt{b}) \} \times \{ (\texttt{b},\texttt{c}),~ \texttt{c}\} ={}& \{ (\texttt{a}, (\texttt{b},\texttt{c})),~ (\texttt{a}, \texttt{c}),~ ((\texttt{a},\texttt{b}),(\texttt{b},\texttt{c})),~ ((\texttt{a},\texttt{b}), \texttt{c}) \}
\end{align*}
The string $\texttt{abc}$ can be formed either by picking $\texttt{a}$ from the left language and $\texttt{bc}$ from the right; or by picking $\texttt{ab}$ and then $\texttt{c}$. The two are indistinguishable, so the result is a language with three strings instead of four. For products, the two cases can be distinguished, and we obtain a four-element type.

Given two types $T$ and $U$, write $T + U$ for their \emph{sum} which is the set all terms of the form $\inl V_1$ and $\inr V_2$, where $V_1$ is in $T$ and $V_2$ is in $U$. Formally, $T + U = \{ \inl V_1 \mid V_1 \in T \} \cup \{ \inr V_2 \mid V_2 \in U \}$.
For example,
\begin{align*}
  & \{\inl \texttt{a}, \inr \texttt{b}\} + \{ [], [\texttt{a}], [\texttt{a},\texttt{a}], ... \} \\
  & \qquad {}= \{ \inl (\inl \texttt{a}), \inl (\inr \texttt{b}), \inr [], \inr [\texttt{a}], \inr [\texttt{a},\texttt{a}], ... \}
\end{align*}
The sum operation is analogous to the union operation $A \cup B$ for languages, but with the difference that it records from which operand a given element comes from, ensuring that no two elements are conflated. Consider the following example, which shows language union on top and type sum in the bottom:
\begin{align*}
  \{\texttt{a}, \texttt{ab}\} \cup \{\texttt{ab}, \texttt{ac}\} ={}& \{\texttt{a}, \texttt{ab}, \texttt{ac}\} \\
  \{\texttt{a},~ (\texttt{a},\texttt{b})\} + \{(\texttt{a},\texttt{b}),~ (\texttt{a},\texttt{c})\} ={}& \{ \inl \texttt{a},~ \inl (\texttt{a},\texttt{b}),~ \inr (\texttt{a},\texttt{b}),~ \inr (\texttt{a},\texttt{c}) \}
\end{align*}
The string $\texttt{ab}$ is in both languages, but the union only contains the string once, resulting in a three-element language. For sums, the elements are tagged by the side they came from and are thus not conflated, resulting in a four-element type.

Finally, if $T$ is a type, then $\List(T)$ is the set of all lists whose elements are from $T$. Formally,
$\List(T) = \{ [V_1, V_2, ..., V_n] \mid n \geq 0, \text{for all $i \leq n\ldotp V_i \in T$} \}$.
For example, if $T$ is as in the previous examples, then
\[ \List(T) = \{ [], [\inl \texttt{a}], [\inr \texttt{b}], [\inl \texttt{a}, \inl \texttt{a}], [\inl \texttt{a}, \inr \texttt{b}], [\inr \texttt{b}, \inl \texttt{a}], ... \} \]
The list operation is analogous to the star operator for languages. The difference between the two is similar to the difference between concatenation and product.

Languages can be understood as string patterns. Types are also a form of string patterns, but where we also care about \emph{how} a given string is in the pattern---this is explained by the terms that flatten to the string. The correspondence between languages and types is summarized in the table below:
\begin{center}
\begin{tabular}{ll|ll}
  \multicolumn{2}{l|}{Strings \& Languages} & \multicolumn{2}{l}{Terms \& Types} \\
  \hline
  All strings & $\Sigma^*$ & All terms & $\Term_\Sigma$ \\
  Languages & $A \subseteq \Sigma^*$ & Types & $T \subseteq T_\Sigma$ \\
  Concatenation & $AB$ & Product & $T \times U$ \\
  Union & $A \cup B$ & Sum & $T + U$ \\
  Star & $A^*$ & List & $\List(T)$
\end{tabular}
\end{center}

\subsection{Regular Expressions as Types}
\label{sec:parsing:regexastypes}

We are now ready to define a parsing semantics for REs. Using the framework of terms and types that we have set up in the previous, we associate every RE $E$ with a type $\Type{E}$ as follows:
\begin{align*}
  \Type{a} ={}& \{a\} & \Type{\epsilon} ={}& \{ \varepsilon \} \\
  \Type{E_1^*} ={}& \List(\Type{E_1}) & \Type{E_1 E_2} ={}& \Type{E_1} \times \Type{E_2} \\
  \Type{E_1 + E_2} ={}& \Type{E_1} + \Type{E_2}
\end{align*}

There is a close connection between the traditional language interpretation and the type interpretation of REs. Namely, for every RE $E$, if $u$ is a string in its language $\Lang{E}$, then there is a term $V$ in its type $\Type{E}$ with flattening $u$, that is $|V| = u$. Vice versa, for any term $T$ in $\Type{E}$, its flattening $|V|$ can also be found in $\Lang{E}$.

The benefit of this change of perspective is that the type interpretation of REs accounts for ambiguity, whereas this aspect is hidden in the language interpretation and only arises as a property of the concrete implementations. For example, consider the two REs $\texttt{a}(\texttt{a}+\texttt{b})^*$ and $(\texttt{ab}+\texttt{a})(\texttt{a}+\texttt{b})^*$. They have the same languages, $\Lang{\texttt{a}(\texttt{a}+\texttt{b})^*} = \Lang{(\texttt{ab}+\texttt{a})(\texttt{a}+\texttt{b})^*}$, but not the same types: \[ \Type{\texttt{a}(\texttt{a}+\texttt{b})^*} \not= \Type{(\texttt{ab}+\texttt{a})(\texttt{a}+\texttt{b})^*}.\] The type interpretation captures the fact that there is only one term with flattening $\texttt{aba}$ in the type of the first RE, while there are two such terms in the type of the second, as illustrated in Figure~\ref{fig:ex:terms}.
\begin{figure}[h]
\begin{center}
\begin{tikzpicture}
\tikzstyle{level 1}=[level distance=1cm, sibling distance=1.55cm]
\tikzstyle{level 2}=[sibling distance=1.3cm]
\tikzstyle{level 3}=[sibling distance=1cm]
\matrix (m) [column sep=0.5cm] {
  \node {$(\cdot,\cdot)$}
  child {child { child {node {$\texttt{a}$}}}}
  child {node {$[\cdots]$}
    child {node {$\inr \cdot$} child {node {$\texttt{b}$}}}
    child {node {$\inl \cdot$} child {node {$\texttt{a}$}}}
  }
  ;
  &
  \node (m-1-2) {\qquad};
  &
  \node {$(\cdot,\cdot)$}
  child {node{$\inl \cdot$}
    child {node{$(\cdot,\cdot)$}
      child {node{$\texttt{a}$}}
      child {node{$\texttt{b}$}}
    }
  }
  child {node{$[\cdots]$}
    child {node{$\inl \cdot$}
      child {node {$\texttt{a}$}}
    }
  }
  ;
  &
  \node {$(\cdot,\cdot)$}
  child {node{$\inr \cdot$}
    child {child{node{$\texttt{a}$}}}
  }
  child {node{$[\cdots]$}
    child {node{$\inr \cdot$}
      child {node {$\texttt{b}$}}
    }
    child {node{$\inl \cdot$}
      child {node {$\texttt{a}$}}
    }
  }
  ;
  \\
  \node[anchor=north] {$(\texttt{\texttt{a}},[\underbrace{\inr \texttt{b}}_{\texttt{\textbackslash{}1}}, \underbrace{\inl\!\!\!\phantom{g} \texttt{a}}_{\texttt{\textbackslash{}1}}])$};
  &
  \node (m-2-2) {};
  &
  \node[anchor=north] {$(\underbrace{\inl (\texttt{\texttt{a}},\texttt{b})}_{\texttt{\textbackslash{}1}}, [\underbrace{\inl\!\!\!{\phantom{g}}\texttt{a}}_{\texttt{\textbackslash{}2}}])$};
  &
  \node[anchor=north] {$(\underbrace{\inr \texttt{\texttt{a}}}_{\texttt{\textbackslash{}1}}, [\underbrace{\inr \texttt{b}}_{\texttt{\textbackslash{}2}}, \underbrace{\inl\!\!\!{\phantom{g}}\texttt{\texttt{a}}}_{\texttt{\textbackslash{}2}}])$};
  \\
};
\draw[dotted] (m-1-2.north) to (m-2-2.south);
\end{tikzpicture}
\end{center}
  \caption{Terms with flattening $\texttt{aba}$ for REs $\texttt{a}(\texttt{a}+\texttt{b})^*$ and $(\texttt{ab}+\texttt{a})(\texttt{a}+\texttt{b})^*$.}
  \label{fig:ex:terms}
\end{figure}

A term fully ``explains'' how its flattened string can be parsed according to the given RE, including information pertaining to the star operator. As such, we will therefore also refer to terms as \emph{parse trees}. The information contained in parse trees transcends the information provided by capturing groups in a regex implementation. As the example illustrates, the groups $\texttt{\textbackslash{}1}$ in the first RE and $\texttt{\textbackslash{}2}$ in the second cannot be assigned unique substrings since they both occur under a star operator, whereas the parse tree just contains a list of all the possible assignments.

The recognition problem introduced in Section~\ref{sec:regular-expressions-in-theory} can now be generalized to the \emph{parsing problem}:
\problembox{Given an RE $E$ and an input string $u$, is there a parse tree $V$ in $\Type{E}$ such that $|V|=u$?}
Unlike the recognition problem, the parsing problem has more than one solution due to the possibility of ambiguous choices for the parse tree $V$. Different strategies for picking such a parse tree are analogous to the solutions to the disambiguation problem for regex matching, and it is possible to give definitions that are compatible with both PCRE~\cite{frca2004} and POSIX~\cite{sulzmann2014} disambiguation.

\section{Parsing as Transduction}
\label{sec:parsing:as-transduction}

The finite automata used as the computational model for the language interpretation of REs are not expressive enough for the type interpretation. Since a finite automaton can only ever accept or reject an input, it does not support the construction of a parse tree.
There is, however, another model called \emph{finite transducers}~\cite{berstel79,mohri1997finite} which can. In order to connect the type interpretation to this machine model, we need to first introduce the concept of \emph{bit-coding}~\cite{nihe2011}.

\subsection{Bit-Coding}
Bit-coding can be seen as a way of writing down a parse tree for an RE as a flat string, but in a way such that the parse tree can easily be recovered again. The coding scheme is based on the observation that if an RE is of the form $E_1 + E_2$, then any of its parse trees must be of one of the two forms $\inl V_1$ or $\inr V_2$. A single number, say $0$ or $1$, can be used to specify which of the respective shapes the parse tree has, and we are left with the problem of finding a code for one of the subtrees $V_1$ or $V_2$. Similarly, every parse tree for an RE $E_1^*$ is a list, and we can again use the symbols $0$ and $1$ to indicate whether the list is non-empty or empty, respectively. In the latter case there is only one possible list (the empty list), and in the first case, we are left with the problem of finding codes for the first element and the rest of the list. For a product $E_1 E_2$, every parse tree is of the form $(V_1, V_2)$, so all we have to do is find codes for the two subtrees. Similarly, for the remaining constructs and $a$, $\epsilon$ there is only one possible parse tree, so no coding is needed to specify which one it is.

Formally, for any parse tree $V$, we define $\Code(V)$ as a string of bits, that is a string in $\{0,1\}^*$, as follows:
\begin{align*}
  \Code(a) ={}& \varepsilon \\ 
  \Code(\varepsilon) ={}& \varepsilon \\
  \Code([V_1, V_2, ..., V_n]) ={}& 0 ~ \Code(V_1) ~ 0 ~ \Code(V_2) ~ ... ~ 0 ~ \Code(V_n) ~ 1 \\
  \Code( (V_1, V_2) ) ={}& \Code(V_1) \Code(V_2) \\
  \Code( \inl V_1) ={}& 0 ~ \Code(V_1) \\
  \Code( \inr V_2) ={}& 1 ~ \Code(V_2)
\end{align*}

For example, the term $(\inr \texttt{a}, [\inr \texttt{b}, \inl \texttt{a}])$ which is a parse tree for $(\texttt{ab}+\texttt{a})(\texttt{a}+\texttt{b})^*$ and depicted as the third tree in Figure~\ref{fig:ex:terms} has the following bit-code:
\begin{align*}
  \Code((\inr \texttt{a}, [\inr \texttt{b}, \inl \texttt{a}])) ={}& \Code(\inr \texttt{a}) ~ \Code([\inr \texttt{b}, \inl \texttt{a}]) \\
  ={}& 1 ~ \Code([\inr \texttt{b}, \inl \texttt{a}]) \\
  ={}& 1 ~ 0 \Code(\inr \texttt{b}) ~ 0 \Code(\inl \texttt{a}) ~ 1 \\
  ={}& 1 ~ 0 1 ~ 00 ~ 1
\end{align*}

A code can easily be decoded again to obtain the original parse tree. That is, for every RE $E$ there is also a function $\Decode_E$ which takes a bit-code and returns the parse tree that we started out with. In other words, we have $\Decode_E(\Code(V)) = V$. We will not need the definition of decoding for this presentation, and refer to Nielsen and Henglein~\cite{nihe2011} for details.

\subsection{Transductions}
By treating parse trees as codes, we can now connect the type interpretation of REs with another type of finite automata called \emph{finite transducers}~\cite{berstel79}. These are finite automata extended such that every transition is now labeled by a pair $a / c$, where $a$ is an \emph{input label} and $c$ is an \emph{output label}. Input labels are symbols from $\Sigma$ as before, or the empty string $\varepsilon$. Output labels are strings over some output alphabet $\Gamma$. A string $u$ is accepted by a transducer if there is a path from the initial to the final state such that the concatenation of all the input labels along the path equals $u$. Furthermore, every such path is associated with a corresponding output string obtained by concatenating all the output labels in the same way. This justifies the name \emph{transducers}, as they model a simple form of string translators. When transducers are used for parsing, we will have $\Gamma = \{0,1\}$ as we will be translating input strings to bit-codes.

It can be shown that every RE has a transducer which accepts the strings in its language and furthermore outputs all the bit-codes of the corresponding parse trees with the same flattening~\cite{nihe2011}. For an example, see the transducer for the RE $(\texttt{ab}+\texttt{a})(\texttt{a}+\texttt{b})^*$ in Figure~\ref{fig:ex:parsing-transducer}. It can be seen that this machine generates two codes for the input $\texttt{aba}$, corresponding to the two parse trees on the right in Figure~\ref{fig:ex:terms}.

\begin{figure}[h]
  \centering
  \begin{tikzpicture}[x=2.5cm,y=1.45cm,>=stealth',auto]
    \node[sstate,initial] at (0,0) (S1) {$1$};
    \node[sstate] at (1,1) (S2) {$2$};
    \node[sstate] at (1,-1) (S3) {$3$};
    \node[sstate] at (2,-1) (S4) {$4$};
    \node[sstate] at (3,0) (S5) {$5$};
    \node[sstate,accepting] at (3,-1.5) (S9) {$9$};
    \node[sstate] at (4,0) (S6) {$6$};
    \node[sstate] at (4,1) (S7) {$7$};
    \node[sstate] at (4,-1) (S8) {$8$};
    \draw[->] (S1) to node {$\varepsilon / 0$} (S2);
    \draw[->] (S1) to node[swap] {$\varepsilon / 1$} (S3);
    \draw[->] (S2) to node {$\texttt{a} / \varepsilon$} (S5);
    \draw[->] (S3) to node {$\texttt{a} / \varepsilon$} (S4);
    \draw[->] (S4) to node {$\texttt{b} / \varepsilon$} (S5);
    \draw[->] (S5) to node[swap] {$\varepsilon / 1$} (S9);
    \draw[->] (S5) to node {$\varepsilon / 0$} (S6);
    \draw[->] (S6) to node[swap] {$\varepsilon / 0$} (S7);
    \draw[->] (S6) to node {$\varepsilon / 1$} (S8);
    \draw[->] (S7) to node[swap] {$\texttt{a} / \varepsilon$} (S5);
    \draw[->] (S8) to node {$\texttt{b} / \varepsilon$} (S5);
  \end{tikzpicture}
  \caption{A bit-coded parsing transducer for the RE $(\texttt{ab}+\texttt{a})(\texttt{a}+\texttt{b})^*$.}
  \label{fig:ex:parsing-transducer}
\end{figure}
Running a non-deterministic transducer is not as straightforward as using the pebble method for automata, since every pebble is now associated with the output string generated along its path. Since there can be an exponential number of different ways to get to a particular state, a pebble strategy will have to limit the number of active pebbles on each state to at most one in order to ensure linear running time. This corresponds to disambiguation of parses when more than one parse of a string is possible.

Bit-coded parsing transducers provide another perspective on the parsing problem which now becomes:
\problembox{Given an RE $E$ and a string $u$, is there an accepting path with input $u$ and output $v$ in the parsing transducer?}

\section{Recognition, Matching and Parsing Techniques}
\label{sec:parsing-techniques}

We have discussed three different problems pertaining to REs: recognition, matching and parsing. The first is formulated in terms of the language semantics of REs, the second arises in concrete implementations and the third is formulated in terms of the type interpretation. The answers to each problem provide increasing amounts of information, as summarized in the following table:
\begin{center}
\begin{tabular}{ll}
  Problem & Solution \\
  \hline
  Recognition & Accept/Reject \\
  Matching & Accept/Reject and disambiguated captures \\
  Parsing & Accept/Reject and disambiguated parse tree
\end{tabular}
\end{center}
In this section we review the work that has been done on techniques for solving the above.

\subsection{Recognition and Matching}
For pure recognition, the NFA (``multi-pebble'') and DFA (``single-pebble'') based techniques described in this chapter are well-known~\cite{kozen1997,alsu2006}. The construction of NFAs from REs is generally attributed to Thompson~\cite{thompson68}, and McNaughton and Yamada~\cite{mnya60}. Instead of automata, one can also use Brzozowski~\cite{brzozowski64} or Antimirov~\cite{antimirov96} derivatives. These are syntactic operators on REs which correspond to removing a single letter from all strings in the underlying language. The recognition problem can then be reduced to taking iterated derivatives and checking whether the resulting RE contains the empty string. Implementations of regex matching not based on automata or derivatives are generally based on backtracking which has already been covered earlier in this chapter.

The first implementation of regex matching appeared in Pike's \texttt{sam} editor~\cite{pike1987}. The method was not based on backtracking, but tracked the locations of capturing groups during running of the NFA. According to Cox~\cite{cox2007}, Pike did not know that his technique was new and thus did not claim it as such. Laurikari~\cite{laurikari2000} later rediscovered it and formalized it using an extension of NFAs with \emph{tags} for tracking captures, and also gave a method for converting tagged NFAs to tagged DFAs.

While NFA based approaches ensure linear time, they are not as fast as DFAs, which on the other hand can get very big for certain REs. Cox~\cite{cox2009} describes a way of constructing DFAs on the fly while running the NFA. The method obtains the performance benefits of DFAs without risking an exponential blowup of the number of states during conversion, and gracefully falls back to using only the NFA when too many DFA states are encountered. It is implemented in the RE2 library~\cite{cox2010,re2} which also supports matching via capturing groups, although the fast DFA technique supports at most one group. Other approaches to augmenting NFAs with information about capturing groups exist~\cite{fihuwi2010,haber2013}, with a particularly elegant one due to Fischer, Huch and Wilke~\cite{fihuwi2010}, implemented in the Haskell~\cite{hudak2007} programming language. It avoids explicitly constructing an NFA by treating the nodes in the RE syntax tree as states in a Glushkov~\cite{glushkov1960} automaton.

It is also possible to perform RE matching without the use of finite automata by applying RE derivatives. This is a popular approach for implementations in functional programming languages where the representation of finite automata can be cumbersome~\cite{ort2009}. Sulzmann and Lu~\cite{sulzmann2012} give an RE matching algorithm by extending Brzozowski and Antimirov derivatives to keep track of partially matched capturing groups added to the syntax of REs, and they give variants of the method for both POSIX and PCRE disambiguation.

\subsection{Parsing}

\subsubsection{Via General Parsing Techniques}
The RE formalism is subsumed by more general language formalisms such as \emph{context-free grammars} (CFG) which are capable of expressing non-regular languages such as $a^nb^n$. Methods for parsing with CFGs can therefore also be applied to solve the RE parsing problem, but due to their generality they cannot take advantage of the limited expressivity. The literature on CFG parsing algorithms is vast~\cite{grune2008}, but they can generally be divided into two categories: \emph{deterministic} and \emph{general} algorithms.

General CFG parsing algorithms include CYK~\cite{younger1967}, Earley~\cite{earley1970} and GLR~\cite{tomita1987}, and they can parse all CFGs regardless of ambiguity, including REs. The result is often a set of all possible parse trees, with disambiguation deferred to the consumer of the algorithm. The disadvantage of using general CFG algorithms for RE parsing is first of all that the worst-case running time is non-linear, a situation which is theoretically impossible to improve~\cite{lee2002}. Furthermore, we are rarely interested in the set of all parse trees and would rather prefer disambiguation to be built in.

Deterministic CFG parsing algorithms include LR($k$)~\cite{knuth1965} and LL($k$)~\cite{lewis1968}, and they guarantee linear time complexity at the expense of only working for a strict subset of CFGs which are \emph{deterministic} (choices are resolved by looking at most $k$ symbols ahead in the input) relative to the strategy employed by the respective algorithms. The result is always a single parse tree, as ambiguity is ruled out by the determinism restriction. This unfortunately also rules out all ambiguous REs, so deterministic CFG parsing will only work for unambiguous RE subclasses such as one-unambiguous REs~\cite{brwo98}.

Ostrand, Paull and Wcyuker~\cite{ostrand1981} restrict themselves to regular CFGs and devise an algorithm for deciding whether such a grammar is FL($k$) for some $k$, where FL($k$) means that at most $k$ next symbols have to be examined in order to resolve any choice. They give two linear time algorithms which can parse any FL($k$) grammar while producing the parse tree on the fly. In the case where unbounded lookahead is required, the latter algorithm still works but may use non-linear time.

Another general language formalism is Ford's~\cite{ford2004} \emph{parsing expression grammars} (PEG), which can also express any RE~\cite{medeiros2014}. Contrary to CFG parsing, PEG parsing can actually be done in linear time~\cite{aho1972,ford2002} and always yields a unique parse tree consistent with the disambiguation policy of PCRE. The known linear time parsing algorithms use quite a lot of memory, however, which is again a consequence of the generality of PEGs.

\subsubsection{Pure RE Parsing}
Most automata-based RE parsing algorithms operate in two separate passes, where the first pass runs over the input string and the second runs over an auxiliary data structure produced during the first pass. We will classify such methods as being either ``forwards-backwards'' or ``backwards-forwards'' depending on the direction of these runs.

Kearns~\cite{kearns91} devised the first known pure RE parsing algorithm which operates by running the NFA in reverse while journaling the sets of active states in each step. If the run succeeds then the journal is traversed again in order to construct the parse tree. It is thus a backwards-forwards algorithm.

Dubé and Feeley~\cite{dufe2000} gave the first method based on NFAs whose transitions are annotated with actions for constructing parse trees. Under this view, NFA runs also produce a parse tree whenever the machine accepts, but since many paths are explored at once in the forward simulation, the problem becomes finding the one that lead to acceptance. Their forwards-backwards algorithm builds a DFA without actions and runs it while journaling the sequence of visited states. If the DFA accepts, the journal can be traversed again to reconstruct a single NFA path using a precomputed lookup table. By executing the actions on this path, the corresponding parse tree is obtained.

Neither of the methods by Kearns or Dubé and Feeley are concerned with implementing a specific formal disambiguation policy. Kearns implements a policy which seems to resemble that of PCRE, but he never proves them equivalent. Dubé and Feeley encode disambiguation in the lookup table which is not uniquely characterized, and so disambiguation is left to the implementation. This situation was resolved by Frisch and Cardelli~\cite{frca2004} who independently rediscovered the backwards-forwards method of Kearns, but also formalized PCRE disambiguation in terms of parse trees and proves that the method actually implements this policy. They also gave a satisfying solution to the problem of dealing with so-called \emph{problematic REs} which cause naïve backtracking implementations to run forever, and thus also pose a problem for a formal account of PCRE disambiguation. Their solution also handles problematic REs, but in a way which gives the same results as backtracking search in all cases where it terminates.

The formalization by Frisch and Cardelli seems to be the first mention of the type interpretation of REs. This interpretation is further investigated by Henglein and Nielsen~\cite{heni2011} who use it to give a sound and complete reasoning system for proving RE containment (is the language of one RE contained in another?). Their system has a computational interpretation as coercions of parse trees and also admits an encoding of other reasoning systems for RE containment~\cite{salomaa66,kozen1994,grabmayer2005}, equipping them with a computational interpretation as well. They also introduce the \emph{bit-coding} of parse trees described in Section~\ref{sec:parsing:regexastypes}. Nielsen and Henglein~\cite{nihe2011} show that the forwards-backwards method of Dubé and Feeley and the backwards-forwards method of Frisch and Cardelli can both be modified to emit bit-codes instead of materializing the parse trees.

Parsing with the POSIX ``leftmost-longest'' disambiguation policy is significantly more difficult than parsing with the PCRE ``greedy'' policy. Okui and Suzuki~\cite{okui2011,okui2013} give the first forwards-backwards algorithm for disambiguated RE parsing using POSIX disambiguation. It runs in linear time, but with a constant that is quadratic in the size of the RE. The correctness proof of the algorithm is also significantly more intricate than the previous methods discussed here, which confirms the impression that the POSIX policy is in fact more difficult to implement than PCRE, at least for automata based techniques. Sulzmann and Lu~\cite{sulzmann2014} formulate POSIX disambiguation as an ordering relation on parse trees and give an alternative forwards-backwards parsing algorithm based on Brzozowski~\cite{brzozowski64} derivatives, as well as a forwards parsing algorithm using only a single pass which produces a bit-code representation of the parse tree.

Borsotti, Breveglieri, Reghizzi and Morzenti~\cite{borsotti2015bsp,borsotti2015ambiguous} recently gave a forwards-backwards parser based on an extension of the Berry-Sethi~\cite{bese86} algorithm for constructing DFAs from REs. The parser can be configured for both POSIX and PCRE disambiguation by only changing the choices made in the second backwards pass, giving a common framework which can accommodate both policies.

\subsection{Connection to Transducers}
It is remarkable that every automaton based method for RE parsing seems to operate in two passes. By applying the interpretation of parsing as transduction from Section~\ref{sec:parsing:as-transduction}, it seems that one should be able to obtain a single-pass parser by turning the non-deterministic parsing transducer into a deterministic one, just as an NFA can be converted to an equivalent DFA. This is however not possible in general, as non-deterministic transducers are strictly more powerful than deterministic ones~\cite*[ex. IV.2.3]{berstel79}. This implies that any deterministic RE parsing machine must be more powerful than finite transducers.

One the other hand, an old result by Elgot and Mezei~\cite{elgot1965}{\cite[Theorem 5.2]{berstel79}} says that every unambiguous transducer can be run in two passes, where each pass is modeled by a deterministic transducer. The first pass runs in the forwards direction, producing an auxiliary string over an intermediate alphabet, and the second pass runs in the opposite direction over this string to produce the reversed output. This is exactly the forwards-backwards model employed by the automata based parsing methods, which can all be seen as rediscoveries of this old result.

\subsection{Disambiguation Policies}
Since most practical REs are ambiguous, any method for RE matching or parsing must employ a disambiguation policy, which furthermore must have a semantics that is transparent to the user. Defining a disambiguation policy which is both efficient to implement and easy to comprehend is not an easy task. The formal definitions by Vansummeren~\cite{vansummeren2006} of various common disambiguation policies, including those employed in PCRE and POSIX, provide a good comparison of their different qualities.

Myers, Oliva and Guimaraes~\cite{myers1998reporting} argue that the PCRE and POSIX disambiguation policies are not intuitive, since they are inherently tied to the structure of the underlying NFA instead of more meaningful semantic criteria such as ``maximize the total length of substrings captured in all capturing groups''. They give a method which in an NFA can select the path that corresponds to either maximizing the length of all captured substrings, or the individual lengths of the leftmost ones.

Although semantic policies are ostensibly more intuitive, they seem to have been largely ignored in most work on RE matching and parsing. Apart from the difficulty of obtaining efficient implementations (Myers' method runs in linear time, but with a large constant overhead), a possible hindrance to adoption could be that most users have familiarized themselves with REs through existing PCRE or POSIX tools, and so this is the behavior that they have come to expect.

\section{Our Contributions}
\label{sec:contributions}

The first two papers of this dissertation are concerned with regular expression based parsing using a PCRE ``greedy'' disambiguation policy. Our approaches are both based on finite automata annotated with bit-codes à la Nielsen and Henglein~\cite{nihe2011} and offer, respectively, improved memory and time usage by lowered constant factors compared to existing methods, as well as a new streaming execution model for parsing.

In Paper~\ref{paper:two-pass} we present a new algorithm for RE parsing which operates in two passes similar to the forwards-backwards algorithms mentioned in the previous section, producing a reversed bit-code representation of the greedy parse tree in the second pass. The first pass runs the NFA in the forwards direction while maintaining an ordered \emph{list} of active states instead of a set, where the ordering of states in the list denote their ranking according to the disambiguation policy. In each step of the forward run we save $k$ bits of information in a \emph{log}, where $k < \frac{1}{3}m$ and $m$ is the number of states in the NFA. In the second pass the log is traversed in opposite order in order to reconstruct the greedy parse tree. Our algorithm is a variant of the method of Dubé and Feeley~\cite{dufe2000} with disambiguation, and using less storage for the log---we only save $k$ bits per input character instead of the full set of active states which requires $m$ bits. We also avoid having to build a DFA and thus avoid the risk of an exponential number of states. We compare the performance of a prototype C implementation with RE2~\cite{cox2010}, Tcl~\cite{Tcl}, Perl~\cite{perl}, GNU \texttt{grep} as well as the implementations by Nielsen and Henglein~\cite{nihe2011} of the methods of Dubé and Feeley~\cite{dufe2000} and Frisch and Cardelli~\cite{frca2004}. It performs well in practice, and is surprisingly competitive with tools that only perform matching such as RE2 and \texttt{grep}.

Paper~\ref{paper:optimal-streaming} takes a new approach and presents a linear time parsing algorithm which also performs bit-coded PCRE disambiguated parsing, but using only a single forward pass. Furthermore, the parse is produced in an \emph{optimally streaming} fashion---bits of the output is produced as early as is semtically possible, sometimes even before the corresponding input symbols have been seen. For REs where an unbounded amount of symbols need to be consumed in order to resolve a choice, such as the RE $\texttt{a}^*\texttt{b} + \texttt{a}^*\texttt{c}$, the algorithm automatically adapts to buffering as many $\texttt{a}$s as needed, and immediately outputs the bit-code as soon as a $\texttt{b}$ or $\texttt{c}$ symbol is encountered. In order to obtain optimal streaming a PSPACE-hard analysis is required, adding a worst-case $O(2^{m \log m})$ preprocessing phase to the algorithm, although this must only be done once for the RE and is independent of the input. The main idea of the method is to maintain a \emph{path tree} from the initial state to all states that can be reached by reading the input read so far, where a branching node in the tree represents the latest point at which two paths diverge. The longest unary branch from the root of the tree thus represents the path prefix that must be followed by \emph{all} viable paths reading a completion of the input seen so far. The path tree model was also used by Ostrand, Paull and Weyuker~\cite{ostrand1981} in their FL($k$) parser, albeit without support for linear-time parsing with unbounded lookahead and with a more primitive condition for resolving choices.

\section{Conclusions and Perspectives}
\label{sec:parsing:conclusions}

We will hope that by the end of reading this chapter, it has become clear that the area of regular expressions still contains interesting problems despite their well-understood language theory and long list of practical applications. By taking a step back to properly identify the core problem that is being solved in practical regex tools, namely \emph{parsing}, we obtain a new perspective from which new and interesting solutions can be uncovered.

We present two new methods for regular expression based parsing. The first improves on previous methods, while the second appears to be the first streaming parsing algorithm for unrestricted regular expressions, and both methods follow a simple disambiguation policy consistent with that found in popular regex implementations such as Perl's. Our work paves the way for new tools with stronger guarantees and greater expressivity than current solutions, as well as new and interesting application areas. Furthermore, a connection is revealed between regular expression based parsing and finite-state transductions.

It would be a mistake to claim that our methods will replace all existing applications of regular expressions for search, extraction and manipulation of data. Existing tools are also appreciated for the features which we deliberately choose not to support, and there continue to be problem areas where the resulting trade-offs are acceptable. On the other hand, the two-pass and streaming regular expression parsing methods offer alternatives for those areas where the performance guarantee or increased expressivity is needed.

Our focus has mainly been on the theoretical aspects of regular expression parsing and little on practical applications, of which we believe there are many. Possible applications include parsing of data formats, streaming protocol implementation, advanced text editing and lexical analysis with \emph{maximal munch}~\cite{reps98}.

In order to enable any application, a considerable amount of effort has to be invested in tools and integration, including software libraries, command-line tools, programming language integration or the design of domain-specific languages. The development of a compiler for the latter is one of the topics of the next chapter, in which we develop a grammar-based programming language for high-performance streaming string processing, based on the streaming parsing algorithm presented in Paper~\ref{paper:optimal-streaming}.

%%% Local Variables:
%%% mode: latex
%%% TeX-master: "thesis"
%%% End:

\chapter{Grammar Based Stream Processing}
\label{chap:streaming-transduction}

In this chapter we will consider two formalisms, \emph{regular grammars} and \emph{parsing expression grammars}, as foundations for specifications of string processing programs. It is our goal to be able to turn such specifications into efficient streaming programs which execute in time proportional to the length of the input string.  ``Streaming'' in this context means that the resulting programs do not need access to the full input string at any time, but instead operate in a single pass from left to right, generating parts of the final result as they go along. Programs of this kind can be used to perform a range of useful tasks, including advanced text substitution, streaming filtering of log files, formatting of data to human-readable form, lexical analysis of programming languages, et cetera.

We first observe that the compact nature of REs cause large specifications to become unwieldy and hard to comprehend. A formalism that scales better is Chomsky's~\cite{chomsky1956,alsu2006} \emph{context-free grammars} (CFG) for specifying linguistic structure using a set of production rules. CFGs have more expressive power than REs, so in order to use them as a replacement of the latter, a syntactic test must be used to discard those that use non-regular features. The regular CFGs have a natural notion of parse trees which is compatible with the transducer based view of RE parsing using greedy disambiguation.

In order to use regular CFGs as specifications of \emph{programs}, we assign a semantics to the parse trees by translating every tree into a sequence of program statements to be executed. This type of specification is called a \emph{syntax-directed translation scheme} (SDT)~\cite{lewis1973}, and is obtained by allowing CFGs to contain program fragments, also called \emph{semantic actions}, embedded within productions. The actions then show up in the parse trees which can be flattened to remove all structure except the sequence of program statements to be executed. This is the first formalism that will be considered by this chapter, and the goal is to apply the streaming RE parsing technique of Paper~\ref{paper:optimal-streaming}.

The restriction to regular languages somewhat limit the possible applications, as it precludes the specification of programs that need to match parentheses or otherwise parse recursive language structures. For this purpose we want to base our program specifications on a more expressive formalism, but without giving up the strong performance guarantees provided by linear time parsing. A candidate for such as formalism is Ford's~\cite{ford2004} \emph{parsing expressing grammars} (PEG) for specifying recursive descent parses with limited backtracking. Every regular CFG parsed using greedy disambiguation corresponds to its interpretation as a PEG, but PEGs can additionally also express recursive parsing rules. This is the second formalism to be considered, and the challenge then becomes to generalize the streaming RE parsing methods to also apply to PEGs.

The rest of the chapter is structured as follows. We define context-free grammars in Section~\ref{st:sec:cfg}, and show how the regular subclass allows for a compact representation of non-deterministic finite automata. In Section~\ref{st:sec:sdt} we describe syntax-directed translation schemes and give a few examples. In Section~\ref{st:sec:peg} we introduce parsing expression grammars as a generalization of regular translation schemes with greedy disambiguation. In Section~\ref{st:sec:related}, we discuss formalisms found in the literature for the specification of string processing programs and their evaluation on commodity hardware. We present our own contributions in Section~\ref{st:sec:contributions}, and offer our conclusions and perspectives for further work in Section~\ref{st:sec:conclusions}.

\section{Context-Free Grammars}
\label{st:sec:cfg}

REs are not a particularly compact way of specifying regular languages. Although REs and finite automata have the same expressive power, there are regular language whose smallest RE description is quadratically bigger than equivalent descriptions using DFAs~\cite[Theorem 23]{ellul2005regular}. Furthermore, since the RE formalism does not include a systematic way of breaking up large REs into more manageable parts, they quickly become unwieldy and hard to comprehend for users. In this section we consider an alternative.

The \emph{context-free grammars} (CFGs) introduced by Chomsky~\cite{chomsky1956} is a formalism for systematically describing formal languages using a set of \emph{production rules}. They are one step above REs in the \emph{Chomsky hierarchy} of increasingly expressive generative language formalisms. Figure \ref{st:fig:ex:cfg} shows an example of a CFG for a simple language.
\begin{figure}
  \centering
\begin{align*}
  \ul{phrase} \to{}& \ul{subject} ~ \ul{verb} ~ \ul{adjectives} \\
  \ul{subject} \to{}& \texttt{he} \mid \texttt{she} \\
  \ul{verb} \to{}& \texttt{was} \mid \texttt{is} \\
  \ul{adjectives} \to{}& \ul{adverb} ~ \ul{adjective} \mid \ul{adverb} ~ \ul{adjective} ~ \texttt{and} ~ \ul{adjectives} \\
  \ul{adverb} \to{}& \ul{verys} \mid \texttt{not} \\
  \ul{verys} \to{}& \texttt{very} ~ \ul{verys} \mid \varepsilon \\
  \ul{adjective} \to{}& \texttt{happy} \mid \texttt{hungry} \mid \texttt{tall}
\end{align*}  
  \caption{A context-free grammar.}
  \label{st:fig:ex:cfg}
\end{figure}
The underlined words in the grammar are called \emph{nonterminal symbols} and act as ``syntactic variables'' in the specification. The letters (written in typewriter font) are called \emph{terminal symbols}. Every line in the CFG is called a \emph{production}, and is of the form $\ul{A} \to \alpha_0 \mid \alpha_1 \mid ... \mid \alpha_{n-1}$, where each $\alpha_i$ is a string (possibly empty) of terminals and nonterminals. The $\alpha_i$ strings are called \emph{alternatives}, as they represent different choices for sentences described by the corresponding nonterminal. They are numbered from left to right starting from zero. The nonterminal to the left of the arrow in the first production is a designated \emph{starting symbol}.

The language described by the CFG contains the following sentences:
\begin{center}
  \texttt{he is tall},

  \texttt{she was very hungry and not happy},

  \texttt{he is tall and very very happy},

  \texttt{she was happy and hungry and tall}
\end{center}
This language can also be described by an RE. However, it is quite big, spanning two lines, and is not very readable:
\[
\begin{array}{l}
(\texttt{he} + \texttt{she})(\texttt{was} + \texttt{is})((\texttt{very})^* + \texttt{not})(\texttt{happy} + \texttt{hungry} + \texttt{tall})\\
(\texttt{and} ((\texttt{very})^* + \texttt{not})(\texttt{happy} + \texttt{hungry} + \texttt{tall}))^*
\end{array}
\]
In particular, note that we have to include duplicate occurrences of most of the words in order to correctly specify that a list of adjectives is separated by the word \texttt{and}. On the other hand, the CFG is self-documenting by having the names of nonterminals describe what kind of sentence structure they define.

The language described by a CFG is determined as the set of all strings of terminal symbols that can be \emph{derived} from the starting symbol. A derivation is a sequence of rewritings of strings containing terminal and nonterminal symbols. For any string $\alpha$ of the form $\beta \ul{A} \delta$, where $\beta$ and $\delta$ are strings of terminals and nonterminals and $\ul{A}$ is a nonterminal with production $\ul{A} \to \gamma_0 \mid \gamma_1 \mid ... \mid \gamma_{n-1}$, we can rewrite $\alpha$ as follows:
\[
  \alpha \Rightarrow \beta \gamma_i \delta
\]
where the alternative $\gamma_i$ is chosen freely among the alternatives in the production for $\ul{A}$. If a string can be rewritten several times, $\alpha_0 \Rightarrow \alpha_1 \Rightarrow ... \Rightarrow \alpha'$, we also write just $\alpha \Rightarrow \alpha'$. Rewriting is a highly non-deterministic process, since neither the expanded non-terminal $\ul{A}$ or the chosen alternative are uniquely determined in each step. Figure \ref{st:fig:derivation} shows an example of how to derive a sentence in the grammar from Figure~\ref{st:fig:ex:cfg}, starting from the starting symbol $\ul{phrase}$, and with the expanded nonterminal highlighted in each step.
\begin{figure}
\begin{center}
\begin{minipage}{0.47\textwidth}
\begin{tikzpicture}
\tikzstyle{level 1}=[sibling distance=2cm,level distance=1cm]
\tikzstyle{level 2}=[sibling distance=1.65cm]
\node {$\ul{phrase}$}
child {node {$\ul{subject}$}
  child {child {child {node {\texttt{he}}}}}
}
child {node {$\ul{verb}$}
  child {child {child {node {\texttt{is}}}}}
}
child {node {$\ul{adjectives}$}
  child {node {$\ul{adverb}$}
    child {node {$\ul{verys}$}
      child {node {$\varepsilon$}}
    }
  }
  child {node {$\ul{adjective}$}
    child {child {node {\texttt{tall}}}}
  }
}
;
\end{tikzpicture}
\end{minipage}
\begin{minipage}{0.25\textwidth}
\begin{align*}
  \highlight{$\ul{phrase}$}
  \Rightarrow_0{}& \highlight{$\ul{subject}$} ~ \ul{verb} ~ \ul{adjectives} \\
  \Rightarrow_0{}& \texttt{he} ~ \highlight{$\ul{verb}$} ~ \ul{adjectives} \\
  \Rightarrow_1{}& \texttt{he} ~ \texttt{is} ~ \highlight{$\ul{adjectives}$} \\
  \Rightarrow_0{}& \texttt{he} ~ \texttt{is} ~ \highlight{$\ul{adverb}$} ~ \ul{adjective} \\
  \Rightarrow_0{}& \texttt{he} ~ \texttt{is} ~ \highlight{$\ul{verys}$} ~ \ul{adjective} \\
  \Rightarrow_1{}& \texttt{he} ~ \texttt{is} ~ \highlight{$\ul{adjective}$} \\
  \Rightarrow_2{}& \texttt{he} ~ \texttt{is} ~ \texttt{tall} \\
\end{align*}
\end{minipage}
\end{center}
\caption{A parse tree and the corresponding derivation.}
\label{st:fig:derivation}
\end{figure}
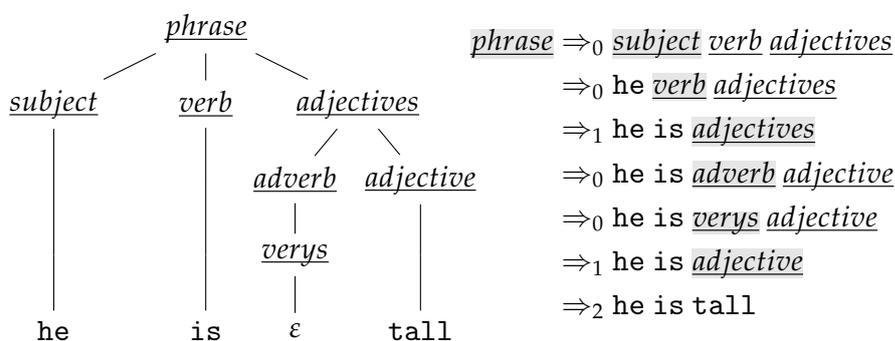

We denote the language of a grammar $G$ with start symbol $\ul{S}$ by $\Lang{G}$, and define it formally as the set of all terminal strings that can be derived from $\ul{S}$:
\[ \Lang{G} = \{ u \in \Sigma^* \mid \text{there is a derivation $\ul{S} \Rightarrow u$ in $G$} \} \]

\subsection{Parse Trees and Codes}

Parse trees for CFGs are naturally defined as pictorial descriptions of how a given string is derived from the starting symbol. As such, a parse tree consists of labeled nodes, where internal nodes are labeled by nonterminals and leaf nodes are labeled by terminals or the empty string $\varepsilon$. The root is always labeled by the starting symbol, and for each internal node with label $\ul{A}$ and child nodes with labels $L_0, L_1, ..., L_{m-1}$, there must be a production $\ul{A} \to \alpha_0 \mid \alpha_1 \mid ... \mid \alpha_{n-1}$ such that $\alpha_i = L_0 L_1 ... L_{m-1}$ for some number $i$. If we assume that no production contains two equal alternatives, then every parse tree uniquely guides the choice of alternatives in derivations, although the order of expanded nonterminals is still nondeterministic. See Figure~\ref{st:fig:derivation} for an example.

We can obtain a true one-to-one correspondence between parse trees and derivations by only considering derivations which choose the next nonterminal to expand in a particular order. For this purpose we only consider \emph{leftmost} derivations, which always expand the leftmost nonterminal before others. The derivation in Figure~\ref{st:fig:derivation} is leftmost, and thus uniquely determines the corresponding parse tree and vice versa.

Every parse tree can be given a serialized code in the same way as we did for RE parse trees in Chapter~\ref{chap:regex-parsing}. Since a parse tree corresponds to a leftmost derivation which performs a series of deterministic expansions, the code can simply be defined as the sequence of numbers which determine the alternatives in each expansion step. For example, the derivation in Figure~\ref{st:fig:derivation} has been annotated with the choice of alternative in each expansion, which leads to the code $0010012$.

\subsection{From CFGs to Transducers}

The motivation for introducing CFGs were as a replacement for REs, allowing us to apply methods specific to RE parsing to parse CFGs. Not every CFG has a corresponding RE describing the same language, as the CFG formalism is significantly more expressive. For example, the simple grammar with only one production $\ul{S} \to \texttt{a} \ul{S} \texttt{b} \mid \varepsilon$ describes the non-regular language consisting of strings of \texttt{a}s followed by exactly the same number of \texttt{b}s. We will have to rule out such grammars to ensure that we only consider the regular CFGs. There is no computer program which can determine for any CFG whether it is regular or not~\cite{bar1961}, but there are simple tests we can use which can verify most regular CFGs as such, but which returns false negatives for some~\cite{anselmo2003finite}.

It can be shown that every regular CFG can be rewritten such that for every production $\ul{A} \to \alpha_0 \mid \alpha_1 \mid ... \mid \alpha_{n-1}$, each $\alpha_i$ is either of the form $a_i \ul{B}_i$ where each $a_i$ is a terminal symbol and $\ul{B}_i$ is a nonterminal symbol, or $\alpha_i = \varepsilon$. Grammars on this form are called \emph{right-regular}, and have a natural interpretation as finite state transducers where each nonterminal identifies a state. If the production for a nonterminal $\ul{A}$ has an alternative $\alpha_i = \varepsilon$, then its state is accepting. For every alternative where $\alpha_i = a_i \ul{B}_i$, there is an outgoing transition $\ul{A} \stackrel{a_i / i}{\to} \ul{B}_i$. See Figure~\ref{st:fig:regular-cfg} for an example.
\begin{figure}
\centering
\begin{minipage}{0.33\linewidth}
\begin{align*}
  \ul{S} \to{}& \texttt{a} \ul{T} \mid \texttt{a} \ul{L} \\
  \ul{T} \to{}& \texttt{b} \ul{L} \mid \texttt{b} \ul{S} \\
  \ul{L} \to{}& \texttt{a} \ul{L} \mid \texttt{b} \ul{L} \mid \varepsilon
\end{align*}
\end{minipage}
\begin{minipage}{0.63\linewidth}
\begin{tikzpicture}[x=0.9cm,auto,>=stealth']
  \node[sstate,initial] (S) at (0,0) {$\ul{S}$};
  \node[sstate] (T) at (3,0) {$\ul{T}$};
  \node[sstate,accepting] (L) at (6,0) {$\ul{L}$};
  \draw[->] (S) to node {$\texttt{a} / 0$} (T);
  \draw[->,bend right] (S) to node[swap] {$\texttt{a} / 1$} (L);
  \draw[->] (T) to node {$\texttt{b} / 0$} (L);
  \draw[->,out=135,in=45] (T) to node[swap] {$\texttt{b} / 1$} (S);
  \draw[->,loop above] (L) to node {$\texttt{a} / 0$} (L);
  \draw[->,loop below] (L) to node {$\texttt{b} / 1$} (L);
\end{tikzpicture}
\end{minipage}
\caption{Regular CFG and its parsing transducer.}
\label{st:fig:regular-cfg}
\end{figure}
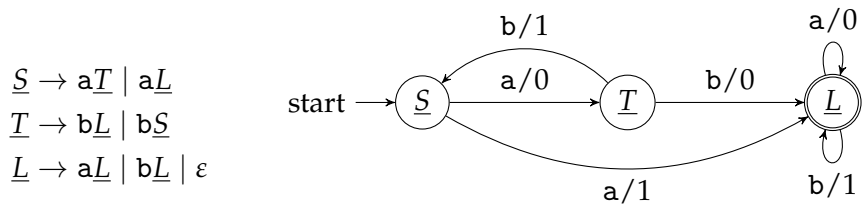

\subsection{Disambiguation}

The problem of ambiguity also arises for CFGs, since there can be more than one parse tree/leftmost derivation for a given string. The problem can be solved in a similar way as for REs by specifying a policy for selecting a single parse tree from a set of candidates. We consider here the greedy disambiguation policy introduced in the previous chapter generalized to CFGs: If there is more than one leftmost derivation for a terminal string, identify the first step at which they made different expansion choices, and pick the one that chose the earliest alternative.

For example, the grammar in Figure~\ref{st:fig:regular-cfg} is ambiguous since the string \texttt{aba} has three leftmost derivations:
\begin{align}
  \label{st:eq:deriv1}
  & \ul{S} \Rightarrow_0 \texttt{a} \ul{T} \Rightarrow_0 \texttt{ab} \ul{L} \Rightarrow_0 \texttt{aba} \ul{L} \Rightarrow_1 \texttt{aba} \\
  \label{st:eq:deriv2}
  & \ul{S} \Rightarrow_0 \texttt{a} \ul{T} \Rightarrow_1 \texttt{ab} \ul{S} \Rightarrow_1 \texttt{aba} \ul{L} \Rightarrow_1 \texttt{aba} \\
  \label{st:eq:deriv3}
  & \ul{S} \Rightarrow_1 \texttt{a} \ul{L} \Rightarrow_1 \texttt{ab} \ul{L} \Rightarrow_0 \texttt{aba} \ul{L} \Rightarrow_0 \texttt{aba}
\end{align}
We see that \eqref{st:eq:deriv1} differs from \eqref{st:eq:deriv2} and \eqref{st:eq:deriv3} in the first and second step, respectively. In both cases, \eqref{st:eq:deriv1} chooses an earlier alternative than the other, so this is the unique greedy derivation.

\section{Syntax-Directed Translation Schemes}
\label{st:sec:sdt}

In a \emph{syntax-directed translation scheme} (SDT)~\cite{alsu2006}, we allow program fragments, also called \emph{semantic actions}, to occur inside the alternatives of each production. The program fragments can be anything that can be executed on the underlying machine, such as manipulation of stateful variables or execution of side-effects. In order to distinguish semantic actions from the terminal and nonterminal symbols, we will write them in braces and highlight them. For example, the action which sets a variable \texttt{x} to value \texttt{"a"} is written \sact{\{x:="a"\}}.

A derivation for an SDT is finished when it has reached a string which only contains interleaved terminal symbols and semantic actions. The parsed string is the substring of terminal symbols, and the substring of semantic actions forms a sequence of program statements to be executed.

We illustate SDTs by an example. Consider the SDT in Figure~\ref{st:fig:sdt} which reads an English noun phrase, reformulates it, and prints the result.
\begin{figure}
  \centering
  \begin{align*}
  \ul{phrase} \to{}& \ul{det} ~ \ul{noun} ~ \ul{wp} ~ \ul{verb} ~ \ul{adj} ~ \sact{\{ p := d+v+a+n; print(p); \}} \\
  \ul{det} \to{}& \texttt{the} ~ \sact{\{ d := "the"; \}} \\ &{} \mid \texttt{a} ~ \sact{\{ d := "a"; \}} \\
  \ul{noun} \to{}& \texttt{man} ~ \sact{\{ n := "man"; \}} \\ &{} \mid \texttt{woman} ~ \sact{\{ n := "woman"; \}} \\
  \ul{wp} \to{}& \texttt{who} \\
  \ul{verb} \to{}& \texttt{is} ~ \sact{\{ v := ""; \}} \\ &{} \mid \texttt{was} ~ \sact{\{ v := "formerly"; \}} \\
  \ul{adj} \to{}& \texttt{happy} ~ \sact{\{ a := "happy"; \}} \\ &{} \mid \texttt{tall} ~ \sact{\{ a := "tall"; \}}
  \end{align*}
  \caption{Example SDT for reformulating simple English phrases.}
  \label{st:fig:sdt}
\end{figure}
If we parse the string \texttt{a man who was happy} using the SDT, we obtain the (greedy) parse tree depicted in Figure~\ref{st:fig:sdt-parse-tree}.
\begin{figure}
\begin{center}
\begin{tikzpicture}
\tikzstyle{level 1}=[sibling distance=1.85cm,level distance=0.9cm]
\tikzstyle{level 2}=[sibling distance=0.65cm,level distance=0.75cm]
\tikzstyle{level 3}=[level distance=0.6cm]
\node {\ul{phrase}}
child {node {\ul{det}}
  child {node {\texttt{a}}}
  child {child {node {\sact{\{d:="a";\}}}}}
}
child {node {\ul{noun}}
  child {node {\texttt{man}}}
  child {child {child {node {\sact{\{n:="man";\}}}}}}
}
child {node {\ul{wp}}
  child {node {\texttt{who}}}
}
child {node {\ul{verb}}
  child {node {\texttt{was}}}
  child {child {node {\sact{\{v:="formerly";\}}}}}
}
child {node {\ul{adj}}
  child {node {\texttt{happy}}}
  child {child {child {node {\sact{\{a:="happy";\}}}}}}
}
child {child {child {child {child {node {\sact{\{p:=d+v+a+n; print(p);\}}}}}}}}
;
\end{tikzpicture}
\end{center}
\caption{Greedy parse of the string \texttt{a man who was happy}, using SDT from Figure~\ref{st:fig:sdt}.}
\label{st:fig:sdt-parse-tree}
\end{figure}
The subsequence of program statements in the leaves forms the following program, which when executed prints the string \texttt{a formerly happy man}:
\[
\begin{array}{l}
  \texttt{d := "a";} \\
  \texttt{n := "man";} \\
  \texttt{v := "formerly";} \\
  \texttt{a := "happy";} \\
  \texttt{p := d+v+a+n;} \\
  \texttt{print(p);}
\end{array}
\]

Many useful string processing programs can be conveniently specified using SDTs. For example, if we needed to collect statistics from a large log of web requests, an SDT could easily be used to parse the log entries and update a database based on the extracted data. We could also use an SDT to read data in a format which is hard to read for humans and automatically format it in a readable report.

For some applications such as implementations of protocols where data arrives in a stream whose total length is unknown, we want to start executing semantic actions as soon a possible, since we may not have enough memory to store the complete stream. Under this execution model, we have to be careful not to execute semantic actions ``prematurely'': if after seeing a prefix of the input stream we decide to execute an action, then that action must be guaranteed to be executed for every complete parse of the input.

This leads us to the first problem that we wish to address in this chapter:
\problembox{How to evaluate the SDT on an input string in a streaming fashion, using at most time proportional to the length of the input?}

\section{Parsing Expression Grammars}
\label{st:sec:peg}

With the restriction to regular SDTs, we lose the ability to express a large number of interesting string processing programs. Regular languages cannot contain unbounded nesting, so this precludes processing languages such as arithmetic expressions, languages containing matching parentheses and nested data formats.

As we pointed out in the previous chapter, we cannot allow specifications based on unrestricted CFGs without losing the guarantee of linear time parsing~\cite{lee2002}, and we would like to avoid restricting ourselves to deterministic CFGs such as LR($k$)~\cite{knuth1965} since they are difficult to write. It seems to be hard to come up with a suitable relaxation of the regular SDTs, so in this section we will step outside the Chomsky hierarchy and instead consider Ford's \emph{parsing expression grammars} (PEG)~\cite{ford2004}.

A PEG is specified as a set of production rules, each of the form $\ul{A} \leftarrow e$, where $\ul{A}$ is a nonterminal as before, and $e$ is a \emph{parsing expression} (PE) generated by the following grammar:
\begin{align*}
  e ::={}& \ul{A} \mid a \mid e_1e_2 \mid e_1 / e_2 \mid !e_1
\end{align*}
A PE can either be a nonterminal $\ul{A}$, a terminal symbol $a$ in $\Sigma$, the empty string $\varepsilon$, a product $e_1 e_2$, an \emph{ordered sum} $e_1 / e_2$, or a \emph{negated expression} $! e_1$, where in all of the previous, $e_1$ and $e_2$ stand for PEs. The rules for associativity of parentheses are the same as for REs, and we write $e_1e_2e_3$ and $e_1 / e_2 / e_3$ for the PEs $e_1(e_2e_3)$ and $e_1 / (e_2 / e_3)$, respectively. See Figure~\ref{st:fig:peg} for an example PEG which parses simple arithmetic expressions with parentheses.
\begin{figure}
\centering
\begin{align*}
  \ul{sum} \leftarrow{}& \ul{factor} ~ \texttt{+} ~ \ul{sum} / \ul{factor} \\
  \ul{factor} \leftarrow{}& \texttt{0} / \ul{digit} ~ \ul{digits} / \texttt{(} ~ \ul{sum} ~ \texttt{)} \\
  \ul{digits} \leftarrow{}& \ul{digit} ~ \ul{digits} / \varepsilon \\
  \ul{digit} \leftarrow{}& \texttt{0} / \texttt{1} / ... / \texttt{9}
\end{align*}
\caption{Example of a simple PEG.}
\label{st:fig:peg}
\end{figure}

\subsection{PEG Semantics}
Although on the surface PEGs resemble CFGs, their semantics are quite different. PEGs do not have a notion of derivations, but instead every parsing expression specifies a recursive backtracking parser which searches for a greedy parse of the input.

The result of a parse is either \emph{success}, in which case zero or more input symbols are consumed, or \emph{failure}, in which case exactly zero input symbols are consumed. If the PE being parsed is a terminal symbol, then the parse succeeds and consumes one symbol if the first symbol in the input matches; otherwise it fails. If the PE is a nonterminal, then parsing proceeds with the PE associated with that nonterminal in the PEG. For sequences $e_1e_2$, the expression $e_1$ is parsed first, and if it succeeds, $e_2$ is parsed with the remainder of the input; otherwise $e_1e_2$ fails. For ordered sums $e_1 / e_2$, the expression $e_1$ is parsed first, and if it succeeds, the whole sum succeeds, disregarding $e_2$. Only if $e_1$ fails is $e_2$ tried. A negation $! e_1$ fails if $e_1$ succeeds; if $e_1$ fails, then $!e_1$ succeeds, but consumes zero symbols.

The behavior for ordered sums means that in the PEG in Figure~\ref{st:fig:peg}, parsing $\ul{factor}$ with input \texttt{0123} will fail: since the first alternative consumes \texttt{0}, the other alternatives are disregarded, leaving the suffix \texttt{123} unhandled. This illustrates the difference with CFGs, where the second alternative would have lead to a successful parse. The backtracking behavior is in this case intentionally used to reject numbers with leading zeros. See Figure~\ref{st:fig:peg-parse} for the parse tree resulting from parsing the input \texttt{(0+1)+46}.

\begin{figure}
\begin{center}
\begin{tikzpicture}
\tikzstyle{level 1}=[sibling distance=3.5cm,level distance=1cm]
\tikzstyle{level 2}=[sibling distance=2.75cm]
\tikzstyle{level 3}=[sibling distance=1.5cm]
\tikzstyle{level 4}=[sibling distance=1cm]
\node {$\ul{sum}$}
child{node {$\ul{factor}$}
  child{child{child{child{child {node{\texttt{(}}}}}}}
  child {node{$\ul{sum}$}
    child {node{$\ul{factor}$}
      child{child{child {node{$\texttt{0}$}}}}
    }
    child{child{child{child {node{$\texttt{+}$}}}}}
    child {node{$\ul{sum}$}
      child {node{$\ul{factor}$}
          child {node {$\ul{digit}$}
            child {node {$\texttt{1}$}}
          }
          child {node {$\ul{digits}$}
            child {node {$\varepsilon$}}
          }
      }
    }
  }
  child{child{child{child{child {node{\texttt{)}}}}}}}
}
child{child{child{child{child{child{node {$\texttt{+}$}}}}}}}
child{node {$\ul{sum}$}
  child{node{$\ul{factor}$}
    child{node{$\ul{digit}$}
      child{child{child{node{\texttt{4}}}}}
    }
    child{node{$\ul{digits}$}
      child{node{$\ul{digit}$}
        child{child{node{\texttt{6}}}}
      }
      child{node{$\ul{digits}$}
        child{child{node{$\varepsilon$}}}
      }
    }
  }
}
;
\end{tikzpicture}
\end{center}  
  \caption{A PEG parse tree for the string \texttt{(0+1)+46}.}
  \label{st:fig:peg-parse}
\end{figure}

Although the semantics of PEGs are formulated as a backtracking parsing process, every PEG can be parsed in time proportional to the input length. One can either apply a dynamic programming approach~~\cite[Theorem 6.4]{aho1972} or apply the memoizing Packrat algorithm due to Ford~\cite{ford2002}. None of these algorithms operate in a streaming fashion, however.

\subsection{Expressivity}
PEGs are equivalent in power to the formalisms TDPL and GTDPL~\cite{ford2004} due to Aho and Ullman~\cite{aho1972}, albeit a lot easier to read.

The expressive power of PEGs and CFGs is incomparable.
The negation operator allows PEGs to parse languages which cannot be described by any CFG. An example of such a language is the following:
\[
  \underbrace{\texttt{aa} \cdots \texttt{a}}_{\text{$n$ times}} \underbrace{\texttt{bb} \cdots \texttt{b}}_{\text{$n$ times}} \underbrace{\texttt{cc} \cdots \texttt{c}}_{\text{$n$ times}}
\]
That is, the strings consisting of \texttt{a}s followed by \texttt{b}s followed by \texttt{c}s, in equal numbers. The PEG recognizing this language crucially depends on the negation operator in order to look ahead in the input string~\cite[Section 3.4]{ford2004}:
\begin{align*}
  \ul{D} \leftarrow{}& !!(\ul{A} ~ !\texttt{b}) ~ \ul{S} ~ \ul{B} ~ !(\texttt{a}/\texttt{b}/\texttt{c}) \\
  \ul{A} \leftarrow{}& \texttt{a} ~ \ul{A} ~ \texttt{b} / \varepsilon \\
  \ul{B} \leftarrow{}& \texttt{b} ~ \ul{B} ~ \texttt{c} / \varepsilon \\
  \ul{S} \leftarrow{}& \texttt{a} ~ \ul{S} / \varepsilon
\end{align*}

On the other hand, since every PEG can be parsed in linear time, then due to the non-linear lower bound of general CFG parsing~\cite{lee2002}, there must exist a CFG describing a language which cannot be parsed by any PEG.\footnote{To the best of our knowledge, finding an example of such a language is an open problem.} However, every \emph{deterministic} CFG can be simulated by PEG~\cite[Theorem 6.1]{aho1972}, including all LL($k$) and LR($k$) grammars.

Although PEGs are incomparable to general CFGs, they \emph{do} have a close connection to the right-regular CFGs. For every right-regular CFG, replace all productions of the form $\ul{A} \to \alpha_0 \mid \alpha_1 \mid ... \mid \alpha_{n-1}$ by PEG rules $\ul{A} \leftarrow \alpha_0 / \alpha_1 / ... / \alpha_{n-1}$, and then replace every occurrence of $\varepsilon$ by a special end-of-input marker \texttt{\#}. It can be shown that for every input string $u$, the greedy leftmost derivation for $u$ in the original CFG will yield the same parse tree as the PEG on input $u \texttt{\#}$. The reason for this is that no ordered sum in the PEG will finish parsing before all of the string has been processed, so all alternatives will be exhausted, resulting in a simulation of the search for the greedy leftmost derivation in the CFG. PEGs can thus be seen as direct generalizations of regular CFGs with greedy leftmost semantics.

It is straightforward to extend PEGs with semantic actions in the same way as we did for CFGs to obtain a generalization of the regular syntax-directed translation schemes. By applying one of the linear time PEG parsing algorithms, we can evaluate such a PEG-based SDT in a non-streaming fashion. This leads to the second problem to be addressed in this chapter:

\problembox{How to evaluate a PEG-based SDT on an input string in a streaming fashion, using at most time proportional to the length of the input?}

\section{String Processing Methods}
\label{st:sec:related}

We discuss formalisms for the specification of string processing programs and methods for evaluating such specifications on commodity hardware.

\subsection{Line-Oriented Stream Processing}
Several methods and tools for streaming text processing rely on a delimiters such as newline symbols to chunk the input stream. Each chunk is processed independently of the following ones, and can be discarded once the next chunk starts processing. The UNIX operating system adopted this model by treating text files as arrays of strings separated by newlines, and as a result all popular UNIX tools for streaming text processing, such as the regex based tools \texttt{sed}~\cite{gnused} and \texttt{awk}/\texttt{gawk}~\cite{gnuawk}, operate using the chunking model. The advantage of this model is that each chunk can be assumed to be small, often a single line in a text file, which means that further pattern matching inside chunks do not have to be streaming. The disadvantage is, as noted by Pike~\cite{pike1987struct}, that \emph{``[...] if the interesting quantum of information isn’t a line, most of the tools [...] don’t help''}, and as a consequence, processing data formats which are not line-oriented is complicated\footnote{\texttt{sed}, \texttt{awk} are Turing-complete and can parse any decidable language, but not without pain.}.

\subsection{Automata Based Methods}
There are several different methods for specification of streaming string processing programs using finite automata and their generalizations. The state machine compiler Ragel~\cite{thurston2003ragel} allows users to specify NFAs whose transitions are annotated by arbitrary program statements from a host programming language. The annotated NFAs are converted to DFAs, and in the case of ambiguity the DFA will simultaneously perform actions from several NFA transitions upon transitioning from one state to the next, even if one of these transitions turns out not to be on a viable path. By contrast, a syntax-directed translation scheme will only perform the actions that occur in the unique final parse tree. For this reason, Ragel is most useful for processing mostly deterministic specifications or for pure recognition.

Methods based on the more expressive transducer model include the Microsoft Research languages \textsc{Bek}\footnote{\url{http://rise4fun.com/Bek}}~\cite{hooimeijer2011} and \textsc{Bex}\footnote{\url{http://rise4fun.com/Bex}}~\cite{veanes2014}, both of which are based on \emph{symbolic transducers}~\cite{veanes2012,dantoni_static_2013}, a compact notation for representing transducers with many similar transitions which can be described using logical theories. Both languages are formalisms for expressing string sanitizers and encoders commonly found in web programming, supporting both synthesis of fast programs as well as automatic checking of common correctness criteria of such specifications. Due to the focus on a limited application domain, both languages are restricted to expressing deterministic transducers only. This trivially ensures linear time execution, but also limits their expressivity. %Another transducer based language is Boomerang~\cite{boomerang2008} for writing bidirectional text transformations between regular languages. A bidirectional translations constructs a modifiable ``view'' of the input string, such that modifications to the view are converted back to updates of the original input. For this reason, all specifications in Boomerang must be unambiguous. We have not been able to determine if the Boomerang compiler generates a backtracking program or whether it does something more sophisticated.

\emph{Streaming string transducers} (SSTs)~\cite{alur2010expressiveness,alur2011streaming} is another extension of DFAs which upon transitioning from one state to another can perform a set of simultaneous copy-free updates to a finite number of string variables. SSTs are deterministic, but are powerful enough to be able express any function describable by non-deterministic transducers, as well as some functions which cannot, such as string reversal. Since they can be run in linear time, they are an interesting model of computation to target for string processing languages. DReX~\cite{adr2015} is a domain specific string processing language based on a combinatory language~\cite{Alur:2014:RCS:2603088.2603151} which can express all string functions describable by SSTs. In order for DReX programs to be evaluated in time proportional to the input length, they must be restricted to an unambiguous subset.

\subsection{Domain-Specific Languages}
There is an abundance of less general solutions which operate within restricted application domains. These include languages for specifying steaming processors for binary~\cite{back2002} and textual~\cite{fisher2005pads,fisher2011pads} data formats, network packets~\cite{mccann2000,madhavapeddy2007,bosshart2014} and wireless protocols~\cite{stewart2015ziria}. Many of these require domain-specific features which are outside the scope of the general grammar based model of SDTs. 

A system which comes close to the SDT model is PADS~\cite{fisher2005pads,fisher2011pads}, a domain-specific language for writing specifications of the physical and textual layouts of ad-hoc data formats from which parsers, statistical tools and streaming string translators to other textual formats or databases can be derived. PADS can be seen as regular SDTs with greedy disambiguation, but extended with extra features such as \emph{data dependencies}---grammar alternatives can be resolved based on semantic predicates on previously parsed data. The parsers generated by a PADS specification operate via backtracking.

\subsection{Parsing Expression Grammars}
Streaming evaluation of PEG-based SDTs will have to rely on a streaming top-down parsing method for PEG. Current practical methods are either based on backtracking~\cite{medeiros2008}, recursive descent with memoization~\cite{ford2002}, or some variant of these using heuristics for optimization~\cite{redziejowski2009,kuramitsu2015}.

There is only one known parsing method which is \emph{streaming}~\cite{mizushima2010}, but it relies on the programmer to manually annotate the grammar with \emph{cut points} to help the parsing algorithm figure out when parts of the parse tree can be written to the output.

For a more in-depth discussion on methods for streaming PEG parsing, we also refer to Section~\ref{sec:discussion2} of Paper~\ref{paper:peg-parsing}.

\section{Our Contributions}
\label{st:sec:contributions}
In Paper~\ref{paper:kleenex} we present \emph{Kleenex}, a language for expressing high-performance streaming string processing programs as regular grammars with embedded semantic actions for string manipulation, and its compilation to efficient C code. Its underlying theory is based on transducer decomposition into oracle and action machines, where an oracle machine corresponds to a bit-coded RE parsing transducer of Chapter~\ref{chap:regex-parsing}, and an action machine is a deterministic transducer which translates bit-codes into sequences of semantic actions to be executed. Based on the optimally streaming RE parsing algorithm of Paper~\ref{paper:optimal-streaming}, the oracle machine, which is non-deterministic and ambiguous, is disambiguated using the greedy policy and converted into a deterministic streaming string transducer, the same machine model employed by DReX. Unlike DReX, we allow unrestricted ambiguity in Kleenex specifications which makes programming in Kleenex easier. By letting the set of semantic actions in Kleenex be copy-free string variable updates, it appears that Kleenex programs are equivalent to the full set of non-deterministic streaming string transducers~\cite{alur2011a}, and thus equivalent in expressive power with DReX.

The generated transducers are translated to efficient C programs which achieve sustained high throughput in the 1Gbps range on practical use cases. The high performance is obtained by avoiding having to compute path trees at run-time---the most expensive part of the streaming algorithm of Paper~\ref{paper:optimal-streaming}---by fully encoding the current path tree structure in the control mechanism of the streaming string transducer. Furthermore, having translated a Kleenex specification to a restricted machine model allows a range of optimizations to be applied, including standard compiler optimizations such as constant propagation~\cite{appel1998} as well as model-specific optimizations such as symbolic representation~\cite{veanes2012}.

In Paper~\ref{paper:peg-parsing} we present a new linear time parsing algorithm for parsing expression grammars. The algorithm is based on a well-known bottom-up tabulation strategy by Aho and Ullman~\cite{aho1972} which is reformulated using least fixed points. Using the method of \emph{chaotic iteration}~\cite{cousot77} for computing least fixed points, we can compute approximations of the parse table, one for each prefix of the input, in an incremental top-down fashion. The approximated parse tables provide enough information for a simple dynamic analysis to predict a prefix of the control flow of all viable parses accepting a completion of the input prefix read so far. The result is a streaming parser which can be used to schedule semantic actions during the parsing process in the same fashion as Kleenex. We evaluate a prototype of the method on selected examples which shows that it automatically adapts to use practically constant space for grammars that do not require lookahead. We also point out directions for further improvements which must be addressed before the algorithm can be used as a basis for an efficient streaming implementation of parsing expression grammars. In particular, the algorithm fails to obtain streaming behavior for strictly right-regular grammars, and it also performs a large amount of superfluous computation, adding a large constant to the time complexity.

\section{Conclusions and Perspectives}
\label{st:sec:conclusions}

In this chapter, we have illustated how syntax-directed translation schemes provide a restricted but expressive formalism which programmers can use to specify streaming string processing programs without having to explicitly deal with orthogonal technical issues related to buffering and disambiguation.

With the Kleenex language, we have demonstrated that streaming regular expression parsing can be used to obtain high-performance implementations of regular syntax-directed translation schemes with greedy disambiguation. Kleenex provides a concise and convenient language for rapid development of streaming string processing programs with predictable high performance. These programs can be used to process many of the common ad-hoc data formats that can be described or approximated by regular grammars, including web request logs, CSV files, HTML documents, JSON files, and more. Kleenex is distinguished from other tools in the same category by allowing unrestricted ambiguity in specifications which are automatically disambiguated using a predictable policy, thus making it easier to combine and reuse Kleenex program fragments without having to worry about compiler errors.

For the cases where the expressivity of Kleenex is not adequate, we show that the foundation of regular grammars can be conservatively extended to the more expressive formalism of parsing expression grammars, thus allowing a larger range of translation schemes to be specified while preserving the input/output-semantics of the regular ones. This however leaves the question of how to evaluate parsing expression grammars in a streaming fashion. We address this issue by providing a streaming linear time algorithm which automatically adapts to constant memory usage in practical use cases, paving the way for a more expressive dialect of Kleenex.

There are several directions for future work on the Kleenex language and its compilation:

\begin{description}
\item[Data-parallel execution] Veanes, Molnar and Mytkowics \cite{veanes2015} show how to implement the symbolic tranducers of \textsc{Bek} and \textsc{Bex} on multi-core hardware in order to hide I/O latencies by processing separate partitions of the input string in parallel. By virtue of also being based on finite state transducers, a similar approach might be applicable to enable Kleenex to run on multi-core hardware as well.
\item[Reducing state complexity] Certain Kleenex specifications have a tendency to result in very large SSTs, which negatively affects both the compile times and the sizes of the produced binary programs. Perhaps we can apply a similar hybrid runtime simulation/compilation technique as used in the RE2~\cite{cox2007} library in order to materialize only the SST states reached during processing of a particular input stream.

We should also point out a result of Roche~\cite{roche1995}, who shows that the number of states in the forwards-backwards decomposition of a transducer can be exponentially smaller than the equivalent representation using a \emph{bimachine}~\cite{schutzenberger1961,berstel79}, another deterministic transducer model. It is future work to see if this also applies to SSTs, and whether it can account for the blowups observed in practice, but if it turns out to be the case then a streaming variant of the forwards-backwards parsing algorithm of Paper~\ref{paper:two-pass} might serve as an alternative, more space economical execution model for Kleenex.
\end{description}

As we also point out in Paper~\ref{paper:peg-parsing}, there are still some issues that need to be addressed before the streaming parsing algorithm for parsing expression grammars can be used as a high-performance execution model in Kleenex:
\begin{description}
\item[Regular grammar parsing] The algorithm fails to be streaming for the purely right-regular grammars, but works as expected for grammars using the non-regular features of parsing expression grammars. This is due to the fact that streaming regular expression parsing relies on orthogonal criteria for detecting when parts of the parse tree can be written to the output, which suggests that we might be able to find a hybrid method which can handle both types of grammars.
\item[Time complexity overhead] In its current form, the algorithm has been optimized for simplicity and performs a large number of computations which are never needed, adding a constant time overhead to the processing of each input symbol. This should be avoidable by integration with a runtime analysis, but requires further study.
\item[Machine models] Can we find a deterministic machine model which can simulate the streaming parsing algorithm such that parts of the expensive computations can be encoded in the control mechanism of the machine? Such a model would necessarily have to generalize the deterministic pushdown automata~\cite{alsu2006,grune2008} used for parsing deterministic context-free languages, but could potentially yield significant speedups.
\end{description}

%%% Local Variables:
%%% mode: latex
%%% TeX-master: "thesis"
%%% End:

}
\putbib[bibliography]
\end{bibunit}

\appendix

% \chapter{Placeholder A}
% \label{paper:two-pass}
% \chapter{Placeholder B}
% \label{paper:optimal-streaming}
% \chapter{Placeholder C}
% \label{paper:kleenex}
% \chapter{Placeholder D}
% \label{paper:peg-parsing}

\begin{bibunit}[abbrv]
\chapter{Two-Pass Greedy Regular Expression Parsing}
\label{paper:two-pass}

{
% \T -> \TC
% \B -> \BC

\newcommand{\figwidth}{0.5\textwidth}

\newcommand{\activeState}[1]{\textbf{\underline{#1}}}
\newcommand{\re}[1]{$#1$}

% Migrated from draft.tex
\newcommand{\setN}{\mathcal{N}}
\newcommand{\N}[1]{\mathcal{N}\langle #1 \rangle}
\newcommand{\Log}{\mathcal{L}}
\newcommand{\itrans}[1]{\stackrel{#1}{\longrightarrow}}
\newcommand{\trans}[3]{#1 \itrans{#2} #3}
\newcommand{\fw}[1]{\mathsf{#1}}
\newcommand{\dom}{\mathrm{dom}}
\newcommand{\bw}[1]{\overline{\mathsf{#1}}}
\newcommand{\bit}[1]{\mathsf{#1}}
\newcommand{\ignoreResult}{\underline{\hspace{1em}}}
\newcommand{\iwalk}[1]{\stackrel{#1}{\leadsto}}
\newcommand{\walk}[3]{#1 \iwalk{#2} #3}
\newcommand{\Read}[1]{\mathsf{read}(#1)}
\newcommand{\Write}[1]{\mathsf{write}(#1)}
\newcommand{\Logged}[1]{\mathsf{log}(#1)}

% Theorems:
%\theorembodyfont{\normalfont}
%\newtheorem{DEF}{Definition}
%\newenvironment{proof*}{\begin{proof}\normalfont}{\qed\end{proof}}

% Bits:
\newcommand{\Bit}[1]{\textsf{#1}}

% Denotational operators:
\newcommand{\bprec}{\prec}
\newcommand{\gprec}{\lessdot}
\newcommand{\BC}[2][\phantom{}]{\mathcal{B}_{#1}[\![ {#2} ]\!]}
\newcommand{\TC}[2][\phantom{,}]{\mathcal{T}_{#1}[\![ {#2} ]\!]}
\newcommand{\Bnp}[2][\phantom{,}]{\mathcal{B}_{#1}^\mathrm{np}[\![ {#2} ]\!]}
\newcommand{\Tnp}[2][\phantom{,}]{\mathcal{T}_{#1}^\mathrm{np}[\![ {#2} ]\!]}
\newcommand{\Lang}[1]{\mathcal{L}[\![ {#1} ]\!]}
\newcommand{\Flat}{\mathsf{flat}}
\newcommand{\Code}{\mathsf{code}}
\newcommand{\Decode}{\mathsf{decode}}
\newcommand{\Lit}[1]{\mathtt{#1}}

% Value syntax:
\newcommand{\fold}[1]{\mathsf{fold}~#1}
\newcommand{\inl}[1]{\mathsf{inl}~#1}
\newcommand{\inr}[1]{\mathsf{inr}~#1}

% Inference rules and judgments:
\newcommand{\Infer}[3][]{\infer[\textsf{#1} :]{#2}{#3}}
\newcommand{\Greedy}[3]{#1 \in #2 \leadsto #3}

% Operations for NFA-simulation:
\newcommand{\Reach}{\mathsf{Reach}}
\newcommand{\Close}{\mathsf{Close}}
\newcommand{\Step}{\mathsf{Step}}
\newcommand{\Path}{\mathsf{Path}}
\newcommand{\Paths}{\mathsf{Paths}}
\newcommand{\LLClose}{\mathsf{LClose}}
\newcommand{\LLCloseLog}{\mathsf{Log}}
\newcommand{\LLTrace}{\mathsf{LTrace}}
\newcommand{\LLStep}{\mathsf{LStep}}
\newcommand{\LLSim}{\mathsf{LSim}}
\newcommand{\LLMerge}{\mathsf{Merge}}

This paper has been published in the following:

\begin{center}
\begin{minipage}{0.9\textwidth}
\small
Niels Bjørn Bugge Grathwohl, Fritz Henglein, Lasse Nielsen and Ulrik Terp Rasmussen. ``Two-Pass Greedy Regular Expression Parsing''. In \emph{Proceedings 18th International Conference on Implementation and Application of Automata (CIAA)}, pages 60-71. Springer, 2013. DOI: \href{http://dx.doi.org/10.1007/978-3-642-39274-0_7}{10.1007/978-3-642-39274-0\_7}.
\end{minipage}
\end{center}

\noindent
The enclosed version contains minor revisions of the published paper in the form of corrections of typos and reformatting to fit the layout of this dissertation. The presentation in Sections~\ref{sec:nfa-sim} and \ref{sec:optimized_algo} has also been improved, but no contributions have been added or removed.

\clearpage

\thispagestyle{plain}
\begin{center}
{\LARGE \textbf{Two-Pass Greedy Regular Expression Parsing}\footnote{The order of authors is insignificant.}}

\vspace{1.5em}

{Niels Bjørn Bugge Grathwohl}, {Fritz Henglein}, {Lasse Nielsen} and {Ulrik Terp Rasmussen}

\vspace{1em}

{Department of Computer Science, University of Copenhagen (DIKU)}
\end{center}

\begin{abstract}
We present new algorithms for producing greedy parses for regular expressions (REs) in a semi-streaming fashion.  Our lean-log algorithm executes in time $O(m n)$ for REs of size $m$ and input strings of size $n$ and outputs a compact bit-coded parse tree representation.  It improves on previous algorithms by: operating in only 2 passes; using only $O(m)$ words of random-access memory (independent of $n$); requiring only $k n$ bits of sequentially written and read log storage,
where $k < \frac{1}{3} m$ is the number of alternatives and Kleene stars in the RE; 
processing the input string as a symbol stream and not requiring it to be stored at all.  Previous RE parsing algorithms do not scale linearly with input size, or require substantially more log storage and employ 3 passes where the first consists of reversing the input, or do not or are not known to produce a greedy parse.  
The performance of our unoptimized C-based prototype indicates that the superior performance of our lean-log algorithm can also be observed in practice; it is also surprisingly competitive with RE tools not performing full parsing, such as Grep.
\end{abstract}

\section{Introduction}

Regular expression (RE) parsing is the problem of producing a parse tree for an input string under a given RE.  In contrast to most regular-expression based tools for programming such as Grep, RE2 and Perl, RE parsing returns not only whether the input is accepted, where a substring matching the RE and/or sub-REs are matched, but a full parse tree.  In particular, for Kleene stars it returns \emph{a list} of all matches, where each match again can contain such lists depending on the star depth of the RE.  

An RE parser can be built using Perl-style backtracking or general context-free parsing techniques.  What the backtracking parser produces is the \emph{greedy} parse amongst potentially many parses.  General context-free parsing and backtracking parsing are not scalable since they have cubic, respectively exponential worst-case running times.   REs can be and often are grammatically ambiguous and can require arbitrary much look-ahead, making limited look-ahead context-free parsing techniques inapplicable.
Kearns \cite{kearns90} describes the first linear-time algorithm for RE parsing.  In a streaming context it consists of 3 passes: reverse the input, perform backward NFA-simulation, and construct parse tree.  Frisch and Cardelli \cite{frca2004} formalize greedy parsing and use the same strategy to produce a greedy parse.  Dub\'{e} and Feeley \cite{dufe2000} and Nielsen and Henglein \cite{nihe2011} produce parse trees in linear time for fixed RE, the former producing internal data structures and their serialized forms, the latter parse trees in bit-coded form; neither produces a greedy parse.

In this paper we make the following contributions:
\begin{enumerate}
\item Specification and construction of symmetric nondeterministic finite automata (NFA) with maximum in- and out-degree 2, whose paths from initial to final state are in one-to-one correspondence with the parse trees of the underlying RE; in particular, the greedy parse for a string corresponds to the lexicographically least path accepting the string.
\item NFA simulation with \emph{ordered state sets}, which gives rise 
to a 2-pass greedy parse algorithm using $\lceil m \lg m \rceil$ bits per input symbol in the original input string, with $m$ the size of the underlying RE.  No input reversal is required.  
\item NFA simulation optimized to require only $k \leq \lceil 1/3 m \rceil$ bits per input symbol, where the input string need not be stored at all and the 2nd pass is simplified.  Remarkably, this \emph{lean-log algorithm} requires fewest log bits, and neither state set nor even the input string need to be stored.
\item An empirical evaluation, which indicates that our prototype implementation of the optimized 2-pass algorithm outperforms also in practice previous RE parsing tools and is sometimes even competitive with 
RE tools performing limited forms of RE matching. 
\end{enumerate}

In the remainder, we introduce REs as types to represent parse trees, define greedy parses and their bit-coding, introduce NFAs with bit-labeled transitions, describe NFA simulation with ordered sets for greedy parsing and finally the optimized algorithm, which only logs join state bits.  We conclude with an empirical evaluation of a straightforward prototype to gauge the competitiveness of full greedy parsing with regular-expression based tools yielding less information for Kleene-stars.

\section{Symmetric NFA Representation of Parse Trees}

% Bit-coding of parses
% Symmetric NFA construction
% Representation theorem

REs are finite terms of the form $0, 1, a, E_1 \times E_2, E_1 + E_2$ or $E_1^*$, where $E_1, E_2$ are REs.

\textbf{Proviso:} For simplicity and brevity we henceforth assume REs that do not contain sub-REs of the form 
$E^*$, where $E$ is nullable (can generate the empty string).
All results reported here can be and have been extended to such problematic REs in the style of Frisch and Cardelli \cite{frca2004}. In particular, our implementation BitC handles problematic REs.

REs can be interpreted as types built from singleton, product, sum, and list type constructors \cite{frca2004,heni2010}:
\begin{align*}
\TC{0} ={}& \emptyset \\
\TC{1} ={}& \{ () \}, \\
\TC{a} ={}& \{ \Lit{a} \}, \\
\TC{E_1 \times E_2} ={}& \{ (V_1, V_2) \mid V_1 \in \TC{E_1}, V_2 \in \TC{E_2} \}, \\
\TC{E_1 + E_2} ={}& \{ \inl{V_1} \mid V_1 \in \TC{E_1} \} \cup \{ \inr{V_2} \mid V_2 \in \TC{E_2} \}, \\
\TC{E_0^\star} ={}& \{[V_1, \ldots, V_n] \mid n \geq 0 \, \wedge \forall 1 \leq i \leq n. V_i \in \TC{E_0} \}
\end{align*}

Its structured values $\TC{E}$ represent the \emph{parse trees} for $E$ such that the regular language $\Lang{E}$ coincides with the strings obtained by flattening the parse trees:
$$\Lang{E} = \{ \Flat(V) \mid V \in \TC{E} \},$$
where the flattening function erases all structure but the leaves:
\begin{align*}
  \Flat(()) ={}& \epsilon \\
  \Flat(a) ={}& \Lit{a} \\
  \Flat((V_1, V_2)) ={}& \Flat(V_1) \Flat(V_2) \\
  \Flat(\inl{V_1}) ={}& \Flat(V_1) \\
  \Flat(\inr{V_2}) ={}& \Flat(V_2) \\
  \Flat([V_1, \ldots, V_n]) ={}& \Flat(V_1) \ldots \Flat(V_n)
\end{align*}

We recall bit-coding from Nielsen and Henglein \cite{nihe2011}. The bit code $\Code(V)$ of a parse tree $V \in \TC{E}$ is a sequence of bits uniquely identifying $V$ within $\TC{E}$; that is, there exists a function $\Decode_E$ such that for all $V\in \TC{E}$, we have $\Decode_E(\Code(V)) = V$:
\begin{align*}
  \Code(()) ={}& \epsilon \\
  \Code(a) ={}& \epsilon \\
  \Code((V_1, V_2)) ={}& \Code(V_1) \, \Code(V_2) \\
  \Code([V_1, \ldots, V_n]) ={}& \Bit{0} \, \Code(V_1) \ldots \Bit{0} \, \Code(V_n) \, \Bit{1} \\
  \Code(\inl{V_1}) ={}& \Bit{0}~\Code(V_1) \\
  \Code(\inr{V_2}) ={}& \Bit{1}~\Code(V_2)
\end{align*}

The definition of $\Decode_E$ is omitted for brevity, but is straightforward.

% 
% \begin{align*}
%   \Decode_1(\varepsilon) ={}& () \\
%   \Decode_a(\varepsilon) ={}& a \\
%   \Decode_{E_1 \times E_2}(d_1 d_2) ={}& (\Decode_{E_1}(d_1), \Decode_{E_2}(d_2)) \\
%   \Decode_{E_1 + E_2}(\Bit{0} d) ={}& \inl{\Decode_{E_1}(d)} \\
%   \Decode_{E_1 + E_2}(\Bit{1} d) ={}& \inr{\Decode_{E_2}(d)}
% \end{align*}

We write $\BC{\ldots}$ instead of $\TC{\ldots}$ whenever we want to refer to the bit codings, rather than the parse trees. We use subscripts to discriminate parses with a specific flattening:
  $\TC[s]{E} = \{V \in \TC{E} \mid \Flat(V) = s\}.$
We extend the notation $\BC[s]{\ldots}$ similarly.

Note that a bit string by itself does not carry enough information to deduce which parse tree it represents.  Indeed this is what makes bit strings a compact representation of strings where the underlying RE is statically known.

The set $\BC{E}$ for an RE $E$ can be compactly represented by an \emph{augmented nondeterministic finite automaton (aNFA)}, a variant of enhanced NFAs \cite{nihe2011} that has in- and outdegree at most 2 and carries a label on each transition.

\begin{figure}[t]
    \scriptsize
    \centering
    \begin{tabular}{|c|c|}
    \hline
    $E$ & $\N{E, q^s, q^f}$ \\
    \hline
    \hline
    $0$ &
    \begin{tikzpicture}[node distance=2cm, auto]
        \node[state,initial] (qs) {$q^s$};
        \node[state,accepting] (qf) [right of=qs] {$q^f$};
    \end{tikzpicture} \\
    \hline
    $1$ &
    \begin{tikzpicture}[node distance=2cm, auto]
        \node[state,initial, accepting] (qs) {$q^s$};
    \end{tikzpicture} {\footnotesize (implies $q^s = q^f$)} \\
    \hline
    $\Lit{a}$ &
    \begin{tikzpicture}[node distance=2cm, auto]
        \node[state,initial] (qs) {$q^s$};
        \node[state,accepting] (qf) [right of=qs] {$q^f$};
        \path[->] (qs) edge node {$\Lit{a}$} (qf);
    \end{tikzpicture} \\
    \hline
    $E_1 \times E_2$ &
    \begin{tikzpicture}[node distance=2.5cm, auto]
        \node[state,initial] (qs) {$q^s$};
        \node[state] (q') [right of=qs] {$q'$};
        \node[state,accepting] (qf) [right of=q'] {$q^f$};

        \path[dashed, ->] (qs) edge node {$\N{E_1,q^s,q'}$} (q');
        \path[dashed, ->] (q') edge node {$\N{E_2,q',q^f}$} (qf);
    \end{tikzpicture}\\
    \hline
  $E_1 + E_2$ &
  \begin{tikzpicture}[node distance=2.5cm, auto]
      \node[state,initial] (qs) {$q^s$};
      \node[state] (qs1) [above right of=qs,xshift=-0.67cm,yshift=-1.33cm] {$q_1^s$};
      \node[state] (qf1) [right of=qs1] {$q_1^f$};
      \node[state] (qs2) [below right of=qs,xshift=-0.67cm,yshift=1.33cm] {$q_2^s$};
      \node[state] (qf2) [right of=qs2] {$q_2^f$};
      \node[state,accepting] (qf) [below right of=qf1,xshift=-0.67cm,yshift=1.33cm] {$q^f$};

      \path[->] (qs) edge node {$\fw{0}$} (qs1)
      edge node [swap] {$\fw{1}$} (qs2)
      (qf1) edge node {$\bw{0}$} (qf)
      (qf2) edge node [swap] {$\bw{1}$} (qf);
      \path[dashed, ->] (qs1) edge node {$\N{E_1,q_1^s,q_1^f}$} (qf1)
      (qs2) edge node {$\N{E_2,q_2^s,q_2^f}$} (qf2);
  \end{tikzpicture} \\
  \hline
  $E_0^\star$ &
  \begin{tikzpicture}[node distance=2cm, auto]
      \node[state,initial] (qs) {$q^s$};
      \node[state] (qs') [right of=qs] {$q'$};
      \node[state] (qs0) [above of=qs',yshift=-1.1cm,xshift=-1.cm] {$q_0^s$};
      \node[state] (qf0) [right of=qs0,xshift=0.5cm] {$q_0^f$};
      \node[state,accepting] (qf) [right of=qs'] {$q^f$};

      \path[->] (qs') edge node [swap] {$\fw{0}$} (qs0)
      edge node {$\fw{1}$} (qf)
      (qf0) edge node [swap] {$\bw{0}$} (qs')
      (qs) edge node {$\bw{1}$} (qs');
      \path[dashed, ->] (qs0) edge node {$\N{E_0,q_0^s,q_0^f}$} (qf0);
  \end{tikzpicture}\\
  \hline
\end{tabular}

%%% Local Variables:
%%% TeX-master: "../thesis"
%%% End:
    \caption{aNFA construction schema.}
    \label{fig:nfa_construction}
\end{figure}
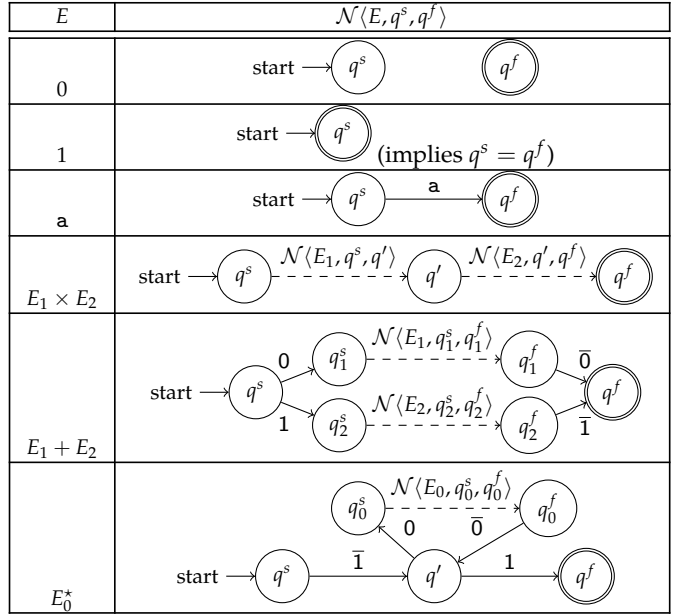

\begin{definition}[Augmented NFA]
  \label{def:nfa}
  An \emph{augmented NFA} (aNFA) is a 5-tuple $M = (Q, \Sigma, \Delta,
  q^s, q^f)$ where $Q$ is the set of states, $\Sigma$ is the input
  alphabet, and $q^s, q^f$ are the start and final states, respectively. The transition
  relation $\Delta \subseteq Q \times (\Sigma~\cup~
  \{\fw{0},\fw{1},\bw{0},\bw{1}\}) \times Q$ contains directed,
  labeled transitions: $(q,\gamma,q') \in \Delta$ is a transition from
  $q$ to $q'$ with label $\gamma$, written $\trans{q}{\gamma}{q'}$.

  We call transition labels in $\Sigma$ \emph{input labels}; labels in
  $\{\fw{0},\fw{1}\}$ \emph{output labels}; and labels in
  $\{\bw{0},\bw{1}\}$ \emph{log labels}.

  We write $\walk{q}{p}{q'}$ if there is a path labeled $p$ 
  from $q$ to $q'$. The
  sequences $\Read{p}$, $\Write{p}$, and $\Logged{p}$ are the subsequences
  of input labels, output labels, and log labels of $p$, respectively.

  We write: $J_M$ for the \emph{join states} $\{ q \in Q \mid \exists
  q_1,q_2.~(q_1,\bw{0},q), (q_2,\bw{1},q) \in \Delta \}$; 
  $S_M$ for the \emph{symbol sources} $\{ q \in Q \mid
  \exists q' \in Q,a \in \Sigma.~(q,a,q') \}$; and
  $C_M$ for the \emph{choice states} $\{ q \in Q \mid \exists
  q_1,q_2.~(q,\fw{0},q_1),(q,\fw{1},q_2) \in \Delta \}$.

  If $M$ is an aNFA, then $\overline{M}$ is the aNFA obtained by
  \emph{flipping} all transitions and exchanging the start and finishing
  states, that is reverse all transitions and interchange output labels with the corresponding log labels. \qed
\end{definition}

Our algorithm for constructing an aNFA from an RE is a standard Thompson-style NFA generation algorithm modified to accomodate output and log labels:

\begin{definition}[aNFA construction]
  \label{def:nfa:construction}
  We write $M=\N{E,q^s,q^f}$ when $M$ is an aNFA constructed according
  to the rules
  in Figure~\ref{fig:nfa_construction}.
\end{definition}

Augmented NFAs are dual under reversal; that is, flipping produces the augmented NFA for the reverse of the regular language.
\begin{proposition}
Let $\overline{E}$ be canonically constructed from $E$ to denote the reverse of $\Lang{E}$, i.e. $\overline{E_1 \times E_2} = \overline{E_2} \times \overline{E_1}$.  Let $M=\N{E,q^s,q^f}$.
Then $\overline{M}=\N{\overline{E},q^f,q^s}$.
\end{proposition}

This is useful since we will be running aNFAs in both forward and backward (reverse) directions.

Well-formed aNFAs---and Thompson-style NFAs in general---are
canonical representations of REs in the sense that
they not only represent their language interpretation, but their type
interpretation:

\begin{theorem}[Representation]
  \label{prop:representation}
  Given an aNFA $M=\N{E,q^s,q^f}$, $M$ outputs the bit-codings of $E$:
  $$\BC[s]{E} = \{\Write{p} \mid \walk{q^s}{p}{q^f} \wedge \Read{p} = s\}.$$
\end{theorem}

\section{Greedy parsing}
The \emph{greedy parse} of a string $s$ under an RE \re{E} is what a backtracking parser returns that tries the left operand of an alternative first and backtracks to try the right alternative only if the left alternative does not yield a successful parse. The name comes from treating the Kleene star \re{E^\star} as \re{E \times E^\star + 1}, which ``greedily'' matches \re{E} against the input as many times as possible.  A ``lazy'' matching interpretation of \re{E^\star} corresponds to treating \re{E^\star} as \re{1 + E \times E^\star}.  (In practice, multiple Kleene-star operators are allowed to make both interpretations available; e.g. \re{E *} and \re{E{**}} in PCRE.) 

Greedy parsing can be formalized by an order $\gprec$
on parse trees, where $V_1 \gprec V_2$ means that $V_1$ is ``more
greedy'' than $V_2$.  The following is adapted from Frisch and Cardelli \cite{frca2004}.

\begin{definition}[Greedy order]
\label{def:structorder:greedy}
The binary relation $\gprec$ is defined inductively on the structure of values as follows:
\[\begin{array}{rcl@{\hskip 1em}c@{\hskip 1em}l}
  (V_1, V_2) & \gprec & (V_1', V_2')
    & \text{if} & V_1 \gprec V_1' \lor (V_1 = V_1' \land V_2 \gprec V_2') \\
  \inl{V_0} & \gprec & \inl{V_0'}
    & \text{if} & V_0 \gprec V_0' \\
  \inr{V_0} & \gprec & \inr{V_0'}
    & \text{if} & V_0 \gprec V_0' \\
  \inl{V_0} & \gprec & \inr{V_0'} & & \\
  {}[V_1, \ldots] & \gprec & [] & & \\
  {}[V_1, \ldots] & \gprec & [V_1', \ldots]
    & \text{if} & V_1 \gprec V_1' \\
  {}[V_1, V_2, \ldots] & \gprec & [V_1, V_2', \ldots] 
    & \text{if} & [V_2, \ldots] \gprec [V_2', \ldots]
\end{array}\]
\end{definition}

The relation $\gprec$ is not a total order; consider for example the incomparable elements
$(\Lit{a}, \inl{()})$ and $(\Lit{b}, \inr{()})$. 
The parse trees of any particular RE are totally ordered, however:
\begin{proposition}
  For each \re{E}, the order $\gprec$ is a strict total
  order on $\TC{E}$.
%  \begin{proof*}
%    By induction on the structure of \re{E}.
%  \end{proof*}
\end{proposition}

In the following, we will show that there is a correspondence between the structural order on values and the lexicographic order on their bit-codings.

\begin{definition}
  \label{def:order:greedy}
  For bit sequences $d, d' \in \{\Bit{0}, \Bit{1}\}^\star$ we
  write $d \bprec d'$ if $d$ is lexicographically strictly less than $d'$;
  that is, $\bprec$ is the least relation satisfying
  \begin{enumerate}
  \item $\epsilon \bprec d$ if $d \not= \epsilon$
  \item $b~d \bprec b'~d'$ if $b < b'$ or $b = b'$ and $d \bprec
    d'$.
  \end{enumerate}
\end{definition}

\begin{theorem}
  \label{prop:lex}
  For all REs $E$ and values $V, V' \in
  \TC{E}$ we have $V \gprec V'$ iff $\Code(V) \bprec \Code(V')$.
%  \begin{proof*}
%    By induction on $V$.
%  \end{proof*}
\end{theorem}

\begin{corollary}
  For any RE $E$ with aNFA $M = \N{E,q^s,q^f}$, and
  for any string $s$, $\min_\gprec \TC[s]{E}$ exists and
  \[\min_\gprec \TC[s]{E}
    = \Decode_E \left(\min_\bprec \left\{ \Write{p} ~\middle|~ \walk{q^s}{p}{q^f} \wedge \Read{p} = s \right\} \right).\]
    \begin{proof}
      Follows from Theorems~\ref{prop:representation} and
      \ref{prop:lex}.
    \end{proof}
\end{corollary}

We can now characterize greedy RE parsing as follows:
Given an RE $E$ and string $s$, find bit sequence $b$
such that there exists a path $\walk{q^s}{p}{q^f}$ from start to finishing state in the
aNFA for $E$ such that:
\begin{enumerate}
\item $\Read{p} = s$,
\item $\Write{p} = b$,
\item $b$ is lexicographically least among all paths satisfying 1 and 2.
\end{enumerate}

This is easily done by a backtracking algorithm that tries
$\Bit{0}$-labeled transitions before $\Bit{1}$-labeled ones.  It is
atrociously slow in the worst case, however: exponential time. How to
do it faster?

\section{NFA-Simulation with Ordered State Sets}
\label{sec:nfa-sim}

Our first algorithm is basically an NFA-simulation.  For reasons of space we only sketch its key idea, which is the basis for the more efficient algorithm in the following section.
  
A standard NFA-simulation consists of computing $\Reach(S, s)$ where
\begin{eqnarray*}
    \Reach(S, \epsilon) & = & S \\
    \Reach(S, a \, s') & = & \Reach(\Close(\Step(S, a)), s') \\
    \Step(S, a) & = & \{ q' \mid q \in S, \trans{q}{a}{q'} \} \\
    \Close(S') & = & \{ q'' \mid q' \in S', \walk{q'}{p}{q''}, \Read{p} = \epsilon \}
\end{eqnarray*}
Checking $q^f \in \Reach(S_0, s)$ where $S_0 = \Close(\{q^s\})$ determines whether $s$ is accepted or not.  But how to construct an accepting \emph{path} and in particular the one corresponding to the greedy parse?  

%When the forward NFA simulation is applied to an input string $s = a_1 \ldots a_n$, it computes a sequence of state sets $S_0, S_1, ..., S_n$ which we write down in a \emph{log}. Given a list of logged state sets, we can reconstruct the set of all paths 

We can \emph{log} the sequence of NFA state sets reached during forward NFA-simulation over an input string $s = a_1 \ldots a_n$. The log thus consists of a list of state sets $S_0, S_1, ..., S_n$, where $S_0$ is defined above, and for each $0 \leq i \leq n-1$, we have $S_{i+1} = \Close(\Step(S_i, a_{i+1}))$.

It is easy to check that every path $\walk{q^s}{p}{q^f}$ with $\Read{p} = s$ is of the form
\[
q^s \iwalk{p_0} q_0 \itrans{a_1} q_0' \iwalk{p_1} q_1 \cdots q_i \itrans{a_{i+1}} q_i' \iwalk{p_{i+1}} q_{i+1} \cdots q_{n-1} \itrans{a_n} q_{n-1}' \iwalk{p_n} q^f
\]
where $p = p_0 a_1 p_1 a_2 p_2 ... a_n p_n$ and $\Read{p_i} = \epsilon$ for all $0 \leq i \leq n$. By definition, each $q_i$ is in the state set $S_i$, so the set of all paths $\{ p \mid q^s \iwalk{p} q^f \wedge \Read{p} = s \}$ can be recovered only from the log, and equals the set $\Paths(n,q^f)$ defined as follows:
\begin{align}
  \nonumber
  \Paths(0,q'') ={}& \{ p \mid q^s \iwalk{p} q'' \mid \Read{p} = \epsilon \} \\
  \label{eqn:path}
  \Paths(i+1,q'') ={}& \{ p' p \mid \exists q \in S_i \ldotp
                      \begin{array}[t]{@{}l}
                        \exists a, q' \ldotp q \itrans{a} q' \iwalk{p} q'' \\
                        {}\wedge \Read{p}=\epsilon \\
                        {}\wedge p' \in \Paths(i,q) \}
                      \end{array}
\end{align}
Using this definition, any single path $p \in \Paths(n,q^f)$ can be recovered in linear time by processing the log in reverse order. In each step $i>0$, we pick some $q \in S_i$ such that the condition in \ref{eqn:path} is satisfied, which can be checked by computing the preimage of the $\epsilon$-closure of $q''$. Note in particular that we do not need the input string for this. $\Write{p}$ gives a bit-coded parse tree, though not necessarily the lexicographically least. We need a way to locally choose $q \in S_i$ such that the lexicographically least path is constructed without backtracking.

We can adapt the NFA-simulation by keeping each state set $S_i$ in a particular order: If $\Reach(\{q^s\}, a_1 \ldots a_i) = \{ q_{i1}, \ldots q_{ij_i} \}$ then order the $q_{ij}$ according to the lexicographic order of the paths reaching them.  Intuitively, the highest ranked state in $S_i$ is on the greedy path if the remaining input is accepted from this state; if not, the second-highest ranked is on the greedy path, if the remaining input is accepted; and so on. Using this, we can resolve the choice of $q$ in \eqref{eqn:path} and define a function which recovers the lexicographically least bit-code $\Path(n,q^f)$ from the log:
\begin{align*}
  \Path(0,q'') ={}& \min_{\bprec} \{ p \mid q^s \iwalk{p} q'' \mid \Read{p} = \epsilon \} \\
  \Path(i+1, q'') ={}& \Path(i,q) \Write{p} \\
  &\quad \text{where $q \in S_i$ is highest ranked such that} \\
  &\quad \exists a, q' \ldotp q \itrans{a} q' \iwalk{p} q'' \wedge \Read{p} = \epsilon
\end{align*}

The NFA-simulation can be refined to construct properly ordered state sequences instead of sets without asymptotic slow-down.  The log, however, is adversely affected by this.  We need $\lceil m \lg m \rceil$ bits per input symbol, for a total of $\lceil m n \lg m \rceil$ bits.

The key property for allowing us to list a state at most once in an ordered state sequence is this:
\begin{lemma}\label{lem:extension_lemma}
    Let $s$, $t_1$, $t_2$, and $t$ be states in an aNFA $M$, and 
    let $p_1$, $p_2$, $q_1$, $q_2$ be paths in $M$ such that
    $\walk{s}{p_1}{t_1}$, $\walk{s}{p_2}{t_2}$, and
    $\walk{t_1}{q_1}{t}$, $\walk{t_2}{q_2}{t }$, where $p_1$ is not a
    prefix of $p_2$. If $\Write{p_1} \bprec
    \Write{p_2}$ then $\Write{p_1q_1} \bprec \Write{p_2q_2}$
\end{lemma}
\begin{proof}
    Application of the lexicographical ordering on paths.
\end{proof}

\section{Lean-log Algorithm}\label{sec:optimized_algo}

We can do better than saving a log where each element is a full sequence of NFA states. Since the join states $J_M$ of an aNFA $M$ become the choice states $C_{\overline{M}}$ of the reverse aNFA $\overline{M}$ we only need to construct one ``direction'' bit for each join state at each input string position.  It is not necessary to record any states in the log at all.  This results in an algorithm that requires only $k$ bits per input symbol for the log, where $k$ is the number of Kleene-stars and alternatives occurring in the RE.  It can be shown that $k \leq \frac{1}{3} m$; in practice we can observe $k <\!\!< m$. % Bille, Thorup exploit this, too.

Instead of writing down state sequences, we write down \emph{log frames} which are partial maps $L : J_M \to \{\bw{0},\bw{1}\}$. The subset of $J_M$ on which $L$ is defined is denoted $\dom(L)$. The empty log frame is $\emptyset$, and the disjoint union of two log frames $L, L'$ is written as $L \cup L'$. The set of all log frames is $\mathsf{Frame}_M$. A modified closure algorithm computes both a state sequence and a log frame:
\begin{align*}
  \Close(q,L) :{}& Q_M \times \mathsf{Frame}_M \to Q_M^* \times \mathsf{Frame}_M \\
  \Close(q,L) ={}&
  \begin{cases}
    (\vec{q}\vec{q'}, L'') & \text{if }\begin{array}[t]{@{}l}
                   q \itrans{\Bit{0}} q_0 \wedge q \itrans{\Bit{1}} q_1 \\
                   \quad {}\wedge \Close(q_0, L) = (\vec{q}, L') \\
                   \quad {}\wedge \Close(q_1, L') = (\vec{q'}, L'')
                 \end{array} \\
    \Close(q',L \cup \{ q' \mapsto t \}) & \text{if }\begin{array}[t]{@{}l}
                     q \itrans{t} q' \wedge t \in \{\bw{0},\bw{1}\} \\
                     \quad {} \wedge q' \not\in \dom(L)
                   \end{array} \\
    (\epsilon, L)
  \end{cases}
\end{align*}
Computing $\Close(q, \emptyset) = (\vec{q}, L)$ results in the sequence of states $\vec{q}$ in the ``frontier'' of the $\epsilon$-closure of $q$, ordered according to their lexicographic order, and a log frame $L$ which uniquely identifies the lexicographically least $\epsilon$-path from $q$ to any state in $\vec{q}$. Note that the algorithm works by backtracking and stops when a join state has previously been encountered. This is sound since the previous encounter must have been via a higher ranked path, and since any extension of the path continues to have higher rank by Lemma~\ref{lem:extension_lemma}.

The closure algorithm is extended to state sequences by applying the statewise closure algorithm in ranking order, using the same log frame:
\begin{align*}
  \Close^* :{}& Q^*_M \times \mathsf{Frame}_M \to Q^*_M \times \mathsf{Frame}_M \\
  \Close^*(\epsilon, L) ={}& L \\
  \Close^*(q~\vec{q}, L) ={}& (\vec{q'} \vec{q''}, L'') \\
  & \text{where $(\vec{q'}, L') = \Close(q, L)$} \\ &\text{and $(\vec{q''}, L'') = \Close^*(\vec{q}, L')$}
\end{align*}

The modified algorithm $\Step : Q_M^* \times \Sigma \to Q_M^*$ is defined on single states $q \in Q_M$ by
\[
  \Step(q, a) =
  \begin{cases}
    q' & \text{if $q \itrans{a} q'$} \\
    \epsilon & \text{otherwise}
  \end{cases}
\]
and extended homomorphically to sequences $Q_M^*$. The forward simulation algorithm is essentially the same process as before, but now explicitly maintains a sequence of log frames $\vec{L}$:
\begin{align*}
  \Reach :{}& Q^*_M \times \Sigma^* \to Q^*_M \times \mathsf{Frame}_M^* \\
  \Reach(\vec{q}, \epsilon) ={}& (\vec{q}, \epsilon) \\
  \Reach(\vec{q}, a~s') ={}& (\vec{q''}, L \vec{L}) \\
  & \text{where $(\vec{q'}, L) = \Close^*(\Step(\vec{q},a), \emptyset)$} \\
  & \text{and $(\vec{q''}, \vec{L}) = \Reach(\vec{q'}, s')$}
\end{align*}

Let $s = a_1 \ldots a_n$. Computing $\Reach(\vec{q_0}, s)$ where $(\vec{q_0}, L_0) = \Close(q^s, \emptyset)$ results in a pair $(q_1 q_2 ... q_m, L_1 ... L_n)$. If for any $1 \leq k \leq m$ we have $q_k = q^f$, then the the lexicographically least path $q^s \iwalk{p} q^f$ with $\Read{p} = s$ exists, and the sequence $L_0 L_1 ... L_n$ can be used to effectively reconstruct its bit-code $\Path(q^f, n)$:
\begin{align*}
  \Path :{}& Q_M \times \{0,1,...,n\} \to \{\fw{0}, \fw{1}\}^* \\
  \Path(q^s, 0) ={}& \epsilon \\
  \Path(q', i) ={}&
  \begin{cases}
    \Path(q, i-1) & \text{if $\exists a \in \Sigma\ldotp q \itrans{a} q'$} \\
    \Path(q_{L_i(q)}, i) & \text{if $q_{\bw{0}} \itrans{\bw{0}} q'$ and $q_{\bw{1}} \itrans{\bw{1}} q'$} \\
    \Path(q, i) b & \text{if $q \itrans{b} q'$ and $b \in \{\fw{0},\fw{1}\}$}
  \end{cases}
\end{align*}

The forward $\Reach$ algorithm keeps the aNFA and the current character in working memory, requiring  $O(m)$ words of random access memory (RAM), writing $n k$ bits to the log, and discarding the input string.  The backward $\Path$ algorithm also requires $O(m)$ words of RAM and reads from the log in reverse write order. The log is thus a 2-phase stack: In the first pass it is only pushed to, in the second pass popped from. 

Both $\Close^*$ and $\Step$ run in time $O(m)$ per input symbol, hence the forward pass requires time $O(m n)$.  Likewise, the backward pass requires time $O(m n)$.

\section{Evaluation}

We have implemented the optimized algorithms in C and in Haskell, and
we compare the performance of the C implementation with the following
existing RE tools:
\begin{description}
  \item[RE2:] Google's RE implementation, available
      from~\cite{re2}.
  \item[Tcl:] The scripting language Tcl~\cite{Tcl}.
  \item[Perl:] The scripting language Perl~\cite{wall2000}.
  \item[Grep:] The UNIX tool \texttt{grep}.
  \item[Rcp:] The implementation of the algorithm ``\emph{DFASIM}''
      from~\cite{nihe2011}.  It is based on Dub\'{e} and Feeley's
      method~\cite{dufe2000}, but altered to produce a bit-coded parse tree.
  \item[FrCa:] The implementation of the algorithm``\emph{FrCa}''
      algorithm used in~\cite{nihe2011}.  It is based on Frisch and
      Cardelli's method from~\cite{frca2004}.
\end{description}

In the subsequent plots, our implementation of the lean-log algorithm is referred to as \emph{BitC}.

The tests have been performed on an Intel Xeon 2.5 GHz machine running GNU/Linux 2.6.  

\begin{figure}[h!]
    \centering
    \begin{subfigure}[b]{0.6\textwidth}
      \includegraphics[width=\textwidth]{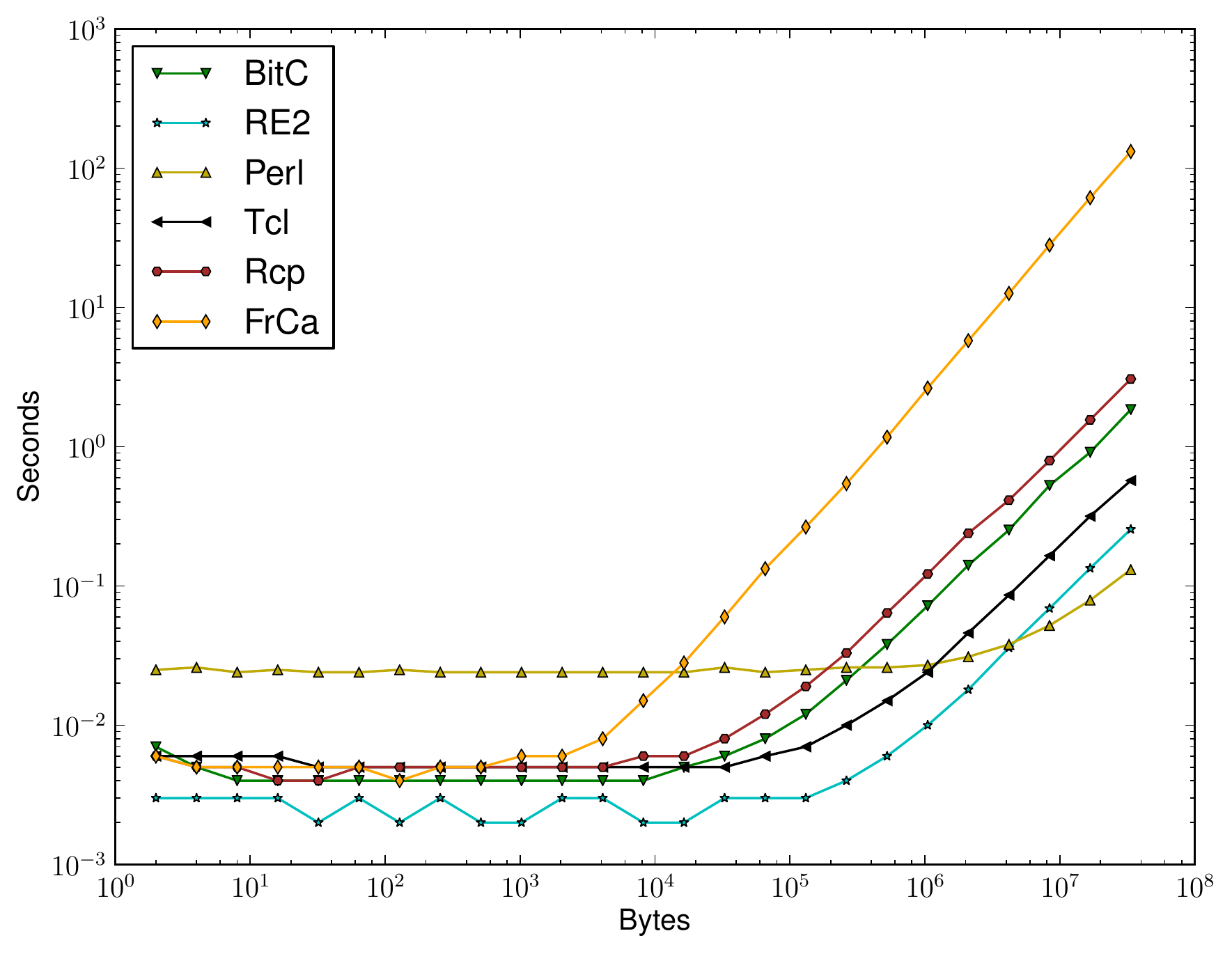}
      \caption{\re{\Lit{a}^\star}, input $\Lit{a}^n$.\label{fig:starrun}}
    \end{subfigure}
    \begin{subfigure}[b]{0.6\textwidth}
      \includegraphics[width=\textwidth]{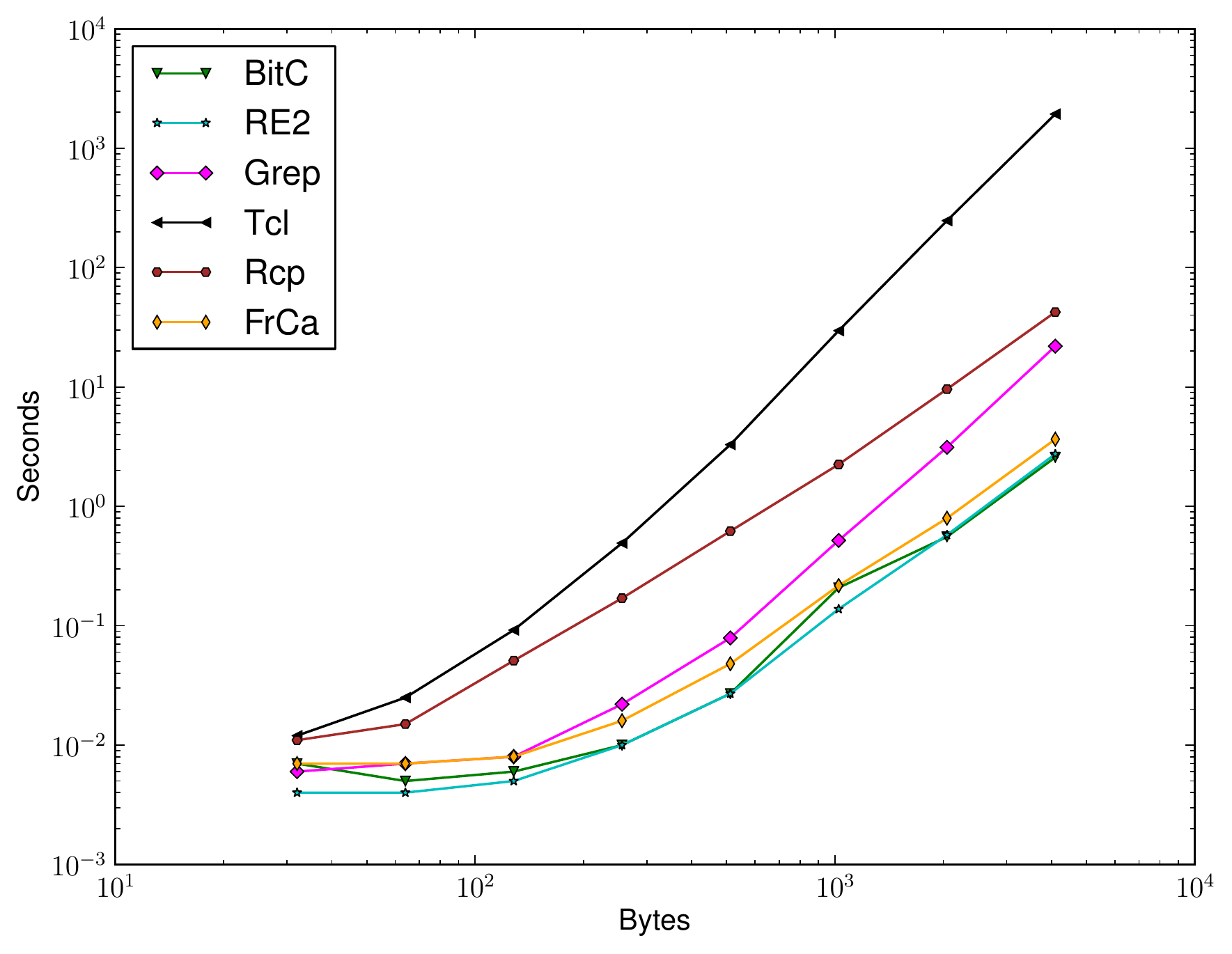}
      \caption{\re{(\Lit{a}|\Lit{b})^\star\Lit{a}(\Lit{a}|\Lit{b})^{n}},
        input $(\Lit{ab})^{n/2}$.\label{fig:dfaworst_var}}
    \end{subfigure}
    \begin{subfigure}[b]{0.6\textwidth}
      \includegraphics[width=\textwidth]{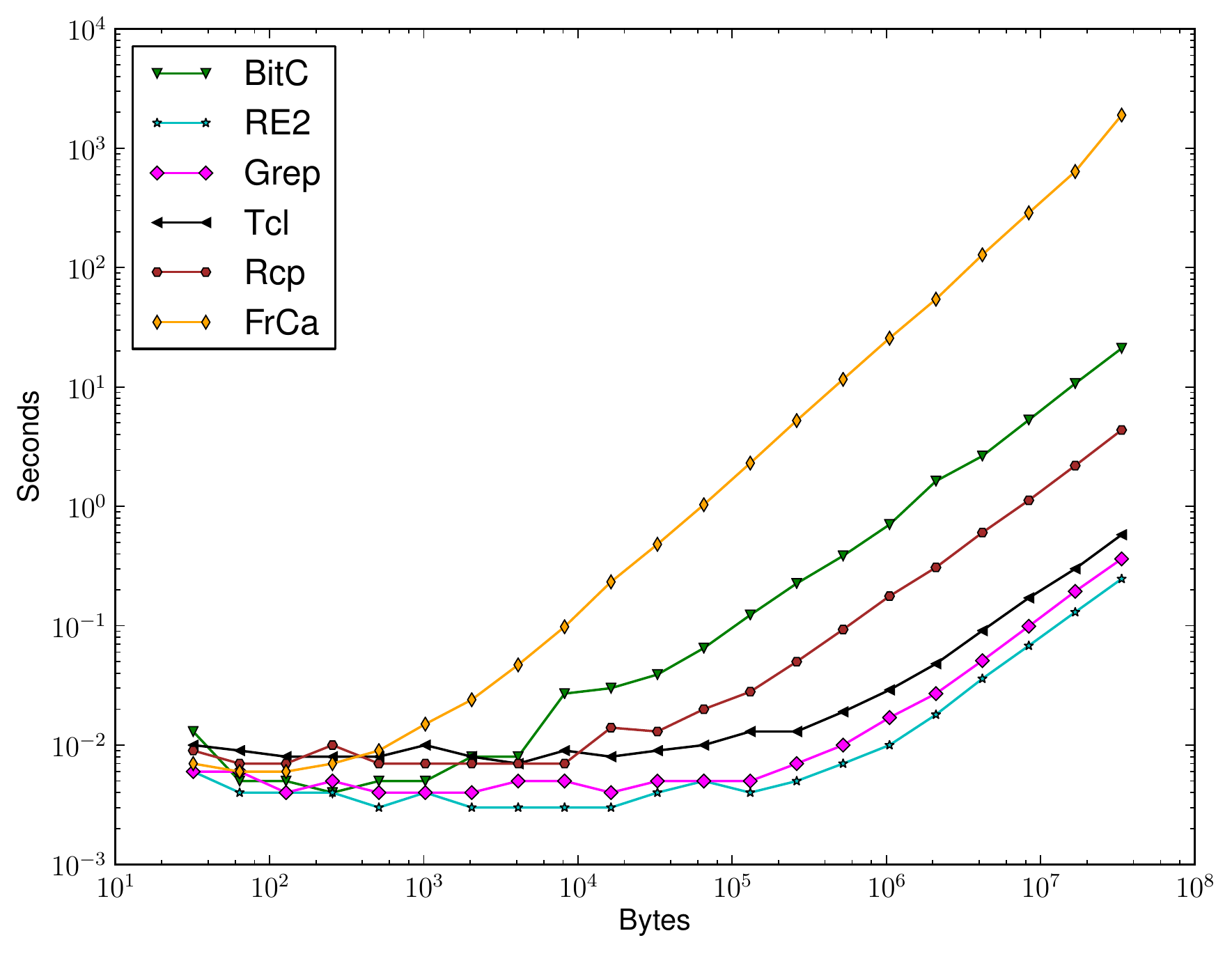}
      \caption{\re{(\Lit{a}|\Lit{b})^\star\Lit{a}(\Lit{a}|\Lit{b})^{25}}, 
        input $(\Lit{ab})^{n/2}$.\label{fig:dfaworst25}}
    \end{subfigure}
    \caption{Comparisons using very simple iteration expressions.}
    \label{fig:runs}
\end{figure}

\begin{figure}[h!]
    \centering
    \begin{subfigure}[b]{0.6\textwidth}
      \includegraphics[width=\textwidth]{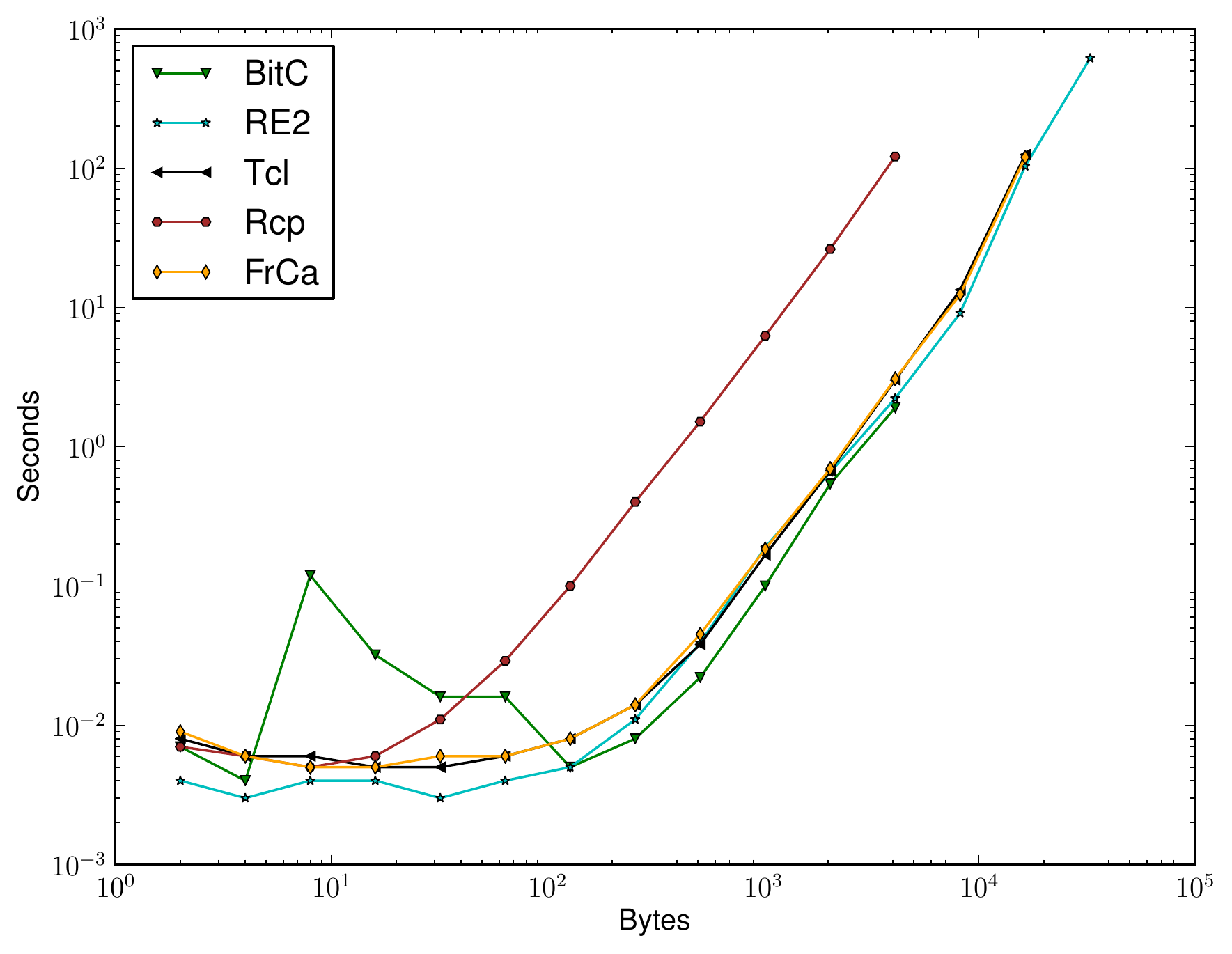}
        \caption{\re{(\Lit{a}?)^n\Lit{a}^n}, input $\Lit{a}^n$.\label{fig:questions}}
    \end{subfigure}
    \begin{subfigure}[b]{0.6\textwidth}
      \includegraphics[width=\textwidth]{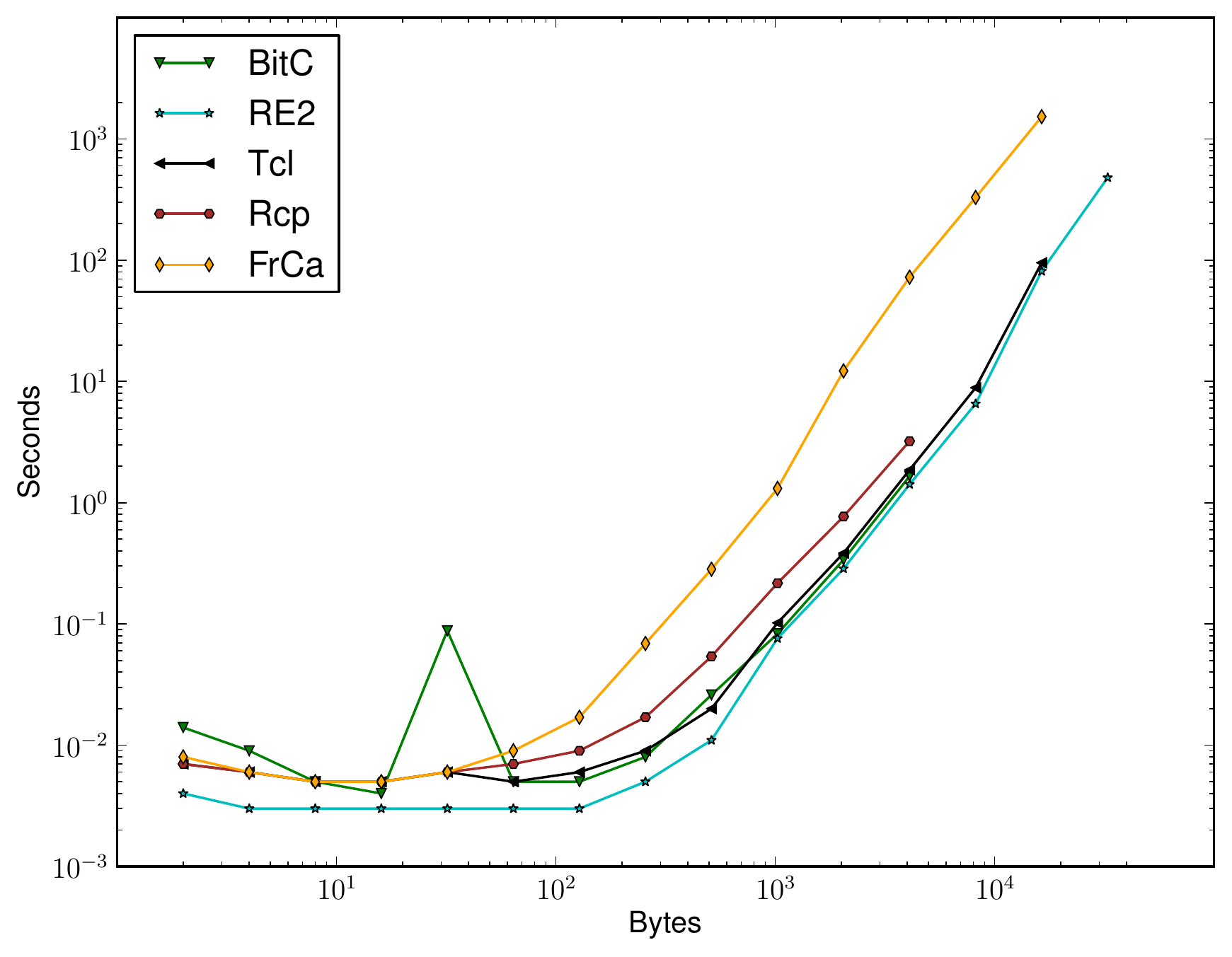}
      \caption{\re{\Lit{a}^n(\Lit{a}?)^n}, input $\Lit{a}^n$.\label{fig:rev_questions}}
    \end{subfigure}
    \caption{Comparison using a backtracking worst case expression, and its reversal.}
\end{figure}

\begin{figure}[h!]
    \centering
    \begin{subfigure}[b]{0.6\textwidth}
      \includegraphics[width=\textwidth]{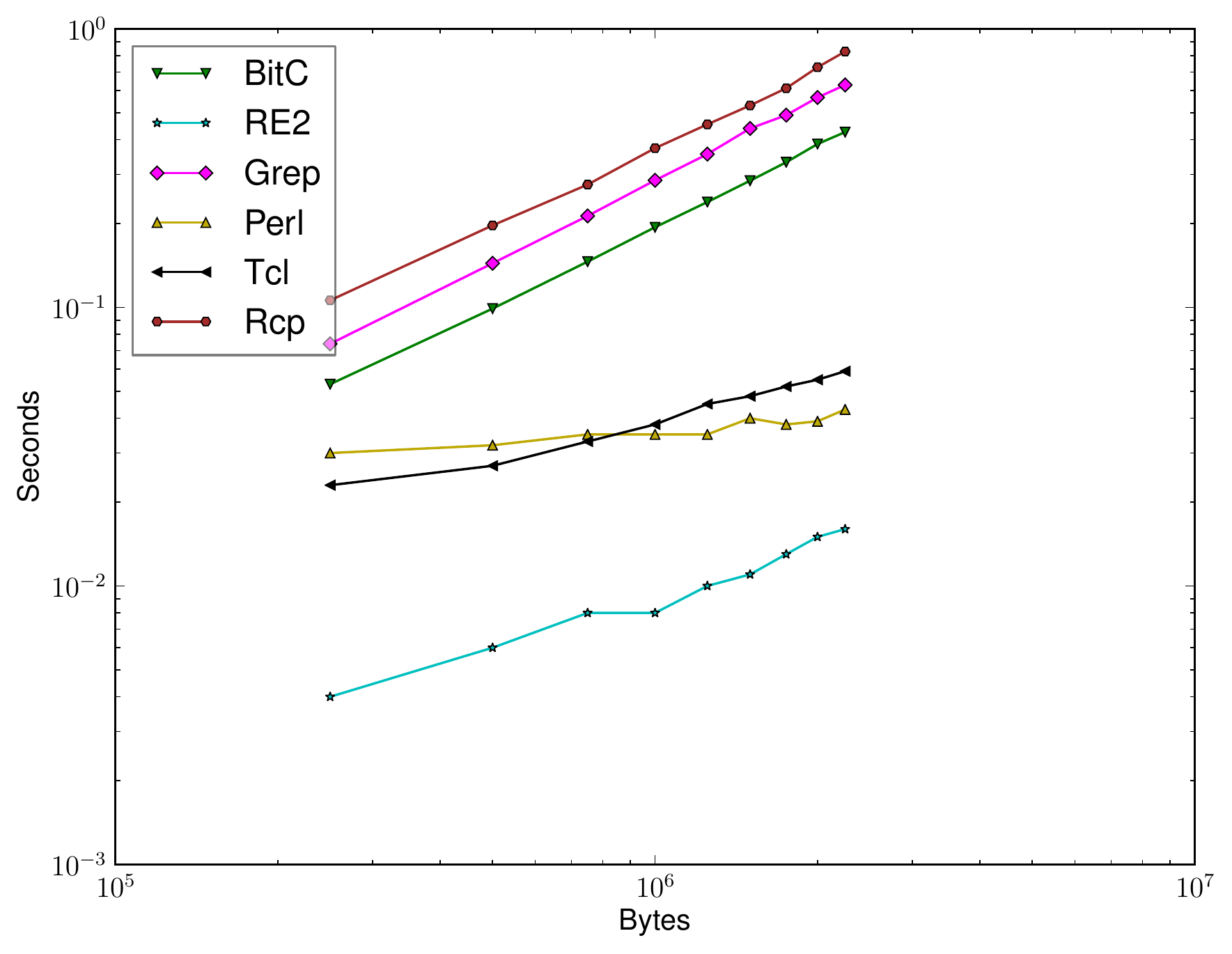}
      \caption{\#4\label{fig:email2}}
    \end{subfigure}
    \begin{subfigure}[b]{0.6\textwidth}
      \includegraphics[width=\textwidth]{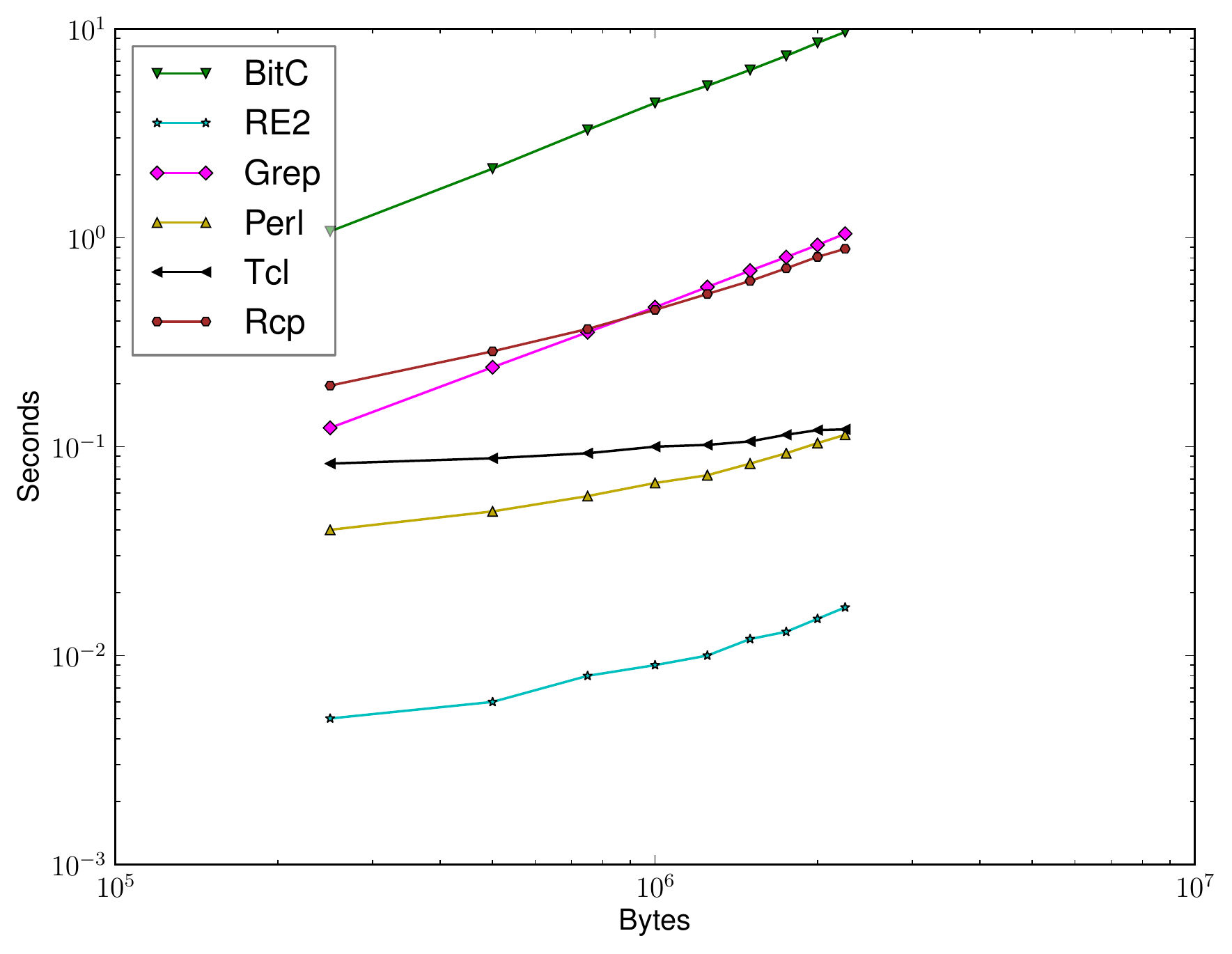}
      \caption{\#7\label{fig:email4}}
    \end{subfigure}
    \begin{subfigure}[b]{0.6\textwidth}
      \includegraphics[width=\textwidth]{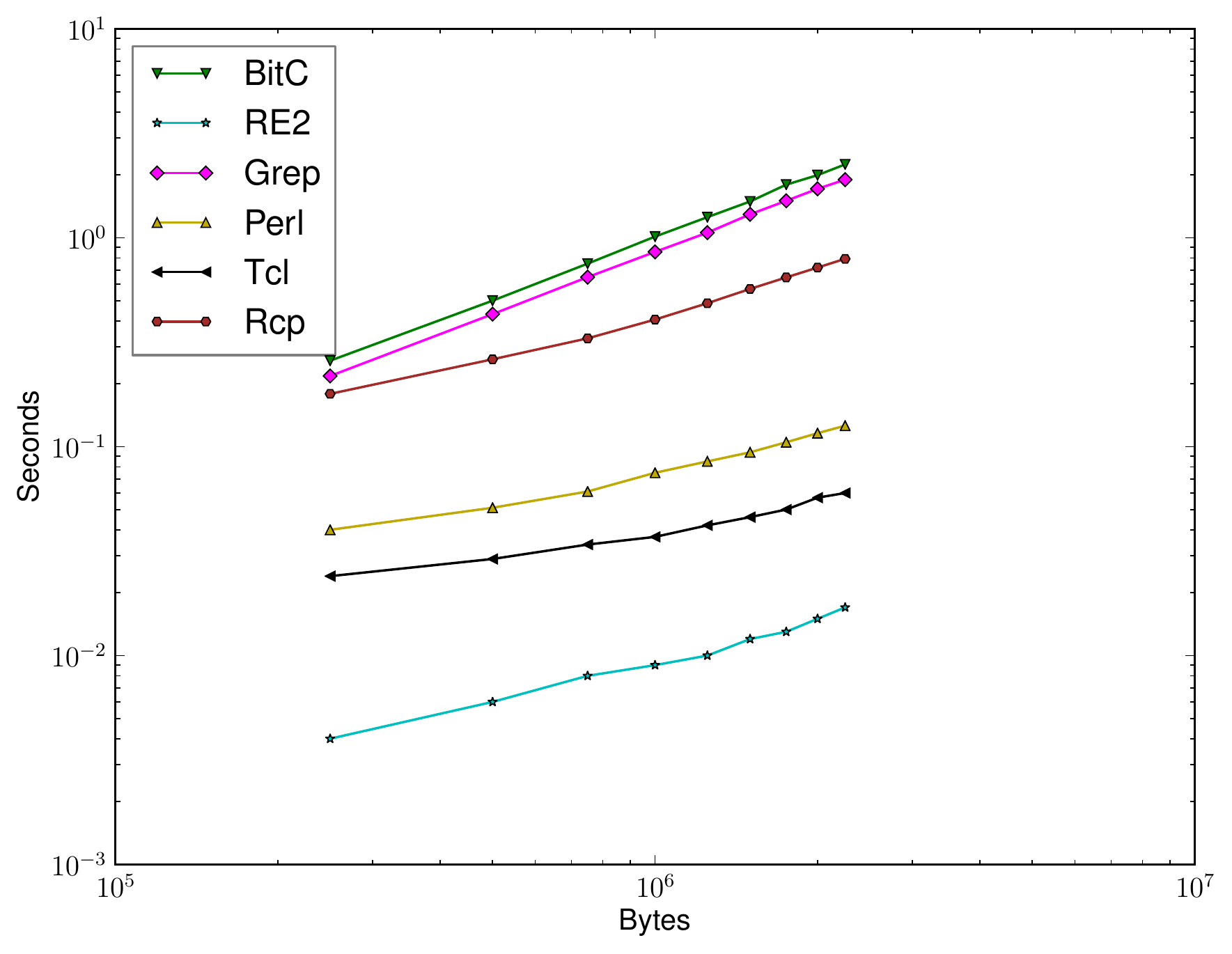}
      \caption{\#8\label{fig:email5}}
    \end{subfigure}
    \caption{Comparison using various e-mail expressions.}
\end{figure}

\subsection{Pathological Expressions}

To get an indication of the ``raw'' throughput for each tool,
\re{\Lit{a}^\star} was run on sequences of $\Lit{a}$s
(Figure~\ref{fig:starrun}).  (Note that the plots use log scales on both
axes, so as to accommodate the dramatically varying running times.)
Perl outperforms the rest, likely due to a strategy where it falls
back on a simple scan of the input.  FrCa stores each position
in the input string from which a match can be made, which in this case
is every position.  As a result, FrCa uses significantly more memory
than the rest, causing a dramatic slowdown.

The expression
\re{(\Lit{a}|\Lit{b})^\star\Lit{a}(\Lit{a}|\Lit{b})^{n}} with the
input $(\Lit{ab})^{n/2}$ is a worst-case for DFA-based methods, as it
results in a number of states exponential in $n$.  Perl has been
omitted from the plots, as it was prohibitively slow.  Tcl, Rcp, and Grep
all perform orders of magnitude slower than FrCa, RE2, and BitC (Figure~\ref{fig:dfaworst_var}),
indicating that Tcl and Grep also use a DFA for this expression.  
If we fix $n$ to $25$, it becomes clear that FrCa is slower than the rest,
likely due to high memory consumption as a result of its storing all
positions in the input string (Figure~\ref{fig:dfaworst25}).  The
asymptotic running times of the others appear to be similar to each other, but
with greatly varying constants.

For the backtracking worst-case expression \re{(\Lit{a}?)^n\Lit{a}^n}
in Figure~\ref{fig:questions}, BitC performs roughly like
RE2.\footnote{The expression parser in BitC failed for the largest
  expressions, which is why they are not on the plot.}  In contrast to
Rcp and FrCa, which are both highly sensitive to the \emph{direction}
of non-determinism, BitC has the same performance for both
\re{(\Lit{a}?)^n\Lit{a}^n} and \re{\Lit{a}^n(\Lit{a}?)^n}
(Figure~\ref{fig:rev_questions}).

\subsection{Practical Examples}

We have run the comparisons with various ``real-life'' examples of
REs taken from~\cite{vht2010}, all of which deal with
expressions matching e-mail addresses. In Figure~\ref{fig:email4}, BitC
is significantly slower than in the other examples.  This can likely
be ascribed to heavy use of bounded repetitions in this expression, 
as they are currently just rewritten into concatenations and alternations in our implementation.

%FH: Isn't this the case for the other tools, too?

%A more efficient implementation would use a
%dedicated construction for repetitions, e.g.\ by adding a simple
%counter instead of $n$ states for an expression like \re{\Lit{a}^n}.

In the other two cases, BitC's performance is roughly like that of
Grep.  This is promising for BitC since Grep performs only RE \emph{matching}, not full \emph{parsing}.  RE2 is consistently ranked
as the fastest program in our benchmarks, presumably due to its aggressive
optimizations and ability to dynamically choose between several
strategies.  Recall that RE2 performs greedy leftmost subgroup matching, not full parsing.  Our present prototype of BitC is coded in less than 1000 lines of C.  It uses only standard libraries and performs no optimizations such as
NFA-minimization, DFA-construction, cached or parallel NFA-simulation,
etc.  This is future work.

\section{Related work}

The known RE parsing algorithms can be divided into four categories.
The first category is 
Perl-style backtracking used in many tools and libraries for RE subgroup matching~\cite{coxbenchmark};
it has an exponential worst case running time, but always produces the
greedy parse and enables some extensions to REs such as
backreferences.  
Another category consists of context-free parsing methods, where the RE is
first translated to a context-free grammar, before a general context-free
parsing algorithm such as Earley's~\cite{earley1970efficient} using cubic time is
applied.  An interesting CFG method is derivatives-based parsing~\cite{might2011parsing}.  While efficient parsers exist for subsets of  unambiguous
context-free languages, this restriction propagates to REs, and
thus these parsers can only be applied for subsets of unambiguous REs.
The third category contains RE scalable parsing algorithms that do not always produce the greedy
parse.  This includes NFA and DFA based algorithms provided by Dub\'{e} and
Feeley~\cite{dufe2000} and Nielsen and
Henglein~\cite{nihe2011}, where the RE is first converted to an NFA with
additional information used to parse strings or to create a DFA preserving the additional information for parsing. 
%While the method presented in this paper is
%uses a Thompson-style NFA generation in order enable the symmetry that
%enables the efficient greedy parsing, these algorithms do not necessarily
%require the same properties, so other techniques such as $\varepsilon$-free
%automata could in theory be applied. 
This category also includes the algorithm by Fischer, Huch and
Wilke~\cite{fihuwi2010}; it is left out of our tests since its Haskell-based implementation often turned out not to be competitive with the other tools.
%As a side note, the algorithm presented by Fischer, Huch, and
%Wilke~\cite{fihuwi2010}, is an elegant version of performing an NFA-simulation
%on a Glushkov automata, and thus extending this method to produce parse trees
%would place it in this category.
The last category consists of the algorithms that scale well and always produce 
greedy parse trees.  Kearns~\cite{kearns90} and Frisch and
Cardelli~\cite{frca2004} reverse the input; perform backwards NFA-simulation, building 
a log of NFA-states reached at each input position; and construct 
the greedy parse tree in a final forward pass over the input.
They require storing the input symbol plus $m$ bits per input symbol for the log.  This can be optimized to storing bits proportional to the number of NFA-states reached at a given input position~\cite{nihe2011}, although the worst case remains the same.  Our lean log algorithm uses only 2 passes,
does not require storing the input symbols and stores only $k < \frac{1}{3} m$ bits per input symbol in the string.

}

\putbib[bibliography]

\end{bibunit}

%%% Local Variables:
%%% mode: latex
%%% TeX-master: "thesis"
%%% End:

\begin{bibunit}[abbrv]
\chapter[Optimally Streaming Greedy Regular Expression Parsing][Optimally Streaming Parsing]{Optimally Streaming Greedy Regular Expression Parsing}
\label{paper:optimal-streaming}

This paper has been published in the following:

\begin{center}
\begin{minipage}{0.9\linewidth}
  \small
  Niels Bjørn Bugge Grathwohl, Fritz Henglein and Ulrik Terp Rasmussen. ``Optimally Streaming Greedy Regular Expression Parsing''. In \emph{Proceedings 11th International Colloquium on Theoretical Aspects of Computing (ICTAC)}. Pages 224-240. Springer, 2014. DOI: \href{http://dx.doi.org/10.1007/978-3-319-10882-7_14}{10.1007/978-3-319-10882-7\_14}.
\end{minipage}
\end{center}
\noindent
The enclosed version has been reformatted to fit the layout of this dissertation.

\clearpage

\thispagestyle{plain}
\begin{center}
{\LARGE \textbf{Optimally Streaming Greedy Regular Expression Parsing}\footnote{The order of authors is insignificant.}}

\vspace{1.5em}

{Niels Bjørn Bugge Grathwohl}, {Fritz Henglein} and {Ulrik Terp Rasmussen}

\vspace{1em}

{Department of Computer Science, University of Copenhagen (DIKU)}
\end{center}

\begin{abstract}
We study the problem of \emph{streaming} regular expression parsing: Given a regular expression and an input stream of symbols, how to output a serialized syntax tree representation as an output stream \emph{during}
input stream processing.

We show that \emph{optimally streaming} regular expression parsing, outputting bits of the output as early as is semantically possible for any regular expression of size $m$ and any input string of length $n$, can be performed in time $O(2^{m \log m} + m n)$ on a unit-cost random-access machine.
This is for the wide-spread \emph{greedy} disambiguation strategy for choosing parse trees of grammatically ambiguous regular expressions.
In particular, for a fixed regular expression, the algorithm's run-time scales linearly with the input string length.
The exponential is due to the need for preprocessing the regular expression to analyze state coverage of its associated NFA, a PSPACE-hard problem, and tabulating all reachable \emph{ordered} sets of NFA-states.

Previous regular expression parsing algorithms operate in multiple phases, \emph{always} requiring processing or storing the whole input string before outputting the first bit of output, not only for those regular expressions and input prefixes where reading to the end of the input is strictly \emph{necessary}.
\end{abstract}

{

% \parse -> \parsef
% \pref -> \prefer

\newcommand{\set}[1]{\ensuremath{\left\{ #1 \right\}}}
\newcommand{\dom}{\mathrm{dom}}

% Annotate bit codes to make them readable.
\newcommand{\annot}[2]{\ensuremath{\underset{#2}{\lit{#1}}}}

\newcommand{\ignoreResult}{\underline{\hspace{1em}}}

% Tree stuff
\newcommand{\treeroot}[1]{\ensuremath{\mathsf{root}(#1)}}
\newcommand{\treechildren}[1]{\ensuremath{\mathsf{children}(#1)}}
\newcommand{\treeparent}[1]{\ensuremath{\mathsf{parent}(#1)}}
\newcommand{\treelabel}[2]{\ensuremath{\mathsf{path}(#1,#2)}}
\newcommand{\treeleaves}[1]{\ensuremath{\mathsf{leaves}(#1)}}
\newcommand{\treeinit}[1]{\ensuremath{\mathsf{init}(#1)}}
\newcommand{\treeempty}{\ensuremath{\mathsf{Tr_{empty}}}}
\newcommand{\reverse}[1]{\ensuremath{\mathsf{reverse}(#1)}}
\newcommand{\treebinarynode}[3]{\ensuremath{#1\langle\bit{0}:#2, \bit{1}:#3\rangle}}
\newcommand{\treeunarynode}[3]{\ensuremath{#1\langle#2:#3\rangle}}
\newcommand{\treenilnode}[1]{\ensuremath{#1\langle\cdot\rangle}}

% Symbols used in path tree examples.
\newcommand{\deathbycoverage}{\lightning}

% literals
\newcommand{\lit}[1]{\ensuremath{\mathtt{#1}}}
% duality operator
\newcommand{\op}[1]{\ensuremath{\overline{#1}}}
\newcommand{\ope}[2]{\ensuremath{\mathsf{op}_{#1}({#2})}}
% complexity classes
\newcommand{\PSPACE}{\mathsf{PSPACE}}
\newcommand{\NPSPACE}{\mathsf{NPSPACE}}
% power set
\newcommand{\pow}[1]{\mathbf{2}^{#1}}

%%% Regular expressions: syntax macros

% symbol constants
\newcommand{\esym}[1]{\ensuremath{#1}}
\newcommand{\elit}[1]{\ensuremath{\lit{#1}}}
% 01-constants
\newcommand{\eone}{\ensuremath{\mathbf{1}}}
\newcommand{\ezero}{\ensuremath{\mathbf{0}}}
% concatenation
\newcommand{\etimes}[2]{\ensuremath{{#1}{#2}}}
% union
\newcommand{\eplus}[2]{\ensuremath{{#1} + {#2}}}
% Kleene star
\newcommand{\estar}[1]{\ensuremath{#1^\star}}
% Set of RE terms
\newcommand{\Exp}{\ensuremath{\mathsf{Exp}}}

%%% Regular expressions: denotations

% language
\newcommand{\lang}[1]{\ensuremath{\mathcal{L}\llbracket {#1} \rrbracket}}
% type
\newcommand{\type}[1]{\ensuremath{\mathcal{T}\llbracket {#1} \rrbracket}}
\newcommand{\typew}[2]{\ensuremath{\mathcal{T}_{#1}\llbracket {#2} \rrbracket}}
\newcommand{\typewnp}[2]{\ensuremath{\mathcal{T}^{\mathsf{np}}_{#1}\llbracket {#2} \rrbracket}}
% bitcode
\newcommand{\bitcode}[1]{\ensuremath{\mathcal{B}\llbracket {#1} \rrbracket}}
\newcommand{\bitcodew}[2]{\ensuremath{\mathcal{B}_{#1}\llbracket {#2} \rrbracket}}
\newcommand{\bitcodewnp}[2]{\ensuremath{\mathcal{B}^{\mathsf{np}}_{#1}\llbracket {#2} \rrbracket}}

%%% Values / parse trees: syntax
\newcommand{\vunit}{\ensuremath{()}}
\newcommand{\vsym}[1]{\ensuremath{#1}}
\newcommand{\vlit}[1]{\ensuremath{\vsym{\lit{#1}}}}
\newcommand{\vinl}[1]{\ensuremath{\mathsf{inl}~{#1}}}
\newcommand{\vinr}[1]{\ensuremath{\mathsf{inr}~{#1}}}
\newcommand{\vprod}[2]{\ensuremath{\langle #1,#2 \rangle}}
\newcommand{\vlist}[1]{\ensuremath{[#1]}}

%%% Values / parse trees: denotations / operators
\newcommand{\vflat}[1]{\ensuremath{|#1|}}
\newcommand{\vcode}[1]{\ensuremath{\ulcorner {#1} \urcorner}}
\newcommand{\vdecode}[1]{\ensuremath{\llcorner {#1} \lrcorner}}
% greedy order on values
\newcommand{\vgreedy}{\lessdot}
\newcommand{\vgreedyg}{\gtrdot}

%%% Bit codes
\newcommand{\bit}[1]{\ensuremath{\mathsf{#1}}}
% log bit
\newcommand{\lbit}[1]{\ensuremath{\overline{\mathsf{#1}}}}
% lexicographic order
\newcommand{\blex}{<_{\mathsf{lex}}}

%%% Positions
\newcommand{\psub}[2]{{#1}|_{#2}}
% empty position (epsilon for now)
\newcommand{\pemp}{\epsilon}

%%% States
\newcommand{\St}{\mathsf{State}}
\newcommand{\DSt}{\mathsf{DState}}
\newcommand{\qin}%{\circ}
                 {\mathsf{in}}
\newcommand{\qout}%{\bullet}
                  {\mathsf{out}}
\newcommand{\qint}%{\dagger}
                  {\mathsf{loop}}
\newcommand{\qinit}{q^{\mathsf{in}}}
\newcommand{\qfinal}{q^{\mathsf{fin}}}

%%% Paths
\newcommand{\plabels}[1]{\mathsf{lab}(#1)}
\newcommand{\pread}[1]{\mathsf{read}(#1)}
\newcommand{\pwrite}[1]{\mathsf{write}(#1)}
% path concatenation / sequencing
\newcommand{\pseq}{\mathrel{;}}
% infix path shorthand notation
\newcommand{\ptrans}[1]{\stackrel{#1}{\rightsquigarrow}}
% single step
\newcommand{\trans}[1]{\stackrel{#1}{\rightarrow}}
% closure operation
\newcommand{\closure}[2][]{\mathsf{closure}_{#1}(#2)}

%% Parsing function
\newcommand{\parsef}[2]{\mathsf{P}_{#1}(#2)}
\newcommand{\fail}{\sharp}

%%% Algorithms
\newcommand{\Accept}{\ensuremath{\mathbf{accept}}}
\newcommand{\Reject}{\ensuremath{\mathbf{reject}}}

%%% Coverage
\newcommand{\covers}{\sqsupseteq}
\newcommand{\cocovers}{\sqsupseteq^\mathsf{op}}

%% Relational symbols
\newcommand{\prefer}{\sqsubseteq}
\newcommand{\ext}{\sqsupseteq}

\tikzstyle{split} = []
\tikzstyle{dummy} = [ inner sep=2pt
                    , fill=white
                    , draw=none
                    ]
\tikzstyle{st} = [ inner sep=3pt
                 ]
\tikzstyle{act} = [ font = \bfseries ]
\tikzstyle{state} = [ draw
                    , circle
                    , inner sep=3pt
                    , minimum size=18pt
                    ]
\tikzstyle{label} = [ font=\ttfamily
                    , inner sep=1pt
                    ]
\tikzstyle{nfa} = [ every node/.style={state}
                  , auto
                  , >=latex
                  ]
\tikzstyle{transitions} = [->, every node/.style={label}]

\section{Introduction}
\label{sec:introduction}

In programming, regular expressions are often used to extract information from an input, which requires an intensional interpretation of regular expressions as denoting parse trees, and not just their ordinary language-theoretic interpretation as denoting strings.

This is a nontrivial change of perspective.
We need to deal with grammatical ambiguity---\emph{which} parse tree to return, not just that it has one---and memory requirements become a critical factor: Deciding whether a string belongs to the language denoted by
$\eplus{\estar{(\lit{ab})}}{\estar{(\eplus{\elit{a}}{\elit{b}})}}$ can be done in constant space, but outputting the first
bit, whether the string matches the first alternative or only the second, may require buffering the whole input string.
This is an instructive case of deliberate grammatical ambiguity to be resolved by the prefer-the-left-alternative policy of greedy disambiguation: Try to match the left alternative; if that fails, return a match according to the right alternative as a fallback.
Straight-forward application of automata-theoretic techniques does not help: $\eplus{\estar{(\lit{ab})}}{\estar{(\eplus{\elit{a}}{\elit{b}})}}$ denotes the same \emph{language} as $\estar{(\eplus{\elit{a}}{\elit{b}})}$, which is unambiguous and corresponds to a small DFA, but is also useless: it doesn't represent any more when a string consists of a sequence of $ab$-groups.

Previous parsing algorithms
\cite{kearns91,dufe2000,frisch2004,nihe2011,sulzmann2012,grathwohl2013two}
require at least one full pass over the input string before outputting any output bits representing the parse tree.
This is the case even for regular expressions requiring only bounded lookahead such as one-unambiguous regular expressions \cite{brwo98}.

In this paper we study the problem of \emph{optimally streaming} parsing.
Consider \[ \eplus{\estar{(\lit{ab})}}{\estar{(\eplus{\elit{a}}{\elit{b}})}}, \] which is ambiguous and in general requires unbounded input buffering, and consider the particular input string \[ \elit{ab} \ldots \elit{ab aa babababab}\ldots. \]
An \emph{optimally} streaming parsing algorithm needs to buffer the prefix $\elit{ab} \ldots \elit{ab}$ in some form because the complete parse might match either of the two alternatives in the regular expression, but once encountering $\elit{aa}$, only the right alternative is possible.
At this point it outputs this information and the output representation for the buffered string as parsed by the second alternative.
After this, it outputs a bit for each input symbol read, with no internal buffering: input symbols are discarded before reading the next symbol.
Optimality means that output bits representing the eventual parse tree must be produced \emph{earliest possible}: as soon as they are semantically determined by the input processed so far under the assumption that the parse will succeed.

\paragraph{Outline.}
In Section~\ref{sec:preliminaries} we recall the \emph{type interpretation} of regular expressions, where a regular expression denotes parse trees, along with the \emph{bit-coding} of parse trees.

In Section~\ref{sec:augmented-automata} we introduce a class of Thompson-style augmented nondeterministic finite automata (aNFAs).
Paths in such an aNFA naturally represent \emph{complete} parse trees, and paths to intermediate states represent \emph{partial} parse trees for prefixes of an input string.

We recall the greedy disambiguation strategy in Section~\ref{sec:disambiguation}, which specifies a deterministic mapping of accepted strings to NFA-paths.

Section~\ref{sec:optimal-streaming} contains a definition of what it means to be an optimally streaming implementation of a parsing function.

We define what it means for a set of aNFA-states to \emph{cover} another state in Section~\ref{sec:coverage}, which constitutes the computationally hardest part needed in our algorithm.

Section~\ref{sec:algorithm} contains the main results.
We present \emph{path trees} as a way of organizing partial parse trees, and based on these we present our algorithm for an optimally streaming parsing function and analyze its asymptotic run-time complexity.

Finally, in Section~\ref{sec:example}, the algorithm is demonstrated by illustrative examples alluding to its expressive power and practical utility.

\section{Preliminaries}
\label{sec:preliminaries}

In the following section, we recall definitions of regular expressions and their interpretation as types \cite{nihe2011}.
\begin{definition}[Regular expression]
A regular expression (RE) over a finite alphabet $\Sigma$ is an expression $E$ generated by the grammar
\[
  E ::= \ezero \mid \eone \mid \esym{a}
    \mid \etimes{E_1}{E_2} \mid \eplus{E_1}{E_2} \mid \estar{E_1}
\]
where $a \in \Sigma$.
\end{definition}
Concatenation (juxtaposition) and alternation ($+$) associates to the right; parentheses may be inserted to override associativity.
Kleene star ($\star$) binds tightest, followed by concatenation and alternation.

The standard interpretation of regular expressions is as descriptions of regular languages.

\begin{definition}[Language interpretation]
Every RE $E$ denotes a language $\lang{E} \subseteq \Sigma^\star$ given as follows:
\begin{align*}
  \lang{\ezero} ={}& \emptyset
  &
  \lang{\etimes{E_1}{E_2}} ={}& \lang{E_1} \lang{E_2}
  &
  \lang{\esym{a}} ={}& \{ a \}
  \\
  \lang{\eone} ={}& \{ \epsilon \}
  &
  \lang{\eplus{E_1}{E_2}} ={}& \lang{E_1} \cup \lang{E_2}
  &
  \lang{\estar{E_1}} ={}& \bigcup_{n\geq 0} \lang{E_1}^n
\end{align*}
where we have $A_1 A_2 = \{w_1 w_2 \mid w_1 \in A_1, w_2 \in A_2\}$, and $A^0 = \{\epsilon\}$ and $A^{n+1} = AA^{n}$.
\end{definition}
\noindent
\textbf{Proviso:} Henceforth we shall restrict ourselves to REs $E$ such that $\lang{E}\neq\emptyset$.

For regular expression parsing, we consider an alternative interpretation of regular expressions as types.
\begin{definition}[Type interpretation]
Let the syntax of \emph{values} be given by
\[
v ::= \vunit
  \mid \vinl{v_1}
  \mid \vinr{v_1}
  \mid \vprod{v_1}{v_2}
  \mid \vlist{v_1, v_2, ..., v_n}
\]
Every RE $E$ can be seen as a \emph{type} describing a set $\type{E}$ of well-typed values:
\begin{align*}
  \type{\ezero} ={}& \emptyset
  &
  \type{\etimes{E_1}{E_2}} ={}& \{ \vprod{v_1}{v_2}
                                   \mid v_1 \in \type{E_1}, v_2 \in \type{E_2} \}
  \\
  \type{\eone} ={}& \{ \vunit \}
  &
  \type{\eplus{E_1}{E_2}} ={}& \{ \vinl{v} \mid v \in \type{E_1} \}
                               \cup \{ \vinr{v} \mid v \in \type{E_2} \}
  \\
  \type{\esym{a}} ={}& \{ \vsym{a} \}
  &
  \type{\estar{E_1}} ={}& \{ \vlist{v_1, \hdots, v_n} \mid n \geq 0
     \land \forall 1 \leq i \leq n. v_i \in \type{E_1} \}
\end{align*}
\end{definition}
We write $\vflat{v}$ for the \emph{flattening} of a value, defined as the word obtained by doing an in-order traversal of $v$ and writing down all the symbols in the order they are visited.
We write $\typew{w}{E}$ for the restricted set $\{ v \in \type{E} \mid \vflat{v} = w \}$.
Regular expression \emph{parsing} is a generalization of the acceptance problem of determining whether a word $w$ belongs to the language of some RE $E$, where additionally we produce a parse tree from $\typew{w}{E}$.
We say that an RE $E$ is \emph{ambiguous} iff there exists a $w$ such that $|\typew{w}{E}| > 1$.

Any well-typed value can be serialized into a sequence of bits.
\begin{definition}[Bit-coding]
Given a value $v \in \type{E}$, we denote its \emph{bit-code} by $\vcode{v} \subseteq \{\bit{0}, \bit{1}\}^\star$, defined as follows:
\begin{align*}
   \vcode{\vunit} ={}& \epsilon
&  \vcode{\vsym{a}} ={}& \epsilon
&  \vcode{\vinl{v}} ={}& \bit{0} \, \vcode{v} \\
   \vcode{\vprod{v_1}{v_2}} ={}& \vcode{v_1} \, \vcode{v_2}
&  \vcode{\vlist{v_1, ..., v_n}} ={}& \bit{0} \, \vcode{v_1}
                                      \, ... \,
                                      \bit{0} \, \vcode{v_n} \, \bit{1}
&  \vcode{\vinr{v}} ={}& \bit{1} \, \vcode{v}
\end{align*}
\end{definition}
We write $\bitcode{E}$ for the set $\{ \vcode{v} \mid v \in \type{E} \}$ and $\bitcodew{w}{E}$ for the set restricted to bit-codes for values with a flattening $w$.
Note that for any RE $E$, bit-coding is an isomorphism when seen as a function $\vcode{\cdot}_E : \type{E} \rightarrow \bitcode{E}$.%, and we will write $\vdecode{\cdot}_E$ for its inverse.

\section{Augmented Automata}
\label{sec:augmented-automata}

In this section we recall from an earlier paper~\cite{grathwohl2013two} the construction of finite automata from regular expressions.
Our construction is similar to that of Thompson~\cite{thompson68}, but augmented with extra annotations on non-deterministic $\epsilon$-transitions.
The resulting state machines can be seen as non-deterministic transducers which for each accepted input string in the language of the underlying regular expression outputs the bit-codes for the corresponding parse trees.

\begin{definition}[Augmented non-deterministic finite automaton]
An \emph{augmented non-deterministic finite automaton} (aNFA) is a tuple $(\St, \delta, \qinit, \qfinal)$, where $\St$ is a finite set of \emph{states}, $\qinit, \qfinal \in \St$ are \emph{initial} and \emph{final} states, respectively, and $\delta \subseteq \St \times \Gamma \times \St$ is a labeled transition relation with labels $\Gamma = \Sigma \uplus \{\bit{0}, \bit{1}, \epsilon \}$.
\end{definition}
Transition labels are divided into the disjoint sets $\Sigma$ (symbol labels); $\{\bit{0},\bit{1}\}$ (bit-labels); and $\{ \epsilon \}$ ($\epsilon$-labels).
$\Sigma$-transitions can be seen as input actions, and bit-transitions as output actions.

\begin{definition}[aNFA construction]
Let $E$ be an RE and define an aNFA $M_E = (\St_E, \delta_E, \qinit_E, \qfinal_E)$ by induction on $E$. We give the definition diagrammatically by cases:
\begin{center}
\begin{tabular}[t]{| c | c |}
\hline
$E$ & $M_E$
\\\hline\hline
$\ezero$ &
\begin{tikzpicture}[nfa,baseline={([yshift=-0.8ex]current bounding box.west)},node distance=1.5cm]
  \node[dummy] (dummy) {};
  \node[right of=dummy](in) {$\qinit$};
  \node[accepting](out) [right=of in] {$\qfinal$};
  \draw[transitions] (dummy) to (in);
\end{tikzpicture}
\\\hline
$\eone$ &
\begin{tikzpicture}[nfa,baseline={([yshift=-0.8ex]current bounding box.west)},node distance=1.5cm]
  \node[dummy] (dummy) {};
  \node[accepting,right of=dummy](in) {$\qinit$};
  \draw[transitions] (dummy) to (in);
\end{tikzpicture} ($\qinit = \qfinal$)
\\\hline
$\esym{a}$ &
\begin{tikzpicture}[nfa,baseline={([yshift=-0.8ex]current bounding box.west)},node distance=1.5cm]
  \node[dummy] (dummy) {};
  \node[right of=dummy] (in) {$\qinit$};
  \node[accepting] (out) [right=of in] {$\qfinal$};
  \draw[transitions] (in) to node {$a$} (out);
  \draw[transitions] (dummy) to (in);
\end{tikzpicture}
\\\hline
$\etimes{E_1}{E_2}$ &
\begin{tikzpicture}[nfa,baseline={([yshift=-0.8ex]current bounding box.west)},node distance=1.5cm]
  \node[dummy] (dummy) {};
  \node[right of=dummy] (in) {$\qinit$};
  \node (q) [right=of in] {$q'$};
  \node[accepting](out) [right=of q] {$\qfinal$};
  \begin{scope}[transitions]
    \draw[dashed] (in) to node {$M_{E_1}$} (q);
    \draw[dashed] (q) to node {$M_{E_2}$} (out);
    \draw (dummy) to (in);
  \end{scope}
\end{tikzpicture}
\\\hline
$\eplus{E_1}{E_2}$ &
\begin{tikzpicture}[nfa,baseline={([yshift=-0.8ex]current bounding box.west)},node distance=1.5cm]
  \node[dummy] (dummy) {};
  \node[right of=dummy] (in) {$\qinit$};
  \node[above right of=in] (in1) {$q_1$};
  \node[below right of=in] (in2) {$q_2$};
  \node[right of=in1,xshift=0.4cm] (out1) {$q_1'$};
  \node[right of=in2,xshift=0.4cm] (out2) {$q_2'$};
  \node[below right of=out1,accepting] (out) {$\qfinal$};
  \begin{scope}[transitions]
    \draw (in) to node {$\bit{0}$} (in1);
    \draw (in) to node[swap] {$\bit{1}$} (in2);
    \draw (out1) to node {$\epsilon$} (out);
    \draw (out2) to node[swap] {$\epsilon$} (out);
    \draw[dashed] (in1) to node {$M_{E_1}$} (out1);
    \draw[dashed] (in2) to node {$M_{E_2}$} (out2);
    \draw (dummy) to (in);
  \end{scope}
\end{tikzpicture}
\\\hline
$\estar{E_1}$ &
\begin{tikzpicture}[nfa,baseline={([yshift=-0.8ex]current bounding box.west)},node distance=1.5cm]
  \node[dummy] (dummy) {};
  \node[right of=dummy](in) {$\qinit$};
  \node[right of=in](loop) {$q'$};
  \node[accepting,right of=loop](out) {$\qfinal$};
  \node[above of=in,yshift=-0.2cm](in1) {$q_1$};
  \node[above of=out,yshift=-0.2cm] (out1) {$q_1'$};
  \begin{scope}[transitions]
    \draw[dashed] (in1) to node {$M_{E_1}$} (out1);
    \draw (in) to node {$\epsilon$} (loop);
    \draw (loop) to node {$\bit{1}$} (out);
    \draw (loop) to node[swap] {$\bit{0}$} (in1);
    \draw (out1) to node[swap] {$\epsilon$} (loop);
    \draw (dummy) to (in);
  \end{scope}
\end{tikzpicture}
\\\hline
\end{tabular}
\end{center}
In the above, the notation \tikz[nfa,baseline=-0.5ex]{\node(q1){$q_1$};\node[right=of q1](q2) {$q_2$};\draw[transitions,dashed] (q1) to node {$M$} (q2);} means that $q_1, q_2$ are initial and final states, respectively, in some (sub-)automaton $M$.
\end{definition}
See Figure~\ref{fig:nfa_ex} for an example.
\begin{figure}[t]
\begin{center}
\begin{tikzpicture}[nfa,node distance=1.5cm]
  \node[dummy] (dummy) {};
  \node(pi)[right of=dummy] {$1$};
  \node(p1l)[right of=pi] {$2$};
  \node(p11i)[above left of=p1l,xshift=-1.1cm,yshift=1.25cm] {$3$};
  \node(p111i)[above right=of p11i,yshift=-0.5cm] {$4$};
  \node(p111o)[right=of p111i] {$5$};
  \node(p112i)[below right=of p11i,yshift=0.5cm] {$7$};
  \node(p112o)[right=of p112i] {$8$};
  \node(p11o)[below right=of p111o,yshift=0.5cm] {$6$};

  \node(p1o)[right=of p1l] {$9$};
  \node(po)[accepting,right=of p1o] {$10$};
  \begin{scope}[transitions]
    \draw (dummy) to (pi);
    \draw (pi) to node[swap] {$\epsilon$} (p1l);
    \draw (p1l) to node[swap] {$\bit{1}$} (p1o);
    \draw (p1o) to node[swap] {$\lit{b}$} (po);
    \draw (p1l) [out=145,in=225] to node {$\bit{0}$} (p11i);
    \draw (p11o) [out=315,in=35] to node {$\epsilon$} (p1l);
    \draw (p11i) to node {$\bit{0}$} (p111i);
    \draw (p111i) to node {$\lit{a}$} (p111o);
    \draw (p11i) to node [swap] {$\bit{1}$} (p112i);
    \draw (p112i) to node {$\lit{b}$} (p112o);
    \draw (p111o) to node {$\epsilon$} (p11o);
    \draw (p112o) to node [swap] {$\epsilon$} (p11o);
  \end{scope}
\end{tikzpicture}
\end{center}
  \caption{Example automaton for the RE $\etimes{\estar{(\eplus{\elit{a}}{\elit{b}})}}{\elit{b}}$}
  \label{fig:nfa_ex}
\end{figure}

\begin{definition}[Path]
A \emph{path} in an aNFA is a finite and non-empty sequence $\alpha \in \St^\star$ of the form $\alpha = p_0 \, p_1 \, ... \, p_{n-1}$ such that for each $i < n$, we have $(p_i, \gamma_i, p_{i+1}) \in \delta_E$ for some $\gamma_i$.
As a shorthand for this fact we might write $p_0 \ptrans{\alpha} p_{n-1}$ (note that a single state is a path to itself).
\end{definition}
Each path $\alpha$ is associated with a (possibly empty) sequence of labels $\plabels{\alpha}$: we let $\pread{\alpha}$ and $\pwrite{\alpha}$ refer to the corresponding subsequences of $\plabels{\alpha}$ filtered by $\Sigma$ and $\{\bit{0},\bit{1}\}$, respectively.
An automaton \emph{accepts} a word $w$ iff we have $\qinit \ptrans{\alpha} \qfinal$ for some $\alpha$ where $\pread{\alpha} = w$. There is a one-to-one correspondence between bit-codes and accepting paths:
\begin{proposition}
For any RE $E$ with aNFA $M_E$, we have for each $w \in \lang{E}$ that
\[
  \{ \pwrite{\alpha} \mid \qinit \ptrans{\alpha} \qfinal \wedge \pread{\alpha} = w \} =
  \bitcodew{w}{E}.
\]
\end{proposition}

\paragraph{Determinization.}
Given a state set $Q$, define its \emph{closure} as the set
\[ \closure{Q} = \{ q' \mid  q \in Q
                     \land \exists \alpha. \pread{\alpha} = \epsilon
                     \land q \ptrans{\alpha} q' \}. \]
For any aNFA $M=(\St,\delta,\qinit,\qfinal)$, let $D(M) = (\DSt_M, I_M, F_M, \Delta_M)$ be the deterministic automaton obtained by applying the standard subset sum construction:
Here, $I_M = \closure{\{\qinit\}}$ is the \emph{initial state}, and $\DSt_M \subseteq \pow{\St}$ is the set of states, defined to be the smallest set containing $I_M$ and closed under the transition function: for all $a \in \Sigma$ and $Q \in \DSt_M$, we have $\Delta_M(Q,a) \in \DSt_M$, where
\[
  \Delta_M(Q, a) = \closure{\{ q' \mid (q, a, q') \in \delta, q \in Q \}}.
\]
The set of \emph{final states} is $F_M = \{Q \in \DSt_M \mid \qfinal \in Q \}.$

\section{Disambiguation}
\label{sec:disambiguation}

A regular expression parsing algorithm has to produce a parse tree for an input word whenever the word is in the language for the underlying RE.
In the case of ambiguous REs, the algorithm has to choose one of several candidates.
We do not want the choice to be arbitrary, but rather a parse tree which is uniquely identified by a \emph{disambiguation policy}.
Since there is a one-to-one correspondence between words in the language of an RE $E$ and accepting paths in $M_E$, a disambiguation policy can be seen as a deterministic choice between aNFA paths recognizing the same string.

We will focus on greedy disambiguation, which corresponds to choosing the first result that would have been found by a backtracking regular expression parsing algorithm such as the one found in the Perl programming language~\cite{wall2000}.
%\review{Added reference to previous work}
The greedy strategy has successfully been implemented in previous work \cite{frisch2004,grathwohl2013two}, and is simpler to define and implement than other strategies such as POSIX~\cite{ieee1003.1-2008,fowlerPOSIX} whose known parsing algorithms are technically more complicated \cite{okui2011,sulzmann2012,sulzmann2014}.

Greedy disambiguation can be seen as picking the accepting path with the lexicographically least bitcode. A well-known problem with backtracking parsing is non-termination in the case of regular expressions with nullable subexpressions under Kleene star, which means that the lexicographically least path is not always well-defined. This problem can easily be solved by not considering paths with non-productive loops, as in \cite{frisch2004}.

\section{Optimal Streaming}
\label{sec:optimal-streaming}
%\review{Changed the use of ``$s$'' to ``$w$'' to be consistent with the notation in the rest of the paper}
In this section we specify what it means to be an \emph{optimally streaming} implementation of a function from sequences to sequences.
%\review{Removed mention of greedy parsing.}
%\paragraph{Notation.}

We write $w \prefer w''$ if $w$ is a \emph{prefix} of $w''$, that is $w w' = w''$ for some $w'$.
%In such case we denote the unique suffix $s'$ such that $s s' = s''$ by $s''/s$.
Note that $\prefer$ is a partial order with greatest lower bounds for nonempty sets: $\bigsqcap L = w$ if $w \prefer w''$ for all $w'' \in L$ and $\forall w'. (\forall w'' \in S. w' \prefer w'') \Rightarrow w' \prefer w$. $\bigsqcap L$ is the longest common prefix of all words in $L$.

% Proviso: In the following, we fix an
%We assume $\lang{E} \neq \emptyset$.

\begin{definition}[Completions]
The set of \emph{completions} $C_E(w)$ of $w$ in $E$ is the set of all words in $\lang{E}$ that have $w$ as a prefix:
$$C_E(w) = \{ w'' \mid w \prefer w'' \wedge w'' \in \lang{E} \}.$$
\end{definition}
Note that $C_E(w)$ may be empty.

\begin{definition}[Extension]
For nonempty $C_E(w)$ the unique \emph{extension} $\hat{w}_E$ of $w$ under $E$ is the longest extension of $w$ with a suffix such that all successful extensions of $w$ to an element of $\lang{E}$ are also extensions of $\hat{w}$:
$$\hat{w}_E = \bigsqcap C_E(w).$$
Word $w$ is \emph{extended} under $E$ if $w = \hat{w}$; otherwise it is unextended.
\end{definition}
Extension is a closure operation: $\hat{\hat{w}} = \hat{w}$; in particular, extensions are extended.

\begin{definition}[Reduction]
For empty $C_E(w)$ the unique \emph{reduction} $\bar{w}_E$ of $w$ under $E$ is the longest prefix $w'$ of $w$ such that $C_E(w') \neq \emptyset$.
\end{definition}

Given parse function $\parsef{E}{\cdot} : \lang{E} \rightarrow \bitcode{E}$ for complete input strings, we can now define what it means for an implementation of it to be optimally streaming:
\begin{definition}[Optimally streaming]
The \emph{optimally streaming} function corresponding to $\parsef{E}{\cdot}$ is
$$O_E(w) = \left\{ \begin{array}{ll}
                  \bigsqcap \{\parsef{E}{w''} \mid w'' \in C_E(w) \} &
                  \mbox{if } C_E(w) \neq \emptyset \\
                  (\bigsqcap O_E(\bar{w})) \fail & \mbox{if } C_E(w) = \emptyset.
                  \end{array} \right.
                  $$
\end{definition}
The first condition expresses that after seeing prefix $w$ the function must output \emph{all} bits that are a common prefix of all bit-coded parse trees of words in $\lang{E}$ that $w$ can be extended to.
The second condition expresses that as soon as it is clear that a prefix has no extension to an element of $\lang{E}$, an indicator $\fail$ of failure must be emitted, with no further output after that.
In this sense $O_E$ is \emph{optimally} streaming: It produces output bits at the semantically earliest possible time during input processing.

It is easy to check that $O_E$ is a streaming function:
$$w \prefer w' \Rightarrow O_E(w) \prefer O_E(w')$$

The definition has the, at first glance, surprising consequence that $O_E$ may output bits for parts of the input it has not even read yet:
\begin{proposition}
\label{prop:closure}
$O_E(w) = O_E(\hat{w})$
\end{proposition}
E.g.\ for $E = (\eplus{\elit{a}}{\elit{a}})(\eplus{\elit{a}}{\elit{a}})$ we have $O_E(\epsilon) = \bit{00}$; that is, $O_E$ outputs $\bit{00}$ off the bat, before reading any input symbols, in anticipation of $\lit{aa}$ being the only possible successful extension.
Assume the input is $\lit{ab}$.
After reading $\lit{a}$ it does not output anything, and after reading $\lit{b}$ it outputs $\fail$ to indicate a failed parse, the total output being $\bit{00}\fail$.

\section{Coverage}
\label{sec:coverage}

Our algorithm is based on simulating aNFAs in lock-step, maintaining a set of partial paths reading the prefix $w$ of the input that has been consumed so far.
In order to be optimally streaming, we have to identify partial paths which are guaranteed not to be prefixes of a greedy parse for a word in $C_E(w)$.

In this section, we define a \emph{coverage relation} which our parsing algorithm relies on in order to detect the aforementioned situation. In the following, fix an RE $E$ and its aNFA $M_E = (\St_E, \delta_E, \qinit_E, \qfinal_E)$.

\begin{definition}[Coverage]
Let $p \in \St_E$ be a state and $Q \subseteq \St_E$ a state set.
We say that \emph{$Q$ covers $p$}, written $Q \covers p$, iff
\begin{equation}
  \label{eq:coverage}
  \{ \pread{\alpha} \mid q \ptrans{\alpha} \qfinal, q \in Q \}
  \supseteq \{ \pread{\beta} \mid p \ptrans{\beta} \qfinal \}
\end{equation}
\end{definition}

Coverage can be seen as a slight generalization of language inclusion.
That is, if $Q \covers p$, then every word suffix read by a path from $p$ to the final state can also be read by a path from one of the states in $Q$ to the final state.

Let $\op{M_e}$ refer to the automaton obtained by reversing the direction of all transitions and swapping the initial and final states.
It can easily be verified that if \eqref{eq:coverage} holds for some $Q,p$, then the following property also holds in the \emph{reverse} automaton $\op{M_E}$:
\begin{equation}
  \label{eq:cocoverage}
  \{ \pread{\alpha} \mid \qinit \ptrans{\alpha} q, q \in Q \}
  \supseteq \{ \pread{\beta} \mid \qinit \ptrans{\alpha} p \}
\end{equation}
If we consider $D(\op{M_E})$, the deterministic automaton generated from $\op{M_E}$, then we see that \eqref{eq:cocoverage} is satisfied iff
\begin{equation}
\label{eq:cocoverage-criterion}
\forall S \in \DSt_{\op{M_E}}.~p \in S \Rightarrow Q \cap S \not= \emptyset
\end{equation}
This is true since a DFA state $S$ is reachable by reading a word $w$ in $D(\op{M_E})$ iff every $q \in S$ is reachable by reading $w$ in $\op{M_E}$.
Since a DFA accepts the same language as the underlying aNFA, this implies that condition \eqref{eq:cocoverage} must hold iff $Q$ has a non-empty intersection with \emph{all} DFA states containing $p$.

The equivalence of \eqref{eq:coverage} and \eqref{eq:cocoverage-criterion} gives us a method to decide $\covers$ in an aNFA $M$, provided that we have computed $D(\op{M})$ beforehand.
Checking \eqref{eq:cocoverage-criterion} for a particular $Q$ and $p$ can be done by intersecting all states of $\DSt_{\op{M_E}}$ with $Q$, using time $O(|Q||\DSt_{\op{M_E}}|) = O(|Q|2^{O(m)})$, where $m$ is the size of the RE $E$.

The exponential cost appears to be unavoidable -- the problem of deciding coverage is inherently hard to compute:
\begin{proposition}
\label{prop:coverage-pspace}
The problem of deciding coverage, that is the set $\{ (E, Q, p) \mid Q \subseteq \St_E \wedge Q \covers p \}$, is PSPACE-hard.
\end{proposition}
\begin{proof}
We can reduce regular expression equivalence to coverage:
Given regular expressions $E$ and $F$, produce an aNFA $M_{E+F}$ for $E + F$ and observe that $M_E$ and $M_F$ are subautomata.
Now observe that there is a path $\qinit_{E+F} \ptrans{\alpha} \qfinal_E$ (respectively $\qinit_{E+F} \ptrans{\beta} \qfinal_F$) in $M_{E+F}$ iff there is a path $\qinit_E \ptrans{\alpha'} \qfinal_E$ with $\pread{\alpha} = \pread{\alpha'}$ in $M_E$ (respectively $\qinit_F \ptrans{\beta'} \qfinal_F$ with $\pread{\beta} = \pread{\beta'}$
in $M_F$).
Hence, we have $\{ \qinit_F \} \covers \qinit_E$ in $M_{E+F}$ iff $\lang{E} \subseteq \lang{F}$.
Since regular expression containment is PSPACE-complete~\cite{stockmeyer1973} this shows that coverage is PSPACE-hard.
\end{proof}

%\todo{Not sure if this section still needs some restructuring. Maybe move PSPACE-theorem to the end of the section?}
Even after having computed a determinized automaton, the decision version of the coverage problem is still NP-complete, which we show by reduction to and from \textsc{Min-Cover}, a well-known NP-complete problem.
Let \textsc{State-Cover} refer to the problem of deciding membership for the language \[\{ (M,D(M),p,k) \mid \exists Q.~|Q| = k \land p \not\in Q \land Q \covers p \text{ in $M$} \}.\]
Recall that \textsc{Min-Cover} is the problem of deciding membership for the language $\{ (X, \mathcal{F}, k) \mid \exists \mathcal{C} \subseteq \mathcal{F}. |\mathcal{C}| = k \land X = \bigcup \mathcal{C} \}$.

\begin{proposition}
\textsc{State-Cover} is NP-complete.
\end{proposition}
\begin{proof}
\textsc{State-Cover} $\Rightarrow$ \textsc{Min-Cover}:
Let $(M, D(M), p, k)$ be given. Define $X = \{ S \in \DSt_{M} \mid p \in S \}$ and $\mathcal{F} = \{ R_q \mid q \in \bigcup X \}$ where $R_q = \{ S \in X \mid q \in S \}$.
Then any $k$-sized set cover $\mathcal{C} = \{ R_{q_1}, ..., R_{q_k} \}$ gives a state cover $Q = \{ q_1, ..., q_k \}$ and vice-versa.

$\textsc{Min-Cover} \Rightarrow \textsc{State-Cover}$:
Let $(X, \mathcal{F},k)$ be given, where $|X| = m$ and $|\mathcal{F}| = n$.
Construct an aNFA $M_{X,\mathcal{F}}$ over the alphabet $\Sigma = X \uplus \{\lit{\$}\}$.
Define its states to be the set $\{\qinit, \qfinal, p\} \cup \{ F_1, ..., F_n \}$, and for each $F_i$, add transitions $F_i \trans{\lit{\$}} \qfinal$ and $\qinit \trans{x_{ij}} F_i$ for each $x_{ij} \in F_i$. Finally add transitions $p \trans{\lit{\$}} \qfinal$ and $\qinit \trans{x} p$ for each $x \in X$.

Observe that $D(M_{X,\mathcal{F}})$ will have states $\{ \{\qinit\}, \{ \qfinal \} \} \cup \{ S_x \mid x \in X \}$ where $S_x = \{ F \in \mathcal{F} \mid x \in F \} \cup \{p\}$, and $\Delta(\{\qinit\}, x) = S_x$.
Also, the time to compute $D(M_{X,\mathcal{F}})$ is bounded by $O(|X| |\mathcal{F}|)$.
Then any $k$-sized state cover $Q = \{ F_1, ..., F_k \}$ is also a set cover.
\end{proof}

\section{Algorithm}
\label{sec:algorithm}

Our parsing algorithm produces a bit-coded parse tree from an input string $w$ for a given RE $E$.
We will simulate $M_E$ in lock-step, reading a symbol from $w$ in each step.
The simulation maintains a set of all partial paths that read the prefix of $w$ that has been consumed so far; there are always only finitely many paths to consider, since we restrict ourselves to paths without non-productive loops.
When a path reaches a non-deterministic choice, it will ``fork'' into two paths with the same prefix.
Thus, the path set can be represented as a tree of states, where the root is the initial state, the edges are transitions between states, and the leaves are the reachable states.

\begin{definition}[Path trees]
  A \emph{path tree} is a rooted, ordered, binary tree with \emph{internal nodes} of outdegrees $1$ or $2$.
Nodes are labeled by aNFA-states and edges by $\Gamma = \Sigma \cup \{\bit{0}, \bit{1}\} \cup \{ \epsilon \}$.
Binary nodes have a pair of $\bit{0}$- and $\bit{1}$-labeled edges (in this order only), respectively.
\end{definition}
We use the following notation:
\begin{itemize}
\item $\treeroot{T}$ is the root node of path tree $T$.
%  \item $\treechildren{T}$ is the set of child nodes of $T$.
\item $\treelabel{n}{c}$ is the path from $n$ to $c$, where $c$ is a descendant of $n$.
\item $\treeinit{T}$ is the path from the root to the first binary node reachable or to the unique leaf of $T$ if it has no binary node.
\item $\treeleaves{T}$ is the \emph{ordered list} of leaf nodes.
\item $\treeempty$ is the empty tree.
\end{itemize}
As a notational convenience, the tree with a root node labeled $q$ and no children is written $\treenilnode{q}$, where $q$ is an aNFA-state.
Similarly, a tree with a root labeled $q$ with children $l$ and $r$ is written $\treebinarynode{q}{l}{r}$, where $q$ is an aNFA-state and $l$ and $r$ are path trees and the edges from $q$ to $l$ and $r$ are labeled $\bit{0}$ and $\bit{1}$, respectively.
Unary nodes are labelled by $\Sigma\cup\set{\epsilon}$ and are written $\treeunarynode{q}{\ell}{c}$, denoting a tree rooted at $q$ with only one $\ell$-labelled child $c$.

In the following we shall use $T_w$ to refer to a path tree created after processing input word $w$ and $T$ to refer to path trees in general, where the input string giving rise to the tree is irrelevant.

\begin{definition}[Path tree invariant]
Let $T_w$ be a path tree and $w$ a word. Define $I(T_w)$ as the proposition that \emph{all} of the following hold:
\begin{enumerate}[label=(\roman*)]
\item The $\treeleaves{T_w}$ have pairwise distinct node labels; all labels are \emph{symbol sources}, that is states with a single symbol transition, or the accept state.\label{state-vector-property}
\item All paths from the root to a leaf read $w$:\label{read-s}
\[\forall n \in \treeleaves{T_w}.~\pread{\treelabel{\treeroot{T_w}}{n}} = w.\]
\item For each leaf $n \in \treeleaves{T_w}$ there exists $w'' \in C_E(w)$ such that the bit-coded parse of $w''$ starts with $\pwrite{\treelabel{\treeroot{T_w}}{n}}$.
\label{soundness}
\item For each $w'' \in C_E(w)$ there exists $n \in \treeleaves{T_w}$ such that the bit-coded parse of $w''$ starts with $\pwrite{\treelabel{\treeroot{T_w}}{n}}$.
\label{completeness}
\end{enumerate}
\end{definition}

%\review{Added explanation of algorithm for invariant.}
The path tree invariant is maintained by Algorithm~\ref{alg:path_tree_invariant}: line 2 establishes part~\ref{state-vector-property}; line 3 establishes part~\ref{read-s}; and lines 4--7 establishes part~\ref{soundness} and \ref{completeness}.

\begin{algorithm}
\caption{Optimally streaming parsing algorithm.}\label{alg:optstream}
\noindent
\textbf{Input:} {An aNFA $M$, a coverage relation $\covers$, and an input stream $S$.}

\noindent
\textbf{Output:} {Greedy leftmost parse tree, emitted in optimally-streaming fashion.}

\begin{algorithmic}[1]
\Function{Stream-Parse}{$M$, $\covers$, $S$}
  \State $w \gets \epsilon$
  \State $(T_\epsilon, \ignoreResult) \gets \textsc{closure}(M,\emptyset,\qinit)$
    \Comment {Initial path tree as output of $\textsc{closure}$}
  \While{$S$ has another input symbol $a$}
     \If{$C_E(wa) = \emptyset$}
        \State \textbf{return} $\pwrite{\treeinit{T_w}}$ followed by $\fail$ and exit.
     \EndIf
     \State $T_{wa} \gets \textsc{Establish-Invariant}(T_w, a, \covers)$
     \State Output new bits on the path to the first binary node in $T_{wa}$, if any.
     \State $w \gets wa$
  \EndWhile
  \If{$\qfinal \in \treeleaves{T_w}$}
     \State \textbf{return} $\pwrite{\treelabel{\treeroot{T_w}}{\qfinal}}$
  \Else
     \State \textbf{return} $\pwrite{\treeinit{T_w}}$ followed by $\fail$
  \EndIf
\EndFunction
\end{algorithmic}
\end{algorithm}

%\review{Reformulated invariant algorithm to be more ``algorithm-y''}
\begin{algorithm}
\caption{Establishing invariant $I(T_{wa})$}\label{alg:path_tree_invariant}

\noindent
\textbf{Input:} {A path tree $T_w$ satisfying $I(T_w)$, a character $a$, and coverage relation $\covers$.}

\noindent
\textbf{Output:} {A path tree $T_{wa}$ satisfying invariant $I(T_{wa})$.}

  \begin{algorithmic}[1]
    \Function{Establish-Invariant}{$T_w$, $a$, $\covers$}
      \State Remove leaves from $T_w$ that do not have a transition on $a$.
      \State Extend $T_w$ to $T_{wa}$ by following all $a$-transitions.
      \For {each leaf $n$ in $T_{wa}$}
        \State $(T', \ignoreResult) \gets \textsc{closure}(M,\emptyset,n)$.
        \State Replace the leaf $n$ with the tree $T'$ in $T_{wa}$.
      \EndFor
      \State \textbf{return} $\textsc{prune}(T_{wa}, \covers)$
    \EndFunction
  \end{algorithmic}
\end{algorithm}

\begin{algorithm}
\caption{Pruning algorithm.}\label{alg:prune}

\noindent\textbf{Input:} {A path tree $T$ and a covering relation $\covers$.}

\noindent\textbf{Output:} {A pruned path tree $T'$ where all leaves are alive.}
  \begin{algorithmic}[1]
    \Function{prune}{$T,\covers$}
      \For{each $l$ in $\reverse{\treeleaves{T}}$}
        \State $S \gets \set{n \mid n \text{ comes before } l \text{ in }\treeleaves{T}}$
        \If {$S \covers l$}
          \State $p \gets \treeparent{l}$
          \State Delete $l$ from $T$
          \State $T \gets \textsc{cut}(T,p)$
        \EndIf
      \EndFor
      \State \textbf{return} $T$
    \EndFunction
    \Function{cut}{$T,n$} \Comment{Cuts a chain of $1$-ary nodes.}
      \If{$|\treechildren{n}| = 0$}
        \State $p \gets \treeparent{n}$
        \State $T' \gets T$ with $n$ removed
        \State \textbf{return} $\textsc{cut}(T',p)$
      \Else
        \State \textbf{return} $T$
      \EndIf
    \EndFunction
  \end{algorithmic}
\end{algorithm}

\begin{algorithm}
  \caption{$\epsilon$-closure with path tree construction.}
  \label{alg:right_closure}
  \begin{algorithmic}[1]
    \Require{An aNFA $M$, a set of visited states $V$, and a state $q$}
    \Ensure{A path tree $T$ and a set of visited states $V'$}
    \Function{closure}{$M,V,q$}
      \If{$q\trans{\bit{0}}q_l$ and $q\trans{\bit{1}}q_r$}
        \State $(T^{l}, V_l) \gets \textsc{closure}(M,V\cup\set{q},q_l)$
          \Comment {Try left option first.}
        \State $(T^{r}, V_{lr}) \gets \textsc{closure}(M,V_l,q_r)$
          \Comment {Use $V_l$ to skip already-visited nodes.}
        \State \textbf{return} $(\treeunarynode{q}{T^l}{T^{r}}, V_{lr})$
      \EndIf
      \If {$q \trans{\epsilon} p$}
        \If {$p \in V$} \Comment{Stop loops.}
          \State \textbf{return} $(\treeempty, V)$
        \Else
          \State $(T', V') \gets \textsc{closure}(M, V\cup\set{q}, p)$
          \State \textbf{return} $(\treeunarynode{q}{\epsilon}{T'}, V')$
        \EndIf
      \Else \Comment{$q$ is a symbol source or the final state.}
        \State \textbf{return} $(\treenilnode{q}, V)$
      \EndIf
    \EndFunction
  \end{algorithmic}
\end{algorithm}

\begin{theorem}[Optimal streaming property]
Assume extended $w$, $C_E(w) \neq \emptyset$.
Consider the path tree $T_w$ after reading $w$ upon entry into the while-loop of the algorithm in Algorithm~\ref{alg:optstream}.
Then $\pwrite{\treeinit{T_w}} = O_E(w)$.
\end{theorem}
In other words, the initial path from the root of $T_w$ to the first binary node in $T_w$ is the longest common prefix of all paths accepting  an extension of $w$.
Operationally, whenever that path gets longer by pruning branches, we output the bits on the extension.
\begin{proof}
%\review{Changed $I(T_s,s)$ to $I(T_s)$.  Changed $s$ to $w$ for consistency with other sects.}
Assume $w$ extended, that is $w = \hat{w}$; assume $C_E(w) \neq \emptyset$, that is there exists $w''$ such that $w \prefer w''$ and $w'' \in \lang{E}$.

Claim: $|\treeleaves{T_w}| \geq 2$ or the unique node in $\treeleaves{T_w}$ is labeled by the accept state.
Proof of claim: Assume otherwise, that is $|\treeleaves{T_w}| = 1$, but its node is not the accept state.
By \ref{state-vector-property} of $I(T_w)$, this means the node must have a symbol transition on some symbol $a$.
In this case, all accepting paths $C_E(wa) = C_E(w)$ and thus $\hat{w} = \hat{wa}$; in particular $\hat{w} \neq w$, which, however, is a contradiction to the assumption that $w$ is extended.

This means we have two cases.
The case $|\treeleaves{T_w}| = 1$ with the sole node being labeled by the accept state is easy: It spells a single path from initial to accept state.
By \ref{read-s} and \ref{soundness} of $I(T_w)$ we have that that path is correct for $w$.
By \ref{completeness} and since the accept state has no outgoing transitions, we have $C_E(w) = \{ w \}$, and the theorem follows for this case.

Let us consider the case $|\treeleaves{T_w}| \geq 2$ then.
Recall that $C_E(w) \neq \emptyset$ by assumption.
By \ref{completeness} of $I(T_w)$ the accepting path of every $w'' \in C_E(w)$ starts with $\treelabel{\treeroot{T_w}}{n}$ for some $n \in \treeleaves{T_w}$, and by \ref{soundness} each path from the root to a leaf is the start of some accept path.
Since $|\treeleaves{T_w}| \geq 2$ we know that there exists a binary node in $T_w$.
Consider the first on the path from the root to a leaf.
It has both $\bit{0}$- and $\bit{1}$-labeled out-edges.
Thus the longest common prefix of $\{ \pwrite{p} \mid n \in \treeleaves{T_w}, p \in \treelabel{\treeroot{T_w}}{n} \}$ is $\pwrite{\treeinit{T_w}}$, the bits on the initial path from the root of $T_w$ to its first binary node.
\end{proof}

%\todo{Not sure what this paragraph means?}
The algorithm, as given, is only optimally streaming for extended prefixes.  It can be made to work for all prefixes by enclosing it in an outer loop that for each prefix $w$ computes $\hat{w}$ and calls the given algorithm with $\hat{w}$.  The outer loop then checks that subsequent symbols match until $\hat{w}$ is reached.
By Proposition~\ref{prop:closure} the resulting algorithm gives the right result for all input prefixes, not only extended ones.

% \paragraph{Reestablishing invariant $I(T,sa)$.}
% We show now how the invariant $I(T_{s'},s')$ for $s' = sa$ in Algorithm~\ref{alg:optstream} is established, given $I(T_s,s)$ holds.
% This is done in the following steps:
% \begin{enumerate}
% \item Mark $\{ n \in \treeleaves{T_s} \mid n \mbox{ does not have transition on } a \}$ as dead.
% \item Extend $T_s$ with all $a$-transitions and the spanning forest (under greedy disambiguation) of the $\epsilon$-closure of $$[ n' \mid n \in \treeleaves{T_s} \wedge n \mbox{ has transition on } a \mbox{ to } n'].$$  (All leaves in $T_s$ are now either internal nodes or marked as dead.) See Algorithm~\ref{alg:right_closure}, which is executed on each node reached by an $a$-transition in the sequence of the nodes in
% $\treeleaves{T_s}$.
% \item Let $[n_1, \ldots, n_k]$ the new leaves after the previous step.
% Mark all $n_i$ such that $\{n_1, \ldots, n_{i-1}\} \covers n_i$ as dead.
% \item Prune all dead branches; that is, remove all branches to dead leaves such that the resulting tree $T_{s'}$ has only live (non-dead) leaves.
% See pruning in Algorithm~\ref{alg:prune} for details.
% \end{enumerate}

% Establishing $I(\epsilon)$ consists of computing the spanning $\epsilon$-tree of the initial state.

\begin{theorem}
The optimally streaming algorithm can be implemented to run in time $O(2^{m \log m} + m n)$, where $m = |E|$ and $n = |w|$.
\end{theorem}
\begin{proof}[Sketch]
%\todo{Re-check this}
As shown in Section~\ref{sec:coverage}, we can decide coverage in time $O(m2^{O(m)})$.
The set of ordered lists $\treeleaves{T}$ for any $T$ reachable from the initial state can be precomputed and covered states marked in it.
(This requires unit-cost random access since there are $O(2^{m \log m})$ such lists.)
The $\epsilon$-closure can be computed in time $O(m)$ for each input symbol, and pruning can be amortized over $\epsilon$-closure computation by charging each edge removed to its addition to a tree path.
\end{proof}

For fixed regular expression $E$ this is linear time in $n$ and thus asymptotically optimal.
An exponential in $m$ as an additive preprocessing cost appears practically unavoidable since we require the coverage relation, which is inherently hard to compute (Proposition~\ref{prop:coverage-pspace}).

\section{Example}
\label{sec:example}

Consider the RE $\estar{(\eplus{\elit{aaa}}{\elit{aa}})}$.
A simplified version of its symmetric position automaton is shown in Figure~\ref{fig:execution_ex}.
The following two observations are requirements for an earliest parse of this expression:
\begin{itemize}
\item After one $\lit{a}$ has been read, the algorithm \emph{must} output a $\bit{0}$ to indicate that one iteration of the Kleene star has been made, but:
\item \emph{five} consecutive $\lit{a}$s determine that the leftmost possibility in the Kleene star choice was taken, meaning that the first \emph{three} $\lit{a}$s are consumed in that branch.
\end{itemize}

The first point can be seen by noting that any parse of a non-zero number of $\lit{a}$s must follow a path through the Kleene star.
This guarantees that \emph{if} a successful parse is eventually performed, it must be the case that at least one iteration was made.

The second point can be seen by considering the situation where only four input $\lit{a}$s  have been read:  It is not known whether these are the only four or more input symbols in the stream.
In the former case, the correct (and only) parse is two iterations with the right alternative, but in the latter case, the first three symbols are consumed in the left branch instead.

These observations correspond intuitively to what ``earliest'' parsing is; as soon as it is impossible that an iteration was \emph{not} made, a bit indicating this fact is emitted, and as soon as the first three symbols must have been parsed in the left alternative, this fact is output.
Furthermore, a $\bit{0}$-bit is emitted to indicate that (at least) another iteration is performed.

Figure~\ref{fig:execution_ex} shows the evolution of the path tree during execution with the RE $\estar{(\eplus{\elit{aaa}}{\elit{aa}})}$ on the input $\lit{aaaaa}$.
% The first path tree represents the paths before any $\lit{a}$ has been read.
% After the first $\lit{a}$, there is a path from the start node to the first split node, and the bit $\bit{0}$ on this path is output (indicated with dashed edges in the tree).
% In the last tree, five $\lit{a}$s have been read, and the bottom ``leg'' of the tree is removed in the pruning step in maintaining invariant $I(\elit{aaaaa})$.
% Consequently, $\pwrite{\treeinit{T}}=\bit{00}$, which can be output.

By similar reasoning as above, after five $\lit{a}$s it is safe to commit to the left alternative after every third $\lit{a}$.
Hence, for the inputs $\lit{aaaaa}(\lit{aaa})^n$, $\lit{aaaaa}(\lit{aaa})^n\lit{a}$, and $\lit{aaaaa}(\lit{aaa})^n\lit{aa}$ the ``commit points'' are placed as follows ($\cdot$ indicate end-of-input):
\begin{center}
  $\annot{a}{\bit{0}} \mid
   \annot{aaaa}{\bit{00}} \mid
   \underbrace{\left(\annot{aaa}{\bit{00}} \mid
   \cdots
   \mid \annot{aaa}{\bit{00}}\right)}_{n\text{ times}} \mid
   \annot{\cdot}{11}$
   \qquad
  $\annot{a}{\bit{0}} \mid
   \annot{aaaa}{\bit{00}} \mid
   \underbrace{\left(\annot{aaa}{\bit{00}} \mid
   \cdots
   \mid \annot{aaa}{\bit{00}}\right)}_{n\text{ times}} \mid
   \annot{a\cdot}{01}$
  \qquad
  $\annot{a}{\bit{0}} \mid
   \annot{aaaa}{\bit{00}} \mid
   \underbrace{\left(\annot{aaa}{\bit{00}} \mid
   \cdots
   \mid \annot{aaa}{\bit{00}}\right)}_{n\text{ times}} \mid
   \annot{aa\cdot}{1011}$
\end{center}

\begin{figure}
  \newcommand{\dist}{0.55cm}
  % step 0
\begin{tikzpicture}[every node/.style={st}, node distance=\dist]
  \node (0a0) {0};
  \node[split,right of=0a0] (0a1) {1};
  \node[split,right of=0a1] (0a2) {2};
  \node[act, right of=0a2] (0a3) {3};
  \node[act, below of=0a3] (0a7) {7};
  \node[act, below of=0a7] (0a11) {11};

  \node[draw=none,above of=0a3] (edummy) {$\epsilon$};

%  \node[right of=0a3] (1a4) {};
%  \node[right of=0a7] (1a8) {};
%
%  \node[draw=none,above right of=1a4] (a1dummy) {};
%
%  \node[right of=1a8] (2a9) {};
%  \node[right of=2a9] (2a10) {};
%  \node[right of=2a10] (2a1) {};
%  \node[right of=2a1] (2a2) {};
%  \node[right of=2a2, yshift=-\dist] (2a7) {};
%  \node[below of=2a7] (2a11) {};
%  \node[right of=1a4, xshift=4*\dist] (2a5) {};
%
%  \node[draw=none,above right of=2a5] (a3dummy) {};
%
%  \node[right of=2a5] (3a6) {};
%  \node[right of=3a6] (3a10) {};
%  \node[right of=3a10] (3a1) {};
%  \node[right of=3a1] (3a2) {};
%  \node[right of=3a2] (3a3) {};
%  \node[below of=3a3] (3a7) {};
%  \node[below of=3a7] (3a11) {};
%  \node[below of=3a11] (3a8) {};

%  \node[draw=none,above right of=3a3] (a4dummy) {};

%  \node[right of=3a3, xshift=3*\dist] (4a4) {};
%  \node[below of=4a4] (4a8) {};
%  \node[right of=3a8] (4a9) {};
%  \node[right of=4a9] (4a10) {};
%  \node[right of=4a10] (4a1) {};
%  \node[right of=4a1, yshift=-\dist] (4a11) {};

%  \node[draw=none,above right of=4a4] (a5dummy) {};

%  \node[right of=4a4, xshift=4*\dist] (5a5) {};
%  \node[right of=4a8] (5a9) {};
%  \node[right of=5a9] (5a10) {};
%  \node[right of=5a10] (5a1) {};
%  \node[right of=5a1] (5a2) {};
%  \node[right of=5a2, yshift=-\dist] (5a7) {};
%  \node[below of=5a7] (5a11) {};

  %%%%%%% Edges
  \draw (0a0) to (0a1)
        (0a1) to (0a2)
        (0a1) to (0a11)
        (0a2) to (0a3)
        (0a2) to (0a7)
        % % 1. a
        % (0a3) to (1a4)
        % (0a7) to (1a8)
        % % 2. a
        % (1a4) to (2a5)
        % (1a8) to (2a9)
        % (2a9) to (2a10)
        % (2a10) to (2a1)
        % (2a1) to (2a2)
        % (2a1) to (2a11)
        % (2a2) to (2a7)
        % % 3. a
        % (2a5) to (3a6) to (3a10) to (3a1)
        % (3a1) to (3a11)
        % (3a1) to (3a2)
        % (3a2) to (3a3)
        % (3a2) to (3a7)
        % (2a7) to (3a8)
        % % 4. a
        % (3a3) to (4a4)
        % (3a7) to (4a8)
        % (3a8) to (4a9) to (4a10) to (4a1) to (4a11)
        % % 5. a
        % (4a4) to (5a5)
        % (4a8) to (5a9) to (5a10) to (5a1) to (5a2) to (5a7)
        % (5a1) to (5a11)
        ;
  \draw[dotted,lightgray] ($(edummy)$) -- ($(edummy) + (0,-1.8)$);
  % \draw[dotted] ($(a1dummy)$) -- ($(a1dummy) + (0,-2.7)$);
  % \draw[dotted] ($(a2dummy)$) -- ($(a2dummy) + (0,-2.7)$);
  % \draw[dotted] ($(a3dummy)$) -- ($(a3dummy) + (0,-2.7)$);
  % \draw[dotted] ($(a4dummy)$) -- ($(a4dummy) + (0,-2.7)$);
  % \draw[dotted] ($(a5dummy)$) -- ($(a5dummy) + (0,-2.7)$);
\end{tikzpicture}
%%% Local Variables:
%%% mode: latex
%%% TeX-master: "../../thesis"
%%% End:
  %!TEX root = ../final.tex

% step 1
\begin{tikzpicture}[every node/.style={st}, node distance=\dist]
  \node (0a0) {0};
  \node[split,right of=0a0] (0a1) {1};
  \node[split,right of=0a1] (0a2) {2};
  \node[right of=0a2] (0a3) {3};
  \node[below of=0a3] (0a7) {7};
  % \node[below of=0a7] (0a11) {11};

  \node[draw=none,above of=0a3] (edummy) {};

  \node[act,right of=0a3] (1a4) {4};
  \node[act,right of=0a7] (1a8) {8};

  \node[draw=none,above of=1a4] (a1dummy) {\texttt{a}};

%  \node[right of=1a8] (2a9) {};
%  \node[right of=2a9] (2a10) {};
%  \node[right of=2a10] (2a1) {};
%  \node[right of=2a1] (2a2) {};
%  \node[right of=2a2, yshift=-\dist] (2a7) {};
%  \node[below of=2a7] (2a11) {};
%  \node[right of=1a4, xshift=4*\dist] (2a5) {};
%
%  \node[draw=none,above  of=2a5] (a3dummy) {};
%
%  \node[right of=2a5] (3a6) {};
%  \node[right of=3a6] (3a10) {};
%  \node[right of=3a10] (3a1) {};
%  \node[right of=3a1] (3a2) {};
%  \node[right of=3a2] (3a3) {};
%  \node[below of=3a3] (3a7) {};
%  \node[below of=3a7] (3a11) {};
%  \node[below of=3a11] (3a8) {};
%
%  \node[draw=none,above  of=3a3] (a4dummy) {};
%
%  \node[right of=3a3, xshift=3*\dist] (4a4) {};
%  \node[below of=4a4] (4a8) {};
%  \node[right of=3a8] (4a9) {};
%  \node[right of=4a9] (4a10) {};
%  \node[right of=4a10] (4a1) {};
%  \node[right of=4a1, yshift=-\dist] (4a11) {};
%
%  \node[draw=none,above  of=4a4] (a5dummy) {};
%
%  \node[right of=4a4, xshift=4*\dist] (5a5) {};
%  \node[right of=4a8] (5a9) {};
%  \node[right of=5a9] (5a10) {};
%  \node[right of=5a10] (5a1) {};
%  \node[right of=5a1] (5a2) {};
%  \node[right of=5a2, yshift=-\dist] (5a7) {};
%  \node[below of=5a7] (5a11) {};

  %%%%%%% Edges
  \draw[dashed] (0a0) to (0a1);
  \draw[dashed] (0a1) to (0a2);
        % (0a1) to (0a11)
  \draw (0a2) to (0a3)
        (0a2) to (0a7)
        % 1. a
        (0a3) to (1a4)
        (0a7) to (1a8)
        % % 2. a
        % (1a4) to (2a5)
        % (1a8) to (2a9)
        % (2a9) to (2a10)
        % (2a10) to (2a1)
        % (2a1) to (2a2)
        % (2a1) to (2a11)
        % (2a2) to (2a7)
        % % 3. a
        % (2a5) to (3a6) to (3a10) to (3a1)
        % (3a1) to (3a11)
        % (3a1) to (3a2)
        % (3a2) to (3a3)
        % (3a2) to (3a7)
        % (2a7) to (3a8)
        % % 4. a
        % (3a3) to (4a4)
        % (3a7) to (4a8)
        % (3a8) to (4a9) to (4a10) to (4a1) to (4a11)
        % % 5. a
        % (4a4) to (5a5)
        % (4a8) to (5a9) to (5a10) to (5a1) to (5a2) to (5a7)
        % (5a1) to (5a11)
        ;
  \draw[dotted,lightgray] ($(edummy)$) -- ($(edummy) + (0,-1.5)$);
  \draw[dotted,lightgray] ($(a1dummy)$) -- ($(a1dummy) + (0,-1.5)$);
  % \draw[dotted] ($(a2dummy)$) -- ($(a2dummy) + (0,-2.7)$);
  % \draw[dotted] ($(a3dummy)$) -- ($(a3dummy) + (0,-2.7)$);
  % \draw[dotted] ($(a4dummy)$) -- ($(a4dummy) + (0,-2.7)$);
  % \draw[dotted] ($(a5dummy)$) -- ($(a5dummy) + (0,-2.7)$);
\end{tikzpicture}

  % step 2
\begin{tikzpicture}[every node/.style={st}, node distance=\dist]
  \node (0a0) {0};
  \node[split,right of=0a0] (0a1) {1};
  \node[split,right of=0a1] (0a2) {2};
  \node[right of=0a2] (0a3) {3};
  \node[below of=0a3] (0a7) {7};
  % \node[below of=0a7] (0a11) {11};

  \node[draw=none,above of=0a3] (edummy) {};

  \node[right of=0a3] (1a4) {4};
  \node[right of=0a7] (1a8) {8};

  \node[dummy,draw=none,above  of=1a4] (a1dummy) {};

  \node[right of=1a8] (2a9) {9};
  \node[right of=2a9] (2a10) {10};
  \node[split,right of=2a10] (2a1) {1};
  \node[split,right of=2a1] (2a2) {2};
  \node[act, right of=2a2, yshift=-\dist] (2a7) {7};
  \node[act, below of=2a7] (2a11) {11};
  \node[act, right of=1a4, xshift=4*\dist] (2a5) {5};

  \node[draw=none,above  of=2a5] (a2dummy) {\texttt{a}};

%   \node[right of=2a5] (3a6) {};
%   \node[right of=3a6] (3a10) {};
%   \node[right of=3a10] (3a1) {};
%   \node[right of=3a1] (3a2) {};
%   \node[right of=3a2] (3a3) {};
%   \node[below of=3a3] (3a7) {};
%   \node[below of=3a7] (3a11) {};
%   \node[below of=3a11] (3a8) {};
% 
%   \node[draw=none,above  of=3a3] (a4dummy) {};
% 
%   \node[right of=3a3, xshift=3*\dist] (4a4) {};
%   \node[below of=4a4] (4a8) {};
%   \node[right of=3a8] (4a9) {};
%   \node[right of=4a9] (4a10) {};
%   \node[right of=4a10] (4a1) {};
%   \node[right of=4a1, yshift=-\dist] (4a11) {};
% 
%   \node[draw=none,above  of=4a4] (a5dummy) {};
% 
%   \node[right of=4a4, xshift=4*\dist] (5a5) {};
%   \node[right of=4a8] (5a9) {};
%   \node[right of=5a9] (5a10) {};
%   \node[right of=5a10] (5a1) {};
%   \node[right of=5a1] (5a2) {};
%   \node[right of=5a2, yshift=-\dist] (5a7) {};
%   \node[below of=5a7] (5a11) {};

  %%%%%%% Edges
  \draw[dashed] (0a0) to (0a1);
  \draw[dashed] (0a1) to (0a2);
        % (0a1) to (0a11)
  \draw (0a2) to (0a3)
        (0a2) to (0a7)
        % 1. a
        (0a3) to (1a4)
        (0a7) to (1a8)
        % 2. a
        (1a4) to (2a5)
        (1a8) to (2a9)
        (2a9) to (2a10)
        (2a10) to (2a1)
        (2a1) to (2a2)
        (2a1) to (2a11)
        (2a2) to (2a7)
        % % 3. a
        % (2a5) to (3a6) to (3a10) to (3a1)
        % (3a1) to (3a11)
        % (3a1) to (3a2)
        % (3a2) to (3a3)
        % (3a2) to (3a7)
        % (2a7) to (3a8)
        % % 4. a
        % (3a3) to (4a4)
        % (3a7) to (4a8)
        % (3a8) to (4a9) to (4a10) to (4a1) to (4a11)
        % % 5. a
        % (4a4) to (5a5)
        % (4a8) to (5a9) to (5a10) to (5a1) to (5a2) to (5a7)
        % (5a1) to (5a11)
        ;
  \draw[dotted,lightgray] ($(edummy)$) -- ($(edummy) + (0,-2.2)$);
  \draw[dotted,lightgray] ($(a1dummy)$) -- ($(a1dummy) + (0,-2.2)$);
  \draw[dotted,lightgray] ($(a2dummy)$) -- ($(a2dummy) + (0,-2.2)$);
  % \draw[dotted] ($(a3dummy)$) -- ($(a3dummy) + (0,-2.7)$);
  % \draw[dotted] ($(a4dummy)$) -- ($(a4dummy) + (0,-2.7)$);
  % \draw[dotted] ($(a5dummy)$) -- ($(a5dummy) + (0,-2.7)$);
\end{tikzpicture}

%%% Local Variables:
%%% mode: latex
%%% TeX-master: "../../thesis"
%%% End:
  %!TEX root = ../final.tex

% step 3
\begin{tikzpicture}[every node/.style={st}, node distance=\dist]
  \node (0a0) {0};
  \node[split,right of=0a0] (0a1) {1};
  \node[split,right of=0a1] (0a2) {2};
  \node[right of=0a2] (0a3) {3};
  \node[below of=0a3] (0a7) {7};
  % \node[below of=0a7] (0a11) {11};

  \node[draw=none,above of=0a3] (edummy) {};

  \node[right of=0a3] (1a4) {4};
  \node[right of=0a7] (1a8) {8};

  \node[dummy,draw=none,above  of=1a4] (a1dummy) {};

  \node[right of=1a8] (2a9) {9};
  \node[right of=2a9] (2a10) {10};
  \node[split,right of=2a10] (2a1) {1};
  \node[split,right of=2a1] (2a2) {2};
  \node[right of=2a2, yshift=-\dist] (2a7) {7};
  % \node[below of=2a7] (2a11) {11};
  \node[right of=1a4, xshift=4*\dist] (2a5) {5};

  \node[dummy,draw=none,above  of=2a5] (a2dummy) {};

  \node[right of=2a5] (3a6) {6};
  \node[right of=3a6] (3a10) {10};
  \node[split,right of=3a10] (3a1) {1};
  \node[split,right of=3a1] (3a2) {2};
  \node[act, right of=3a2] (3a3) {3};
  \node[act, below of=3a3] (3a7) {7};
  \node[act, below of=3a7] (3a11) {11};
  \node[act, below of=3a11] (3a8) {8};
  \node[below of=3a6, yshift=-2*\dist,outer sep=0pt,inner sep=0pt,minimum size=0pt,fill=black] (3adummy) {};

  \node[draw=none,above  of=3a3] (a3dummy) {\texttt{a}};

%  \node[right of=3a3, xshift=3*\dist] (4a4) {};
%  \node[below of=4a4] (4a8) {};
%  \node[right of=3a8] (4a9) {};
%  \node[right of=4a9] (4a10) {};
%  \node[right of=4a10] (4a1) {};
%  \node[right of=4a1, yshift=-\dist] (4a11) {};
%
%  \node[draw=none,above  of=4a4] (a5dummy) {};
%
%  \node[right of=4a4, xshift=4*\dist] (5a5) {};
%  \node[right of=4a8] (5a9) {};
%  \node[right of=5a9] (5a10) {};
%  \node[right of=5a10] (5a1) {};
%  \node[right of=5a1] (5a2) {};
%  \node[right of=5a2, yshift=-\dist] (5a7) {};
%  \node[below of=5a7] (5a11) {};

  %%%%%%% Edges
  \draw[dashed] (0a0) to (0a1);
  \draw[dashed] (0a1) to (0a2);
        % (0a1) to (0a11)
  \draw (0a2) to (0a3)
        (0a2) to (0a7)
        % 1. a
        (0a3) to (1a4)
        (0a7) to (1a8)
        % 2. a
        (1a4) to (2a5)
        (1a8) to (2a9)
        (2a9) to (2a10)
        (2a10) to (2a1)
        (2a1) to (2a2)
        % (2a1) to (2a11)
        (2a2) to (2a7)
        % 3. a
        (2a5) to (3a6) to (3a10) to (3a1)
        (3a1) to (3a11)
        (3a1) to (3a2)
        (3a2) to (3a3)
        (3a2) to (3a7)
        (2a7) to (3adummy) to (3a8)
        % % 4. a
        % (3a3) to (4a4)
        % (3a7) to (4a8)
        % (3a8) to (4a9) to (4a10) to (4a1) to (4a11)
        % % 5. a
        % (4a4) to (5a5)
        % (4a8) to (5a9) to (5a10) to (5a1) to (5a2) to (5a7)
        % (5a1) to (5a11)
        ;
  \draw[dotted,lightgray] ($(edummy)$) -- ($(edummy) + (0,-2.2)$);
  \draw[dotted,lightgray] ($(a1dummy)$) -- ($(a1dummy) + (0,-2.2)$);
  \draw[dotted,lightgray] ($(a2dummy)$) -- ($(a2dummy) + (0,-2.2)$);
  \draw[dotted,lightgray] ($(a3dummy)$) -- ($(a3dummy) + (0,-2.2)$);
  % \draw[dotted] ($(a4dummy)$) -- ($(a4dummy) + (0,-2.7)$);
  % \draw[dotted] ($(a5dummy)$) -- ($(a5dummy) + (0,-2.7)$);
\end{tikzpicture}

%%% Local Variables:
%%% mode: latex
%%% TeX-master: "../../thesis"
%%% End:
  %!TEX root = ../final.tex

% step 4
\begin{tikzpicture}[every node/.style={st}, node distance=\dist]
  \node (0a0) {0};
  \node[split,right of=0a0] (0a1) {1};
  \node[split,right of=0a1] (0a2) {2};
  \node[right of=0a2] (0a3) {3};
  \node[below of=0a3] (0a7) {7};
  % \node[below of=0a7] (0a11) {11};

  \node[draw=none,above of=0a3] (edummy) {};

  \node[right of=0a3] (1a4) {4};
  \node[right of=0a7] (1a8) {8};

  \node[dummy,draw=none,above  of=1a4] (a1dummy) {};

  \node[right of=1a8] (2a9) {9};
  \node[right of=2a9] (2a10) {10};
  \node[split,right of=2a10] (2a1) {1};
  \node[split,right of=2a1] (2a2) {2};
  \node[right of=2a2, yshift=-\dist] (2a7) {7};
  % \node[below of=2a7] (2a11) {11};
  \node[right of=1a4, xshift=4*\dist] (2a5) {5};

  \node[dummy,draw=none,above  of=2a5] (a2dummy) {};

  \node[right of=2a5] (3a6) {6};
  \node[right of=3a6] (3a10) {10};
  \node[split,right of=3a10] (3a1) {1};
  \node[split,right of=3a1] (3a2) {2};
  \node[right of=3a2] (3a3) {3};
  \node[below of=3a3] (3a7) {7};
  % \node[below of=3a7] (3a11) {11};
  \node[below of=3a11] (3a8) {8};
  \node[below of=3a6, yshift=-2*\dist,outer sep=0pt,inner sep=0pt,minimum size=0pt,fill=black] (3adummy) {};

  \node[dummy,draw=none,above  of=3a3] (a3dummy) {};

  \node[act, right of=3a3, xshift=3*\dist] (4a4) {4};
  \node[act, below of=4a4] (4a8) {8};
  \node[right of=3a8] (4a9) {9};
  \node[right of=4a9] (4a10) {10};
  \node[split,right of=4a10] (4a1) {1};
  \node[act, right of=4a1, yshift=-\dist] (4a11) {11};

  \node[draw=none,above  of=4a4] (a4dummy) {\texttt{a}};

  \node[right of=4a4, xshift=4*\dist] (5a5) {};
  \node[right of=4a8] (5a9) {};
  \node[right of=5a9] (5a10) {};
  \node[right of=5a10] (5a1) {};
  \node[right of=5a1] (5a2) {};
  \node[right of=5a2, yshift=-\dist] (5a7) {};
  \node[below of=5a7] (5a11) {};

  %%%%%%% Edges
  \draw[dashed] (0a0) to (0a1);
  \draw[dashed] (0a1) to (0a2);
        % (0a1) to (0a11)
  \draw (0a2) to (0a3)
        (0a2) to (0a7)
        % 1. a
        (0a3) to (1a4)
        (0a7) to (1a8)
        % 2. a
        (1a4) to (2a5)
        (1a8) to (2a9)
        (2a9) to (2a10)
        (2a10) to (2a1)
        (2a1) to (2a2)
        % (2a1) to (2a11)
        (2a2) to (2a7)
        % 3. a
        (2a5) to (3a6) to (3a10) to (3a1)
        % (3a1) to (3a11)
        (3a1) to (3a2)
        (3a2) to (3a3)
        (3a2) to (3a7)
        (2a7) to (3adummy) to (3a8)
        % 4. a
        (3a3) to (4a4)
        (3a7) to (4a8)
        (3a8) to (4a9) to (4a10) to (4a1) to (4a11)
        % 5. a
        % (4a4) to (5a5)
        % (4a8) to (5a9) to (5a10) to (5a1) to (5a2) to (5a7)
        % (5a1) to (5a11)
        ;
  \draw[dotted,lightgray] ($(edummy)$) -- ($(edummy) + (0,-2.7)$);
  \draw[dotted,lightgray] ($(a1dummy)$) -- ($(a1dummy) + (0,-2.7)$);
  \draw[dotted,lightgray] ($(a2dummy)$) -- ($(a2dummy) + (0,-2.7)$);
  \draw[dotted,lightgray] ($(a3dummy)$) -- ($(a3dummy) + (0,-2.7)$);
  \draw[dotted,lightgray] ($(a4dummy)$) -- ($(a4dummy) + (0,-2.7)$);
  % \draw[dotted] ($(a5dummy)$) -- ($(a5dummy) + (0,-2.7)$);
\end{tikzpicture}
  %!TEX root = ../final.tex

% step 5
\begin{tikzpicture}[every node/.style={st}, node distance=\dist]
  \node (0a0) {0};
  \node[split,right of=0a0] (0a1) {1};
  \node[split,right of=0a1] (0a2) {2};
  \node[right of=0a2] (0a3) {3};
  % \node[below of=0a3] (0a7) {7};
  % \node[below of=0a7] (0a11) {11};

  \node[draw=none,above of=0a3] (edummy) {};

  \node[right of=0a3] (1a4) {4};
  % \node[right of=0a7] (1a8) {8};

  \node[dummy,draw=none,above of=1a4] (a1dummy) {};

  % \node[right of=1a8] (2a9) {9};
  % \node[right of=2a9] (2a10) {10};
  % \node[right of=2a10] (2a1) {1};
  % \node[right of=2a1] (2a2) {2};
  % \node[right of=2a2, yshift=-\dist] (2a7) {7};
  % % \node[below of=2a7] (2a11) {11};
  \node[right of=1a4, xshift=4*\dist] (2a5) {5};

  \node[dummy,draw=none,above of=2a5] (a2dummy) {};

  \node[right of=2a5] (3a6) {6};
  \node[right of=3a6] (3a10) {10};
  \node[split,right of=3a10] (3a1) {1};
  \node[split,right of=3a1] (3a2) {2};
  \node[right of=3a2] (3a3) {3};
  \node[below of=3a3] (3a7) {7};
  % \node[below of=3a7] (3a11) {11};
  % \node[below of=3a11] (3a8) {8};

  \node[dummy,draw=none,above of=3a3] (a3dummy) {};

  \node[right of=3a3, xshift=3*\dist] (4a4) {4};
  \node[below of=4a4] (4a8) {8};
  % \node[right of=3a8] (4a9) {9};
  % \node[right of=4a9] (4a10) {10};
  % \node[right of=4a10] (4a1) {1};
  % \node[right of=4a1, yshift=-\dist] (4a11) {11};

  \node[dummy,draw=none,above of=4a4] (a4dummy) {};

  \node[act, right of=4a4, xshift=4*\dist] (5a5) {5};
  \node[right of=4a8] (5a9) {9};
  \node[right of=5a9] (5a10) {10};
  \node[split,right of=5a10] (5a1) {1};
  \node[split,right of=5a1] (5a2) {2};
  \node[act, right of=5a2, yshift=-\dist] (5a7) {7};
  \node[act, below of=5a7] (5a11) {11};

  \node[dummy,draw=none,above of=5a5] (a5dummy) {\texttt{a}};

  %%%%%%% Edges
  \draw[dashed] (0a0) to (0a1);
  \draw[dashed] (0a1) to (0a2);
        % (0a1) to (0a11)
  \draw[dashed] (0a2) to (0a3)
        % (0a2) to (0a7)
        % 1. a
        (0a3) to (1a4)
        % (0a7) to (1a8)
        % 2. a
        (1a4) to (2a5)
        % (1a8) to (2a9)
        % (2a9) to (2a10)
        % (2a10) to (2a1)
        % (2a1) to (2a2)
        % (2a1) to (2a11)
        % (2a2) to (2a7)
        % 3. a
        (2a5) to (3a6) to (3a10) to (3a1)
        % (3a1) to (3a11)
        (3a1) to (3a2);
  \draw
        (3a2) to (3a3)
        (3a2) to (3a7)
        % (2a7) to (3a8)
        % 4. a
        (3a3) to (4a4)
        (3a7) to (4a8)
        % (3a8) to (4a9) to (4a10) to (4a1) to (4a11)
        % 5. a
        (4a4) to (5a5)
        (4a8) to (5a9) to (5a10) to (5a1) to (5a2) to (5a7)
        (5a1) to (5a11)
        ;
  \draw[dotted,lightgray] ($(edummy)$) -- ($(edummy) + (0,-2.3)$);
  \draw[dotted,lightgray] ($(a1dummy)$) -- ($(a1dummy) + (0,-2.3)$);
  \draw[dotted,lightgray] ($(a2dummy)$) -- ($(a2dummy) + (0,-2.3)$);
  \draw[dotted,lightgray] ($(a3dummy)$) -- ($(a3dummy) + (0,-2.3)$);
  \draw[dotted,lightgray] ($(a4dummy)$) -- ($(a4dummy) + (0,-2.3)$);
  \draw[dotted,lightgray] ($(a5dummy)$) -- ($(a5dummy) + (0,-2.3)$);
\end{tikzpicture}

  \raisebox{12.5cm}[0pt][0pt]{\hspace{19em}%!TEX root = ../final.tex

\begin{tikzpicture}[ nfa, node distance=1.2cm ]
  \node[draw=none] (dummy) {};
  \node[right of=dummy] (1) {0};
  \node[right of=1]          (2) {1};
  \node[accepting,right of=2] (12) {11};

  \node[above left of=2,yshift=1.5cm,xshift=-1.8cm] (3) {2};

  \node[above right of=3] (4)   {3};
  \node[right of=4]     (5)  {4};
  \node[right of=5]    (6)  {5};
  \node[right of=6]    (7)  {6};

  \node[below right of=3]  (9) {7};
  \node[right of=9,xshift=0.6cm]      (10) {8};
  \node[right of=10,xshift=0.6cm]      (11) {9};

  \node[below right of=7] (8) {10};

  \begin{scope}[transitions]
    \draw (dummy) to (1);
    \draw (1) to node [swap] {$\epsilon$} (2);
    \draw (2) to node [swap] {$\bit{1}$} (12);
    \draw[out=135,in=270] (2) to node {$\bit{0}$} (3);
    \draw[out=270,in=45] (8) to node {$\epsilon$} (2);

    \draw (3) to node {$\bit{0}$} (4);
    \draw (3) to node {$\bit{1}$} (9);
    \draw (7) to node {$\epsilon$} (8);
    \draw (11) to node {$\epsilon$} (8);

    \draw (4) to node {$\lit{a}$} (5);
    \draw (5) to node {$\lit{a}$} (6);
    \draw (6) to node {$\lit{a}$} (7);
    \draw (9) to node {$\lit{a}$} (10);
    \draw (10) to node {$\lit{a}$} (11);

    % Make a box around the whole automaton with the fit library.
    % \node[draw,fit=(1) (3) (7) (8),inner sep=9pt] {};
  \end{scope}

\end{tikzpicture}}
  \raisebox{10cm}[0pt][0pt]{\hspace{29em}\begin{minipage}{2cm}
    \begin{align*}
    \set{5} & \covers 8 \\
    \set{4} & \covers 7 \\
    \set{8} & \covers 5 \\
    \set{7} & \covers 4 \\
    \set{5} & \covers 3 \\
    \end{align*}
  \end{minipage}
  }

  \caption[Example of streaming algorithm on RE $\estar{(\eplus{\elit{aaa}}{\elit{aa}})}$.]{Example run of the algorithm on the regular expression $E=\estar{(\eplus{\elit{aaa}}{\elit{aa}})}$ and the input string $\lit{aaaaa}$.
The dashed edges represent the partial parse trees that can be emitted: thus, after one $\lit{a}$ we can emit a $\bit{0}$, and after five $\lit{a}$s we can emit $\bit{00}$ because the bottom ``leg'' of the tree has been removed in the pruning step.
The automaton for $E$ and its associated minimal covering relation are shown in the inset.}
  \label{fig:execution_ex}
\end{figure}
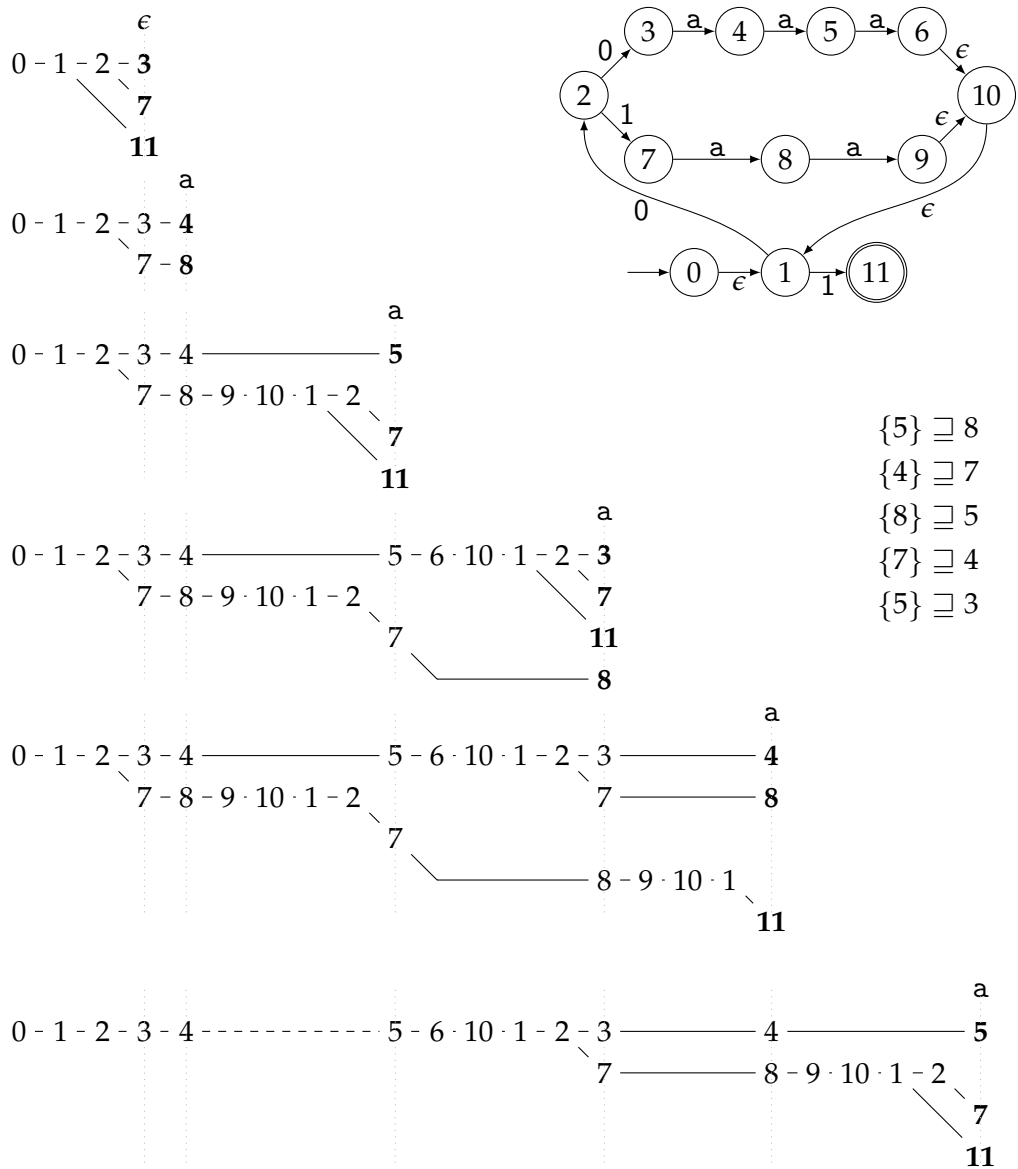

\begin{figure}
  \newcommand{\dist}{0.65cm}
  %!TEX root = ../final.tex

\begin{tikzpicture}[every node/.style={st}, node distance=\dist]
  \node (e1) {1};
  \node[right of=e1,xshift=\dist] (e2) {2};
  \node[right of=e2] (e3) {3};
  \node[act, right of=e3,xshift=\dist] (e4) {4};
  \node[act, below of=e4] (e8) {8};
  \node[act, below of=e8] (e13) {13};
  \node[act, below of=e13] (e18) {18};
  \node[act, below of=e18] (e20) {20};
  \node[left of=e8] (e7) {7};
  \node[left of=e18] (e17) {17};
  \node[left of=e17, xshift=-\dist] (e10) {10};
%%%% first a
  \node[right of=e18] (a19) {19};
  \node[right of=a19] (a17) {17};
  \node[act, right of=a17] (a18) {18};
  \node[act, below of=a18] (a20) {20};
  \node[act, above of=a18, yshift=2*\dist] (a5) {5};
%%%% second a
  \node[right of=a5] (aa6) {6};
  \node[right of=aa6] (aa3) {3};
  \node[act, right of=aa3, xshift=\dist] (aa4) {4};
  \node[act, below of=aa4] (aa8) {8};
  \node[act, below of=aa8] (aa13) {13};
  \node[act, below of=aa13] (aa18) {18};
  \node[act, below of=aa18] (aa20) {20};
  \node[left of=aa8] (aa7) {7};
  \node[left of=aa18] (aa17) {17};
  \node[right of=a18] (aa19) {19};
%%%% third a
  \node[right of=aa18] (aaa19) {19};
  \node[right of=aaa19] (aaa17) {17};
  \node[act, right of=aaa17] (aaa18) {18};
  \node[act, below of=aaa18] (aaa20) {20};
  \node[act, above of=aaa18, yshift=2*\dist] (aaa5) {5};
%%%% first z
  \node[right of=aaa20] (z21) {21};
  \node[act, right of=z21] (z22) {22};
  \node[act, below of=z22] (z25) {25};
%%%% first b
  \node[right of=z25] (b26) {26};
  \node[right of=b26] (b24) {24};
  \node[act, right of=b24] (b12) {12};

%%%%%%%%%%%%
% Dummys for drawing ``death lines''
  \begin{scope}[node distance=0.8*\dist]
    \node[right of=e8] (e8dead) {};
    \node[right of=e13] (e13dead) {};
    \node[right of=e20] (e20dead) {};

    \node[right of=a20] (a20dead) {};

    \node[right of=aa8] (aa8dead) {};
    \node[right of=aa13] (aa13dead) {};
    \node[right of=aa20] (aa20dead) {};

    \node[right of=aaa5] (aaa5dead) {};
    \node[right of=aaa18] (aaa18dead) {};

    \node[right of=z22] (z22dead) {};
  \end{scope}
%%%%%%%%%%%%

  \draw (e1) to (e2)
        (e2) to (e3)
        (e3) to (e4)
        (e3) to (e7)
        (e7) to (e8)
        (e7) to (e13)
        (e1) to (e10)
        (e10) to (e17)
        (e17) to (e18)
        (e17) to (e20)
%% First a
        (e4) to (a5)
        (e18) to (a19)
        (a19) to (a17)
        (a17) to (a18)
        (a17) to (a20)
%% Second a
        (a5) to (aa6)
        (aa6) to (aa3)
        (aa3) to (aa4)
        (aa3) to (aa7)
        (aa7) to (aa8)
        (aa7) to (aa13)
        (a18) to (aa19)
        (aa19) to (aa17)
        (aa17) to (aa18)
        (aa17) to (aa20)
%% Third a
        (aa4) to (aaa5)
        (aa18) to (aaa19)
        (aaa19) to (aaa17)
        (aaa17) to (aaa18)
        (aaa17) to (aaa20)
%% First z
        (aaa20) to (z21)
        (z21) to (z22)
        (z21) to (z25)
%% First b
        (z25) to (b26)
        (b26) to (b24)
        (b24) to (b12)
        ;

%%%% ``Death lines''
  \draw (e8) to (e8dead)
        (e13) to (e13dead)
        (e20) to (e20dead)
        (a20) to (a20dead)
        (aa8) to (aa8dead)
        (aa13) to (aa13dead)
        (aa20) to (aa20dead)
        (aaa5) to (aaa5dead)
        (aaa18) to (aaa18dead)
        (z22) to (z22dead)
        ;

  \begin{scope}[]
    \draw[shift=(e8dead)] ($(-0.1,-0.1)$) -- ($(0.1,+0.1)$);
    \draw[shift=(e13dead)] ($(-0.1,-0.1)$) -- ($(0.1,+0.1)$);
    \draw[shift=(e20dead)] ($(-0.1,-0.1)$) -- ($(0.1,+0.1)$);
    \draw[shift=(a20dead)] ($(-0.1,-0.1)$) -- ($(0.1,+0.1)$);
    \draw[shift=(aa8dead)] ($(-0.1,-0.1)$) -- ($(0.1,+0.1)$);
    \draw[shift=(aa13dead)] ($(-0.1,-0.1)$) -- ($(0.1,+0.1)$);
    \draw[shift=(aa20dead)] ($(-0.1,-0.1)$) -- ($(0.1,+0.1)$);
    \draw[shift=(aaa5dead)] ($(-0.1,-0.1)$) -- ($(0.1,+0.1)$);
    \draw[shift=(aaa18dead)] ($(-0.1,-0.1)$) -- ($(0.1,+0.1)$);
    \draw[shift=(z22dead)] ($(-0.1,-0.1)$) -- ($(0.1,+0.1)$);
  \end{scope}

  \begin{scope}[draw=none]
    \node[above of=e4] (edummy) {$\epsilon$};
    \node[above of=a5] (a1dummy) {\texttt{a}};
    \node[above of=aa4] (a2dummy) {\texttt{a}};
    \node[above of=aaa5] (a3dummy) {\texttt{a}};
    \node[above of=z22, yshift=4*\dist] (zdummy) {\texttt{z}};
    \node[above of=b12, yshift=5*\dist] (bdummy) {\texttt{b}};
  \end{scope}

  \begin{scope}[dotted, lightgray]
    \newcommand{\h}{-7*\dist}
    \draw[dotted,lightgray] ($(edummy)$) -- ($(edummy) + (0,\h)$);
    \draw[dotted,lightgray] ($(a1dummy)$) -- ($(a1dummy) + (0,\h)$);
    \draw[dotted,lightgray] ($(a2dummy)$) -- ($(a2dummy) + (0,\h)$);
    \draw[dotted,lightgray] ($(a3dummy)$) -- ($(a3dummy) + (0,\h)$);
    \draw[dotted,lightgray] ($(zdummy)$) -- ($(zdummy) + (0,\h)$);
    \draw[dotted,lightgray] ($(bdummy)$) -- ($(bdummy) + (0,\h)$);
  \end{scope}

\end{tikzpicture}

%%% Local Variables:
%%% mode: latex
%%% TeX-master: "../../thesis"
%%% End:
  %!TEX root = ../final.tex

\begin{tikzpicture}[every node/.style={st}, node distance=\dist]
  \node (e1) {1};
  \node[right of=e1,xshift=2*\dist] (e2) {2};
  \node[right of=e2] (e3) {3};
  \node[act, right of=e3,xshift=\dist] (e4) {4};
  \node[act, below of=e4] (e8) {8};
  \node[act, below of=e8] (e13) {13};
  \node[act, below of=e13] (e18) {18};
  \node[act, below of=e18] (e20) {20};
  \node[left of=e8] (e7) {7};
  \node[left of=e18] (e17) {17};
  \node[left of=e17, xshift=-\dist] (e10) {10};
%%%% first a
  \node[right of=e18] (a19) {19};
  \node[right of=a19] (a17) {17};
  \node[act, right of=a17] (a18) {18};
  \node[act, below of=a18] (a20) {20};
  \node[act, above of=a18, yshift=2*\dist] (a5) {5};
%%%% second a
  \node[right of=a5] (aa6) {6};
  \node[right of=aa6] (aa3) {3};
  \node[act, right of=aa3, xshift=\dist] (aa4) {4};
  \node[act, below of=aa4] (aa8) {8};
  \node[act, below of=aa8] (aa13) {13};
  \node[act, below of=aa13] (aa18) {18};
  \node[act, below of=aa18] (aa20) {20};
  \node[left of=aa8] (aa7) {7};
  \node[left of=aa18] (aa17) {17};
  \node[right of=a18] (aa19) {19};
%%%% first z
  \node[act, right of=aa8] (z9) {9};
  \node[act, right of=aa13] (z14) {14};
%%%% first b
  \node[right of=z14] (b15) {15};
  \node[right of=b15] (b11) {11};
  \node[act, right of=b11] (b12) {12};

%%%%%%%%%%%%
% Dummys for drawing ``death lines''
  \begin{scope}[node distance=0.8*\dist]
    \node[right of=e8] (e8dead) {};
    \node[right of=e13] (e13dead) {};
    \node[right of=e20] (e20dead) {};

    \node[right of=a20] (a20dead) {};

    \node[right of=aa4] (aa4dead) {};
    \node[right of=aa18] (aa18dead) {};
    \node[right of=aa20] (aa20dead) {$\deathbycoverage$}; % COVERED TO DEATH!

    \node[right of=z9] (z9dead) {};
  \end{scope}
%%%%%%%%%%%%

  \draw (e1) to (e2)
        (e2) to (e3)
        (e3) to (e4)
        (e3) to (e7)
        (e7) to (e8)
        (e7) to (e13)
        (e1) to (e10)
        (e10) to (e17)
        (e17) to (e18)
        (e17) to (e20)
%% First a
        (e4) to (a5)
        (e18) to (a19)
        (a19) to (a17)
        (a17) to (a18)
        (a17) to (a20)
%% Second a
        (a5) to (aa6)
        (aa6) to (aa3)
        (aa3) to (aa4)
        (aa3) to (aa7)
        (aa7) to (aa8)
        (aa7) to (aa13)
        (a18) to (aa19)
        (aa19) to (aa17)
        (aa17) to (aa18)
        (aa17) to (aa20)
%% First z
        (aa8) to (z9)
        (aa13) to (z14)
%% First b
        (z14) to (b15)
        (b15) to (b11)
        (b11) to (b12)
        ;
%%%% ``Death lines''
  \draw (e8) to (e8dead)
        (e13) to (e13dead)
        (e20) to (e20dead)
        (a20) to (a20dead)
        (aa4) to (aa4dead)
        (aa20) to (aa20dead)
        (z9) to (z9dead)
        ;

  \begin{scope}[]
    \draw[shift=(e8dead)] ($(-0.1,-0.1)$) -- ($(0.1,+0.1)$);
    \draw[shift=(e13dead)] ($(-0.1,-0.1)$) -- ($(0.1,+0.1)$);
    \draw[shift=(e20dead)] ($(-0.1,-0.1)$) -- ($(0.1,+0.1)$);
    \draw[shift=(a20dead)] ($(-0.1,-0.1)$) -- ($(0.1,+0.1)$);
    \draw[shift=(aa4dead)] ($(-0.1,-0.1)$) -- ($(0.1,+0.1)$);
    \draw[shift=(z9dead)] ($(-0.1,-0.1)$) -- ($(0.1,+0.1)$);
  \end{scope}

  \begin{scope}[draw=none]
    \node[above of=e4] (edummy) {$\epsilon$};
    \node[above of=a5] (a1dummy) {\texttt{a}};
    \node[above of=aa4] (a2dummy) {\texttt{a}};
    \node[above of=z9, yshift=\dist] (zdummy) {\texttt{z}};
    \node[above of=b12, yshift=2*\dist] (bdummy) {\texttt{b}};
  \end{scope}

  \begin{scope}[dotted,lightgray]
    \newcommand{\h}{-7*\dist}
    \draw ($(edummy)$) -- ($(edummy) + (0,\h)$);
    \draw ($(a1dummy)$) -- ($(a1dummy) + (0,\h)$);
    \draw ($(a2dummy)$) -- ($(a2dummy) + (0,\h)$);
    \draw ($(zdummy)$) -- ($(zdummy) + (0,\h)$);
    \draw ($(bdummy)$) -- ($(bdummy) + (0,\h)$);
  \end{scope}

\end{tikzpicture}
  %!TEX root = ../final.tex

% NFA for E = (aa)*(za+zb) + a*z(a+b)

\newcommand{\ndist}{1.2cm}
\begin{tikzpicture}[ nfa, node distance=\ndist ]

  \node[draw=none] (dummy) {};
  \node[right of=dummy] (1) {1};
    \node[right of=1,yshift=\ndist] (2) {2};
    \node[right of=2] (3) {3};
      \node[above left of=3,xshift=-0.33*\ndist,yshift=0.2*\ndist] (4) {4};
      \node[right of=4] (5) {5};
      \node[right of=5] (6) {6};
    \node[right of=3] (7) {7};
      \node[above right of=7] (8) {8};
      \node[right of=8] (9) {9};
      \node[right of=9] (10) {10};
    \node[below right of=10] (11) {11};
      \node[below right of=7] (13) {13};
      \node[right of=13] (14) {14};
      \node[right of=14] (15) {15};

    \node[right of=1,yshift=-1.2*\ndist] (16) {16};
    \node[right of=16] (17) {17};
    \node[above left of=17, xshift=0.2*\ndist, yshift=0.2*\ndist] (18) {18};
    \node[right of=18] (19) {19};
    \node[right of=17] (20) {20};
    \node[right of=20] (21) {21};
    \node[above right of=21] (22) {22};
    \node[right of=22] (23) {23};
    \node[below right of=23] (24) {24};
    \node[below right of=21] (25) {25};
    \node[right of=25] (26) {26};

  \node[accepting, right of=24, yshift=\ndist] (12) {12};

  \begin{scope}[transitions]
    \newcommand{\0}{$\bit{0}$}
    \newcommand{\1}{$\bit{1}$}
    \newcommand{\bz}{$\epsilon$}
    \newcommand{\bo}{$\epsilon$}
    \draw (dummy) to (1);
    \draw (1) to node {\0} (2);
    \draw[swap] (2) to node {\bo} (3);
    \draw[swap] (3) to node {\0} (4);
    \draw (4) to node {a} (5);
    \draw (5) to node {a} (6);
    \draw[swap] (6) to node {\bz} (3);
    \draw[swap] (3) to node {\1} (7);
    \draw (7) to node {\0} (8);
    \draw (8) to node {z} (9);
    \draw (9) to node {a} (10);
    \draw (10) to node {\bz} (11);
    \draw[swap] (7) to node {\1} (13);
    \draw (13) to node {z} (14);
    \draw (14) to node {b} (15);
    \draw[swap] (15) to node {\bo} (11);
    \draw (11) to node {\bz} (12);
    \draw[swap] (1) to node {\1} (16);
    \draw[swap] (16) to node {\bo} (17);
    \draw[swap] (17) to node {\0} (18);
    \draw (18) to node {a} (19);
    \draw[swap] (19) to node {\bz} (17);
    \draw[swap] (17) to node {\1} (20);
    \draw (20) to node {z} (21);
    \draw (21) to node {\0} (22);
    \draw (22) to node {a} (23);
    \draw (23) to node {\bz} (24);
    \draw[swap] (21) to node {\1} (25);
    \draw (25) to node {b} (26);
    \draw[swap] (26) to node {\bo} (24);
    \draw[swap] (24) to node {\bo} (12);
  \end{scope}
\end{tikzpicture}

%%% Local Variables:
%%% mode: latex
%%% TeX-master: "../../thesis"
%%% End:

  \raisebox{2.7cm}[0pt][0pt]{\hspace{10.8cm}\begin{minipage}{2cm}
    \begin{align*}
      \set{14} & \covers 25 \\
      \set{9} & \covers 22 \\
      \set{8,13} & \covers 20 \\
      \set{4,5} & \covers 18 \\
      \set{25} & \covers 14 \\
      \set{20} & \covers 13 \\
      \set{22} & \covers 9 \\
      \set{20} & \covers 8 \\
      \set{18} & \covers 5 \\
      \set{18} & \covers 4
    \end{align*}
  \end{minipage}}

  \caption[Example of streaming algorithm on $\eplus{\estar{(\elit{aa})}(\eplus{\elit{za}}{\elit{zb}})}{\estar{\elit{a}}\elit{z}(\eplus{\elit{a}}{\elit{b}})}$.]{Example run of the algorithm on $E=\eplus{\estar{(\elit{aa})}(\eplus{\elit{za}}{\elit{zb}})}{\estar{\elit{a}}\elit{z}(\eplus{\elit{a}}{\elit{b}})}$.
  Note that state $20$ is covered by \emph{the combination of} states $8$ and $13$.
  The earliest time the algorithm can commit is when a \texttt{z} is encountered, which determines if the number of \texttt{a}s is even or odd.
  The top shows the path tree on the input \texttt{aaazb}.
  There is a ``trunk'' from state $1$ to state $21$ after reading \texttt{z}, as the rest of the branches have been pruned (not shown).
  This path corresponds to choosing the right top-level alternative.
  In the second figure, we see that if the \texttt{z} appears after an even number of \texttt{a}s, a binary-node-free path from $1$ to $7$ emerges.
  Due to the cover $\set{8,13}\covers 20$, the branch starting from $20$ is not expanded further, even though there could be a \texttt{z}-transition on it.
  This is indicated with $\deathbycoverage$.
  Overall, the resulting parse tree corresponds to the leftmost option in the sum.
  }
  \label{fig:execution_ex_nontrivial_coverage}
\end{figure}
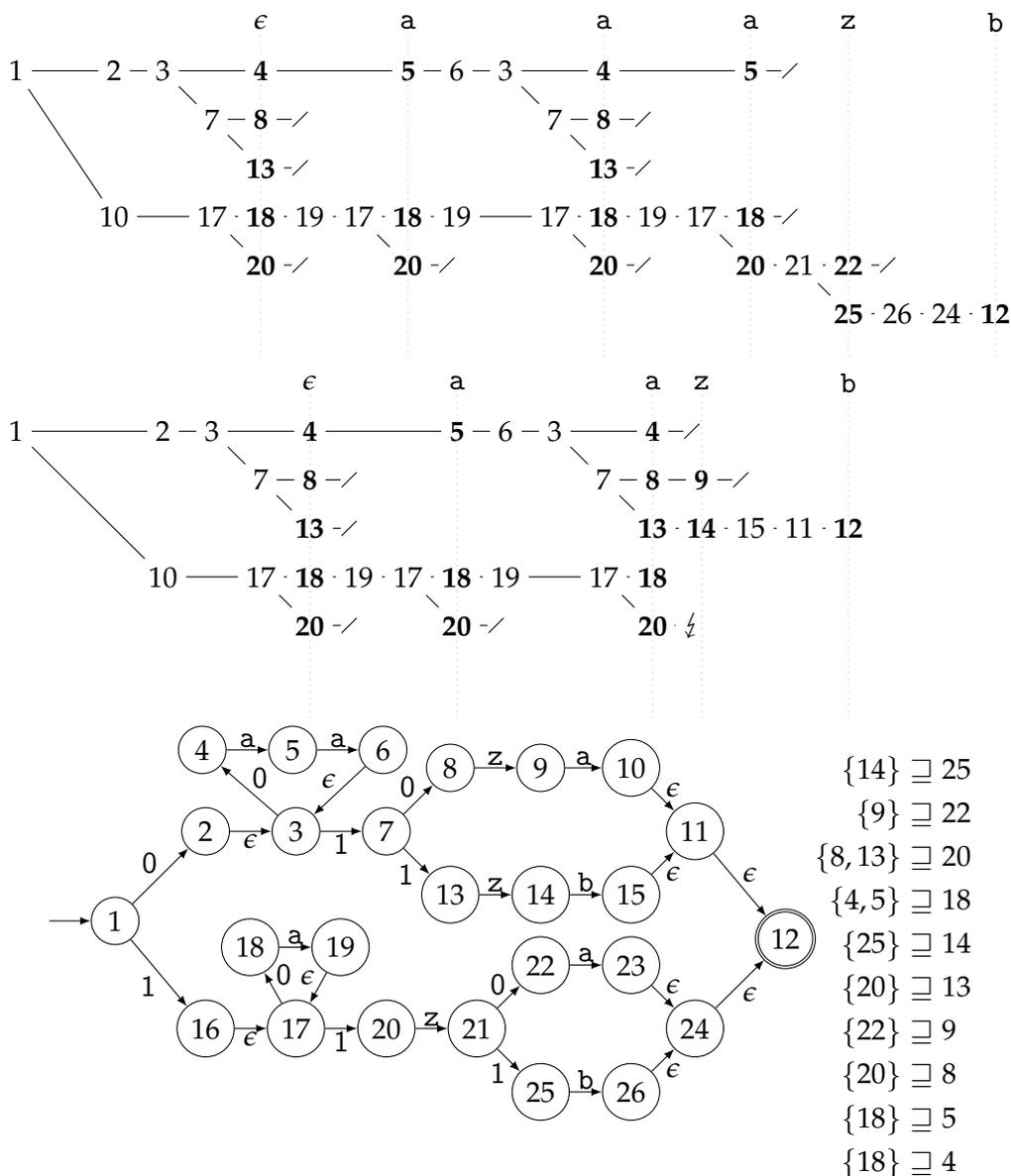

\paragraph{Complex coverage.}
\label{par:complex_coverage}

The previous example does not exhibit any non-trivial coverage, i.e., situations where a state $n$ is covered by $k>1$ other states.
One can construct an expression that contains non-trivial coverage relations by observing that if each symbol source $s$ in the aNFA is associated with the RE representing the language recognized from $s$, coverage can be expressed as a set of (in)equations in Kleene algebra.
Thus, the coverage $\set{n_0,n_1}\covers n$ becomes $RE(n_0)+RE(n_1)\geq RE(n)$ in KA, where $RE(\cdot)$ is the function that yields the RE from a symbol source in an aNFA.

Any expression of the form $x_1zy_1 + x_2zy_2 + x_3z(y_1+y_2)$ satisfies the property that two subterms cover a third.
If the coverage is to play a role in the algorithm, however, the languages denoted by $x_1$ and $x_2$ must not subsume that of $x_3$, otherwise the part starting with $x_3$ would never play a role due to greedy leftmost disambiguation.

Choose $x_1=x_2=\estar{(\elit{aa})}$, $x_3=\estar{\elit{a}}$, $y_1=\elit{a}$, and $y_2=\elit{b}$.
Figure~\ref{fig:execution_ex_nontrivial_coverage} shows the expression $$ \estar{(\elit{aa})}\elit{za} + \estar{\elit{aa}}\elit{zb} + \estar{\elit{a}}z\eplus{\elit{a}}{\elit{b}} = \estar{(\elit{aa})}(\eplus{\elit{za}}{\elit{zb}})+\estar{a}\elit{z}(\eplus{\elit{a}}{\elit{b}}). $$
The earliest point where any bits can be output is when the \texttt{z} is reached.
Then it becomes known whether there was an even or odd number of \texttt{a}s.
Due to the coverage $\set{8,13}\covers 20$ state $20$ is pruned away on the input \texttt{aazb}, thereby causing the path tree to have a large trunk that can be output.

\paragraph{CSV files.}
\label{par:csv_files}

The expression $\estar{(\estar{(\eplus{\elit{a}}{\elit{b}})} \estar{(\elit{;}\estar{(\eplus{\elit{a}}{\elit{b}})})}\elit{n})}$ defines the format of a simple semicolon-delimited data format, with data consisting of words over $\set{\elit{a},\elit{b}}$ and rows separated by the newline character, $\elit{n}$.
Our algorithm emits the partial parse trees after each letter has been parsed, as illustrated on the example input below:
\begin{center}
  \begin{tabular}{cc}
    \begin{minipage}{1cm}
      \begin{verbatim}a;ba;a
b;;b
\end{verbatim}
    \end{minipage}
    &
    \qquad$\annot{a\vphantom{;}}{\bit{000}} \mid
           \annot{;}{\bit{10}} \mid
           \annot{b\vphantom{;}}{\bit{01}} \mid
           \annot{a\vphantom{;}}{\bit{00}} \mid
           \annot{;}{\bit{10}} \mid
           \annot{a\vphantom{;}}{\bit{00}} \mid
           \annot{n\vphantom{;}}{\bit{11}} \mid
           \annot{b\vphantom{;}}{\bit{001}} \mid
           \annot{;}{\bit{10}} \mid
           \annot{;}{\bit{10}} \mid
           \annot{a\vphantom{;}}{\bit{00}} \mid
           \annot{n\vphantom{;}}{\bit{11}} \mid
           \annot{\cdot\vphantom{;}}{\bit{1}}$
  \end{tabular}
\end{center}

Due to the star-height of three, many widespread implementations would not be able to meaningfully handle this expression using only the RE engine.
Capturing groups under Kleene stars return either the first or last match, but not a \emph{list} of matches---and certainly not a list of lists of matches! Hence, if using an implementation like Perl's~\cite{wall2000}, one is forced to rewrite the expression by removing the iteration in the outer Kleene star and reintroduce it as a looping construct in Perl.

\section{Related and Future Work}
\label{sec:future_work}
%\todo[inline]{Not finished}
Parsing regular expressions is not new~\cite{grathwohl2013two,frisch2004,dufe2000,nihe2011,sulzmann2014}, and streaming parsing of XML documents has been investigated for more than a decade in the context of \textsc{XQuery} and \textsc{XPath}---see, e.g.,~\cite{debarbieux2013,gupta2003,wu2013}.
However, \emph{streaming regular expression} parsing appears to be new.
%\todo{how new?}

In earlier work~\cite{grathwohl2013two} we described a compact ``lean log'' format for storing intermediate information required for two-phase regular expression parsing.
The algorithm presented here may degenerate to two passes, but requires often just one pass in the sense being effectively streaming, using only $O(m)$ work space, independent of $n$.  The preprocessing of the regular expression and the intermediate data structure during input string processing are more complex, however.
It may be possible to merge the two approaches using a tree of lean log frames with associated counters, observing that edges in the path tree that are \emph{not} labeled $\bit{0}$ or $\bit{1}$ are redundant.
This is future work.

}

\putbib[bibliography]
\end{bibunit}

%%% Local Variables:
%%% mode: latex
%%% TeX-master: "thesis"
%%% End:

\begin{bibunit}[abbrv]
\chapter[Kleenex: High-Performance Stream Processing][Kleenex]{Kleenex: High-Performance Grammar Based Stream Processing}
\label{paper:kleenex}

\begin{center}
\begin{minipage}{0.9\linewidth}
Bjørn Bugge Grathwohl, Fritz Henglein, Ulrik Terp Rasmussen, Kristoffer Aalund Søholm and Sebastian Paaske Tørholm. ``Kleenex: Compiling Nondeterministic Transducers to Deterministic Streaming Transducers''. In \emph{Proceedings of the 43rd Annual ACM SIGPLAN-SIGACT Symposium on Principles of Programming Languages (POPL)}. Pages 284-297. ACM, 2016. DOI: \href{http://dx.doi.org/10.1145/2837614.2837647}{10.1145/2837614.2837647}.
\end{minipage}
\end{center}

\noindent
The enclosed paper has been renamed. It has also been reformatted to fit the layout of this dissertation. A confusing typo have been corrected in Section~\ref{sec:simulation} ($|u|$ was consistently used instead of the correct $\overline{u}$).

\clearpage

\thispagestyle{plain}
\begin{center}
{\LARGE \textbf{Kleenex: High-Performance Grammar Based Stream Processing}\footnote{The order of authors is insignificant.}}

\vspace{1.5em}

{Niels Bjørn Bugge Grathwohl}$^*$, {Fritz Henglein}$^*$, {Ulrik Terp Rasmussen}$^*$, {Kristoffer Aalund Søholm}$^\dagger$ and {Sebastian Paaske Tørholm}$^\dagger$

\vspace{1em}

{$^*$Department of Computer Science, University of Copenhagen (DIKU)}

\vspace{0.5em}

$^\dagger${Jobindex, Denmark}
\end{center}

\begin{abstract}
We present and illustrate Kleenex, a language for expressing general nondeterministic finite transducers, and its novel compilation to streaming string transducers with
%essentially optimal streaming behavior,
worst-case linear-time performance and sustained high throughput. Its underlying theory is based on transducer decomposition into oracle and action machines: the oracle machine performs  streaming greedy disambiguation of the input; the action machine performs the output actions.  
In use cases Kleenex achieves consistently high throughput rates around the 1 Gbps range on stock hardware. It performs well, especially in complex use cases, in comparison to both specialized and related tools such as \texttt{awk}, \texttt{sed}, RE2, Ragel and regular-expression libraries.
\end{abstract}

{
% \C -> \Cal
% \tone -> \tunit

\tikzstyle{state}=[draw,circle,minimum size=17pt,inner sep=0pt]
\tikzstyle{oft} = [ every node/.style={state,scale=0.8}
                  , >=stealth'
                  , auto
                  ]
\tikzstyle{dummy} = [ inner sep=2pt
                    , fill=white
                    , draw=none
                    ]
\tikzstyle{label} = []
\tikzstyle{transition} = [->, every node/.style={label}]

\input{p3-popl2016/highlight-style}
\DefineVerbatimEnvironment%
    {KleenexVerb}{Verbatim}%
    {commandchars=\\\{\},fontsize=\small}
\CustomVerbatimCommand{\KleenexInline}{Verb}{commandchars=\\\{\},fontsize=\small}
% Custom highlighting!
\makeatletter
\expandafter\def\csname PY@tok@esc\endcsname{\def\PY@tc##1{\textcolor[rgb]{1.00,0.13,1.00}{##1}}}
\makeatother

\newcommand{\longdoublearrow}[2]{\!\!\!\xymatrix{{}\ar@{->>}[r]^{#1}_{#2}&{}}\!\!\!}

\newcommand{\RPathtree}{\mathcal{R}}
\newcommand{\Pathtree}{\mathrm{P}}
\newcommand{\Leaves}{\mathrm{leaves}}
\newcommand{\Nat}{\mathbb{N}}
\newcommand{\dom}{\mathrm{dom}}
\newcommand{\rng}{\mathrm{rng}}
\newcommand{\sem}[1]{\mathcal{R}[\![ {#1} ]\!]}
\newcommand{\gsem}[1]{\mathcal{G}[\![ {#1} ]\!]}
\newcommand{\fsem}[1]{\mathcal{F}[\![ {#1} ]\!]}
\newcommand{\lsem}[1]{\mathcal{L}[\![ {#1} ]\!]}
\newcommand{\powerset}[1]{\mathcal{P}(#1)}
\newcommand{\IsSat}{\mathit{IsSat}}
\newcommand{\byteclass}[1]{\texttt{[#1]}}
\newcommand{\adjtop}{\top} %adjoined top element
\newcommand{\code}[1]{\mathsf{#1}}
\newcommand{\outp}[1]{\mathrm{outp}}

\newcommand{\iffdef}{\ensuremath{\stackrel{\mathrm{def}}{\iff}}}
\newcommand{\trans}[2]{\xrightarrow{#1{\scriptscriptstyle /}#2}}
\newcommand{\translbl}[2]{{#1}{\scriptscriptstyle \vert}{#2}}
\renewcommand{\path}[2]{\mathrel{\xrightarrow{#1{\scriptscriptstyle /}#2}{}}}
\newcommand{\np}{\mathsf{np}}
\newcommand{\pathnp}[2]{\mathrel{\xrightarrow{#1{\scriptscriptstyle /}#2}_{\np}}}
\newcommand{\minnp}{\mathsf{min}}
\newcommand{\pathmin}[2]{\mathrel{\xrightarrow{#1{\scriptscriptstyle /}#2}_{\minnp}}}
\newcommand{\supp}{\mathrm{supp}}
\newcommand{\qinit}{q^{-}}
\newcommand{\qfinal}{q^{f}}

\newcommand{\outputs}[2]{#1 \downarrow #2}
\newcommand{\grammarrule}[2]{#1 \longrightarrow #2}
\newcommand{\suppressoutput}{\texttt{\textasciitilde}}
\newcommand{\prog}[1]{\texttt{#1}}
\newcommand{\plot}[1]{\centering\includegraphics[width=1\columnwidth]{p3-popl2016/benchplots/crop_#1}}

\newcommand{\Act}{\mathsf{A}}
\newcommand{\Code}{\mathsf{C}}
\newcommand{\Two}{\mathbf{2}}
\newcommand{\Zero}{\emptyset}
\newcommand{\Reach}[1]{\mathrm{Reach}({#1})}
\newcommand{\Cal}[1]{\mathcal{#1}}

\newcommand{\tact}[1]{\texttt{"}#1\texttt{"}}
\newcommand{\tsup}[1]{\suppressoutput #1}
\newcommand{\tdef}[1]{\texttt{/} #1 \texttt{/}}
\newcommand{\tstar}[1]{#1\texttt{*}}
\newcommand{\tplus}[1]{#1\texttt{+}}
\newcommand{\tquest}[1]{#1\texttt{?}}
\newcommand{\tunit}{\texttt{1}}
\newcommand{\trangeexact}[2]{#1\texttt{\{}#2\texttt{\}}}
\newcommand{\trangemin}[2]{#1\texttt{\{}#2\texttt{,\}}}
\newcommand{\trangemax}[2]{#1\texttt{\{}\texttt{,}#2\texttt{\}}}
\newcommand{\trangebetwn}[3]{#1\texttt{\{}#2\texttt{,}#3\texttt{\}}}
\newcommand{\tset}{\texttt{<-}}
\newcommand{\tappend}{\texttt{+=}}
\newcommand{\tcapture}[2]{#1~\texttt{@}~#2}
\newcommand{\toutput}[1]{\texttt{!} #1}
\newcommand{\taction}[1]{\texttt{[}\,#1\,\texttt{]}}
\newcommand{\desug}[1]{\mathcal{D} \sem{ #1 }}

% Decoration for things that are "symbolic"
\newcommand{\symb}[1]{\widehat{#1}}

% Decoration for things that are "abstract"
\newcommand{\abs}[1]{\overline{#1}}

\newcommand{\pre}{\mathsf{pre}}

\newcommand{\const}{\mathsf{c}}

% Implementation names
\newcommand{\gnuawk}{\texttt{AWK}}
\newcommand{\gnugrep}{\texttt{grep}}
\newcommand{\gnused}{\texttt{sed}}
\newcommand{\gnucut}{\texttt{cut}}

% Save space in list
\newcommand{\compresslist}{     %
    \setlength{\itemsep}{0pt}   %
    \setlength{\parskip}{0pt}   %
    \setlength{\parsep}{0pt}    %
}

% RE example abbreviations
\newcommand{\for}{\mathtt{for}}
\newcommand{\wsp}{\mathtt{\backslash w}}

\section{Introduction}

%We present Kleenex, a declarative language for specifying string transformations, and its compilation to high-performance streaming programs.  
A Kleenex program consists of a context-free grammar, restricted to guarantee regularity, with embedded side-effecting semantic actions.  
%Its semantics is the composition of actions corresponding to the parse tree of an input under greedy disambiguation.  
%Greedy disambiguation corresponds semantically, but avoiding its exponential run-time cost, to using a parser that follows the left alternative of a choice first and backtracks to the right alternative only if it does not lead to a successful parse.
%The compiled programs execute in worst-case linear time in the length of the input with a constant factor proportional to 
%the size of transducer associated with the Kleenex source.  They furthermore execute in a streaming fashion: 
%Informally, an output action is executed as soon as it is uniquely determined from the input prefix read so far.
%In particular, input grammars with finite-lookahead disambiguation execute in constant space.

We illustrate Kleenex by an example.
Consider a large text file containing unbounded numerals, which we want to make more readable by inserting separators; e.g.\ ``12742'' is to be replaced by ``12,742'').
% It is not immediately obvious how to express this task declaratively in a regular expression based tool such as \texttt{sed}, since the number of separators to insert will have to vary with the length of each numeral.
% Alternatively, one may resort to using a scripting language to solve the task in a more imperative fashion.
In Kleenex, this transformation can be specified as follows:
 
\begin{KleenexVerb}
\PY{n+nf}{main}  \PY{o}{:=} (\PY{n+nf}{num} \PY{l+s+sx}{/[^0-9]/} | \PY{n+nf}{other})\PY{o}{*}
\PY{n+nf}{num}   \PY{o}{:=} \PY{n+nf}{digit}\PYZob{}\PY{n+nf}{1},\PY{n+nf}{3}\PYZcb{} (\PY{l+s}{\PYZdq{},\PYZdq{}} \PY{n+nf}{digit}\PYZob{}\PY{n+nf}{3}\PYZcb{})\PY{o}{*}
\PY{n+nf}{digit} \PY{o}{:=} \PY{l+s+sx}{/[0-9]/}
\PY{n+nf}{other} \PY{o}{:=} \PY{l+s+sx}{/./}
\end{KleenexVerb}
This is the complete program.  
The program defines a set of nonterminals, with \KleenexInline!\PY{n+nf}{main}! being the start symbol. The constructs \KleenexInline!\PY{l+s+sx}{/[0-9]/}!, \KleenexInline!\PY{l+s+sx}{/[^0-9]/}! and \KleenexInline!\PY{l+s+sx}{/./}! specify matching a single digit, any non-digit and any symbol, respectively, and echoing the matched symbol to the output. The construct \KleenexInline!\PY{l+s}{\PYZdq{},\PYZdq{}}! reads nothing and outputs a single comma. The star \KleenexInline!\PY{o}{*}! performs the inner transformation zero or more times; the repetition \KleenexInline!\PYZob{}\PY{n+nf}{1},\PY{n+nf}{3}\PYZcb{}! performs it between 1 and 3 times.  Finally, the \KleenexInline!|! operator denotes prioritized choice, with priority given to the left alternative.
An example of its execution is as follows:
\begin{center}
\begin{tabular}{|l|l|} \hline
Input read so far & \ldots and output produced so far\\ \hline 
Surf & Surf \\
Surface:\textvisiblespace  & Surface:\textvisiblespace \\
Surface:\textvisiblespace{}14479 & Surface:\textvisiblespace \\
Surface:\textvisiblespace{}1447985 & Surface:\textvisiblespace \\
Surface:\textvisiblespace{}144798500\textvisiblespace & Surface:\textvisiblespace{}144,798,500\textvisiblespace \\
Surface:\textvisiblespace{}144798500\textvisiblespace{}km\^{}2 & Surface:\textvisiblespace{}144,798,500\textvisiblespace{}km\^{}2 \\ \hline
\end{tabular}
\end{center}
The example highlights the following:
\begin{description}
\item[Ambiguity by design.] Any string is \emph{accepted} by this program, since any string matching \KleenexInline!\PY{n+nf}{num} \PY{l+s+sx}{/[^0-9]/}! also matches
\KleenexInline!(\PY{n+nf}{other})\PY{o}{*}!.  Greedy disambiguation forces the \KleenexInline!\PY{n+nf}{num} \PY{l+s+sx}{/[^0-9]/}! transformation to be tried first, however, and only if that fails do we fall back to echoing the input verbatim to the output using \KleenexInline!\PY{n+nf}{other}!.
\item[Streaming output.] The program almost always detects the earliest possible time an output action can be performed. Any non-digit symbol is written to the output immediately, and as soon as the first non-digit symbol after a sequence of digits is read, the resulting numeral with separators is written to the output stream. The first of a sequence of digits is not output right away, however.  
Employing a strategy that \emph{always} outputs as early as possible would require solving a $\mathsf{PSPACE}$-hard problem. 
\end{description}

%Disambiguation of Kleenex programs is done at the level of transducers.
A Kleenex program is first compiled to a possibly ambiguous \emph{(finite-state) transducer}. Any transducer can be decomposed into two transducers: an \emph{oracle machine}, which maps an input string to a bit-coded representation of the transducer paths accepting the input, and a deterministic \emph{action machine}, which translates such a bit-code to the corresponding sequence of output actions in the original transducer.  
The greedy leftmost path in the oracle machine corresponds to the lexicographically least bit-code of paths accepting a given input; consequently, disambiguation reduces to computing this bit-code for a given input.
To compute it, the oracle machine is simulated in a streaming fashion.  This generalizes NFA simulation to not just yield a single-bit output---accept or reject---but also the lexicographically least path witnessing acceptance.  The simulation algorithm maintains a \emph{path tree} from the initial state to all the oracle machine states reachable by the input prefix read so far. A branching node represents both sides of an alternative where both are still viable. The output actions on the (possibly empty) path segment from the initial state to the first branching node can be performed based on the input prefix processed so far without knowing which of the presently reached states will eventually accept the rest of the input.  This algorithm generalizes \emph{greedy regular expression parsing}~\cite{ghnr2013,grathwohl2014a} to \emph{arbitrary} right-regular grammars. Regular expressions correspond to certain well-structured oracle machines via their McNaughton-Yamada-Thompson construction.  The simulation algorithm \emph{automatically} results in constant memory space consumption for grammars that are deterministic modulo finite lookahead, e.g.~one-unambiguous regular expressions~\cite{brwo98}.  For arbitrary transducers the simulation requires linear space in the size of the input in the worst case.  No algorithm can guarantee constant space consumption: the number of unique path trees computed by the streaming algorithm is potentially unbounded due to the possibility of arbitrarily much lookahead required to determine which of two possible alternatives will eventually succeed.  Unbounded lookahead is the reason that not all unambiguous transducers can be determinized to a finite state machine~\cite{schutzenberger1977,berstel79}.

By identifying path trees with the same ordered leaves and underlying branching structure, we obtain an equivalence relation with finite index.  That is, a path tree can be seen as a rooted full binary tree together with an association of output strings with tree edges, and the set of reachable rooted full binary trees of an oracle machine can can be precomputed analogous to the NFA state sets reachable in an NFA. 
We can thus compile an oracle machine to a \emph{streaming string transducer}~\cite{alur2011streaming,alur2010expressiveness,adr2015}, a deterministic machine model with (unbounded sized) string registers and affine (copy-free) updates associated with each transition: a path tree is represented as an abstract state and the contents of a finite set of registers, each containing a bit sequence coding a path segment of the represented path tree.   Upon reading an input, the state is changed and the registers are updated in-place to represent the subsequent path tree.  This yields a both asymptotically and practically very efficient implementation: the example shown earlier compiles to an efficient C program that operates with sustained high throughput in the 1 Gbps range on stock desktop hardware.

The semantic model of context-free grammars with unbridled ``regular'' ambiguity and embedded semantic actions is flexible and the above implementation technology is quite general.  For example, the action transducer is not constrained to producing output in the string monoid, but can be extended to any monoid.  By considering the monoid of affine register updates, Kleenex can code all nondeterministic streaming string transducers~\cite{alur2011a}.

\subsection{Contributions}

This paper makes the following novel contributions:
\begin{itemize}
\compresslist
\item A \emph{streaming} algorithm for \emph{nondeterministic finite state transducers (FST)}, which emits the lexicographically least output sequence generated by all accepting paths of an input string based on decomposition into an input-processing \emph{oracle machine} and an output-effecting \emph{action machine}.  It runs in $O(m n)$ time for transducers of size $m$ and inputs of size $n$.
%; output symbols are usually emitted as soon as they are uniquely determined by an input prefix,
%This generalizes optimally streaming parsing for regular expressions \cite{grathwohl2014a}, retaining the same performance bounds. 
\item An effective determinization of FSTs into a subclass of \emph{streaming string transducers (SST)} \cite{alur2010expressiveness}, finite state machines with copy-free updating of string registers when entering a new state upon reading an input symbol.  
%\todo{valid claim? see todo.}The number of registers required adapts to the number of output actions in the FST: The fewer output actions the fewer registers.
%In particular, without \todo{clarify}special-casing, no registers are generated---yielding a deterministic finite automaton (DFA).  
\item An expressive declarative language, \emph{Kleenex}, for specifying FSTs with full support for and clear semantics of unrestricted nondeterminism by greedy disambiguation.  
A basic Kleenex program is a context-free grammar with embedded semantic output actions, but syntactically restricted to ensure that the input is regular.\footnote{This avoids the $\Omega(M(n))$ lower bound for context-free grammar parsing, where $M(n)$ is the complexity of multiplying $n \times n$ matrices \cite{lee2002fast}.} Basic Kleenex programs can be functionally composed into pipelines.  
The central technical aspect of Kleenex is its semantic support for unbridled nondeterminism and its effective determinization and compilation to SSTs, which both highlights and complements the significance of SSTs as a deterministic machine model.
\item An implementation, including empirically evaluated optimizations, of Kleenex that generates SSTs and deterministic finite-state machines, each rendered as standard single-threaded C-code that is eventually compiled to x86 machine code.  The optimizations illustrate the design and implementation flexibility obtained by the underlying theories of FSTs and SSTs. 
\item Use cases that illustrate the expressive power of Kleenex, and a performance comparison with related tools, including Ragel~\cite{ragel}, RE2~\cite{re2} and specialized string processing tools.
These document Kleenex's consistently high performance (typically around 1 Gbps, single core, on stock hardware) even when compared to less expressive tools with special-cased algorithms and to tools with no or limited support for nondeterminism.
\end{itemize}

%
%
%
%streaming optimal streaming works for any transducer \\
%$O(m n)$ worst-case bound; linear in both $m$ and $n$ \\
%determinization to subclass of SST, $O(k n)$ runtime, adaptive to $k$ where $k$ is number of non-$\epsilon$ output actions \\
%expressive language, handles nested Kleene-star \\
%many use cases, \\
%efficient and adaptive in practice \\
%expressive (well-founded) \\
%semantically well-founded, purely declarative, reasonable \\
%stock hardware, straightforward implementation, 
%    multistriding, automata optimizations, special-casing of 
%    algorithms to automata classes
%

\subsection{Overview of paper}

In Section~\ref{sec:transducers} we introduce normalized transducers with explicit deterministic and nondeterministic $\epsilon$-transitions.  Kleenex and its translation to such transducers is defined in Section~\ref{sec:kleenex}.  We then devise an efficient streaming transducer simulation (Section~\ref{sec:simulation}) and its determinization (Section~\ref{sec:determinization}) to streaming string transducers.
In Section \ref{sec:benchmarks} we briefly describe the compilation to C-code and some optimizations, and we then empirically evaluate the implementation on a number of simple benchmarks and more realistic use cases (Section~\ref{sec:usecases}).
We conclude with a discussion of related and possible future work (Section~\ref{sec:discussion}).

We assume basic knowledge of automata~\cite{kozen1997}, compilation~\cite{alsu2006}, and algorithms~\cite{clrs2009}.  Basic results in these areas are not explicitly cited. 

\section{Transducers}
\label{sec:transducers}

An \emph{alphabet} $A$ is a finite set; e.g.\ the binary alphabet $\Two = \{ \code{0}, \code{1} \}$ and the empty alphabet $\Zero = \{ \}$. $A^*$ denotes the free monoid generated by $A$, that is the strings over $A$ with concatenation, expressed by juxtaposition, and the empty string $\varepsilon$ as neutral element. 
%Note that $\Zero^* = \{ \varepsilon \}$.  
We write $A[x,\ldots]$ for extending $A$ with additional elements $x, \ldots$ not in $A$.

%Every function $m: A \rightarrow M$, where $A$ is an alphabet and $M$ a monoid
%extends uniquely to a monoid homomorphism, also denoted $m : A^* \rightarrow M$.
%
%Henceforth, let $\Sigma, \Gamma$ denote alphabets not containing the three designated symbols $\epsilon, \epsilon_0, \epsilon_1$.  Their use in transducers will become clear shortly.
%\begin{eqnarray*}
%\overline{a} & = & a, \mbox{ if } a \in \Sigma \\
%\overline{\epsilon} & = & \varepsilon \\
%\overline{\epsilon_0} & = & \varepsilon \\
%\overline{\epsilon_1} & = & \varepsilon
%\end{eqnarray*}
 
\begin{definition}[Finite state transducer]
\label{def:fst}
A \emph{finite state transducer (FST)} $\mathcal{T}$ over $\Sigma$ and $\Gamma$ is a tuple $(\Sigma, \Gamma, Q, q^{-}, q^f, E)$ where
\begin{itemize} \compresslist
\item $\Sigma$ and $\Gamma$ are alphabets;
\item $Q$ is a finite set of \emph{states};
\item $\qinit, \qfinal \in Q$ are the \emph{initial} and \emph{final} states, respectively; 
\item $E : Q \times \Sigma[\epsilon] \times \Gamma[\epsilon] \times Q$
is the \emph{transition relation}.
\end{itemize}
Its \emph{size} is the cardinality of its transition relation: $|T| = |E|$.

$\Cal T$ is \emph{deterministic} if for all $q \in Q, a \in \Sigma[\epsilon]$ we have
\begin{eqnarray*}
(q, a, b', q') \in E \wedge (q, a, b'', q'') \in E & \Rightarrow & b' = b'' \wedge q' = q'' \\(q, \epsilon, b', q') \in E \wedge (q, a, b'', q'') \in E & \Rightarrow &\epsilon = a
\end{eqnarray*}

The \emph{support} of a state is the set of symbols it has transitions on: $$\supp(q) = \{ a \in \Sigma[\epsilon] \mid \exists q', b \ldotp (q, a, b, q') \in E \}.$$ 
\end{definition}
\noindent Deterministic FSTs with no $\epsilon$-transitions and $\supp(q) = \Sigma$ for all $q$ are \emph{Mealy machines}.  Conversely, every deterministic FST is easily turned into a Mealy machine by adding a failure state and transitions to it.

We write $q \trans{a}{b} q'$ whenever $(q,a,b,q') \in E$, and $E$ is understood from the context. 
A \emph{path} in $\mathcal{T}$ is a possibly empty sequence of transitions
\[
q_0 \trans{a_1}{b_1} q_1 \trans{a_2}{b_2} \ldots \trans{a_n}{b_n} q_n
\]
It has \emph{input} $u=a_1 a_2 \ldots a_n$ and \emph{output} $v=b_1 b_2 \ldots b_n$. 
We write $q_0 \trans{u}{v} q_n$ if there exists such a path. 

\begin{definition}[Relational semantics, input language]
\label{def:relsem}
FST $\mathcal{T}$ denotes the binary relation
$$\sem{\mathcal{T}} = \{ (\overline u, \overline v) \mid q^- \trans{u}{v} q^f \}$$
where the \emph{$\epsilon$-erasure} $\overline{\cdot} : \Sigma[\epsilon]^* \rightarrow \Sigma^*$ is $\overline\epsilon = \varepsilon$ and $\overline a = a$ for all $a \in \Sigma$, extended homomorphically to strings.  
Its \emph{input language} is 
$$\lsem{\mathcal{T}} = \{ s \mid \exists t \, . \, (s, t) \in \sem{\mathcal{T}} \}.$$
Two FSTs are \emph{equivalent} if they have the same relational semantics.
\end{definition}
The class of relations denotable by FSTs are the \emph{rational relations}; their input languages are the \emph{regular languages}~\cite{berstel79}.
%the smallest class of binary relations containing the empty relation and all singleton relations that are closed under union, relational composition and relational iteration.

\begin{definition}[Normalized FST]
A \emph{normalized finite state transducer} over $\Sigma$ and $\Gamma$ is a deterministic FST over $\Sigma[\epsilon_0, \epsilon_1]$ and $\Gamma$ such that for all $q \in Q$, $q$ is:
\begin{itemize} \compresslist
\item a \emph{choice state}: $\supp(q) = \{ \epsilon_0, \epsilon_1 \}$ and $q \neq q^f$, or
\item a \emph{skip state}: $\supp(q) = \{ \epsilon \}$ and $q \neq q^f$, or
\item a \emph{symbol state}: $\supp(q) = \{ a \}$ for some $a \in \Sigma$ and $q \neq q^f$, or
\item the \emph{final state}: $\supp(q) = \{ \}$ and $q = q^f$
\end{itemize}
\end{definition}
We say that $q$ is a \emph{resting state} 
%and write $q \downarrow$ 
if $q$ is either a symbol state or the final state. 
%In normalized FSTs, any path that ends in a resting state can be decomposed as follows:
%\begin{proposition}
%  \label{prop:normalized-decomp}
%  If $\Cal T$ is normalized, then $p \path{uv}{z} r \downarrow$ if and only if there exists a $q \in Q$ such that $z = xy$ and $p \path{u}{x} q \downarrow$ and $q \path{v}{y} r \downarrow$.
%\end{proposition}

The relational semantics $\sem{\Cal T}$ of a normalized FST is the same as in Definition~\ref{def:relsem}, where $\epsilon$-erasure is extended by $\overline\epsilon_0 = \overline\epsilon_1 = \varepsilon$. 

\begin{proposition}
For every FST of size $m$ there exists an equivalent normalized FST of size at most $3m$.  
Conversely, for every normalized FST of size $m$ there exists an equivalent FST of the same size.
\end{proposition}
\begin{proof} (Sketch) For each state $q$ with $k>1$ outgoing transitions, add $k$ new states $q^{(1)}, \ldots, q^{(k)}$, replace the $i$-th outgoing transition $(q, a, b, q')$ by
$(q^{(i)}, a, b, q')$ and add a full binary tree of $\epsilon_0$- and $\epsilon_1$-transitions for reaching each $q^{(i)}$ from $q$.  In the converse direction, replace $\epsilon_0$ and $\epsilon_1$ by $\epsilon$. 
\end{proof}
Normalized FSTs are useful by limiting transition outdegree to $2$, having explicit $\epsilon$-transitions and classifying them into deterministic ($\epsilon$) and ordered nondeterministic ones ($\epsilon_0, \epsilon_1$). 

\textbf{Proviso.} Henceforth we will call normalized FSTs simply \emph{transducers}.

Let $|{\cdot}| : \Sigma[\epsilon_0, \epsilon_1, \epsilon] \rightarrow \Two[\epsilon]$ be defined by $|\epsilon_0| = \code{0}, |\epsilon_1| = \code{1}$ and $|a| = \epsilon$ for
all $a \in \Sigma[\epsilon]$.
\begin{definition}[Oracle and action machines]
\label{def:oracle_action_machines}
Let $\Cal T$ be a transducer.  The \emph{oracle machine} $\Cal{T}^\Code$ is defined as $\Cal T$, but with each transition $(q, a, b, q')$ replaced by
$(q, a, |a|, q')$.  
Its \emph{action machine} $\Cal{T}^\Act$ is $\Cal T$, but with each transition $(q, a, b, q')$ replaced by $(q, |a|, b, q')$.  
\end{definition}
The oracle machine is a transducer over $\Sigma$ and $\Two$; the action machine a deterministic FST over $\Two$ and $\Gamma$.
Each transducer can be canonically decomposed into its oracle and action machines:
\begin{proposition}
\label{prop:transducer_decomposition}
$\sem{\Cal T} = \sem{\Cal{T}^\Act} \circ \sem{\Cal{T}^\Code}$
\end{proposition}
\noindent where $\circ$ denotes relational composition.  Note that the oracle machine is independent of the outputs in the original transducer; in particular, a transducer where only the outputs are changed has the same oracle machine.  
%A bit string emitted by the oracle machine uniquely identifies a complete path from initial to any resting state \todo{final state -- semantics doesn't account for incremental computation yet} in the transducer:
%\begin{proposition}
%  \label{prop:uniqueness}
%  For all $q \in Q$, if $q {\downarrow}$ and $q^- \trans{u}{v} q$ and $q^- \trans{u'}{v} q$ then $u=u'$.
%\end{proposition}
Intuitively, the action machine starts at the initial state the original transducer, automatically follows transitions from resting and skip states, and uses the bit string from the oracle machine as an oracle---hence the name---to choose which transition to take from a choice state; in this process it emits the outputs it traverses.  

\begin{figure*}
    \tikzset{resting/.style={very thick,fill=black!10}}
\tikzset{restingtrans/.style={very thick}}
\newcommand{\drawnormalizedfst}%
{\begin{tikzpicture}[>=stealth',x=0.7cm,y=0.5cm,baseline=(current bounding box.west)]
\begin{scope}[state/.append style={minimum size=0.45cm,inner sep=0}]
\node[state,initial above,initial distance=0.18cm,\bstyle] (s1) at (0,0) {$1$};
\node[state,\astyle] (s2) at (2,0) {$2$};
\node[state,\bstyle] (s3) at (4,0) {$3$};
\node[state,\astyle] (s4) at (6,0) {$4$};
\node[state,\bstyle] (s5) at (8,0) {$5$};
\node[state,\astyle] (s6) at (10,0) {$6$};
\node[state,\bstyle] (s7) at (12,0) {$7$};
\node[state] (s8) at (14,0) {$8$};
\node[state,\astyle] (s9) at (14,-2) {$9$};
\node[state,\astyle] (s10) at (16.25,-2) {$10$};
\node[state,\astyle] (s11) at (16.25,0) {$11$};
\node[state,accepting,\fstyle] (s12) at (0,-2) {$12$};
\node[state,\astyle] (s13) at (12,-2) {$13$};
\end{scope}
\begin{scope}[lab/.style={auto,inner sep=1pt}]
\foreach \from/\to in {1/2, 3/4, 5/6, 7/8}
  \draw[->,\bstyletrans] (s\from) to node[lab] {\zlabel} (s\to);
\draw[->,\bstyletrans] (s1) to node[lab] {\olabel} (s12);
\draw[rounded corners,->,\bstyletrans] (s3) -- (4.5,-1) -- node[lab,swap] {\olabel} (7.5,-1) -- (s5);
\draw[rounded corners,->,\bstyletrans] (s5) -- (8.5,-1) -- node[lab,swap] {\olabel} (11.5,-1) -- (s7);
\draw[->,\bstyletrans] (s7) to node[lab] {\olabel} (s13);
\foreach \from/\to in {2/3, 4/5, 6/7, 10/11}
  \draw[->,\astyletrans] (s\from) to node[lab] {\alabel} (s\to);
\draw[->,\astyletrans] (s9) to node[lab,swap] {\alabel} (s10);
\draw[rounded corners,->,\astyletrans] (s13) -- node[lab] {\nlabel} (4,-2) -- (s1);
\draw[rounded corners,->,\astyletrans] (s11) -- (15.75,1) -- node[lab,swap] {\alabel} (12.5,1) -- (s7);
\draw[->] (s8) to node[lab] {\clabel} (s9);
\end{scope}
\end{tikzpicture}}

\begin{minipage}{0.2\linewidth}
\begin{KleenexVerb}
\PY{n+nf}{main} \PY{o}{:=} (\PY{n+nf}{num} \PY{l+s+sx}{/\PY{esc}{\PYZbs{}n}/})\PY{o}{*}
\PY{n+nf}{num} \PY{o}{:=} \PY{n+nf}{digit}\PYZob{}\PY{n+nf}{1},\PY{n+nf}{3}\PYZcb{} (\PY{l+s}{\PYZdq{},\PYZdq{}} \PY{n+nf}{digit}\PYZob{}\PY{n+nf}{3}\PYZcb{})\PY{o}{*}
\PY{n+nf}{digit} \PY{o}{:=} \PY{l+s+sx}{/a/}
\end{KleenexVerb}
\end{minipage}

%\begin{minipage}{0.28\linewidth}
%\begin{align*}\footnotesize
%  M \mathtt{:=}& N_1 \mathtt{ | } \epsilon \\
%  N_1 \mathtt{:=}& \mathtt{a "a"} ((\mathtt{a "a"} N_2) \mathtt{ | } N_2) \\
%  N_2 \mathtt{:=}& (\mathtt{a "a"} N_3) \mathtt{ | } N_3 \\
%  N_3 \mathtt{:=}& (\mathtt{"," aaa } N_3) \mathtt{ | } (\texttt{\textbackslash{}n"\textbackslash{}n"} M)
%\end{align*}
%\end{minipage}
\begin{center}
{\newcommand{\zlabel}{$\epsilon_0 / \epsilon$}
\newcommand{\olabel}{$\epsilon_1 / \epsilon$}
\newcommand{\alabel}{$\texttt{a} / \texttt{a}$}
\newcommand{\nlabel}{$\texttt{\textbackslash n} / \texttt{\textbackslash n}$}
\newcommand{\clabel}{$\epsilon / \texttt{,}$}
\newcommand{\bstyle}{}\newcommand{\bstyletrans}{}
\newcommand{\astyle}{}\newcommand{\astyletrans}{}
\newcommand{\fstyle}{}
\drawnormalizedfst}
\end{center}

\hrule

\begin{center}
{\newcommand{\zlabel}{$\epsilon_0 / \code{0}$}
\newcommand{\olabel}{$\epsilon_1 / \code{1}$}
\newcommand{\alabel}{$\texttt{a} / \epsilon$}
\newcommand{\nlabel}{$\texttt{\textbackslash n} / \epsilon$}
\newcommand{\clabel}{$\epsilon / \epsilon$}
\newcommand{\bstyle}{}\newcommand{\bstyletrans}{}
\newcommand{\astyle}{resting}\newcommand{\astyletrans}{restingtrans}
\newcommand{\fstyle}{}
\drawnormalizedfst}
{\newcommand{\zlabel}{$\code{0}/\epsilon$}
\newcommand{\olabel}{$\code{1}/\epsilon$}
\newcommand{\alabel}{$\epsilon/\texttt{a}$}
\newcommand{\nlabel}{$\epsilon/\texttt{\textbackslash n}$}
\newcommand{\clabel}{$\epsilon / \texttt{,}$}
\newcommand{\bstyle}{resting}\newcommand{\bstyletrans}{restingtrans}
\newcommand{\astyle}{}\newcommand{\astyletrans}{}
\newcommand{\fstyle}{}
\drawnormalizedfst}
\end{center}

%%% Local Variables:
%%% mode: latex
%%% TeX-master: "../thesis"
%%% End:
    \caption[Kleenex program with transducer, oracle and action machines.]{Top: a Kleenex program and its associated transducer. The program accepts a list of newline-separated numbers (simplified to unary numbers with digit \texttt{a}) and inserts thousands separators. Bottom: The corresponding oracle and action machines.
}
\label{fig:decimal_example}
\end{figure*}
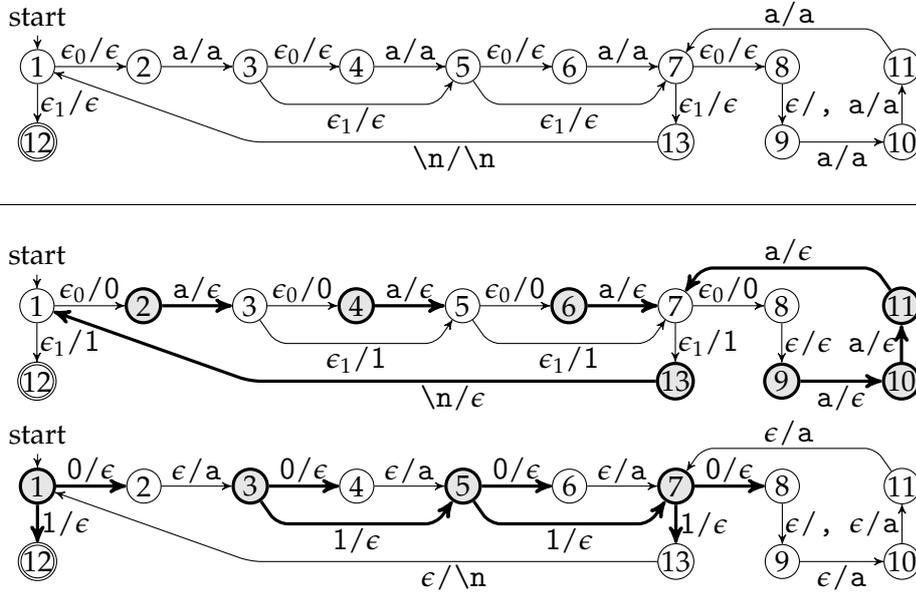

\begin{example}
Figure~\ref{fig:decimal_example} shows a Kleenex program (see Section~\ref{sec:kleenex}), the associated transducer and its decomposition into oracle and action machines.
\end{example}

Observe that if there is a path $q \trans{u}{v} q'$ then
$u$ uniquely identifies the path from $q$ to $q'$ in a transducer and, furthermore, in an oracle machine so does $v$.

We write $q \pathnp{u}{v} q''$ if the path $q \trans{u}{v} q''$ does not contain an \emph{$\epsilon$-loop}, that is a subpath $q' \trans{u'}{v'} q'$ where 
$\overline{u'} = \varepsilon$.  Paths without $\epsilon$-loops are called 
\emph{nonproblematic} paths~\cite{frca2004}.

\begin{definition}[Greedy semantics]
The \emph{greedy semantics} of a transducer $T$ is
$\gsem{\Cal T} = \sem{\Cal{T}^\Act} \circ \gsem{\Cal{T}^\Code}$ where
\begin{eqnarray*}
\gsem{\Cal{T}^\Code} & = & \{ (\overline u, \overline v) \mid q^- \pathnp{u}{v} q^f \wedge \\
                   & & \forall u', v'. \, q^- \pathnp{u'}{v'} q^f \wedge \overline u = \overline{u'} \Longrightarrow \overline v \leq \overline{v'} \}
\end{eqnarray*}
and $\leq$ denotes the lexicographic ordering on bit strings. 
\end{definition}
Given input string $s$, the greedy semantics chooses the lexicographically least path in the transducer accepting $s$ and outputs the corresponding output symbols encountered along the path.  The restriction to nonproblematic paths ensures that there are only finitely many paths accepting $s$ and thus the lexicographically least amongst them exists, if $s$ is accepted at all.
We write $q \pathmin{u}{v} q'$ if $q \trans{u}{v} q'$ is the lexicographically least nonproblematic path from $q$ to $q'$.

A transducer $\Cal T$ over $\Sigma$ and $\Gamma$ is \emph{single-valued} if $\sem{\Cal T}$ is a partial function from $\Sigma^*$ to $\Gamma^*$.  
\begin{proposition}
Let $\Cal T$ be a transducer over $\Sigma$ and $\Gamma$.
\begin{itemize}
\item $\gsem{\Cal T}$ is a partial function from $\Sigma^*$ to $\Gamma^*$. 
\item $\gsem{\Cal T} = \sem{\Cal T}$ if $\Cal T$ is single-valued.
\end{itemize} 
\end{proposition}
The greedy semantics 
can be thought of as a \emph{disambiguation policy} for transducers that
conservatively extends the standard semantics for single-valued transducers to a deterministic semantics for arbitrary transducers.

%Why the restriction to nonproblematic paths in the above definition? Consider the following transducer $\mathcal{T}$:
%\begin{center}
%\begin{tikzpicture}[>=stealth',auto,every node/.style={scale=0.8}]
%\node[state,initial,xshift=-0.5cm] (q0) {$q_1$};
%\node[state,accepting,right=of q0,xshift=0.75cm] (q1) {$q_2$};
%\draw[->] (q0) to node {$a/\code{1}$} (q1);
%\draw[loop,->,looseness=5] (q0) to node[swap] {$\epsilon/\code{0}$} (q0);
%\end{tikzpicture}
%\end{center}
%Then $\min \{ v \mid q_1 \path{a}{v} q_2 \}$ is not well-defined, as evidenced by the following infinitely descending chain of outputs:
%$\code{1} \geq \code{01} \geq \code{001} \geq \code{0001} \geq ...$. Operationally, such a chain corresponds to a non-terminating backtracking search.
%On the other hand, the number of nonproblematic paths with a given input label is always finite, ensuring well-foundedness of the lexicographic order. Every problematic path has a corresponding nonproblematic path with the same input label; consequently, $\dom(\sem{\mathcal{T}}_{\leq}) = \dom(\sem{\mathcal{T}})$.

%It should be noted that the refined semantics and ordering of transitions does not buy us anything in terms of expressive power.
%\begin{proposition}
%\label{prop:functionalization}
%For any FST $\mathcal{T}$, there exists an FST $\mathcal{T}'$ such that $\sem{\mathcal{T}}_\leq = \sem{\mathcal{T}'}$.
%\end{proposition}

\section{Kleenex}
\label{sec:kleenex}

Kleenex\footnote{Kleenex is a contraction of \emph{Kleene} and \emph{expression} in recognition of the fundamental contributions by Stephen Kleene to language theory.}
is a language for compactly and conveniently expressing transducers.
%and, more generally, nondeterministic action machines including nondeterministic streaming string transducers. 
%Syntactically a Kleenex program is a context-free grammar with embedded semantic actions.
%Semantically, it denotes a function that transforms an input string from a regular language into a sequence of actions determined by the greedy leftmost parse of the input, with the caveat that the grammar be regular.  

\subsection{Core Kleenex}

Core Kleenex is a grammar for directly coding transducers.  

%For the remainder of the section we fix input and action alphabets $\Sigma$ and $\Gamma$, 
%and we assume that $\Sigma \subseteq \Gamma$.

\begin{definition}[Core Kleenex syntax]
\label{def:syntax}
A \emph{Core Kleenex} program is a nonempty list $p = d_0 d_1 \hdots d_n$ of \emph{definitions} $d_i$, each of the form $N \texttt{:=}\ t$, where $N$ is an identifier and $t$ is generated by the grammar
\begin{align*}
  t &::= \varepsilon \mid N \mid a~N' \mid \texttt{"} b \texttt{"} ~ N' \mid N_0 \texttt{|} N_1
\end{align*}
where $a \in \Sigma$ and $b \in \Gamma$ for given alphabets $\Sigma, \Gamma$, e.g.\ some character set. $N$ ranges over some set of identifiers.  The identifiers occurring in $p$ are called \emph{nonterminals}.  There must be at most one definition of each nonterminal, and every occurrence of a nonterminal must have a definition.
\end{definition}

\begin{definition}[Core Kleenex transducer semantics]
\label{def:kleenex-semantics}
The \emph{transducer associated with Core Kleenex program $p$} for nonterminal $N \in \mathcal{N}$ is 
$$\Cal{T}_p(N) = (\Sigma, \Gamma, \mathcal N[q^f], N, q^f, E)$$ where
$\mathcal N$ is the set of nonterminals in $p$, and
$E$ consists of transitions constructed from each production in $p$ as follows:
\[ \begin{array}{|l|l|} \hline
N \mathtt{:=} \, \varepsilon & N \trans{\epsilon}{\epsilon} q^f \\ \hline
N \mathtt{:=} \, N' & N \trans{\epsilon}{\epsilon} N' \\ \hline
N \mathtt{:=} \, a~N' & N \trans{a}{\epsilon} N' \\ \hline
N \mathtt{:=} \, \texttt{"} b \texttt{"} ~ N' & N \trans{\epsilon}{b} N' \\ \hline
N \mathtt{:=} \, N' \texttt{|} N''& N \trans{\epsilon_0}{\epsilon} N' \mbox{ and } \\
                                & N \trans{\epsilon_1}{\epsilon} N'' \\ \hline
\end{array} \]
The \emph{semantics} of $p$ is the greedy semantics of its associated transducer: $\gsem{p} = \gsem{\Cal{T}_p}(N_0)$ where $N_0$ is a designated start nonterminal. (By convention, this
is $\texttt{main}$.)

\end{definition}
%The semantics of Kleenex is equivalent to parsing the input using a backtracking 
%parser, always trying the left alternative first and backtracking to the second alternative
%only if that fails, and emitting the embedded output characters once a parse has been found.
%The disadvantage is that this takes exponential time in the worst case and requires 
%random access to the input.  For large inputs, say hundreds of gigabytes, this is problematic for several reasons.  The key point of this paper is that an equivalent transducer-based semantics and its streaming implementation can guarantee linear time execution and ``earliest-possible'' outputting without the need for storing the input string. 

\subsection{Standard Kleenex}

We extend the syntax of right-hand sides in Kleenex productions with arbitrary concatenations of the form and $N' N''$ and slightly simplify the remaining rules as follows:
\begin{align*}
  t &::= \varepsilon \mid N \mid a \mid \texttt{"} b \texttt{"} \mid N_0 \texttt{|} N_1
\mid N' N''
\end{align*}

Let $p$ be such a \emph{Standard Kleenex} program.  Its \emph{dependency graph} $G_p = (\mathcal N, D)$ consists of its nonterminals $\mathcal N$ and the \emph{dependencies}
\[D = \{ N \rightarrow N' \mid N' \mbox{ occurs in the definition of } N \mbox{ in } p \}.\] Define the \emph{strict dependencies} $D_s = \{ N \rightarrow N'  \mid (N \texttt{:=} N' N'') \in p \}$.

\begin{definition}[Well-formedness]
A Standard Kleenex program $p$ is \emph{well-formed} if no strong component of $G_p$ contains a strict
dependency.  
\end{definition}
Well-formedness ensures that the underlying grammar is non-self-embedding  
\cite{anselmo2003finite}, and thus its input language is regular.  

\begin{definition}[Kleenex syntax and semantics]
Let $p$ be a well-formed Kleenex program with nonterminals $\mathcal N$.  Define the transitions $E \subseteq \mathcal N^* \times \Sigma[\epsilon_0, \epsilon_1, \epsilon] \times \Gamma[\epsilon] \times \mathcal N^*$ as 
follows:
\[ \begin{array}{|l|l|} \hline
\mbox{For rule $d$} & \mbox{add these transitions for all } X \in \mathcal N^* \mbox{ to } E \\ \hline \hline
N \mathtt{:=} \, \varepsilon & N X \trans{\epsilon}{\epsilon} X \\ \hline
N \mathtt{:=} \, N' & N X \trans{\epsilon}{\epsilon} N' X \\ \hline
N \mathtt{:=} \, a & N X \trans{a}{\epsilon} X \\ \hline
N \mathtt{:=} \, \texttt{"} b \texttt{"} & N X \trans{\epsilon}{b} X \\ \hline
N \mathtt{:=} \, N' \, N'' & N X \trans{\epsilon}{\epsilon} N' N'' X \\ \hline
N \mathtt{:=} \, N' \texttt{|} N'' & N X \trans{\epsilon_0}{\epsilon} N' X \mbox{ and } \\
                                & N X \trans{\epsilon_1}{\epsilon} N'' X \\ \hline                                
\end{array} \]
Let $\Reach{N} = \{ \vec N_k \mid N \trans{.}{.} \ldots \trans{.}{.} \vec N_k \}$ be the nonterminal sequences reachable from $N$ along transitions in $E$.
The \emph{transducer $\Cal{T}_p$ associated with $p$} is $(\Sigma, \Gamma, R, N, \varepsilon, E|_R)$ 
where $R = \Reach{N}$ for designated start symbol $N$ and $E|_R$ is $E$ restricted to $R$.  The \emph{(greedy) semantics}
of $p$ is the greedy semantics of $\Cal{T}_p$: $\gsem{p} = \gsem{\Cal{T}_p}$.
\end{definition}
The following proposition justifies calling $\Cal{T}_p$ a transducer.
\begin{proposition}
Let $p$ be a well-formed Standard Kleenex program, with $\Cal{T}_p$ as defined above.
Then $R$ is finite, and $\Cal{T}_p$ is a transducer, that is normalized FST.
\end{proposition}
\begin{proof} (Sketch)
$\Reach{N}$ consists of all the nonterminal suffixes of sentential forms of left-most derivations of $p$ considered as a context-free grammar.  In well-formed Kleenex programs, their maximum length is bounded by $|\mathcal{N}|$.
It is easy to check that every state in $R$ is either a resting, skip, choice or final state.
%Let $k$ be the strong component rank of a nonterminal ($k=0$ means $N$ doesn't have right-hand side of
%form $N'N''$, and the same holds recursively true for all the nonterminal on $N$'s right-hand side, if any).
%
\end{proof}
\noindent Observe that the transducer associated with a Kleenex program can be exponentially bigger than the program itself.

Since a transducer has a straightforward representation in Core Kleenex, the construction of $\Cal{T}_p$ provides a translation of a well-formed Standard Kleenex program into Core Kleenex. For example, the Kleenex program on the left translates into the Core Kleenex program on the right:
\begin{align*}
  \begin{array}{@{}r@{}l}
    M \texttt{ := }{}& M' \texttt{|} N \\
    M' \texttt{ := }{}& N N_a \\
    N_a \texttt{ := }{} & a \\
    N \texttt{ := }{}& N' \texttt{|} N_\varepsilon \\
    N' \texttt{ := }{} & N_b N \\
    N_b \texttt{ := }{} & b \\
    N_\varepsilon \texttt{ := }{}& \varepsilon
  \end{array}
                       \Longrightarrow{}&
  \begin{array}{r@{}l@{}}
  M \texttt{ := }{}& N' \texttt{|} N \\
  M' \texttt{ := }{}& N' \texttt{|} N_a \\
  N_a \texttt{ := }{} & a N_\varepsilon \\ 
  N' \texttt{ := }{} & b M' \\
  N \texttt{ := }{}& N' \texttt{|} N_\varepsilon \\
  N_\varepsilon \texttt{ := }{}& \varepsilon
  \end{array}
\end{align*}

\subsection{The Full Surface Language}

The full surface syntax of \emph{Kleenex} is obtained by extending Standard Kleenex with the following term-level constructors, none of which increase the expressive power:
\begin{align*}
  t ::= \hdots &\mid \tact{v} \mid \tdef{e} \mid \tsup{t} \mid t_0 t_1 
  \mid t_0\texttt{|}t_1 \mid \tstar{t} \mid \tplus{t} \mid \tquest{t} \\
               &\mid \trangeexact{t}{n} \mid \trangemin{t}{n} \mid \trangemax{t}{m} \mid \trangebetwn{t}{n}{m}
\end{align*}
where $v \in \Gamma^*$, $n, m\in\Nat$, and $e$ is a \emph{regular expression}.
The terms $t_0 t_1$ and $t_0\texttt{|}t_1$ desugar into $N_0 N_1$ and $N_0\texttt{|}N_1$, respectively,
with additional productions $N_0 \texttt{ := } t_0$ and $N_1 \texttt{ := } t_1$ for new nonterminals $N_0, N_1$. 
The term $\tact{v}$ is shorthand for a sequence of outputs. 

Regular expressions are special versions of Kleenex terms without nonterminals.
They desugar to terms that output the matched input string, i.e.~$\tdef{e}$ desugars by adding an output symbol $\tact{a}$ after every input symbol $a$ in $e$.
For example, the regular expression $\tdef{\texttt{a*|b\{n,m\}|c?}}$ becomes \[\tstar{(a\tact{a})} \texttt{|} \trangebetwn{(b\tact{b})}{n}{m} \texttt{|} \tquest{(c\tact{c})},\] which can then be further desugared.

A \emph{suppressed} subterm $\tsup{t}$ desugars into $t$ with all output symbols removed, including any that might have been added in $t$ by the above construction.
For example, $\tsup{(\tact{b} \tdef{a})}$ desugars into $\tsup{(\tact{b} \, a \, \tact{a})}$, which further desugars into $a$.

The operators $\tstar{\cdot}, \tplus{\cdot}$ and $\tquest{\cdot}$ desugar to their usual meaning as regular operators, as do the repetition operators $\trangeexact{\cdot}{n}$, $\trangemin{\cdot}{n}$, $\trangemax{\cdot}{m}$, and $\trangebetwn{\cdot}{n}{m}$. Note that they all desugar into their \emph{greedy} variants where matching a subexpression is preferred over skipping it. For example:
\begin{align*}
  \begin{array}{@{}r@{}l}
    M \texttt{ := }{}& \tplus{(a \, \tact{b})}
  \end{array}
                       \Longrightarrow{}&
  \begin{array}{r@{}l@{}}
  M \texttt{ := }{}& (a \, \tact{b})N' \\
  N' \texttt{ := }{}& a \, \tact{b} N' \texttt{|} \varepsilon
  \end{array}
\end{align*}
\emph{Lazy} variants can be encoded by making $\varepsilon$ the left rather than the right choice of an alternative.

\subsection{Register Update Actions}
\label{sec:kleenex-register-update-actions}

By viewing $\Gamma$ as an alphabet of \emph{effects}, we can extend the expressivity of Kleenex beyond rational functions \cite{berstel79}.  Let $X$ be a computable set, and assume that there is an effective partial action $\Gamma \times X \to X$. It is simple to define a deterministic machine implementing the function $\Gamma^* \times X \to X$ by successively applying a list of actions to some starting state $X$. Any Kleenex program then denotes a function $\Sigma^* \times X \to X$ by composing its greedy semantics with such a machine. If we can implement the pure transducer part in a streaming fashion, then a state $X$ can be maintained on-the-fly by interpreting output actions as soon as they become available.

Let $X = (\Gamma^*)^+ \times (\Gamma^*)^n$ for some $n$, representing a non-empty stack of output strings and $n$ string registers. The transducer output alphabet is extended to $\Gamma[\mathsf{push},\mathsf{pop}_0, ..., \mathsf{pop}_n, \mathsf{write}_0, ..., \mathsf{write}_n]$, with actions defined by
\begin{align*}
  (t \vec{w}, v_0, ..., v_n) \cdot a ={}& ((ta) \vec{w},v_0, ..., v_n) & \text{ ($a \in \Gamma$)} \\
  (\vec{w}, v_0, ..., v_n) \cdot \mathsf{push} ={}& ((\varepsilon) \vec{w}, v_0, ..., v_n) \\
%  (t, v_0, ..., v_n) \cdot \mathsf{pop}_i ={}& (t, v_0, ..., v_n) \\
  (t \vec{w}, v_0, ..., v_i, ..., v_n) \cdot \mathsf{pop}_i ={}& (\vec{w}, v_0, ..., t, ..., v_n) & \text{($|\vec{w}|>0$)}\\
  (t \vec{w}, v_0, ..., v_n) \cdot \mathsf{write}_i ={}& ((t v_i) \vec{w}, v_0, ..., v_n)
\end{align*}
The bottom stack element can only be appended to and models a designated output register---popping it is undefined. The stack and the variables can be used to perform complex string interpolation. To access the extended actions, we extend the surface language:
\begin{align*}
    t ::= \hdots{} & \mid \tcapture{R}{t} \mid \toutput{R} \\
                   & \mid \taction{R~\tset{}~ (R\mid\tact{v})^\star} %\\ 
                     \mid \taction{R ~\tappend{}~ (R \mid \tact{v})^\star}
\end{align*}
where $R$ ranges over \emph{register names} standing for indices.

The term $\tcapture{R}{t}$ desugars to $\tact{\mathsf{push}} ~ t ~ \tact{\mathsf{pop}_R}$, and the term $\toutput{R}$ desugars to $\tact{\mathsf{write}_R}$. The term $\taction{R ~\tset{}~ x_1 ... x_m}$ desugars to $\tact{\mathsf{push}} t_1' ... t_m' \tact{\mathsf{pop}_R}$, where $t_i' = \mathsf{write}_{R_i}$ if $x_i = R_i$, and $t_i' = x_i$ otherwise. Finally, $\taction{R ~\tappend{}~ \vec{x}}$ desugars to $\taction{R ~\tset{}~ R~\vec{x}}$.

Thus all streaming string transducers (see Section~\ref{sec:determinization}) can be coded. As an example, the following program swaps two input lines by storing them in registers \texttt{a} and \texttt{b} and outputting them in reverse order:
\begin{KleenexVerb}
\PY{n+nf}{main} \PY{o}{:=} \PY{n+nf}{a}\PY{o}{@}\PY{n+nf}{line} \PY{n+nf}{b}\PY{o}{@}\PY{n+nf}{line} \PY{o}{!}\PY{n+nf}{b} \PY{o}{!}\PY{n+nf}{a}
\PY{n+nf}{line} \PY{o}{:=} \PY{l+s+sx}{/[^\PY{esc}{\PYZbs{}n}]*\PY{esc}{\PYZbs{}n}/}
\end{KleenexVerb}
where the first line above desugars to
\begin{align*}
  \mathtt{main} \texttt{ := }& \tact{\mathsf{push}}~\texttt{line}~\tact{\mathsf{pop}_a}~\tact{\mathsf{push}}~\texttt{line}~\tact{\mathsf{pop}_b}\\ & \quad \tact{\mathsf{write}_b}~\tact{\mathsf{write}_a}
\end{align*}

\section{Streaming Simulation}
\label{sec:simulation}

As we have seen, every Kleenex program has an associated transducer, which can be
split into oracle and action machines.  The action machine is a straightforwardly implemented deterministic FST.  The oracle machine is nondeterministic, however: The key challenge is how to (deterministically) find and output the lexicographically least path that accepts a given input string.
In this section we develop an efficient oracle machine simulation algorithm that inputs a stream of symbols and streams the output bits almost as early as possible during input processing.

\subsection{Path Trees}

Given an oracle machine $\Cal{T}^\Code$ as in Definition~\ref{def:oracle_action_machines}, consider input $s$ such that  
$q^- \pathmin{u}{v} q^f$ where $\overline u = s$.  Recall that $q \pathmin{u}{v} q'$ uniquely identifies a path from 
$q$ to $q'$ in $\Cal{T}^\Code$, which is furthermore asserted to be the lexicographically minimal amongst all nonproblematic paths from $q$ to $q'$.

\begin{proposition}[Path decomposition]
\label{prop:path_decomposition}
Assume $q^- \pathmin{u}{v} q^f$.
For every prefix $s'$ of $\overline u$ there exist unique $u', v', u'', v'', q'$ such that $q^- \pathmin{u'}{v'} q' \pathmin{u''}{v''} q^f$, $q'$ is a resting state, $\overline{u'}=s'$, $u' u'' = u$ and $v' v'' = v$.
\end{proposition} 
\begin{proof}
Let $u'$ be the longest prefix of $u$ such that $\overline{u'}=s'$ and let $q^- \pathnp{u'}{v'} q'$
be the path from $q$ determined by $u'$.  (Such a prefix must exist.)  
Claim: This is the $q'$ in the proposition.
\begin{enumerate}
\item $q'$ is a resting state.  If it were not, we could transition on $\epsilon, \epsilon_0$ or $\epsilon_1$ resulting in a longer prefix $w$ with $\overline w = s'$.
\item $q^- \pathmin{u'}{v'} q'$ and $q' \pathmin{u''}{v''} q^f$.  
If any of these subpaths were not lexicographically minimal, we could replace it with one that is lexicographically less, resulting in a path from $q^-$ to $q^f$ that is lexicographically less than $q^- \pathnp{u}{v} q^f$, contradicting our assumption $q^- \pathmin{u}{v} q^f$. \qedhere \end{enumerate}
\end{proof}

After reading input prefix $s'$ we need to find the above $q^- \pathmin{u'}{v'} q'$
where $\overline{u'}=s'$.  Since we do not know the remaining input yet, however, we maintain \emph{all} paths $q^- \pathmin{u'}{v'} q'$ for any resting state $q'$ such that $\overline{u'}=s'$.

\begin{definition}[Path tree]
Let $\Cal{T}^\Code$ be given.
Its \emph{path tree} $\Pathtree(s)$ for $s$ is the set of paths 
$\{ q^- \pathmin{u}{v} q' \mid \overline{u}=s \}$. 
\end{definition}

Consider a transducer as a directed labeled graph where the nodes are transducer states indexed by
the strings reaching them, \[ \{ q_s \mid \exists u, v.\, q^- \trans{u}{v} q \wedge \overline{u} = s \},\] and
the edges are the corresponding transitions, \[\{ q_s \trans{a}{b} q'_{s \overline{a}} \mid q \trans{a}{b} q' \}. \]
It can be seen that $\Pathtree(s)$ is a subgraph that forms a non-full rooted edge-labeled binary tree.  The \emph{stem} of $\Pathtree(s)$ is the longest path in this tree from $q_\varepsilon^-$ to some $q_{s'}$ for a prefix $s'$ of $s$ only involving nodes with at most one child.  The \emph{leaves} of $\Pathtree(s)$ are the states $q$ such that $q_s$ is reachable, in lexicographic
order of the paths reaching them from $q_\varepsilon^-$.

\begin{example}
Recall the oracle machine for the decimal converter in the lower left of Figure~\ref{fig:decimal_example}.  Its path tree for input \texttt{a} is shown in the upper left of Figure~\ref{fig:path_tree}.  The nodes are subscripted with the length of the input prefix rather the input prefix itself. 
Note that the leaf states are listed from top to bottom in lexicographic order of their paths reaching them.  This means that the top state is the prime candidate for being $q'$ in Proposition~\ref{prop:path_decomposition}. If the remainder of the input is not accepted from it, though, the other leaf states take over in the given order.
\end{example}

\begin{landscape}
{\tikzstyle{corner} = [outer sep=0pt,inner sep=0pt,minimum size=0pt,fill=black]
\tikzstyle{treelab} = [auto,font=\tiny,inner sep=1pt]
\newcommand{\cL}{$\epsilon_{\mathsf{0}} / \code{0}$}
\newcommand{\cR}{$\epsilon_{\mathsf{1}} / \code{1}$}
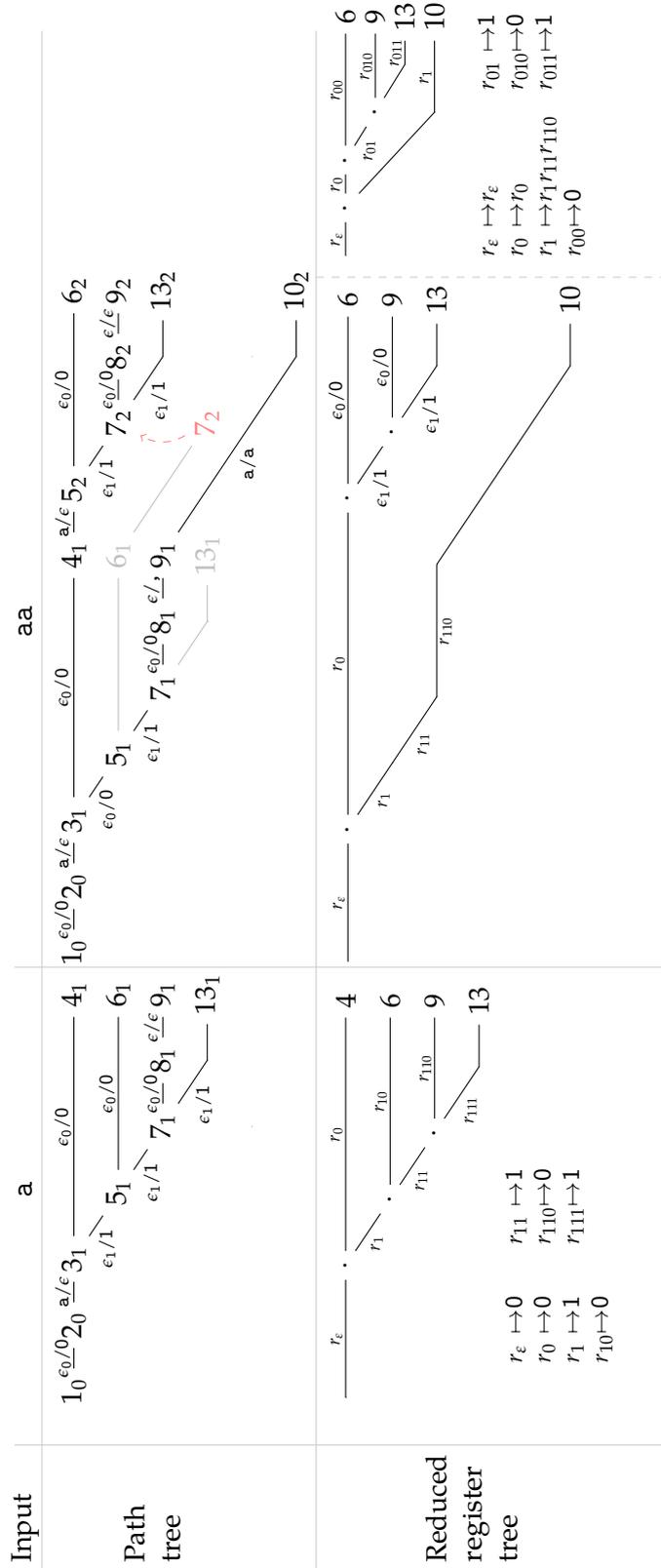
\begin{figure*}
    \centering
        \begin{tikzpicture}[x=6.7cm,y=3.7cm]
        \node[anchor=north east] (1a_brutto) at (0.025,0) {\begin{tikzpicture}[x=0.9cm,y=0.6cm,every node/.style={anchor=center}]
    \node (s1) at (0,0) {$1_0$};
    \node (s2) at (1,0) {$2_0$};
    \node (s3) at (2,0) {$3_1$};
    \node (s4) at (6,0) {$4_1$};
    \node (s5) at (3,-1) {$5_1$};
    \node (s6) at (6,-1) {$6_1$};
    \node (s7) at (4,-2) {$7_1$};
    \node (s8) at (5,-2) {$8_1$};
    \node (s9) at (6,-2) {$9_1$};
%    \node (s12) at (6,-4) {$12$};
    \node[corner] (s12dummy) at (4,-4) {};
    \node (s13) at (6,-3) {$13_1$};
    \node[corner] (s13dummy) at (5,-3) {};

    \draw (s1) to node[treelab] {\cL} (s2) to node[treelab] {$\texttt{a} / \epsilon$} (s3) to node[treelab] {\cL} (s4)
    (s3) to node[swap,treelab] {\cR} (s5) to node[treelab] {\cL} (s6)
    (s5) to node[swap,treelab] {\cR} (s7) to node[treelab] {\cL} (s8) to node[treelab] {$\epsilon / \epsilon$} (s9)
    (s7) to node[swap,treelab] {\cR} (s13dummy) to (s13)
%    (s1) to node[swap,treelab] {\cR} (s12dummy) to (s12);
    ;
\end{tikzpicture}

%%% Local Variables:
%%% mode: latex
%%% TeX-master: "../../thesis"
%%% End:
};
%        \node[anchor=north] (trans_brutto) at (1,-0.5) {$\Longrightarrow$};
        \node[anchor=north west] (2a_brutto) at (-0.05,0) {\tikzstyle{dead} = [gray!50]
\begin{tikzpicture}[x=0.9cm,y=0.6cm,every node/.style={anchor=center}]
    \node (s1) at (0,0) {$1_0$};
    \node (s2) at (1,0) {$2_0$};
    \node (s3) at (2,0) {$3_1$};
    \node (s4) at (6,0) {$4_1$};
    \node (s5) at (3,-1) {$5_1$};
    \node (s5') at (7,0) {$5_2$};
    
    \node (s6) at (10,0) {$6_2$};
    \node (s7) at (4,-2) {$7_1$};
    \node (s7') at (8,-1) {$7_2$};
    \node (s8) at (5,-2) {$8_1$};
    \node (s8') at (9,-1) {$8_2$};
    \node (s9) at (6,-2) {$9_1$};
    \node (s9') at (10,-1) {$9_2$};
    \node (s10) at (10,-5) {$10_2$};
    \node[corner] (s10dummy) at (9,-5) {};
    
    \node (s13') at (10,-2) {$13_2$};
    \node[corner] (s13'dummy) at (9,-2) {};

    \draw (s1) to node[treelab] {\cL} (s2) to node[treelab] {$\texttt{a} / \epsilon$} (s3) to node[treelab] {\cL} (s4) to node[treelab] {$\texttt{a} / \epsilon$} (s5') to node[treelab] {\cL} (s6)
    (s5') to node[swap,treelab] {\cR} (s7') to node[treelab] {\cL} (s8') to node[treelab] {$\epsilon / \epsilon$} (s9')
    (s7') to node[swap,treelab] {\cR} (s13'dummy) to (s13')
    (s3) to node[swap,treelab] {\cL} (s5) to node[swap,treelab] {\cR} (s7) to node[treelab] {\cL} (s8) to node[treelab] {$\epsilon / \texttt{,}$} (s9) to node[swap,treelab] {$\texttt{a} / \texttt{a}$} (s10dummy) to (s10)
    ;

    \begin{scope}[every node/.style={dead,anchor=center}, every path/.style={dead}, killed/.style={red!50}]
        \node (d6) at (6,-1) {$6_1$};
        \node[killed] (d7) at (8,-3) {$7_2$};
%        \node[killed] (d8) at (9,-3) {$8_2$};
%        \node[killed] (d9) at (10,-3) {$9_2$};
        \node (d13) at (6,-3) {$13_1$};
        \node[corner] (d13dummy) at (5,-3) {};
%        \node[killed] (d13') at (10,-4) {$13_2$};
        \node[corner] (d13'dummy) at (9,-4) {};

        \draw (d6) to (d7) % to (d8) to (d9)
%        (d7) to (d13'dummy) to (d13')
        (s7) to (d13dummy) to (d13)
        (s5) to (d6)
        ;

        \draw[->,bend left,dashed,killed] (d7) to (s7');
%        \draw[->,bend right=90,dashed,gray!70] (d9) to (s9');
%        \draw[->,bend right=90,dashed,gray!70] (d13') to (s13');
    \end{scope}
\end{tikzpicture}

%%% Local Variables:
%%% mode: latex
%%% TeX-master: "../kleenex-popl16"
%%% End:
};

        \node[anchor=north east] (1a) at (0,-1) {\begin{tikzpicture}[x=0.9cm,y=0.6cm,every node/.style={anchor=center}]
    \node[corner] (s1) at (0,0) {};
    \node (s3) at (2,0) {$\cdot$};
    \node (s4) at (6,0) {$4$};
    \node (s5) at (3,-1) {$\cdot$};
    \node (s6) at (6,-1) {$6$};
    \node (s7) at (4,-2) {$\cdot$};
    \node (s9) at (6,-2) {$9$};
%    \node (s12) at (6,-4) {$12$};
%    \node[corner] (s12dummy) at (4,-4) {};
    \node (s13) at (6,-3) {$13$};
    \node[corner] (s13dummy) at (5,-3) {};

    \node[anchor=north,shape=rectangle,inner sep=0pt] (regs) at (2,-3.5) {
      \footnotesize
      \begin{minipage}{3cm}
          $\begin{array}{l@{}c@{}lcl@{}c@{}l}
            r_\varepsilon & \mapsto & \code{0} && r_{11} & \mapsto & \code{1} \\
            r_{0} & \mapsto & \code{0} && r_{110} & \mapsto & \code{0} \\
            r_{1} & \mapsto & \code{1} && r_{111} & \mapsto & \code{1} \\
            r_{10} & \mapsto & \code{0} && &&
          \end{array}$
      \end{minipage}
    };
    
    \draw (s1) to node[treelab] {$r_\varepsilon$} (s3) to node[treelab] {$r_{0}$} (s4)
    (s3) to node[swap,treelab] {$r_{1}$} (s5) to node[treelab] {$r_{10}$} (s6)
    (s5) to node[swap,treelab] {$r_{11}$} (s7) to node[treelab] {$r_{110}$} (s9)
    (s7) to node[swap,treelab] {$r_{111}$} (s13dummy) to (s13);
%    (s1) to node[swap,treelab] {$r_1$} (s12dummy) to (s12);
\end{tikzpicture}

%%% Local Variables:
%%% mode: latex
%%% TeX-master: "../kleenex-popl16"
%%% End:
};
        \node[anchor=north west] (2a) at (-0.01,-1) {\begin{tikzpicture}[x=0.9cm,y=0.6cm,every node/.style={anchor=center}]
    \node[corner] (s1) at (0,0) {};
%    \node (s1') at (7,-4) {$\cdot$};
%    \node (s2') at (10,-4) {$2$};
    \node (s3) at (2,0) {$\cdot$};
    \node (s5') at (7,0) {$\cdot$};
    \node (s6) at (10,0) {$6$};
    \node[corner] (s5corner) at (3,-1) {};
    \node[corner] (s7) at (4,-2) {};
    \node (s7') at (8,-1) {$\cdot$};
    \node (s9') at (10,-1) {$9$};
    \node[corner] (s9) at (6,-2) {};
    \node (s10) at (10,-5) {$10$};
    \node[corner] (s10dummy) at (9,-5) {};
%    \node[corner] (s10dummy2) at (6,-2) {};
%    \node (s12) at (10,-5) {$12$};
%    \node[corner] (s12dummy) at (8,-5) {};
    % \node[corner] (s13dummy2) at (6,-3) {};
    % \node[corner] (s13dummy) at (5,-3) {};
    \node (s13') at (10,-2) {$13$};
    \node[corner] (s13'dummy) at (9,-2) {};

     \draw (s1) to node[treelab] {$r_\varepsilon$} (s3) to node[treelab] {$r_0$} (s5') to node[treelab] {\cL} (s6)
     (s5') to node[swap,treelab] {\cR} (s7') to node[treelab] {\cL} (s9')
     (s7') to node[swap,treelab] {\cR} (s13'dummy) to (s13')
     (s3) to node[swap,treelab] {$r_1$} (s5corner) to node[swap,treelab] {$r_{11}$} (s7) to node [swap,treelab] {$r_{110}$} (s9) to (s10dummy) to (s10)
     % (s7) to (s13dummy) to node[swap,treelab] {$\code{1}$} (s13dummy2) to (s1') to node[treelab] {$\code{0}$} (s2')
     % (s1') to (s12dummy) to node[swap,treelab] {$\code{1}$} (s12);
     ;
 \end{tikzpicture}

%%% Local Variables:
%%% mode: latex
%%% TeX-master: "../kleenex-popl16"
%%% End:
};

        \node[anchor=north west] (2ar) at (2a.north east) {\begin{tikzpicture}[x=0.65cm,y=0.4cm,every node/.style={anchor=center}]
    \node[corner] (s1) at (0,0) {};
    \node (s3) at (1,0) {$\cdot$};
    \node (s5') at (2,0) {$\cdot$};
    \node (s6) at (5,0) {$6$};
%    \node[corner] (s5corner) at (2,-1) {};
%    \node[corner] (s7) at (4,-2) {};
    \node (s7') at (3,-1) {$\cdot$};
    \node (s9') at (5,-1) {$9$};
%    \node[corner] (s9) at (6,-2) {};
    \node (s10) at (5,-3) {$10$};
%    \node[corner] (s10dummy) at (3,-3) {};

    \node (s13') at (5,-2) {$13$};
    \node[corner] (s13'dummy) at (4,-2) {};
    \node[corner] (s10'dummy) at (3,-3) {};

     \draw (s1) to node[treelab] {$r_\varepsilon$} (s3) to node[treelab] {$r_0$} (s5') to node[treelab] {$r_{00}$} (s6)
     (s5') to node[swap,treelab] {$r_{01}$} (s7') to node[treelab] {$r_{010}$} (s9')
     (s7') to (s13'dummy) to node[treelab] {$r_{011}$} (s13')
     (s10) to node[swap,treelab] {$r_1$} (s10'dummy) to (s3);
     ;
 \end{tikzpicture}

%%% Local Variables:
%%% mode: latex
%%% TeX-master: "../kleenex-popl16"
%%% End:
};
        \node[anchor=north west] (trans) at ($(2ar.south west)$) {\footnotesize
          $\begin{array}{@{}l@{}c@{}lc@{}l@{}c@{}l@{}}
            r_\varepsilon &\mapsto& r_\varepsilon && r_{01} &\mapsto& \code{1} \\
            r_0 &\mapsto& r_0 && r_{010} &\mapsto& \code{0} \\
            r_1 &\mapsto& r_1 r_{11} r_{110} && r_{011} &\mapsto& \code{1} \\
            r_{00} &\mapsto& \code{0} && &&
          \end{array}$
        };

%        \node (name1) at ($(1a.south west) + (0.1,0)$) {$[A_1]$};
        \node (dummy2) at ($(2a.west) + (0.1,0)$) {};
%        \node (name2) at (name1 -| dummy2.west) {$[A_2]$};

        \node[anchor=south] (id_a) at (1a_brutto.north) {\texttt{a}};
        \node[anchor=south] (id_aa) at (2a_brutto.north) {\texttt{aa}};

        \node (leftpoint) at ($(1a_brutto.west) + (-0.18,0)$) {};
        \node (desc1) at (id_a -| leftpoint) {
          \begin{minipage}{3.5em}
              Input
          \end{minipage}
        };
        \node (desc2) at (1a_brutto.west -| leftpoint) {
          \begin{minipage}{3.5em}
              Path tree
          \end{minipage}
        };
        \node (desc3) at (1a.west -| leftpoint) {
          \begin{minipage}{3.5em}
              Reduced register tree
          \end{minipage}
        };

        \draw[gray!40] (-1.22,-1) to (1.9,-1);
        \draw[gray!40] (0,0.1) to (0,-2.3);
        \draw[gray!40] (-1.22,0) to (1.9,0);
        \draw[gray!40] (-0.97,0.1) to (-0.97,-2.3);
        \draw[gray!40,dashed] (1.4,-1) to (1.4,-2.3);
        
    \end{tikzpicture}
%%% Local Variables:
%%% mode: latex
%%% TeX-master: "../../thesis"
%%% End:
    \caption[Path tree example.]{
      Path trees for the decimal conversion oracle in Figure~\ref{fig:decimal_example}.
      Above left: path tree reading \texttt{a}. Subscripts denote the number of input symbols read when the given state was visited.
      Above right: path tree reading \texttt{aa}. Failing paths are shown in gray.
      Below left: reduced register tree reading \texttt{a}, with register valuation.
      Below middle: extension of register tree after reading an additional \texttt{a}.  Note mix of registers and bits, and that bottom branch is now labeled by a sequence of registers.
      Below right: the path tree and register \emph{update} after reading \texttt{aa}. The registers $r_1$, $r_{11}$, and $r_{110}$ are all on the same unary path and are concatenated.
    }
    \label{fig:path_tree}
\end{figure*}
}
\end{landscape}

\subsection{Basic Simulation Algorithm}

The basic streaming simulation algorithm works as follows:
\begin{algorithm}
\caption{Basic streaming algorithm}
\label{alg:basic_streaming}
Let $s = a_1 \ldots a_n \in \Sigma^*$ be the input string.
\begin{algorithmic}[1]
\For{$i = 1$ \textbf{to} $n$}
  \If{$\Pathtree(a_1 ... a_i) = \emptyset$}
    \State \textbf{terminate} with failure (input rejected)
  \EndIf
  \If{$\mathrm{stem}(\Pathtree(a_1 ... a_i))$ longer than $\mathrm{stem}(\Pathtree(a_1 ... a_{i-1}))$}
    \State emit the output bits on the stem extension
  \EndIf
\EndFor
\If{$\Pathtree(a_1 ... a_n)$ contains path to $q^f$}
  \State \parbox[t]{\dimexpr\linewidth-\algorithmicindent\relax}{if path tree contains at least one branch, emit output bits on path from highest binary ancestor to $q^f$\strut}
  \State \textbf{terminate} with success (input accepted)
\Else
  \State \textbf{terminate} with failure (input rejected)
\EndIf
\end{algorithmic}
%\begin{enumerate} \compresslist
%\item \textbf{for} $i \in 1 \ldots n$ \textbf{do}
%\begin{itemize}
%\item \textbf{if} $\Pathtree(a_1 \ldots a_i) = \emptyset$ 
%\item \textbf{then} terminate with failure (input rejected)
%\item \textbf{else if} the stem of $\Pathtree(a_1 \ldots a_i)$ is longer than the stem of $\Pathtree(a_1 \ldots a_{i-1})$
%\item \textbf{then} emit the output bits on the stem extension.
%\end{itemize}
%\item Terminate with success (input accepted).
%\end{enumerate}
\end{algorithm}

The critical step in the algorithm is incrementally computing the path tree for $s' a$ from
the path tree for $s'$.
\begin{algorithm}
\caption{Incremental path tree computation}
\label{alg:incremental_path_trees}
Let $\Pathtree$ be $\Pathtree(s')$ for some prefix $s'$ of the input string, and let $[q_0,...,q_n]$ be its leaves in lexicographic order of the paths reaching them. Upon reading $a$, incrementally compute
$\Pathtree(s' a)$ as follows.

\begin{algorithmic}[1]
\For{$q = q_0$ \textbf{to} $q_n$}
  \State \parbox[t]{\dimexpr\linewidth-\algorithmicindent\relax}{compute $\Pathtree_q(a)$, the path tree of lexicographically least $(u/v)$ paths with $\overline{u} = a$ from $q$ to resting states, but excluding resting states that have been reached in a previous iteration\strut}
  \If{$\Pathtree_q(a)$ is non-empty}
    \State replace leaf node $q$ in $\Pathtree$ by $\Pathtree_q(a)$
  \Else
    \State \parbox[t]{\dimexpr\linewidth-\algorithmicindent-\algorithmicindent\relax}{prune branch from lowest binary ancestor to leaf node $q$; if binary ancestor does not exist, then \textbf{terminate} with failure (input rejected) \label{pruning}\strut}
  \EndIf
\EndFor
\end{algorithmic}
\end{algorithm}

\begin{example}
The upper right in Figure~\ref{fig:path_tree} shows $\Pathtree(\mathtt{aa})$ 
for the decimal converter.  Observe how it arises from $\Pathtree(\mathtt{a})$ by 
extending leaf states 4 and 9, which have an \texttt{a}-transition, 
and building the $\epsilon$-closure as a binary tree. It prunes
branches either because they reach a state already reached by a lexicographical
lower path (state 6) or because the leaf does not have transition on \texttt{a} 
(state 13).  The algorithm outputs \texttt{0} after reading the first \texttt{a} since
\texttt{0} is the sequence of output bits on the stem of the path tree.
It does not output anything after reading the second \texttt{a} since $\Pathtree(aa)$
has the same stem as $\Pathtree(a)$.
\end{example}

%\begin{algorithm}[Incremental path tree computation]
%\label{alg:incremental_path_trees}
%Let $\Pathtree$ be $\Pathtree(s')$ for some prefix $s'$ of the input string. Upon reading $a$, incrementally compute
%$\Pathtree(s' a)$ as follows. (Iteration over $\Leaves(\Pathtree)$ is in lexicographic order.)
%
%\begin{enumerate} \compresslist
%\item \textbf{for} $q \in \Leaves(\Pathtree)$ \textbf{do}
%\begin{itemize}
%\item \textbf{if} $\Cal T^\Code$ contains transition $q \trans{a}{b} q'$
%\item \textbf{then} replace leaf node $q$ with the path tree of 
%lexicographically least $\epsilon$-paths from $q$ to resting states, but excluding resting states that have been reached in a previous iteration; 
%\item \textbf{else} 
%\begin{enumerate}
%\item prune the branch from nearest binary ancestor to leaf node $q$; \label{pruning}
%\item \textbf{if} binary ancestor does not exist \\ \textbf{then} terminate with failure (input rejected)
%\end{enumerate} 
%\end{itemize}
%\end{enumerate}
%\end{algorithm
%}

\begin{definition}[Optimal streaming]
Let $f$ be a partial function from $\Sigma^*$ to $\Gamma^*$, $s \in \Sigma^*$.
Let $T(s) = \{ f(s s') \mid s' \in \Sigma^* \wedge s s' \in \dom{f} \}$. 
The \emph{output $f^\#(s)$ determined by $f$ for $s$} is  
the longest common prefix of $T(s)$ if $T(s)$ is nonempty; otherwise it is undefined.
The partial function $f^\#$ is called the \emph{optimally streaming version} of $f$. 
An \emph{optimally streaming algorithm} for $f$ is an algorithm that implements $f^\#$: 
It emits output symbols as soon as they are semantically determined by the 
input prefix read so far.
\end{definition}

Let transducer $\Cal T$ be given.  Write $\lsem{q}$ for $\lsem{\Cal T'}$ where $\Cal T'$ is $\Cal T$, but with 
$q$ as initial state instead of $q^-$.  A state $q$ is \emph{covered} by $\{q_1, \ldots, q_k\}$
if $\lsem{q} \subseteq \lsem{q_1} \cup \ldots \cup \lsem{q_k}$.   A path tree $\Pathtree(s)$ with 
lexicographically ordered leaves $[q_1, \ldots, q_n]$ is \emph{cover-free} if no $q_i$ is covered by $\{ q_1, \ldots, q_{i-1} \}$.
$\Cal T$ is \emph{cover-free} if $\Pathtree(s)$ is cover-free for all $s \in \Sigma^*$.

\begin{theorem}
\label{thm:linear_time_streaming}
Let $\Cal T$ be cover-free.  Then Algorithm~\ref{alg:basic_streaming} with Algorithm~\ref{alg:incremental_path_trees} for
incremental path tree recomputation is an optimally streaming algorithm for $\gsem{\Cal T^\Code}$ that runs in time $O(m n)$, where $m=|\Cal T^\Code|$ and $n$ is the length of the input string.
\end{theorem}
\begin{proof} (Sketch)
Algorithm~\ref{alg:incremental_path_trees}  
can be implemented to run in time $O(m)$ since it visits each transition in $\Cal T^\Code$
at most once and pruning can be amortized: every deallocation of an edge can be charged to its allocation.  Algorithm~\ref{alg:basic_streaming} invokes Algorithm~\ref{alg:incremental_path_trees} $n$ times. Optimal streaming follows from a generalization of the proof of optimal streaming for regular expression parsing \cite{grathwohl2014a}.
\end{proof}
The algorithm can be made optimally streaming for all oracle transducers by also pruning leaf states that are covered by other leaf states in Step~\ref{pruning} of Algorithm~\ref{alg:incremental_path_trees}.  Coverage is $\mathsf{PSPACE}$-complete, however.  
Eliding the coverage check does not seem to make much of a difference to the streaming behavior in practice.  

\section{Determinization}
\label{sec:determinization}

NFA simulation maintains a set of NFA states.  This is the basis of compiling
an NFA into a DFA: precompute and number the set of \emph{all} NFA state sets reachable by \emph{any} input from the initial NFA state, observing
that there are only finitely many such sets.  
In the transducer simulation in Section~\ref{sec:simulation} path trees play the role of NFA state sets. The corresponding 
determinization idea does not work for transducers, however: $\{ \Pathtree(s) \mid s \in \Sigma^* \}$ is in general infinite.  For example, for the oracle machine in Figure~\ref{fig:decimal_example}, the trees $\Pathtree(\mathtt{a}^n)$ all have the same stem, but contain paths with bit strings of length proportional to $n$.  This is inherently so.
A single-valued transducer can be transformed effectively~\cite{beal2002,vanoord2001} into a form of deterministic finite-state transducer if its relational semantics is \emph{subsequential}~\cite{schutzenberger1977,berstel79}, but nondeterministic finite state transducers in general are properly more expressive than their deterministic counterparts.  We can factor a path tree into its underlying full binary 
tree and the labels associated with the edges, though.  Since there are only finitely many different such trees, we can achieve determinization to transducers with registers storing the potentially unbounded label data.  

% \subsection{Streaming string transducers}

%A \emph{streaming string transducer}~\cite{alur2010expressiveness} (SST) is a deterministic model of computation that generalizes subsequential transducers by allowing copy-free updates to a finite set of word registers. It turns out every transducer can be 
%compiled into an equivalent SST. 

\begin{definition}[Streaming String Transducer~\cite{alur2010expressiveness}]
  A deterministic \emph{streaming string transducer} (SST) over alphabets $\Sigma, \Gamma$ is a tuple $\mathcal{S} = (X, Q, q^{-}, F, \delta^1, \delta^2)$ where
\begin{itemize}\compresslist
\item $X$ is a finite set of \emph{register variables};
\item $Q$ is is a finite set of \emph{states};
\item $F$ is a partial function $Q \to (\Gamma \cup X)^*$ mapping each \emph{final state} $q \in \dom(F)$ to a word $F(q) \in (\Gamma \cup X)^*$ such that each $x \in X$ occurs at most once in $F(q)$;
\item $\delta^1$ is a transition function $Q \times \Sigma \to Q$;
\item $\delta^2$ is a \emph{register update} function $Q \times \Sigma \to (X \to (\Gamma \cup X)^*)$ such that for each $q \in Q$, $a \in \Sigma$ and $x \in X$, there is at most one occurrence of $x$ in the multiset of strings $\{ \delta^2(q,a)(y) \mid y \in X \}$.
\end{itemize}
\end{definition}
%The semantics is defined as follows.
\noindent
A \emph{configuration} of an SST $\mathcal{S} = (X,Q,q^{-},F,\delta^1,\delta^2)$ is a pair $(q, \rho)$ where $q \in Q$ is a state, and $\rho : X \to \Gamma^*$ is a \emph{valuation}. A valuation extends to a monoid homomorphism $\widehat{\rho} : (X \cup \Gamma)^* \to \Gamma^*$ by setting $\rho(x) = x$ for $x \in \Gamma$. The initial configuration is $(q^{-}, \rho^{-})$ where $\rho^{-}(x) = \epsilon$ for all $x \in X$.

A configuration steps to a new one given an input symbol:
$\delta((q,\rho), a) ={} (\delta^1(q,a), \widehat{\rho} \circ \delta^2(q,a))$.
The transition function extends to a transition function on words $\delta^*$ by $\delta^*((q,\rho),\epsilon) = (q,\rho)$ and $\delta^*((q,\rho),au) = \delta^*(\delta((q,\rho), a), u).$

Every SST $\mathcal{S}$ denotes a partial function $\fsem{\mathcal{S}} : \Sigma^* \to \Gamma^*$ where for any $u \in \Sigma^*$ such that $\delta^*((q^{-},\rho^{-}),u) = (q',\rho')$, we define
\[
\fsem{\mathcal{S}}(u) =
\begin{cases}
\widehat{\rho'}(F(q')) & \text{if $q' \in \dom(F)$} \\
\text{undefined} & \text{otherwise}
\end{cases}
\]

%\subsection{Tabulation}

In the following, let $X = \{ r_p \mid p \in \Two^* \}$ be a set of registers.
\begin{definition}[Reduced register tree]
  Let $\Pathtree$ be a path tree. Its \emph{reduced register tree} $\RPathtree(\Pathtree)$ is a pair $(R_\Pathtree, \rho_\Pathtree)$ where $\rho_\Pathtree$ is a valuation $X \to \Two^*$ and $R_\Pathtree$ is a full binary tree with state-labeled leaves, obtained from $\Pathtree$ by first contracting all unary branches and concatenating edge labels; then replacing each edge label $(u/v)$ by a single register symbol $r_p$, where $p$ denotes the unique path from the root to the edge destination node, and setting $\rho_\Pathtree(r_p) = v$.
\end{definition}

The set $\{ R_{\Pathtree(s)} \mid s \in \Sigma^* \}$ is finite: it is bounded by the number of full binary trees with up to $|Q|$ leaves 
times the number of possible permutations of the leaves.

Let $R$ be $R_\Pathtree$ and $a \in \Sigma$ a symbol, and apply Algorithm~\ref{alg:incremental_path_trees} to $R$. The result is a non-full binary tree with edges labeled either by a register or by a $(u/v)$ pair. By reducing the tree again and treating registers as output labels, we get a pair $(R_a, \kappa_{R,a})$ where $\kappa_{R,a} : X \to (\Two \cup X)^*$ is a register update.

\begin{example}
  Consider the bottom left tree in Figure~\ref{fig:path_tree}. This is the reduced register tree obtained from the path tree above it. The evaluation map $\rho$ can be seen below it, where register subscripts denote their position in the register tree. In the middle is the result of extending the register tree using Algorithm~\ref{alg:incremental_path_trees}. Reducing this again yields the tree on the right. The update map $\kappa$ is shown below it---note that the range of this map is mixed register/bit sequences.
\end{example}

\begin{proposition}
  \label{prop:abstract_step}
  Let $\Cal{T}^\Code$ be given, and let $\Pathtree = \Pathtree(s)$, $\Pathtree' = \Pathtree(sa)$, $(R, \rho) = \RPathtree(\Pathtree)$ and $(R', \rho') = \RPathtree(\Pathtree')$ for some $s$ and $a$.
Then $R' = R_a$ and $\rho' = \widehat{\rho} \circ \kappa_{R, a}$.
\end{proposition}

\begin{theorem}
\label{thm:SST_construction}
Let $\Cal{T}^\Code$ be an oracle machine of size $m$. There is an SST $\mathcal{S}$ with $O(2^{m \log m})$ states 
such that $\fsem{\Cal S} = \gsem{\Cal T}$.
\end{theorem}
\begin{proof}
Let $Q_{\Cal S} = \{R_{\Pathtree(s)} \mid s \in \Sigma^* \} \cup \{ R_0 \}$ and $q_{\Cal S}^{-} = R_0$, where $R_0$ is the single-leaf binary tree with leaf $q^{-}_{\Cal T}$.
The set of registers $X_{\Cal S}$ is the finite subset of register variables occurring in $Q_{\Cal S}$. The transition maps are given by $\delta^1_{\Cal S}(R,a) = R_a$ and $\delta^2_{\Cal S}(R,a) = \kappa_{R,a}$. For any $R \in Q_{\Cal S} - \{R_0\}$, define the final output $F_{\Cal S}(R)$ to be the sequence of registers on the path from the root to the final state $q^f_{\Cal T}$ in $R$ if $R$ contains it as a leaf; otherwise let $F_{\Cal S}(R)$ be undefined. Let $F_{\Cal S}(R_0) = \overline{v}$ if $q^{-}_{\Cal T} \pathmin{\epsilon}{v} q^f$ for some $v$; otherwise let $F_{\Cal S}(R_0)$ be undefined.  

Correctness follows by showing $\delta^*((R_0,\rho^{-}),u)=\RPathtree(\Pathtree(u))$ for all $u \in \Sigma^+$. We prove this by induction, applying Proposition~\ref{prop:abstract_step} in each step. For the case $u=\varepsilon$ correctness follows by the definition of $F_{\Cal S}(R_0)$. 

The upper bound follows from the fact that there are at most $C_{k-1} (k-1)! = O(2^{m \log m})$ full binary trees with $k$ pairwise distinct leaves where $k$ is the number of resting states in $\Cal T^\Code$ and $C_{k-1}$ is the $(k-1)$-st \emph{Catalan} number.
\qedhere
\end{proof}

\begin{example}
The oracle machine in Figure~\ref{fig:decimal_example} yields the SST in Figure~\ref{fig:decimal-sst}. The states $1$ and $2$ are identified by the left and right reduced trees, respectively, in the bottom of Figure~\ref{fig:path_tree}.
\end{example}

\begin{corollary}
The SST $\mathcal{S}$ for $\Cal T^\Code$ can be implemented to execute in time $O(m n)$ where $m = |\Cal T^\Code|$.
\end{corollary}
\begin{proof} (Sketch)
Use a data structure for imperatively extending a string register, $r \,\texttt{:=}\, r s$, in amortized time $O(n)$ where $n$ is the size of $s$,
independent of the size of the string stored in $r$.
The result then follows from the fact that the steps in Algorithm~\ref{alg:incremental_path_trees} can
be implemented in the same amortized time.
\end{proof}
In practice, the compiled version of the SST is much more efficient---roughly one to two orders of magnitude faster---than streaming simulation since it compiles away the interpretive overhead of explicitly managing the binary trees underlying path trees and employs machine word-level parallelism by operating on bit strings in fewer registers rather than many edges each labeled by at most one bit.

\begin{landscape}
\begin{figure*}
  {\begin{tikzpicture}[>=stealth',x=1.5cm,y=0.9cm]
\begin{scope}[state/.append style={minimum size=0.6cm,inner sep=0}]
\node[state,initial,initial where=below] (a0) at (0.6,0) {$0$};
\node[state] (a1) at (2,0) {$1$};
\node[state] (a2) at (4,0) {$2$};
\node[state] (a3) at (6,0) {$3$};
\node[state] (a4) at (8,0) {$4$};
\node[state] (a5) at (10,0) {$5$};
\node[state,accepting by double,accepting by arrow,accepting where=right,accepting text={$r_\varepsilon r_1$}] (ae) at (12,0) {$6$};
\end{scope}
\begin{scope}[lab/.style={auto,inner sep=1pt}]
\newcommand{\alabel}[1]{$\texttt{a}\left|{#1}\right.$}
\newcommand{\nlabel}[1]{$\texttt{\textbackslash n}\left| {#1} \right.$}
\newcommand{\kOneE}{\footnotesize \begin{array}{@{}l@{}c@{}l@{}}
  r_\varepsilon &\mapsto& r_\varepsilon r_1 r_{11} r_{111} \\
  r_0 &\mapsto& \code{0} \\
  r_1 &\mapsto& \code{1}
\end{array}}
\newcommand{\kTwoE}{\footnotesize \begin{array}{@{}l@{}c@{}l@{}}
  r_\varepsilon &\mapsto& r_\varepsilon r_0 r_{01} r_{011} \\
  r_0 &\mapsto& \code{0} \\
  r_1 &\mapsto& \code{1}
\end{array}}
\newcommand{\kThreeE}{\footnotesize \begin{array}{@{}l@{}c@{}l@{}}
  r_\varepsilon &\mapsto& r_\varepsilon r_0 r_{00} r_{001} \\
  r_0 &\mapsto& \code{0} \\
  r_1 &\mapsto& \code{1}
\end{array}}
\newcommand{\kFourE}{\footnotesize \begin{array}{@{}l@{}c@{}l@{}}
  r_\varepsilon &\mapsto& r_\varepsilon r_1 r_{11} \\
  r_0 &\mapsto& \code{0} \\
  r_1 &\mapsto& \code{1}
\end{array}}
\newcommand{\kFiveE}{\footnotesize \begin{array}{@{}l@{}c@{}l@{}}
  r_\varepsilon &\mapsto& r_{\varepsilon} r_0 r_{01} r_{011} \\
  r_0 &\mapsto& \code{0} \\
  r_1 &\mapsto& \code{1}
\end{array}}
\newcommand{\kZeroOne}{\footnotesize \begin{array}{@{}l@{}c@{}l@{}}
  r_\varepsilon &\mapsto& r_\varepsilon \code{0} \\
  r_0 &\mapsto& \code{0} \\
  r_1 &\mapsto& \code{1} \\
  r_{10} &\mapsto& \code{0} \\
  r_{11} &\mapsto& \code{1} \\
  r_{110} &\mapsto& \code{0} \\
  r_{111} &\mapsto& \code{1}
\end{array}}
\newcommand{\kOneTwo}{\footnotesize \begin{array}{@{}l@{}c@{}l@{}}
%  r_\varepsilon &\mapsto& r_\varepsilon \\
%  r_0 &\mapsto& r_0 \\
  r_1 &\mapsto& r_1 r_{11} r_{110} \\
  r_{00} &\mapsto& \code{0} \\
  r_{01} &\mapsto& \code{1} \\
  r_{010} &\mapsto& \code{0} \\
  r_{011} &\mapsto& \code{1}
\end{array}}
\newcommand{\kTwoThree}{\footnotesize \begin{array}{@{}l@{}c@{}l@{}}
%  r_\varepsilon &\mapsto& r_\varepsilon \\
%  r_0 &\mapsto& r_0 \\
%  r_{00} &\mapsto& r_{00} \\
  r_{01} &\mapsto& r_{01} r_{010} \\
  r_{000} &\mapsto& \code{0} \\
  r_{001} &\mapsto& \code{1}
%  r_1 &\mapsto& r_1
\end{array}}
\newcommand{\kThreeFour}{\footnotesize \begin{array}{@{}l@{}c@{}l@{}}
%  r_\varepsilon &\mapsto& r_\varepsilon \\
%  r_0 &\mapsto& r_0 \\
  r_{00} &\mapsto& r_{00} r_{000} \\
%  r_{01} &\mapsto& r_{01} \\
%  r_1 &\mapsto& r_1 \\
  r_{10} &\mapsto& \code{0} \\
  r_{11} &\mapsto& \code{1}
\end{array}}
\newcommand{\kFourFive}{\footnotesize \begin{array}{@{}l@{}c@{}l@{}}
%  r_\varepsilon &\mapsto& r_\varepsilon \\
%  r_0 &\mapsto& r_0 \\
%  r_{00} &\mapsto& r_{00} \\
%  r_{01} &\mapsto& r_{01} \\
  r_{010} &\mapsto& \code{0} \\
  r_{011} &\mapsto& \code{1} \\
  r_1 &\mapsto& r_1 r_{10}
\end{array}}
\newcommand{\kFiveThree}{\footnotesize \begin{array}{@{}l@{}c@{}l@{}}
%  r_\varepsilon &\mapsto& r_\varepsilon \\
%  r_0 &\mapsto& r_0 \\
%  r_{00} &\mapsto& r_{00} \\
  r_{000} &\mapsto& \code{0} \\
  r_{001} &\mapsto& \code{1} \\
  r_{01} &\mapsto& r_{01} r_{010} \\
%  r_1 &\mapsto& r_1
\end{array}}
\newcommand{\kEOne}{\footnotesize \begin{array}{@{}l@{}c@{}l@{}}
  r_\varepsilon &\mapsto& r_\varepsilon r_0 \\
  r_0,r_{10}, r_{110} &\mapsto& \code{0} \\
  r_1, r_{11}, r_{111} &\mapsto& \code{1}
\end{array}}
\draw[->] (a0) to node[lab] {\alabel{\kZeroOne}} (a1);
\draw[->] (a1) to node[lab,pos=0.6] {\alabel{\kOneTwo}} (a2);
\draw[->] (a2) to node[lab] {\alabel{\kTwoThree}} (a3);
\draw[->] (a3) to node[lab] {\alabel{\kThreeFour}} (a4);
\draw[->] (a4) to node[lab,pos=0.45] {\alabel{\kFourFive}} (a5);
\draw[rounded corners,->] (a5) -- (10,1.75) -- node[lab,swap] {\alabel{\kFiveThree}} (6,1.75) -- (a3);
\draw[rounded corners,->] (ae) -- (12,3.5) -- node[lab,swap] {\alabel{\kEOne}} (2,3.5) -- (a1);

\draw[rounded corners,->] (a1) -- node[lab,anchor=west,yshift=0.07cm] {\nlabel{\kOneE}} (2,-2.2) -- (12,-2.2) -- (ae);
\draw[rounded corners] (a2) -- node[lab,anchor=west,yshift=0.07cm] {\nlabel{\kTwoE}} (4,-2.2) -- (6,-2.2);
\draw[rounded corners] (a3) -- node[lab,anchor=west,yshift=0.07cm] {\nlabel{\kThreeE}} (6,-2.2) -- (8,-2.2);
\draw[rounded corners] (a4) -- node[lab,anchor=west,yshift=0.07cm] {\nlabel{\kFourE}} (8,-2.2) -- (10,-2.2);
\draw[rounded corners] (a5) -- node[lab,anchor=west,xshift=-0.03cm,yshift=0.07cm] {\nlabel{\kFiveE}} (10,-2.2) -- (11,-2.2);
\end{scope}
\end{tikzpicture}}
%%% Local Variables:
%%% mode: latex
%%% TeX-master: "../thesis"
%%% End:
  \caption{SST constructed from the oracle machine in Figure~\ref{fig:decimal_example}.}
  \label{fig:decimal-sst}
\end{figure*}
\end{landscape}

\section{Implementation and Benchmarks}
\label{sec:benchmarks}

% \begin{figure}
% \centering
% \input{fig_compilation_paths}
% \caption{Compilation paths. $1$-LA is symbolic SST construction with single-symbol transitions; $k$-LA is with up to $k$ symbols of lookahead for some $k$ determined by the program. ``inline'' indicates deforestation, while ``pipeline'' indicates that action and oracle machines are composed at runtime.}
% \label{fig:architecture}
% \end{figure}

Our implementation\footnote{Source code and benchmarks available at \url{http://kleenexlang.org/}} compiles the action machine and the oracle SST to machine code via C.
%The possible compilation paths can be seen in Figure~\ref{fig:architecture}.
We have implemented several optimizations which are orthogonal to the underlying principles behind our compilation from Kleenex via transducers to SSTs:
\paragraph{Inlining of output actions} The action machine and the oracle SST need to be composed. We can do this at runtime by piping the SST output to the action machine, or we can apply a form of deforestation~\cite{wadler1990} to inline the output actions directly into the SST. This is straightforward since the machines are deterministic.

\paragraph{Constant propagation} The SSTs generated by the construction underlying Theorem~\ref{thm:SST_construction} typically contain many constant-valued registers (e.g.\ most registers in Figure~\ref{fig:decimal-sst} are constant). We eliminate these using constant propagation: compute reaching definitions by solving a set of data-flow constraints.

\paragraph{Symbolic representation} A more succinct SST representation is obtained by using a symbolic representation of transitions where input symbols are replaced by \emph{predicates} and output symbols by \emph{terms} indexed by input symbols. This is a straightforward extension of similar representations for automata~\cite{watson1996} and transducers~\cite{vanoord2001,veanes2012,veanes2014,veanes2015}. Our implementation uses simple predicates in the form of byte ranges, and simple output terms represented by byte-indexed lookup tables. We refer the reader to the cited literature for the technical details of symbolic transducers.

\paragraph{Finite lookahead} Symbolic FSTs with bounded lookahead have been shown to reduce the state space when representing string encoders~\cite{dantoni_static_2013,veanes2014,veanes2015}. We have implemented a form of finite lookahead in our SST representation. Opportunities for lookahead is detected by the compiler, and arise in the case where the program contains a string constant with length above one. In this case a lookahead transition is used to check once and for all if the string constant is matched by the input instead of creating an SST state for each symbol. This may in some cases reduce the size of the generated code since we avoid tabulating all states of the whole program for every prefix of the string constant.

We have run comparisons with different combinations of the following tools:
\begin{description}\compresslist
    \item[RE2,] Google's regular expression C++ library~\cite{re2}.
    \item[RE2J,] a recent re-implementation of RE2 in Java~\cite{re2j}.
    \item[GNU \gnuawk{} and GNU \gnused{},] programming languages and tools for text processing and extraction~\cite{gnu}.
    \item[Oniglib,] a regular expression library written in C++ with support for different character encodings~\cite{oniguruma}.
    \item[Ragel,] a finite state machine compiler with multiple language backends~\cite{ragel}.
\end{description}

In addition, we implemented test programs using the standard regular expression libraries in the scripting languages Perl~\cite{wco2000}, Python~\cite{lutz2010}, and Tcl~\cite{welch2003}.

The benchmark suite, Kleenex programs, and version numbers of libraries used can be found at \url{http://kleenexlang.org}.

\paragraph{Meaning of plot labels}
Kleenex plot labels indicate the compilation path, and follow the format
\texttt{[<0|3>[-la] | woACT] [clang|gcc]}. 
\texttt{0}/\texttt{3} indicates whether constant propagation was disabled/enabled. 
\texttt{la} indicates whether lookahead was enabled. 
\texttt{clang}/\texttt{gcc}  indicates which C compiler was used. 
The last part indicates that custom register updates are disabled, in which case we generate a single fused SST as described in Section \ref{sec:bench-with-without-splitting}.
These are only run with constant propagation and lookahead enabled.

\paragraph{Experimental setup}
The benchmark machine runs Linux, has 32 GB RAM and an eight-core Intel Xeon E3-1276 3.6 GHz CPU with 256 KB L2 cache and 8 MB L3 cache.
Each benchmark program was run 15 times, after first doing two warm-up rounds.
% All programs ran on a single core. 
%Version numbers of libraries, etc.\ can be seen on the project homepage.
All C and C++ files have been compiled with \texttt{-O3}.  
%All tested versions of Kleenex are single-core.

\paragraph{Difference between Kleenex and the other implementations}
Unless otherwise stated, the structure of all the non-Kleenex implementations is a loop that reads input line by line and applies an action to the line.  
Hence, in these implementations there is an interplay between the regular expression library used and the external language, e.g., RE2 and C++.  
In Kleenex, line breaks do not carry any special significance, so the multi-line programs can be formulated entirely within Kleenex.

\paragraph{Ragel optimization levels}
Ragel is compiled with three different optimization levels: T1, F1, and G2.
``T1'' and ``F1'' means that the generated C code should be based on a lookup-table, and ``G2'' means that it should be based on C \texttt{goto} statements.

\paragraph{Kleenex compilation timeout}

On some plots, some versions of the Kleenex programs are not included. 
This is because the C compiler times out (after 30 seconds).
As we fully determinize the transducers, the resulting C code can explode in some cases. 
%Minimization of transducers and SSTS is an area for future research.  
The two  
worst-case exponential blow-ups in generating transducers from Kleenex and then generating SSTs implemented in C code from transducers are \emph{inherent}, though, and as such can be considered a \emph{feature} of Kleenex: 
tools based on finite machines with no or limited nondeterminism support such as Ragel would 
require \emph{hand-coding} a potentially huge machine that Kleenex generates \emph{automatically}.\footnote{We have found it excessively difficult to employ Ragel in some use cases with a natural nondeterministic specification.}  

\subsection{Baseline}
\label{sec:baseline-benchmarks}

The following two programs are intended to give a baseline impression of the performance of Kleenex programs.

% Programs:
%   baseline: rot13
%             flip_ab
%             patho2
%   rewriting: thousandsep
%              csv_project3
%              irc
%   play: apache_log
%         ini2json (pipeline needed, no plot)
%         syntax highlighter for kleenex! (no plot)
%         issuu_json2sql
%         bibtex (DRex-stuff... no plot)
%   with/without bytesplitting:  isodatetime2json (only kleenex)
%                                csv_project3  (only kleenex )
%   
% \paragraph{\prog{rot13}}

\paragraph{\prog{flip\_ab}}

The program \prog{flip\_ab} swaps ``a''s and ``b''s on all its input lines.
In Kleenex it looks like this:
\begin{KleenexVerb}
\PY{n+nf}{main} \PY{o}{:=} (\PY{l+s}{\PYZdq{}b\PYZdq{}} \PY{c+c1}{\PYZti{}/a/} | \PY{l+s}{\PYZdq{}a\PYZdq{}} \PY{c+c1}{\PYZti{}/b/} | \PY{l+s+sx}{/\PY{esc}{\PYZbs{}n}/})\PY{o}{*}
\end{KleenexVerb}

We made a corresponding implementation with Ragel, using a \texttt{while}-loop in C to get each new input line and feed it to the automaton code generated by Ragel.

Implementing this functionality with regular expression libraries in the other tools would be an unnatural use of them, so we have not measured those. 

The performance of the two implementations run on input with an average line length of 1000 characters is shown in Figure~\ref{fig:bench-flip_ab}.

\begin{figure}[t]
    \plot{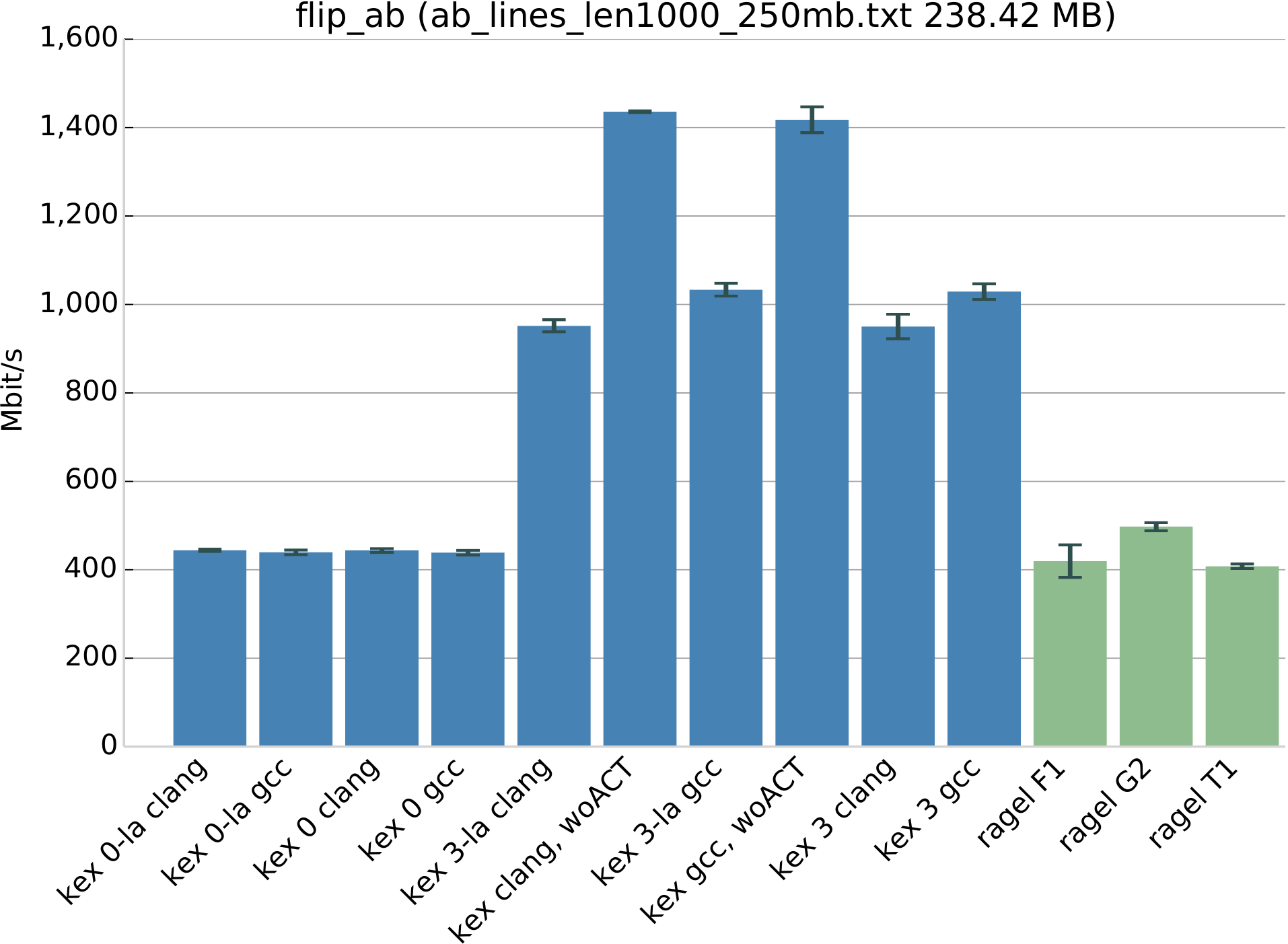}
    \caption{\prog{flip\_ab} run on lines with average length 1000.}
    \label{fig:bench-flip_ab}
\end{figure}

\paragraph{\prog{patho2}}

The program \prog{patho2} forces Kleenex to wait until the very last character of each line has been read before it can produce any output:
\begin{KleenexVerb}
\PY{n+nf}{main} \PY{o}{:=} ((\PY{c+c1}{\PYZti{}/[a-z]*a/} | \PY{l+s+sx}{/[a-z]*b/})\PY{o}{?} \PY{l+s+sx}{/\PY{esc}{\PYZbs{}n}/})\PY{o}{+}
\end{KleenexVerb}

In this benchmark, the constant propagation makes a big difference, as Figure~\ref{fig:bench-patho2} shows. 
Due to the high degree of interleaving and the lack of keywords, in this program the lookahead optimization has reduced overall performance.

This benchmark was not run with Ragel because Ragel requires the programmer to do all disambiguation manually when writing the program; the C code that Ragel generates does not handle ambiguity in a for us predictable way.

\begin{figure}[t]
    \plot{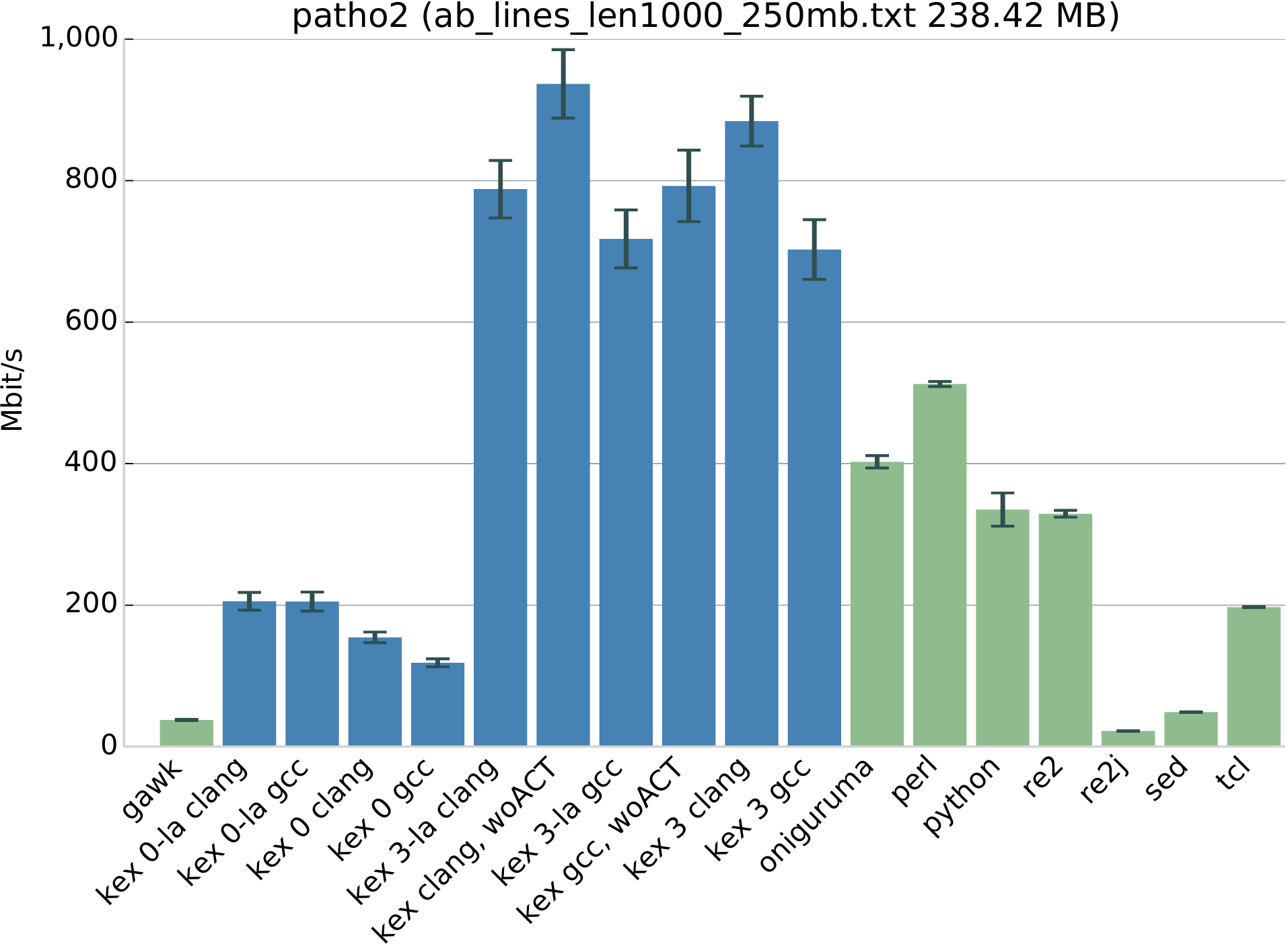}
    \caption{\prog{patho2} run on lines with average length 1000.}
    \label{fig:bench-patho2}
\end{figure}

\subsection{Rewriting}
\label{sec:benchmark-rewriting}

\paragraph{Thousand separators}

The following Kleenex program inserts thousand separators in a sequence of digits:
\begin{KleenexVerb}
\PY{n+nf}{main}  \PY{o}{:=} (\PY{n+nf}{num} \PY{l+s+sx}{/\PY{esc}{\PYZbs{}n}/})\PY{o}{*}
\PY{n+nf}{num}   \PY{o}{:=} \PY{n+nf}{digit}\PYZob{}\PY{n+nf}{1},\PY{n+nf}{3}\PYZcb{} (\PY{l+s}{\PYZdq{},\PYZdq{}} \PY{n+nf}{digit}\PYZob{}\PY{n+nf}{3}\PYZcb{})\PY{o}{*}
\PY{n+nf}{digit} \PY{o}{:=} \PY{l+s+sx}{/[0-9]/}
\end{KleenexVerb}
We evaluated the Kleenex implementation along with two other implementations using Perl and Python.
The performance can be seen in Figure~\ref{fig:bench-thousand-sep}.
Both Perl and Python are significantly slower than all of the Kleenex implementations; the problem is tricky to solve with regular expressions unless one reads the input right-to-left.

\begin{figure}[t]
    \plot{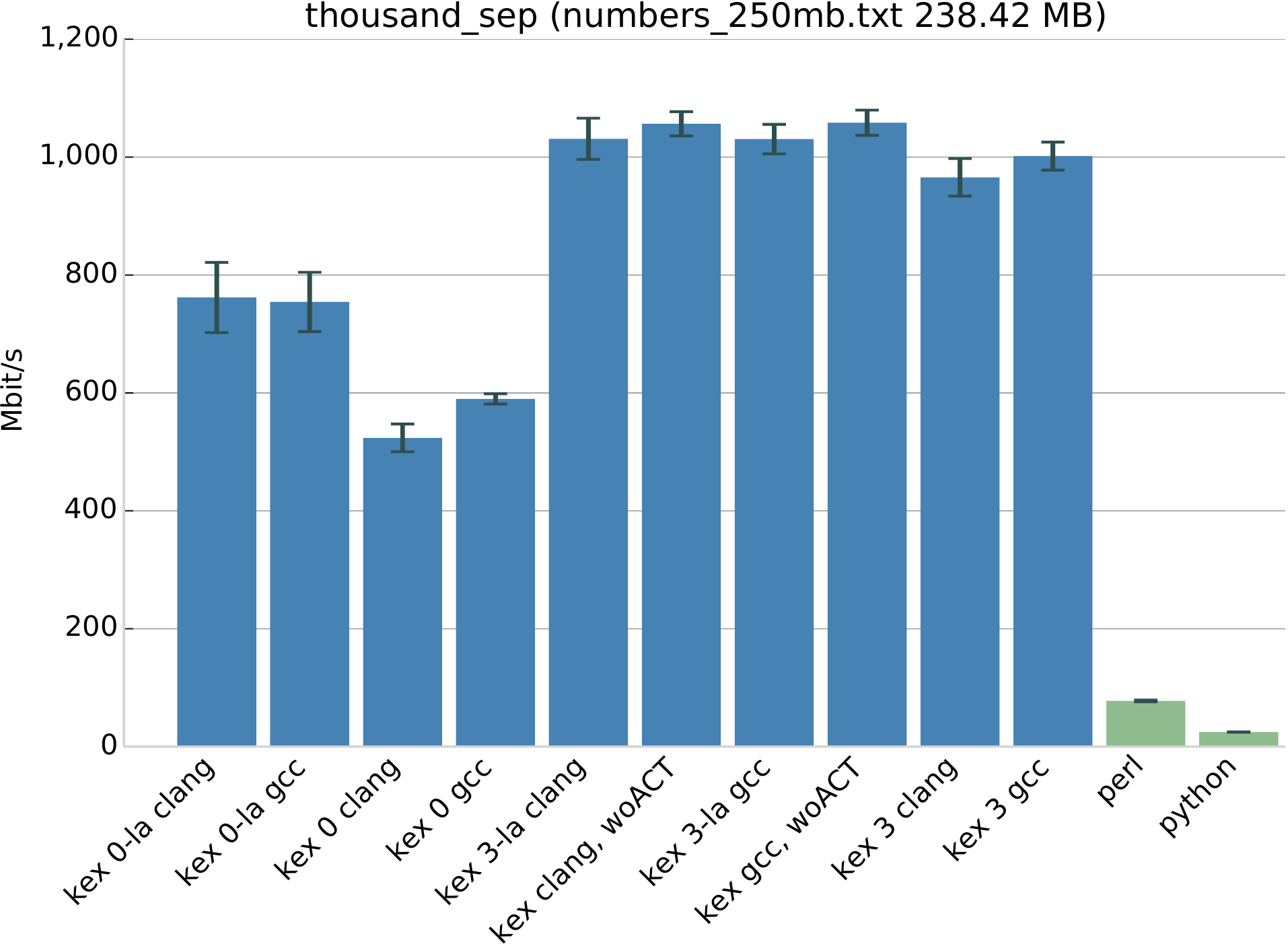}
    \caption{Inserting separators in random numbers of average length 1000.}
    \label{fig:bench-thousand-sep}
\end{figure}

\paragraph{IRC protocol handling}

The following Kleenex program parses the IRC protocol as specified in RFC 2812.\footnote{\url{https://tools.ietf.org/html/rfc2812}}
It follows roughly the output style described in part 2.3.1 of the RFC.
Note that the Kleenex source code and the BNF grammar in the RFC are almost identical.
Figure~\ref{fig:bench-irc} shows the throughput on 250 MiB data.

\begin{KleenexVerb}
\PY{n+nf}{main} \PY{o}{:=} (\PY{n+nf}{message} | \PY{l+s}{\PYZdq{}Malformed line: \PYZdq{}} \PY{l+s+sx}{/[^\PY{esc}{\PYZbs{}r}\PY{esc}{\PYZbs{}n}]*\PY{esc}{\PYZbs{}r}?\PY{esc}{\PYZbs{}n}/})\PY{o}{*}
\PY{n+nf}{message} \PY{o}{:=} (\PY{c+c1}{\PYZti{}/:/} \PY{l+s}{\PYZdq{}Prefix: \PYZdq{}} \PY{n+nf}{prefix} \PY{l+s}{\PYZdq{}\PY{esc}{\PYZbs{}n}\PYZdq{}}  \PY{c+c1}{\PYZti{}/ /})\PY{o}{?}
           \PY{l+s}{\PYZdq{}Command: \PYZdq{}} \PY{n+nf}{command} \PY{l+s}{\PYZdq{}\PY{esc}{\PYZbs{}n}\PYZdq{}}
           \PY{l+s}{\PYZdq{}Parameters: \PYZdq{}} \PY{n+nf}{params}\PY{o}{?} \PY{l+s}{\PYZdq{}\PY{esc}{\PYZbs{}n}\PYZdq{}}
           \PY{c+c1}{\PYZti{}crlf}
\PY{n+nf}{command} \PY{o}{:=} \PY{n+nf}{letter}\PY{o}{+} | \PY{n+nf}{digit}\PYZob{}\PY{n+nf}{3}\PYZcb{}
\PY{n+nf}{prefix} \PY{o}{:=} \PY{n+nf}{servername} | \PY{n+nf}{nickname} ((\PY{l+s+sx}{/!/} \PY{n+nf}{user})\PY{o}{?} \PY{l+s+sx}{/@/} \PY{n+nf}{host} )\PY{o}{?}
\PY{n+nf}{user} \PY{o}{:=} \PY{l+s+sx}{/[^\PY{esc}{\PYZbs{}n}\PY{esc}{\PYZbs{}r} @]/}\PY{o}{+} \PY{c+c1}{// Missing \PY{esc}{\PYZbs{}x00}}
\PY{n+nf}{middle} \PY{o}{:=} \PY{n+nf}{nospcrlfcl} ( \PY{l+s+sx}{/:/} | \PY{n+nf}{nospcrlfcl} )\PY{o}{*}
\PY{n+nf}{params} \PY{o}{:=} (\PY{c+c1}{\PYZti{}/ /} \PY{n+nf}{middle} \PY{l+s}{\PYZdq{}, \PYZdq{}})\PYZob{},\PY{n+nf}{14}\PYZcb{} ( \PY{c+c1}{\PYZti{}/ :/} \PY{n+nf}{trailing} )\PY{o}{?}
        | ( \PY{c+c1}{\PYZti{}/ /} \PY{n+nf}{middle} )\PYZob{}\PY{n+nf}{14}\PYZcb{} ( \PY{l+s+sx}{/ /} \PY{l+s+sx}{/:/}\PY{o}{?}  \PY{n+nf}{trailing} )\PY{o}{?}
\PY{n+nf}{trailing} \PY{o}{:=} (\PY{l+s+sx}{/:/} | \PY{l+s+sx}{/ /} | \PY{n+nf}{nospcrlfcl})\PY{o}{*}
\PY{n+nf}{nickname} \PY{o}{:=} (\PY{n+nf}{letter} | \PY{n+nf}{special}) 
            (\PY{n+nf}{letter} | \PY{n+nf}{special} | \PY{n+nf}{digit})\PYZob{},\PY{n+nf}{10}\PYZcb{}
\PY{n+nf}{host} \PY{o}{:=} \PY{n+nf}{hostname} | \PY{n+nf}{hostaddr}
\PY{n+nf}{servername} \PY{o}{:=} \PY{n+nf}{hostname}
\PY{n+nf}{hostname} \PY{o}{:=} \PY{n+nf}{shortname} ( \PY{l+s+sx}{/\PY{esc}{\PYZbs{}.}/} \PY{n+nf}{shortname})\PY{o}{*}
\PY{n+nf}{hostaddr} \PY{o}{:=} \PY{n+nf}{ip4addr}
\PY{n+nf}{shortname} \PY{o}{:=} (\PY{n+nf}{letter} | \PY{n+nf}{digit}) (\PY{n+nf}{letter} | \PY{n+nf}{digit} | \PY{l+s+sx}{/-/})\PY{o}{*} 
             (\PY{n+nf}{letter} | \PY{n+nf}{digit})\PY{o}{*}
\PY{n+nf}{ip4addr} \PY{o}{:=} (\PY{n+nf}{digit}\PYZob{}\PY{n+nf}{1},\PY{n+nf}{3}\PYZcb{} \PY{l+s+sx}{/\PY{esc}{\PYZbs{}.}/} )\PYZob{}\PY{n+nf}{3}\PYZcb{} \PY{n+nf}{digit}\PYZob{}\PY{n+nf}{1},\PY{n+nf}{3}\PYZcb{}
\end{KleenexVerb}

\begin{figure}[t]
    \plot{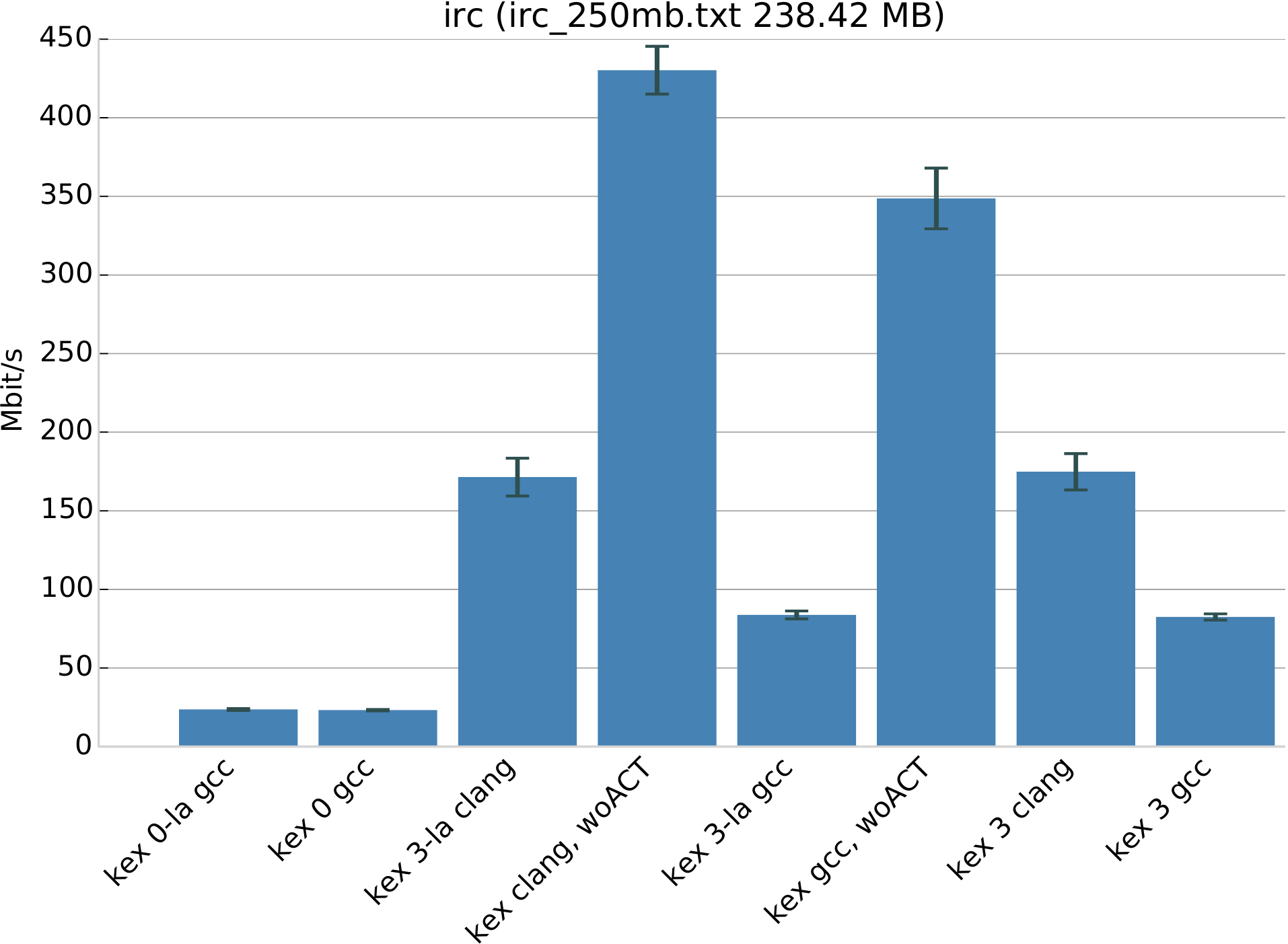}
    \caption{Throughput when parsing 250 MiB random IRC data.}
    \label{fig:bench-irc}
\end{figure}

\paragraph{CSV rewriting}

The program \prog{csv\_project3} deletes all columns but the $2$nd and $5$th from a CSV file:
\begin{KleenexVerb}
\PY{n+nf}{main} \PY{o}{:=} (\PY{n+nf}{row} \PY{l+s+sx}{/\PY{esc}{\PYZbs{}n}/})\PY{o}{*}
\PY{n+nf}{col}  \PY{o}{:=} \PY{l+s+sx}{/[^,\PY{esc}{\PYZbs{}n}]*/}
\PY{n+nf}{row}  \PY{o}{:=} \PY{c+c1}{\PYZti{}(col /,/)} \PY{n+nf}{col} \PY{l+s}{\PYZdq{}\PY{esc}{\PYZbs{}t}\PYZdq{}} \PY{c+c1}{\PYZti{}/,/} \PY{c+c1}{\PYZti{}(col /,/)}
        \PY{c+c1}{\PYZti{}(col /,/)} \PY{n+nf}{col} \PY{c+c1}{\PYZti{}/,/}      \PY{c+c1}{\PYZti{}col}
\end{KleenexVerb}
Various specialized tools that can handle this transformation are included in Figure~\ref{fig:bench-csv_project3}; GNU \gnucut{} is a command that splits its input on certain characters, and GNU \gnuawk{} has built-in support for this type of transformation.

Apart from \gnucut{}, which is very fast for its own use case, a Kleenex implementation is the fastest.
The performance of Ragel is slightly lower, but this is likely due to the way the implementation produces output. In a Kleenex program, output strings are automatically put in an output buffer which is flushed routinely, whereas a programmer has to manually handle buffering when writing a Ragel program.

\begin{figure}[t]
    \plot{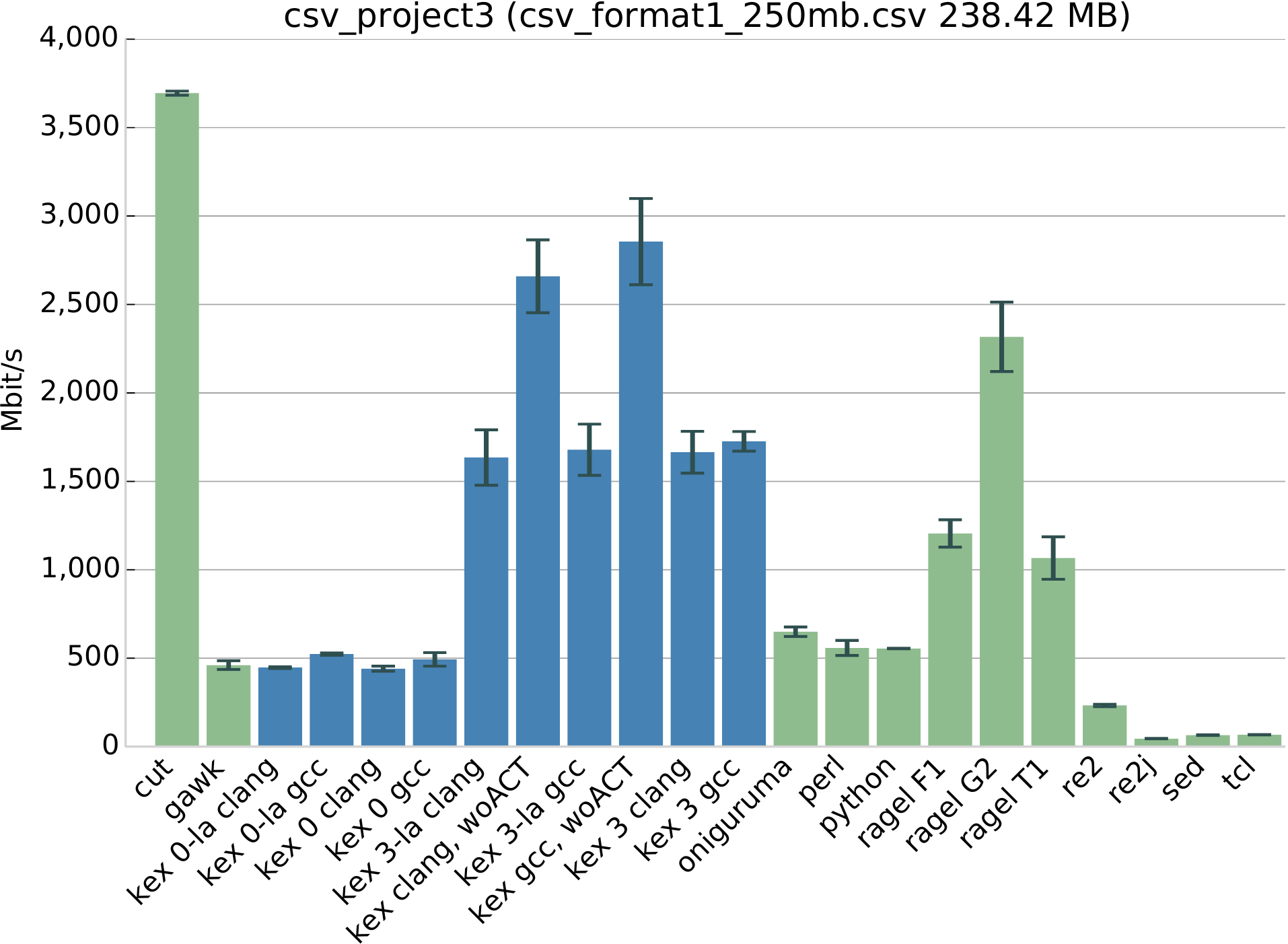}
    \caption[Benchmark for program \prog{csv\_project3}.]{\prog{csv\_project3} reads in a CSV file with six columns and outputs columns two and five.
``gawk'' is GNU \gnuawk{} that uses the native \gnuawk{} way of splitting up lines.
``cut'' is a tool from GNU coreutils that splits up lines.}
    \label{fig:bench-csv_project3}
\end{figure}

\subsection{With or Without Action Separation}
\label{sec:bench-with-without-splitting}

One can choose to use the machine resulting from fusing the oracle and action machines when compiling Kleenex.
Doing so results in only one process performing both disambiguation and outputting, which in some cases is faster and in other cases slower.
Figures~\ref{fig:bench-csv_project3}, \ref{fig:bench-issuu_json2sql}, and \ref{fig:bench-iso_datetime_to_json} illustrate both situations.
It depends on the structure of the problem whether it pays off to split up the work into two processes; if all the work happens in the oracle machine and the action machine does nearly nothing, then the added overhead incurred by the process context switches becomes noticeable.
On the other hand, in cases where both machines perform much work, the fact that two CPU cores can be utilized in parallel speeds up execution. This is more likely once Kleenex has support for actions that can perform arbitrary computations, e.g.\ in the form of embedded C code.

\section{Use Cases}
\label{sec:usecases}

We briefly touch upon various use cases---natural application scenarios---for Kleenex.

\paragraph{JSON logs to SQL}

We have implemented a Kleenex program that transforms a JSON log file into an SQL insert statement. 
The program works on the logs provided by Issuu.\footnote{The line-based data set consists of 30 compressed parts; part one is available from \url{http://labs.issuu.com/anodataset/2014-03-1.json.xz}}

The Ragel version we implemented outperforms Kleenex by about 50\% (Figure~\ref{fig:bench-issuu_json2sql}), indicating that further optimizations of our SST construction should be possible.

\begin{figure}[t]
    \plot{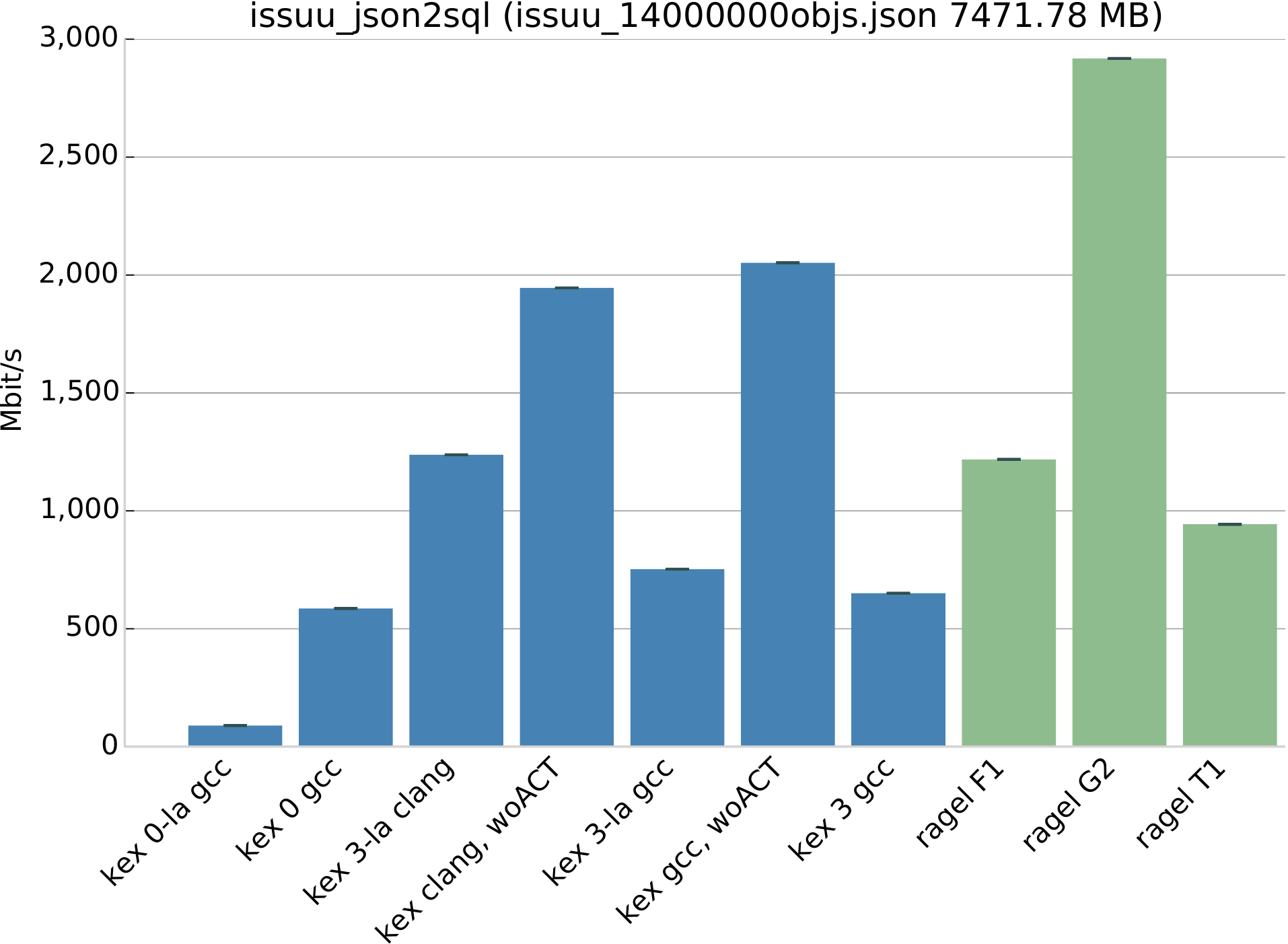}
    \caption[JSON to SQL benchmark.]{The speeds of transforming JSON objects to SQL INSERT statements using Ragel and Kleenex.}
    \label{fig:bench-issuu_json2sql}
\end{figure}

\paragraph{Apache CLF to JSON}

The Kleenex program below rewrites Apache CLF\footnote{\url{https://httpd.apache.org/docs/trunk/logs.html\#common}} log files into a list of JSON records:
\begin{KleenexVerb}
\PY{n+nf}{main} \PY{o}{:=} \PY{l+s}{\PYZdq{}[\PYZdq{}} \PY{n+nf}{loglines}\PY{o}{?} \PY{l+s}{\PYZdq{}]\PY{esc}{\PYZbs{}n}\PYZdq{}}
\PY{n+nf}{loglines} \PY{o}{:=} (\PY{n+nf}{logline} \PY{l+s}{\PYZdq{},\PYZdq{}} \PY{l+s+sx}{/\PY{esc}{\PYZbs{}n}/})\PY{o}{*} \PY{n+nf}{logline} \PY{l+s+sx}{/\PY{esc}{\PYZbs{}n}/}
\PY{n+nf}{logline} \PY{o}{:=} \PY{l+s}{\PYZdq{}\PYZob{}\PYZdq{}} \PY{n+nf}{host} \PY{c+c1}{\PYZti{}sep} \PY{c+c1}{\PYZti{}userid} \PY{c+c1}{\PYZti{}sep} \PY{c+c1}{\PYZti{}authuser} \PY{n+nf}{sep} 
               \PY{n+nf}{timestamp} \PY{n+nf}{sep} \PY{n+nf}{request} \PY{n+nf}{sep} \PY{n+nf}{code} \PY{n+nf}{sep}
               \PY{n+nf}{bytes} \PY{n+nf}{sep} \PY{n+nf}{referer} \PY{n+nf}{sep} \PY{n+nf}{useragent} \PY{l+s}{\PYZdq{}\PYZcb{}\PYZdq{}}
\PY{n+nf}{host} \PY{o}{:=} \PY{l+s}{\PYZdq{}\PY{esc}{\PYZbs{}\PYZdq{}}host\PY{esc}{\PYZbs{}\PYZdq{}}:\PY{esc}{\PYZbs{}\PYZdq{}}\PYZdq{}} \PY{n+nf}{ip} \PY{l+s}{\PYZdq{}\PY{esc}{\PYZbs{}\PYZdq{}}\PYZdq{}}
\PY{n+nf}{userid} \PY{o}{:=} \PY{l+s}{\PYZdq{}\PY{esc}{\PYZbs{}\PYZdq{}}user\PY{esc}{\PYZbs{}\PYZdq{}}:\PY{esc}{\PYZbs{}\PYZdq{}}\PYZdq{}} \PY{l+s+sx}{/-/} \PY{l+s}{\PYZdq{}\PY{esc}{\PYZbs{}\PYZdq{}}\PYZdq{}}
\PY{n+nf}{authuser} \PY{o}{:=} \PY{l+s}{\PYZdq{}\PY{esc}{\PYZbs{}\PYZdq{}}authuser\PY{esc}{\PYZbs{}\PYZdq{}}:\PY{esc}{\PYZbs{}\PYZdq{}}\PYZdq{}} \PY{l+s+sx}{/[^ \PY{esc}{\PYZbs{}n}]+/} \PY{l+s}{\PYZdq{}\PY{esc}{\PYZbs{}\PYZdq{}}\PYZdq{}}
\PY{n+nf}{timestamp} \PY{o}{:=} \PY{l+s}{\PYZdq{}\PY{esc}{\PYZbs{}\PYZdq{}}date\PY{esc}{\PYZbs{}\PYZdq{}}:\PY{esc}{\PYZbs{}\PYZdq{}}\PYZdq{}} \PY{c+c1}{\PYZti{}/\PY{esc}{\PYZbs{}[}/} \PY{l+s+sx}{/[^\PY{esc}{\PYZbs{}n}\PY{esc}{\PYZbs{}]}]+/} \PY{c+c1}{\PYZti{}/]/} \PY{l+s}{\PYZdq{}\PY{esc}{\PYZbs{}\PYZdq{}}\PYZdq{}}
\PY{n+nf}{request} \PY{o}{:=} \PY{l+s}{\PYZdq{}\PY{esc}{\PYZbs{}\PYZdq{}}request\PY{esc}{\PYZbs{}\PYZdq{}}:\PYZdq{}} \PY{n+nf}{quotedString}
\PY{n+nf}{code} \PY{o}{:=} \PY{l+s}{\PYZdq{}\PY{esc}{\PYZbs{}\PYZdq{}}status\PY{esc}{\PYZbs{}\PYZdq{}}:\PY{esc}{\PYZbs{}\PYZdq{}}\PYZdq{}} \PY{n+nf}{integer} \PY{l+s}{\PYZdq{}\PY{esc}{\PYZbs{}\PYZdq{}}\PYZdq{}}
\PY{n+nf}{bytes} \PY{o}{:=} \PY{l+s}{\PYZdq{}\PY{esc}{\PYZbs{}\PYZdq{}}size\PY{esc}{\PYZbs{}\PYZdq{}}:\PY{esc}{\PYZbs{}\PYZdq{}}\PYZdq{}} (\PY{n+nf}{integer} | \PY{l+s+sx}{/-/}) \PY{l+s}{\PYZdq{}\PY{esc}{\PYZbs{}\PYZdq{}}\PYZdq{}}
\PY{n+nf}{referer} \PY{o}{:=} \PY{l+s}{\PYZdq{}\PY{esc}{\PYZbs{}\PYZdq{}}url\PY{esc}{\PYZbs{}\PYZdq{}}:\PYZdq{}} \PY{n+nf}{quotedString}
\PY{n+nf}{useragent} \PY{o}{:=} \PY{l+s}{\PYZdq{}\PY{esc}{\PYZbs{}\PYZdq{}}agent\PY{esc}{\PYZbs{}\PYZdq{}}:\PYZdq{}} \PY{n+nf}{quotedString}
\PY{n+nf}{sep} \PY{o}{:=} \PY{l+s}{\PYZdq{},\PYZdq{}} \PY{c+c1}{\PYZti{}/[\PY{esc}{\PYZbs{}t} ]+/}
\PY{n+nf}{quotedString} \PY{o}{:=} \PY{l+s+sx}{/"([^"\PY{esc}{\PYZbs{}n}]|\PY{esc}{\PYZbs{}\PYZbs{}}")*"/}
\PY{n+nf}{integer} \PY{o}{:=} \PY{l+s+sx}{/[0-9]+/}
\PY{n+nf}{ip} \PY{o}{:=} \PY{n+nf}{integer} (\PY{l+s+sx}{/\PY{esc}{\PYZbs{}.}/} \PY{n+nf}{integer})\PYZob{}\PY{n+nf}{3}\PYZcb{}
\end{KleenexVerb}
This is a re-implementation of a Ragel program.\footnote{\url{https://engineering.emcien.com/2013/04/5-building-tokenizers-with-ragel}}
Figure~\ref{fig:bench-apache_log} shows the benchmark results.
The versions compiled with clang are not included, as the compilation timed out after 30 seconds.
Curiously, the non-optimized Kleenex program is the fastest in this case.

\begin{figure}[t]
    \plot{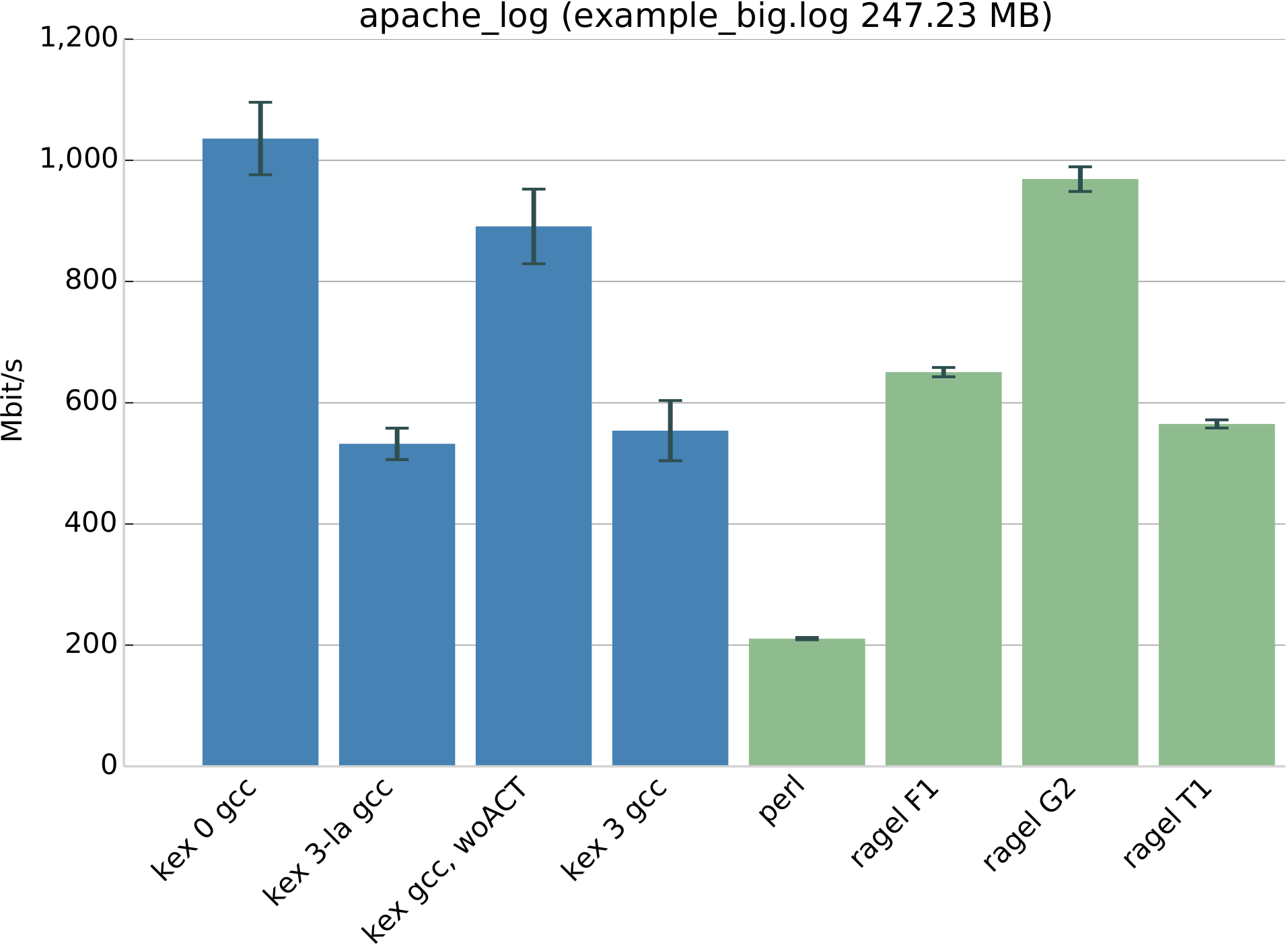}
    \caption[Apache Log to JSON benchmark.]{Speed of the conversion from the Apache Common Log Format to JSON.}
    \label{fig:bench-apache_log}
\end{figure}

\paragraph{ISO date/time objects to JSON}

Inspired by an example in~\cite{goyvaerts2009}, the program \prog{iso\_datetime\_to\_json} converts date and time stamps in an ISO standard format to a JSON object.
Figure~\ref{fig:bench-iso_datetime_to_json} shows the performance.

\begin{figure}[t]
    \plot{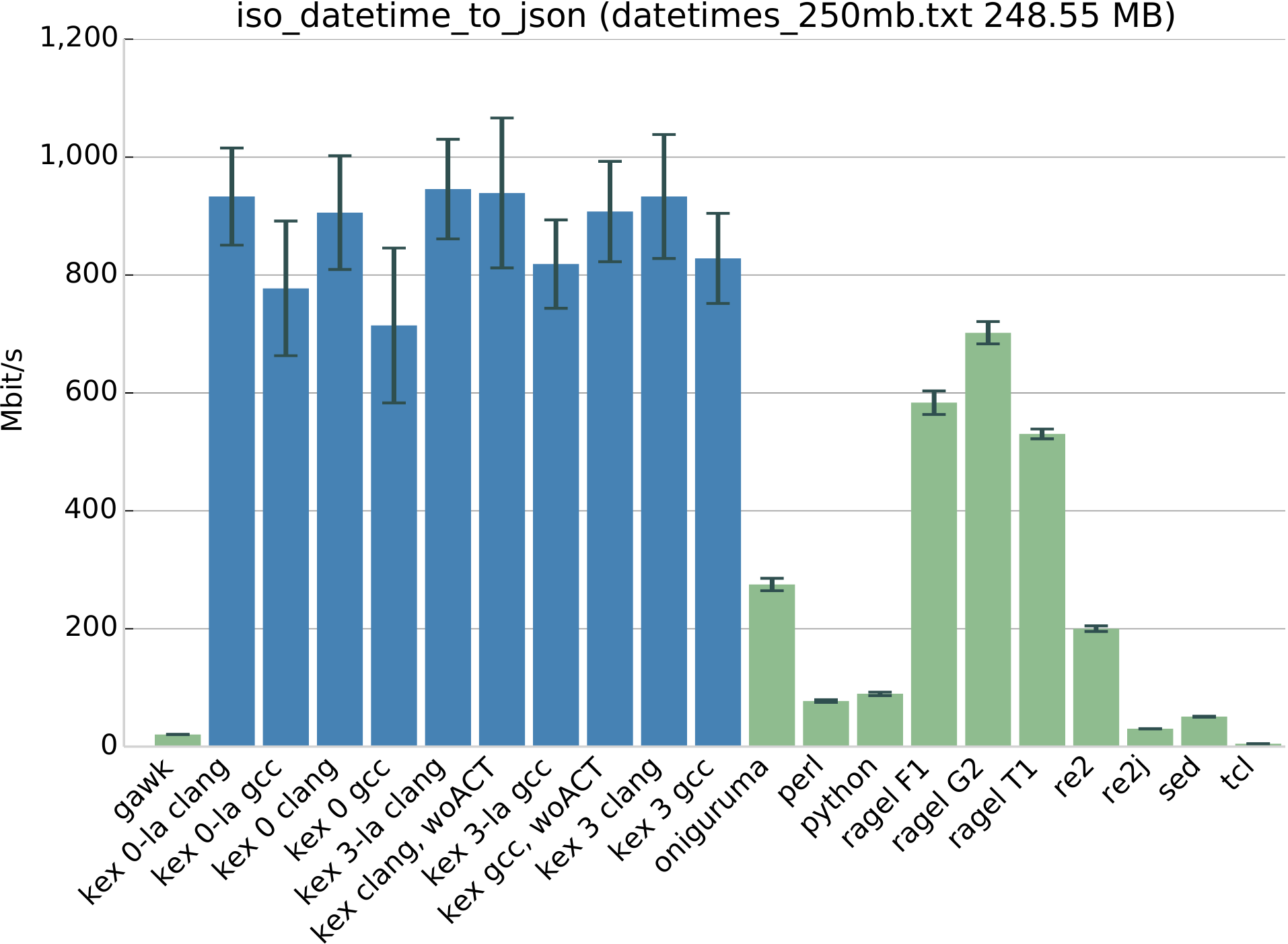}
    \caption[ISO time stamps to JSON benchmark.]{The performance of the conversion of ISO time stamps into JSON format.}
    \label{fig:bench-iso_datetime_to_json}
\end{figure}

\paragraph{HTML comments}

The following Kleenex program finds HTML comments with basic formatting commands and renders them in HTML after the comment.
For example, \texttt{<!-- doc: *Hello* world -->} becomes \texttt{<!-- doc: *Hello* world --><div> <b>Hello</b> world </div>}.

\begin{KleenexVerb}
\PY{n+nf}{main} \PY{o}{:=} (\PY{n+nf}{comment} | \PY{l+s+sx}{/./})\PY{o}{*}
\PY{n+nf}{comment} \PY{o}{:=} \PY{l+s+sx}{/<!-- doc:/} \PY{n+nf}{clear} \PY{n+nf}{doc}\PY{o}{*} \PY{o}{!}\PY{n+nf}{orig} \PY{l+s+sx}{/-->/}
           \PY{l+s}{\PYZdq{}<div>\PYZdq{}} \PY{o}{!}\PY{n+nf}{render} \PY{l+s}{\PYZdq{}</div>\PYZdq{}}
\PY{n+nf}{doc} \PY{o}{:=} \PY{c+c1}{\PYZti{}/\PY{esc}{\PYZbs{}*}/} \PY{n+nf}{t}\PY{o}{@}\PY{l+s+sx}{/[^*]*/} \PY{c+c1}{\PYZti{}/\PY{esc}{\PYZbs{}*}/} 
       [ \PY{n+nf}{orig} \PY{o}{+=} \PY{l+s}{\PYZdq{}*\PYZdq{}} \PY{n+nf}{t} \PY{l+s}{\PYZdq{}*\PYZdq{}} ] [ \PY{n+nf}{render} \PY{o}{+=} \PY{l+s}{\PYZdq{}<b>\PYZdq{}} \PY{n+nf}{t} \PY{l+s}{\PYZdq{}</b>\PYZdq{}} ]
     | \PY{n+nf}{t}\PY{o}{@}\PY{l+s+sx}{/./} [ \PY{n+nf}{orig} \PY{o}{+=} \PY{n+nf}{t} ] [ \PY{n+nf}{render} \PY{o}{+=} \PY{n+nf}{t} ]
\PY{n+nf}{clear} \PY{o}{:=} [ \PY{n+nf}{orig}  \PY{o}{<-} \PY{l+s}{\PYZdq{}\PYZdq{}} ] [ \PY{n+nf}{render} \PY{o}{<-} \PY{l+s}{\PYZdq{}\PYZdq{}} ]
\end{KleenexVerb}

% \paragraph{URL parsing}
% 
% Kleenex allows one to naturally follow the URL specification given in RFC1738.\footnote{\url{http://www. ietf.org/rfc/rfc1738.txt}}
% We implemented a URL parser by directly following the BNF-grammar in the RFC.
% %; its code, along with all other benchmark and example programs can be found on the homepage.

\paragraph{Syntax highlighting}

Kleenex can be used to write syntax highlighters; in fact, the Kleenex syntax in this paper was highlighted using a Kleenex program.

\section{Discussion}
\label{sec:discussion}

We discuss related and future work by building Kleenex conceptually up from regular expression matching via regular expressions as types for bit-coded parsing to transducers and eventually grammars with embedded actions.

\paragraph{Regular Expression Matching.}
Regular expression \emph{matching} has different meanings in the literature.  

For \emph{acceptance testing}, the subject of \emph{automata theory} where only a single bit is output, NFA-simulation and DFA-construction are classical techniques. Bille and Thorup \cite{bith2009} improve on Myers' \cite{myers92} log-factor improved classical NFA-simulation for regular expressions, based on tabling.  They design an $O(k n)$  algorithm \cite{bith2010} with word-level parallelism, where $k \leq m$ is the number of strings occurring in an RE.  The tabling technique may be promising in practice; the algorithms have not been implemented and evaluated empirically, though.  

In \emph{subgroup matching} as in PCRE \cite{pcre2010}, an input is not only classified as accepting or not, but a substring is returned for each sub-RE of interest. Subgroup matching exposes ambiguity in the RE. 
Subgroup matching is often implemented by backtracking over alternatives, which implements \emph{greedy} disambiguation.\footnote{Committing to the left alternative before checking that the remainder of the input is accepted is the essence of \emph{parsing expression grammars} \cite{ford2004parsing}.}  Backtracking may result in exponential-time worst case behavior, however, even in the absence of inherently hard matching with backreferences \cite{aho90}.  Considerable human effort is usually expended to engineer REs used in practice to perform well anyway.  More recently, REs designed to force exponential run-time behavior are used in algorithmic attacks, though \cite{sumi2014,DBLP:journals/corr/RathnayakeT14}.  Some subgroup matching libraries have guaranteed worst-case linear-time performance based on automata-theoretic techniques, notably Google's RE2 \cite{re2}.  Intel's Hyperscan \cite{hyperscan} is also described as employing automata-theoretic techniques.  A key point of Kleenex is implementing the natural backtracking semantics without actually performing backtracking and without requiring storage of the input.
 
Myers, Oliva and Guimaraes \cite{myers1998reporting} and Okui, Suzuki \cite{okui2011} describe a $O(m n)$, respectively $O(m^2 n)$ \emph{POSIX}-disambiguated matching algorithms.  Sulzmann and Lu \cite{sulzmann2012} use Brzozowski \cite{brzozowski64} and Antimirov derivatives \cite{antimirov96} for Perl-style subgroup matching for greedy and POSIX disambiguation.  
Borsotti, Breveglieri, Reghizzi, and Morzenti \cite{borsotti2015bsp,borsotti2015ambiguous} have devised a Berry-Sethi based parser generator that can be configured for greedy or POSIX disambiguation.

\paragraph{Regular expression parsing.} Full RE \emph{parsing}, also called RE matching \cite{frca2004}, generalizes subgroup matching to return a full parse tree.  The set of parses are exactly the elements of a regular expression read as a \emph{type} \cite{frca2004,heni2011}: Kleene-star is the (finite) list type constructor, concatenation the Cartesian product, alternation the sum type and an individual character the singleton type containing that character. A \emph{(McNaughton-Yamada-)Thompson NFA} \cite{mnya60,thompson68} represents an RE in a strong sense: the complete paths---paths from initial to final state---are in one-to-one correspondence with the parses \cite{ghnr2013,graulund2015}.
A Thompson NFA equipped with $0$, $1$ outputs \cite{ghnr2013} is a certain kind of oracle machine. The bit-code it generates can also be computed directly from the RE underlying the Thompson automaton \cite{heni2011,nihe2011}.
The \emph{greedy RE parsing problem} produces the lexicographically least bit-code for a string matching a given RE.  Kearns \cite{kearns91}, Frisch and Cardelli \cite{frca2004} devise 3-pass linear-time \emph{greedy} RE parsing; they require 2 passes over the input, the first consisting of reversing the entire input, before generating output in the third pass. 
%D ub\'{e} and Feeley \cite{dufe2000}
Grathwohl, Henglein, Nielsen, Rasmussen devise a two-pass \cite{ghnr2013} and an optimally streaming \cite{grathwohl2014a} greedy regular expression parsing algorithm.  The algorithm works for all NFAs, indeed transducers, not just Thompson NFAs. 

%Streaming guarantees that line-by-line RE matching can be coded as a single RE matching problem.
Sulzman and Lu \cite{sulzmann2014} remark that POSIX is notoriously difficult to implement correctly and show how to use Brzozowski derivatives \cite{brzozowski64} for POSIX RE parsing.

%Here it is critical that Thompson NFAs have $\epsilon$-transitions since equivalent $\epsilon$-free automata require $\Omega(m \log m)$ transitions \cite{schnitger2006regular} and standard
%$\epsilon$-free NFA-constructions \cite{glushkov61,antimirov96,ilie2003follow} even $\Omega(m^2)$.)

\paragraph{Regular expression implementation optimizations.} There are specialized RE matching tools and techniques too numerous to review comprehensively.  We mention a few employing automaton optimization techniques potentially applicable to Kleenex, but presently unexplored.  Yang, Manadhata, Horne, Rao, Ganapathy \cite{yang2012fast} propose an OBDD representation for subgroup matching and apply it to intrusion detection REs; the cycle counts per byte appear a bit high, but are reported to be competitive with RE2.  
Sidhu and Prasanna \cite{sidhu2001} implement NFAs directly on an FPGA, essentially performing NFA-simulation in parallel; it outperforms GNU \gnugrep.
Brodie, Taylor, Cytron \cite{btc2006} construct a multistride DFA, which
processes multiple input symbols in parallel, and devise a compressed implementation on stock FPGA, also achieving very high throughput rates.  Likewise, Ziria employs tabled multistriding to achieve high throughput \cite{stewart2015ziria}.
Navarro and Raffinot \cite{navarro2001compact} show how to code DFAs compactly for efficient simulation.

%Might, Darais, Spiewak \cite{might2011parsing} context-free parsing with derivatives. 
%Caron, Champarnaud, Mignot \cite{caron2011partial}, derivatives for extended REs

%Owens, Reppy, Turon \cite{ort2009}, rediscovering Brzozowski derivatives for constructing DFAs.

%Nielsen, Henglein \cite{nihe2011}, bit-coded greedy RE parsing.  
 
%\subsection{Ambiguity}
%
%REs may be ambiguous, which is irrelevant for acceptance testing, but  problematic for subgroup matching and parsing since the output depends on which amongst possibly multiple matches is to be returned.  
%Br\"{u}ggemann-Klein \cite{brueggemann93} provides an efficient $O(m^2)$ RE ambiguity testing algorithm. 
%Vansummeren \cite{vansummeren2006} illustrates differences between POSIX, first/longest and greedy matches.
%%Clark and Cormack \cite{clco97} disambiguation policy for searching text, first-longest (POSIX) not found to be good.
%Colcombet \cite{colcombet2012} analyzes notions of (non)determinism of automata.  

\paragraph{Finite state transducers.}
From RE parsing it is a surprisingly short distance to the implementation of arbitrary nondeterministic finite state transducers (FSTs) \cite{berstel79,mohri1997finite}.  In contrast to the situation for \emph{automata}, nondeterministic transducers are strictly more powerful than deterministic transducers; this, together with observable ambiguity,  highlights why RE parsing is more challenging than RE acceptance testing.

As we have noted, efficient RE parsing algorithms operate on arbitrary NFAs, not only those corresponding to REs.  Indeed, REs are not a particularly convenient or compact way of specifying regular languages: they can be represented by \emph{certain} small NFAs with low tree width \cite{johnson2001directed}, but may be inherently quadratically bigger than automata, even for DFAs \cite[Theorem 23]{ellul2005regular}.  This is why Kleenex employs well-formed context-free grammars, which are much more compact than regular expressions.  

\paragraph{Streaming string transducers.} We have shown in this paper that the greedy semantics of arbitrary FSTs can be compiled to a \emph{subclass} of streaming string transducers (SSTs).
SSTs extensionally correspond to regular transductions, functions implementable by 2-way deterministic finite-state transducers \cite{alur2010expressiveness}, MSO-definable string transductions \cite{enho2001} and a combinator language analogous to regular expressions \cite{Alur:2014:RCS:2603088.2603151}.  The implementation
techniques used in Kleenex appear to be directly applicable to all
SSTs, not just the ones corresponding to FSTs. 

DReX~\cite{adr2015} is a combinatory functional language for expressing all SST-definable transductions.  Kleenex without register operations is expressively more restrictive; with copy-less register operations it appears to compactly code exactly the nondeterministic SSTs and thus SSTs.  Programs in DReX must be unambiguous by construction while programs in Kleenex may be nondeterministic and ambiguous, which is greedily disambiguated.

\paragraph{Symbolic transducers.} Veanes, Molnar, Mytkowics \cite{veanes2015} employ symbolic transducers \cite{veanes2012,dantoniveanes2014} in the implementation of the Microsoft Research languages BEK\footnote{\url{http://research.microsoft.com/en-us/projects/bek}} and BEX\footnote{\url{http://research.microsoft.com/en-us/projects/bex}} for multicore execution.  These techniques can be thought of as synthesizing code that implements the transition function of a finite state machine not only efficiently, but also compactly.  Tabling in code form (switch statement) or data form (lookup in array) is the standard implementation technique for the transition function.  It is efficient when applicable, but not compact enough for large alphabets and multistrided processing.  Kleenex employs basic symbolic transition. Compact coding of multistrided transitions is likely to be crucial for exploiting word-level parallelism---processing 64 bits at a time---in practice.

\paragraph{Parallel transducer processing.} Allender and Mertz \cite{alme2015} show that the functions computable by cost register automata \cite{alur2013regular}, which generalize the string monoid used in SSTs to admit arbitrary monoids and more general algebraic structures, are in NC and thus inherently parallelizable.  This appears to be achievable by performing relational FST-composition by matrix multiplication on the matrix representation of FSTs \cite{berstel79}, which can be performed by parallel reduction.  This requires in principle running an FST from all states, not just the input state, on input string fragments.  Mytkowicz, Musuvathi, Schulte \cite{mytkowicz2014data} observe that there is often a small set of cut states sufficient to run each FST.  This promises to
be an interesting parallel harness for a suitably adapted Kleenex implementation running on fragments of very large inputs.

\paragraph{Syntax-directed translation schemes.} A Kleenex program is an example of a \emph{syntax-directed translation scheme (SDTS)} or a domain-specific stream processing language such as PADS \cite{fisher2005pads,fisher2011pads} and Ziria \cite{stewart2015ziria}.  In these the underlying grammar is typically deterministic modulo short lookahead so that semantic actions can be executed immediately when encountered during parsing.

Kleenex is restricted to non-self-embedding grammars to avoid the matrix-multiplication lower bound on general context-free parsing \cite{lee2002fast}; it supports full nondeterminism without lookahead restriction, though.
A key contribution of Kleenex is that semantic actions are scheduled no earlier than semantically permissible and no later than necessary.

\section{Conclusions}
\label{sec:conclusions}

We have presented Kleenex, a convenient language for specifying nondeterministic finite state transducers, and its compilation to machine code implementing streaming string transducers.

Kleenex is comparatively expressive and performs consistently well.  For complex regular expressions with nontrivial amounts of output it is almost always better than industrial-strength text processing tools such as RE2, Ragel, \gnuawk, \gnused~ and RE-libraries of Perl, Python and Tcl in the evaluated use cases.

We believe Kleenex's clean semantics, streaming optimality, algorithmic generality, worst-case guarantees and absence of tricky code and special casing provide a useful basis for 
\begin{itemize}\compresslist
\item extensions, specifically visibly push-down transducers \cite{raskin2008,talbot2014}, restricted versions of backreferences and approximate regular expression matching\cite{myers1989approximate,wu1992agrep};
\item known, but so far unexplored optimizations, such as multistriding, automata minimization and symbolic representation, hybrid FST simulation and SST construction; 
\item massively parallel (log-depth, linear work) processing.
\end{itemize}

% \acks
% 
% This work has been partially supported by The Danish Council for Independent Research under Project 11-106278, ``Kleene Meets Church: Regular Expressions and Types''; see \url{http://diku.dk/kmc}.
% We would like to thank Alexandra Silva and Nate Foster for their critical questions and comments and the anonymous referees for their detailed reviews, which have given rise to numerous changes in the final version. 
% We thank Issuu for releasing their data set to the research community.
% The work by the Jobindex authors was performed while being Master's students at DIKU.
% 
% The order of authors is insignificant; please list all authors---or none---when citing this paper.

}

\putbib[bibliography]
\end{bibunit}

%%% Local Variables:
%%% mode: latex
%%% TeX-master: "thesis"
%%% End:

\begin{bibunit}[abbrv]
\chapter[PEG Parsing Using Tabling and Dynamic Analysis][PEG Parsing Using Progressive Tabling]{PEG Parsing in Less Space Using Progressive Tabling and Dynamic Analysis}
\label{paper:peg-parsing}

The following paper is unpublished at the time of writing, but is planned for submission. The manuscript and the majority of the development was done by the author of this dissertation, with parts of the theory developed in collaboration with Fritz Henglein.

{
%%%%%%%%%%%%%%%%%%%%%%%%%%%%%%%%%%%%%%%%%%%%%%%%%%%%%%%%%%%%%%%%%%%%%%
%% Macros
%%%%%%%%%%%%%%%%%%%%%%%%%%%%%%%%%%%%%%%%%%%%%%%%%%%%%%%%%%%%%%%%%%%%%%

% Highlighting
\newcommand{\highlight}[1]{\colorbox{gray!20}{#1}}

%% Name of the exeuction model. TBD
\newcommand{\TT}{PTP} % PTP == Progressive Tabular Parsing

%% TODO-notes
\newcommand{\fixme}[1]{%
   \todo[inline%
        ,caption={}%
        ,size=\small%
        ,bordercolor=black!50%
        ,backgroundcolor=orange!40]%
        {#1}}

%% Concrete faces
\newcommand{\C}[1]{\ensuremath{\mathcal{#1}}}     % mathcal shorthand

%% Semantic faces
\newcommand{\Set}[1]{\mathsf{#1}}                 % Named sets
\newcommand{\Ctor}[1]{\mathsf{#1}}                % Term constructors
\newcommand{\Lit}[1]{\texttt{#1}}                 % Character literals

%% Sets
\newcommand{\Nat}{\mathbb{N}}
\newcommand{\Two}{\mathbf{2}}

%% Special symbols
\newcommand{\eof}{\#}

%% Encoding
\newcommand{\enc}[1]{\ulcorner {#1} \urcorner}

%% GTDPLO expressions
\newcommand{\GExpr}{\Set{GExpr}}
\newcommand{\geps}{\epsilon}
\newcommand{\gfail}{\Ctor{f}}
\newcommand{\gout}[1]{\Ctor{"} {#1} \Ctor{"}}
\newcommand{\gto}{{\leftarrow}} % production relation

%% Walk states
\newcommand{\waccept}{\Ctor{accept}}
\newcommand{\wreject}{\Ctor{reject}}

%% GTDPLO relations
\newcommand{\gparse}{\Rightarrow} % Binary parse relation

%% Lists
%\newcommand{\app}{\mathop{{+}\!\!{+}}}
\newcommand{\app}{}
\newcommand{\cons}{{::}}
\newcommand{\nil}{\varepsilon}
\newcommand{\hole}{\bullet}

%% Partial results
%\newcommand{\pendft}[3]{\setlength{\fboxsep}{0.12em}\fbox{${#1}[{#2},{#3}]$}}
%\newcommand{\pendf}[3]{\setlength{\fboxsep}{0.12em}\fbox{${#1}[{#2},{#3}]$}}
\newcommand{\pends}[2]{{#1} \lapp {#2}}
% the "unknown" partial result
\newcommand{\pendnext}[1]{\setlength{\fboxsep}{0.12em}\fbox{${#1}\Ctor{?}$}}
% a reference to a (non-terminal, offset) cell
\newcommand{\noderef}[2]{\setlength{\fboxsep}{0.12em}\fbox{${#1}$}_{#2}}
% lifted append. Used to indicate that an output prefix should be appended to
% the result of cell reference (\noderef) upon success.
\newcommand{\lapp}{\mathop{{+}\!\!{+}}}

% PEG pretty-printing
\newcommand{\ul}[1]{\underline{\smash{#1}}\vphantom{#1}}
\newcommand{\nonterminal}[1]{\mathop{\ul{#1}}}
\newcommand{\Prange}[2]{{[}\Pin{#1} ... \Pin{#2}{]}}
\newcommand{\Pin}[1]{\texttt{#1}}
\newcommand{\spacesym}{space}
\newcommand{\newlinesym}{newline}

\clearpage

\thispagestyle{plain}
\begin{center}
{\LARGE \textbf{PEG Parsing in Less Space Using Progressive Tabling and Dynamic Analysis\footnote{The order of authors is insignificant.}}}

\vspace{1.5em}

{{Fritz Henglein} and {Ulrik Terp Rasmussen}}

\vspace{1em}

{Department of Computer Science, University of Copenhagen (DIKU)}
\end{center}

\begin{abstract}
Tabular top-down parsing and its lazy variant, Packrat, are linear-time execution models for the TDPL family of recursive descent parsers with limited backtracking.
By tabulating the result of each (nonterminal, offset)-pair, we avoid exponential work due to backtracking at the expense of always using space proportional to the product of the input length and grammar size.
Current methods for limiting the space usage relies either on manual annotations or on static analyses which are sensitive to the syntactic structure of the grammar.

We present \emph{progressive tabular parsing} (PTP), a new execution model which progressively computes parse tables for longer prefixes of the input and simultaneously generates a leftmost expansion of the parts of the parse tree that can be resolved. Table columns can be discarded on-the-fly as the expansion progresses through the input string, providing best-case constant and worst-case linear memory use. Furthermore, semantic actions are scheduled before the parser has seen the end of the input. The scheduling is conservative in the sense that no action has to be ``undone'' in the case of backtracking.

The time complexity is $O(dmn)$ where $m$ is the size of the parser specification, $n$ is the size of the input string, and $d$ is either a configured constant or the maximum parser stack depth.

For common data exchange formats such as JSON, we demonstrate practically constant space usage, and without static annotation of the grammar.
\end{abstract}

\section{Introduction}

Parsing of computer languages has been a topic of research for several decades, leading to a large family of different parsing methods and formalisms. Still, with each solution offering varying degrees of expressivity, flexibility, speed and memory usage, and often at a trade-off, none of them can be regarded as an ideal general approach to solving to all parsing problems. For example, compiler writers often specify their languages in a declarative formalism such as context-free grammars (CFG), relying LL($k$) or LR($k$) parser generators to turn their specifications into executable parsers. The resulting parsers are often fast, but with the downsides that a separate lexical preprocessing is needed, and that the programmer is required to mold the grammar into a form that is deterministic for the chosen parser technology. Such solutions require a large investment in time, as identifying the sources of non-determinism in a grammar can be quite difficult. A user who needs to write an ad-hoc parser will thus not find that the amount of time invested makes up for the apparent benefits.

Aho and Ullman's TDPL/GTDPL languages~\cite{aho1972}, which were later popularized as Parsing Expression Grammars (PEG)~\cite{ford2004}, provide a formal foundation for the specification of recursive-descent parsers with limited backtracking. They do away with the problem of non-determinism by always having, by definition, a single unique parse for every accepted input. The syntax of PEGs resembles that of CFGs, but where a CFG is a set of generative rules specifying its language, a PEG is a set of rules for a backtracking \emph{recognizer}, and its language is the set of strings recognized. This ensures unique parses, but with the downside that it can sometimes be quite hard to determine what language a given PEG represents.
Recognition can be performed in linear time and space by an algorithm which computes a table of results for every (nonterminal, input offset)-pair~\cite{aho1972}, although it seems to never have been used in practice, probably due to its large complexity constants. Ford's Packrat parsing~\cite{ford2002} reduces these constants by only computing the table entries that are needed to resolve the actual parse. However, the memory usage of Packrat is $\Theta(mn)$ for PEGs of size $m$ and inputs of size $n$, which can be prohibitively expensive for large $m$ and $n$, and completely precludes applying it in a streaming context where input is potentially infinite. Heuristics for reducing memory usage~\cite{kuramitsu2015,redziejowski2009} still store the complete input string, and even risks triggering exponential time behavior. One method~\cite{mizushima2010} can remove both table regions and input prefixes from memory during runtime, but relies on manual annotations and/or a static analysis which does not seem to perform well beyond LL languages~\cite{redziejowski2016}.

In this paper, we present \emph{progressive tabular parsing} (PTP), a new execution model for the TDPL family of languages. The method is based on the tabular parsing of Aho and Ullman, but avoids computing the full parse table at once. We instead start by computing a table with a single column based on the first symbol in the input. For each consecutive symbol, we append a corresponding column to the table and update all other entries based on the newly added information. We continue this until the end of the input has been reached and the full parse table has been computed. During this process, we have access to partial parse tables which we use to guide a leftmost expansion of the parse tree for the overall parse. Whenever a prefix of the input has been uniquely parsed by this process, the prefix and its corresponding table columns can be removed from memory. The result is a linear-time parsing algorithm which still uses $O(mn)$ memory in the worst case, but $O(m)$ in the best case. Since we have access to the partial results of every nonterminal during parsing, a simple dynamic analysis can use the table to rule out alternative branches and speculatively expand the parse tree before the corresponding production has been fully resolved. The speculation is conservative and never has to undo an expansion unless the whole parse turns out to fail. The analysis changes the time complexity to $O(dmn)$ for a configurable constant $d$ bounded by the maximum stack depth of the parser, but preliminary experiments suggests that it pays for itself in practice by avoiding the computation of unused table entries.

The method can be formulated elegantly using least fixed points of monotone table operators in the partial order of tables with entrywise comparison, and where unresolved entries are considered a bottom element in the partial order. The computation of parse tables is then an instance of \emph{chaotic iteration}~\cite{cousot77} for computing least fixed points using a work set instead of evolving all entries in parallel. The work set is maintained such that we obtain meaningful partial parse tables as intermediate results which can be used by the dynamic analysis. Linear time is obtained by using an auxiliary data structure to ensure that each table entry is added to the work set at most once.

Our evaluation demonstrates that PTP dynamically adapts its memory usage based on the amount of lookahead required to resolve productions. The complexity constant due to indiscriminately computing all entries of the parse table can be quite large, but we are confident that this problem can be alleviated in the same way as Packrat reduced the constants for conventional tabular parsing.
We believe that our general formulation of PTP offers a solid foundation for further development of both static and dynamic analyses for improving performance.

To summarize, we make the following contributions:

\begin{itemize}
\item \emph{Progressive tabular parsing} (PTP), a new execution model for the TDPL family of parsing formalisms. The execution of a program proceeds by progressively computing parse tables, one for each prefix of the input, using the method of \emph{chaotic iteration} for computing least fixed points. Meanwhile, a leftmost expansion of the parse tree is generated in a streaming fashion using the parse table as an oracle. Table columns are discarded on-the-fly as soon as the method detects that a backtracking parser would never have to return to the corresponding part of the input.

\item An algorithm for computing progressive parse tables in an incremental fashion. It operates in amortized time $O(mn)$ for grammars of size $m$ and inputs of size $n$, and produces $n$ progressive approximations of the parse table. The algorithm implements the execution model in $O(mn)$ time and space. We show that for certain grammars and inputs, as little as $O(m)$ space is consumed.

\item A configurable dynamic analysis which can dramatically improve the streaming behavior of parsers by allowing a longer trace to be generated earlier in the parse.
%Existing methods largely rely on manual program annotations or offline static analysis to obtain the same results.
The dynamic analysis changes the time complexity to $O(dmn)$ where $d$ is either a configured constant or the maximum parser stack depth.

\item An evaluation of a prototype of the algorithm which demonstrates that
  \begin{enumerate*}[label=\alph*)]
    \item for an unannotated JSON parser written in the PEG formalism, memory usage is practically constant,
    \item for parsers of non-LL languages, the algorithm adjusts memory usage according to the amount of lookahead required,
    \item however, ambiguous tail-recursive programs trigger worst-case behavior.
  \end{enumerate*}
\end{itemize}

The rest of the paper is organized as follows. The GTDPL and PEG parsing formalisms are introduced in Section~\ref{sec:parsing-formalism}, together with a notion of parse trees and a definition \emph{streaming parsing}. In Section~\ref{sec:tabulation} we recall the linear-time tabular parsing method, but defined using least fixed points. We extend this in Section~\ref{sec:streaming-parsing} to obtain an approximation of the full parse table based on a prefix of the full input string. In the same section, we define the streaming generation of execution traces based on dynamic analysis of approxmation tables, which we then use to present the \emph{progressive tabular parsing} method. In Section~\ref{sec:algorithm2} we define---and prove correct---an amortized linear-time algorithm for computing all progressive table approximations for all consecutive prefixes of an input string. A prototype implementation is evaluated on three different parsing programs in Section~\ref{sec:evaluation}, where we also point out the main challenges towards a high-performance implementation. We conclude with a discussion of related and future work in Section~\ref{sec:discussion2}.

\section{Parsing Formalism}
\label{sec:parsing-formalism}

The \emph{generalized top-down parsing language} (GTDPL) is a language for specifying top-down parsing algorithms with limited backtracking~\cite{aho1972,birman1970}. It has the same recognition power as the \emph{top-down parsing language} (TDPL), from which it was generalized, and  \emph{parsing expression grammars} (PEG)~\cite{ford2004}, albeit using a smaller set of operators.

The top-down parsing formalism can be seen as a recognition-based alternative to declarative formalisms used to describe machine languages, such as context-free grammars (CFGs). A CFG constitutes a set of generative rules that characterize a language, and the presence of ambiguity and non-determinism poses severe challenges when such a specification must be turned into a deterministic parsing algorithm. In contrast, every GTDPL/PEG by definition denotes a deterministic \emph{program} which operates on an input string and returns with an outcome indicating failure or success. The recognition power of CFGs and GTDPL/PEG is incomparable. There are GTDPLs which recognize languages that are not context-free~\cite{aho1972}, e.g. the language $\{a^nb^nc^n \mid n \geq 0\}$. On the other hand, GTDPL recognition is linear-time~\cite{aho1972} and CFG recognition is super-linear~\cite{lee2002}, which suggests that there exists a context-free languages that cannot be recognized by any GTDPL.\footnote{To the best of our knowledge, no such language is known.}

Let $\Sigma$ be a finite alphabet, and $\C{N}$ a finite set of \emph{nonterminal symbols}.

\begin{definition}[Program]
A GTDPL program (henceforth just \emph{program}) is a tuple $P = (\Sigma, V, S, R)$ where
\begin{enumerate}
  \item $\Sigma$ is a finite input alphabets; and
  \item $V$ is a finite set of \emph{nonterminal} symbols; and
  \item $S \in V$ is the starting nonterminal; and
  \item $R = \{ A_0 \gto g_0, ..., A_{m-1} \gto g_{m-1} \}$ is a non-empty finite set of numbered \emph{rules}, where each $A_i$ is in $V$ and each $g_i \in \GExpr$ is an \emph{expression} generated by the grammar
  \[  \GExpr \ni g ::= \geps \mid \gfail \mid a \mid A[B,C] \]
  where $A,B,C \in V$, $a \in \Sigma$. Rules are unique: $i \not= j$ implies $A_i \not= A_j$.
\end{enumerate}
\end{definition}
\noindent
Define the \emph{size} $|P|$ of a program to be the cardinality of its rule set $|R| = m$.
When $P$ is understood, we will write $A \gto g$ for the assertion $A \gto g \in R$.
By uniqueness of rule definitions, we can write $i_A$ for the unique index of a rule $A_i \gto g_i$ in $R$. If $g_i$ is of the form $B[C,D]$ we call it a \emph{complex expression}, otherwise we call it a \emph{simple expression}.

The intuitive semantics of a production $A \gto B[C,D]$ is to first try parsing the input with $B$. If this succeeds, parse the remainder with $C$, otherwise backtrack and parse from the beginning of the input with $D$. For this reason we call $B$ the \emph{condition} and $C$ and $D$ the \emph{continuation branch} and \emph{failure branch}, respectively.

Given sets $X, Y$, write $X + Y$ for their disjoint union $\{0\} \times X \cup \{1\} \times Y$.

\begin{definition}[Operational semantics]
  \label{def:operational-semantics}
  Let $P = (\Sigma, V, S, R)$ be a program and define a matching relation $\gparse_P$ from $V \times \Sigma^*$ to results $r \in \Sigma^* + \{\gfail\}$. That is, it relates pairs of the form $(A, u) \in V \times \Sigma^*$ to either the failure value $\gfail$, or a result value $v \in \Sigma^*$ indicating success, where $v$ is the suffix of $u$ that remains unconsumed. We leave out the subscript $P$ when it is clear from the context.

Let $\gparse_P$ be generated by the following rules:
\[
\begin{prooftree}
\Infer[left label=(1),right label=($A \gto \geps$)]0{ (A,u) \gparse_P u }
\end{prooftree}
\qquad
\begin{prooftree}
\Infer[left label=(2),right label=($A \gto \gfail$)]0{ (A,u) \gparse_P \gfail }
\end{prooftree}
\]
\[
\begin{prooftree}
\Infer[left label=(3i),right label=($A \gto a$)]0{(A, au) \gparse_P u}
\end{prooftree}
\]
\[
\begin{prooftree}
\Infer[left label=(3ii),right label=($A \gto a$ and $a$ not prefix of $u$)]0{(A, u) \gparse_P \gfail}
\end{prooftree}
\]
\[
\begin{prooftree}
\Hypo{ (B, u) \gparse_P v }
\Hypo{ (C, v) \gparse_P r }
\Infer[left label=(4i),right label={($A \gto B[C,D]$)}]2{(A,u) \gparse_p r}
\end{prooftree}
\]
\[
\begin{prooftree}
\Hypo{ (B, u) \gparse_P \gfail }
\Hypo{ (D, u) \gparse_P r }
\Infer[left label=(4ii),right label={($A \gto B[C,D]$)}]2{(A,u) \gparse_p r}
\end{prooftree}
\]
\end{definition}
The proof derivations generated by the rules will be denoted by subscripted variations of the letter $\C{D}$.

Write $(A, u) \not\gparse_P$ when there does not exist an $r$ such that $(A,u) \gparse_P r$.
Say that $A$ \emph{matches} $u$ when $(A,u) \gparse_P v$ for $v \in \Sigma^*$ (note that $A$ does not have to consume all of the input). The \emph{language} recognized by a nonterminal $A$ is the set $L_P(A) = \{ u \in \Sigma^* \mid \exists v \in \Sigma^* \ldotp (A, u) \gparse_P v \}$. The language \emph{rejected} by $A$ is the set $\overline{L}_P(A) = \{ u \in \Sigma^* \mid (A,u) \gparse_P \gfail \}$. We say that $A$ \emph{handles} $u$ when $u \in L_P(A) \cup \overline{L}_P(A)$. The program $P$ is \emph{complete} if the start symbol $S$ handles all strings $u \in \Sigma^*$.

The following two properties are easily shown by induction.
\begin{proposition}[Suffix output]
  If $(A, u) \gparse_P (s, w)$, then $w$ is a suffix of $u$ ($\exists v\ldotp u = v \app w$).
\end{proposition}

\begin{proposition}[Determinacy]
  If $(A, u) \gparse_P r_1$ by $\C{D}_1$ and $(A, u) \gparse_P r_2$ by $\C{D}_2$, then $\C{D}_1 = \C{D}_2$ and $r_1 = r_2$.
\end{proposition}

We recall the following negative decidability results proved by Ford for the PEG formalism~\cite{ford2004}. Since any GTDPL can be converted to an equivalent PEG and vice-versa, they hold for GTDPL as well.
\begin{proposition}
  \label{prop:undec:emptiness}
  It is undecidable whether $L_P(A) = \emptyset$ and whether $L_P(A) = \Sigma^*$.
\end{proposition}
\begin{proposition}
  \label{prop:undec:completeness}
  It is undecidable whether a program is complete.
\end{proposition}

\subsection{Parsing Expression Grammars}
Having only a single complex operator, GTDPL offers a minimal foundation which simplifies the developments in later sections. The drawback is that it is very hard to determine the language denoted by a given GTDPL program. In order to make examples more readable, we will admit programs to be presented with expressions from the extended set $\Set{PExpr}$ defined as follows:
\[ \Set{PExpr} \ni e ::= g \in \Set{GExpr} \mid e_1 e_2 \mid e_1 / e_2 \mid e^* \mid {!e_1} \]
This corresponds to the subset of \emph{predicate-free parsing expressions} extended with the ternary GTDPL operator. A program $P$ with productions in $\Set{PExpr}$ is called a \emph{PEG program}, and desugars to a pure GTDPL program by adding productions $E \gto \geps$ and $F \gto \gfail$ and replacing every non-conforming production as follows:
\[
  \begin{array}{l@{}c@{}lcl@{}c@{}l}
    A &\gto& e_1 e_2 & \longmapsto & A &\gto& B[C, F] \\
    &&&& B &\gto& e_1 \\
    &&&& C &\gto& e_2 \\
    \hline
    A &\gto& e_1 / e_2 & \longmapsto & A &\gto& B[E, C] \\
    &&&& B &\gto& e_1 \\
    &&&& C &\gto& e_2 \\
    \hline
    A &\gto& e_1^* & \longmapsto & A &\gto& B[A, E] \\
    &&&& B &\gto& e_1 \\
    \hline
    A &\gto& {!e_1} & \longmapsto & A &\gto& B[F,E] \\
    &&&& B &\gto& e_2
  \end{array}
\]
The desugaring embeds the semantics of PEG in GTDPL~\cite{ford2004}, so there is no need to introduce semantic rules for parsing expressions. Note that although parsing expressions resemble regular expressions, the recognizers that they denote may not recognize the same languages as their usual set-theoretic interpretation. For example, the expression $a^* a$ recognizes the empty language!

\subsection{Parse Trees}
We are usually interested in providing a parse tree instead of just doing recognition, e.g. for the purpose of executing semantic actions associated with parsing decisions. Unlike generative frameworks, any program uniquely matches an input via a unique derivation $\C{D}$, which we therefore could take as our notion of parse tree. However, for space complexity reasons we will employ a more compact notion for which we also define a bit coding for the purpose of providing a definition of streaming parsing.

A \emph{parse tree} $\C{T}$ is an ordered tree where each leaf node is labeled by the empty string or a symbol in $\Sigma$, and each internal node is labeled by a nonterminal subscripted by a symbol from $\Two \cup \{\varepsilon\}$ where $\Two = \{0, 1\}$.

\begin{definition}[Parse trees and codes]
  \label{def:parse-tree}
  For any $A \in V$, $u,v \in \Sigma^*$, and derivation $\C{D} :: (A,u) \gparse_P v$, define simultaneously a \emph{parse tree} $\C{T}_{\C{D}}$ and a \emph{parse code} $\C{C}_{\C{D}} \in \Two^*$ by recursion on $\C{D}$:
\end{definition}
\begin{enumerate}
\item If $A \gto \geps$, respectively $A \gto a$, then $\C{T}_{\C{D}}$ is a node labeled by $A_\varepsilon$ with a single child node labeled by $\varepsilon$, respectively $a$. Let $\C{C}_{\C{D}} = \varepsilon$.
\item If $A \gto B[C,D]$ and $\C{D}_1 :: (B,u) \gparse_P u'$ we must have $\C{D}_2 :: (C, u') \gparse_P v$. Let $\C{T}_{\C{D}}$ be a node $A_0$ with subtrees $\C{T}_{\C{D}_1}$ and $\C{T}_{\C{D}_2}$. Let $\C{C}_{\C{D}} = 0 ~ \C{C}_{\C{D}_1} \C{C}_{\C{D}_2}$.
\item If $A \gto B[C,D]$ and $\C{D}_1 :: (B,u) \gparse_P \gfail$, then we must have $\C{D}_2 :: (D, u') \gparse_P v$. Create a node labeled by $A_1$ with a single subtree $\C{T}_{\C{D}_2}$. Let $\C{C}_{\C{D}} = 1 ~ \C{C}_{\C{D}_2}$.
\end{enumerate}
The size of a parse tree $|\C{T}|$ is the number of nodes in it.
Note that only the parts of a derivation counting towards the successful match contribute to its parse tree, while failing subderivations are omitted.
This ensures that parse trees have size proportional to the input, in contrast to derivations which can grow exponentially in the worst case.
\begin{restatable}[Linear tree complexity]{proposition}{proplinearparsetrees}
  Fix a program $P$. For all $A \in V$ and $u,v \in \Sigma^*$ and derivations $\C{D} :: (A,u) \gparse_P v$ we have
  $|\C{T}(\C{D})| = O(|u|)$.
\end{restatable}

Parse trees and parse codes both provide injective codings of the subset of derivations with non-failing results.
\begin{restatable}[Injectivity]{proposition}{propinjectivity}
  Fix a program $P$ and symbol $A \in V$. For all $u_1,u_2,v_1,v_2 \in \Sigma^*$ and derivations $\C{D}_1 :: (A, u_1) \gparse_P v_1$ and $\C{D}_2 :: (A, u_2) \gparse_P v_2$, if $\C{D}_1 \not= \C{D}_2$, then $\C{T}_{\C{D}_1} \not= \C{T}_{\C{D}_2}$ and $\C{C}_{\C{D}_1} \not= \C{C}_{\C{D}_2}$.
\end{restatable}
It is easy to check that a code can be used to construct the corresponding parse tree in linear time, regardless of the size of the underlying derivation. In general, a code can be viewed as an oracle which guides a leftmost expansion of the corresponding parse tree. Any prefix of a code can thus be seen as a partially expanded parse tree. During expansion, we maintain a stack of nodes that are not yet expanded. If the top node is simple it can be expanded deterministically, and if it is complex the next code symbol determines its expansion; its child nodes are pushed on the stack.

\begin{example}
\label{ex:gtdpl1}
Consider the PEG program $S \gto (a^* b / \geps) a^*$, which desugars into:
\begin{align*}
S \gto{}& L[R, F] &
L \gto{}& P[E, E] &
P \gto{}& A[P, B] &
R \gto{}& A[R, E] \\
A \gto{}& a &
B \gto{}& b &
E \gto{}& \geps &
F \gto{}& \gfail
\end{align*}
We have derivations $\C{D} :: (S,aa) \gparse \varepsilon$ and $\C{D}' :: (S,aaba) \gparse \varepsilon$. Visualized below is, from left to right: the trees $\C{T}_{\C{D}}$, $\C{T}_{\C{D}'}$, and the partial tree expanded from the prefix $000$ of the code $\C{C}_{\C{D}'}$. The leftmost nonterminal leaf is the next to be expanded.
\begin{center}
\begin{tikzpicture}[level distance=0.85cm]
\tikzstyle{level 1}=[sibling distance=1cm]
\tikzstyle{level 2}=[sibling distance=0.7cm]
\tikzstyle{level 3}=[sibling distance=0.5cm]
\node {$S_0$}
child {
  node {$L_1$}
  child {
    node {$E$}
    child {
      child { child { node {$\underline{\varepsilon}$} } }
    }
  }
}
child {
  node {$R_0$}
  child {
    node {$A$}
    child {
      child { child { node {$\underline{a}$} } }
    }
  }
  child {
    node {$R_0$}
    child {
      node {$A$}
      child {
        child { node {$\underline{a}$} }
      }
    }
    child {
      node {$R_1$}
      child {
        node {$E$}
        child {
          node {$\underline{\varepsilon}$}
        }
      }
    }
  }
};
\end{tikzpicture}\qquad
\begin{tikzpicture}[level distance=0.8cm]
\tikzstyle{level 1}=[sibling distance=1.5cm]
\tikzstyle{level 2}=[sibling distance=1cm]
\tikzstyle{level 3}=[sibling distance=0.75cm]
\tikzstyle{level 4}=[sibling distance=0.5cm]
\node {$S_0$}
child {
  node {$L_0$}
  child {
    node {$P_0$}
    child {
      node {$A$}
      child {
        child { child { node {$\underline{a}$} } }
      }
    }
    child {
      node {$P_0$}
      child {
        node {$A$}
        child {
          child { node {$\underline{a}$} }
        }
      }
      child {
        node {$P_1$}
        child {
          node {$B$}
          child {
            node {$\underline{b}$}
          }
        }
      }
    }
  }
  child {
    node {$E$}
    child {
      child { child { child { node {$\underline{\varepsilon}$} } } }
    }
  }
}
child {
  node {$R_0$}
  child {
    node {$A$}
    child {
      child { child { child { node {$\underline{a}$} } } }
    }
  }
  child {
    node {$R_1$}
    child {
      node {$E$}
      child {
        child { child { node {$\underline{\varepsilon}$} } }
      }
    }
  }
};
\end{tikzpicture}\qquad
\begin{tikzpicture}[level distance=0.8cm]
\tikzstyle{level 1}=[sibling distance=0.5cm]
\tikzstyle{level 2}=[sibling distance=0.5cm]
\tikzstyle{level 3}=[sibling distance=0.5cm]
\node {$S_0$}
child {
  node {$L_0$}
  child {
    node {$P_0$}
    child {
      node {$A$}
      child {
        child { child { node {$\underline{a}$} } }
      }
    }
    child {
      node {$\underline P$}
    }
  }
  child {
    node {$\underline E$}
  }
}
child {
  node {$\underline R$}
};
\end{tikzpicture}
\end{center}

%%% Local Variables:
%%% mode: latex
%%% TeX-master: "../thesis"
%%% End:

The parse codes are $\C{C}_{\C{D}} = 01001$ and $\C{C}_{\C{D}'} = 0000101$, respectively. Observe that codes correspond to the subscripts of the internal nodes in the order they would be visited by an in-order traversal, reflecting the leftmost expansion order.
\end{example}

\subsection{Streaming Parsing}

Using parse codes, we can define \emph{streaming parsing}.
\begin{definition}[Streaming parsing function]
\label{def:streaming-parsing}
Let $\eof \not\in\Sigma$ be a special end-of-input marker. A \emph{streaming parsing function} for a program $P$ is a function $f : \Sigma^*(\eof \cup \varepsilon) \to \Two^*$ which for every input prefix $u \in \Sigma^*$ satisfies the following:
\begin{enumerate}
\item it is monotone: For all $v \in \Sigma^*$, $f(uv) = f(u) c'$ for some $c' \in \Two^*$.
\item it computes code prefixes: For all $v \in \Sigma^*$ and matching derivations $\C{D} :: (A, uv) \gparse_P w$ ($w \in \Sigma^*$), we have $\C{C}_{\C{D}} = f(u) c'$ for some $c' \in \Two^*$.
\item it completes the code: if there exists a matching derivation $\C{D} :: (A,u) \gparse_P w$, then $\C{C}_{\C{D}} = f(u \eof)$.
\end{enumerate}
\end{definition}
In the rest of this chaper, we develop an algorithm which implements a streaming parsing function as defined above.
The code prefix produced allows consumers to perform parsing actions (e.g. construction of syntax trees, evaluation of expressions, printing, etc.) before all of the input string has been consumed. Monotonicity ensures that no actions will have to be ``un-done'', with the caveat that further input might cause the whole parse to be rejected.

\section{Tabulation of Operational Semantics}
\label{sec:tabulation}

In the following we fix a program $P = (\Sigma, V, S, R)$.

We will be working with various constructions defined as least fixed points of monotone operators on partially ordered sets.
A partial order is a pair $(X, \sqsubseteq)$ where $X$ is a set and $\sqsubseteq$ is a reflexive, transitive and antisymmetric relation on $X$. Given two elements $x,y \in X$, we will write $x \sqsubset y$ when $x \sqsubseteq y$ and $x \not= y$.

For any set $X$, let $(X, \sqsubseteq)$ be the discrete partial order, the smallest partial order on $X$ (i.e. $x \sqsubseteq x'$ implies $x = x'$). Write $X_\bot$ for the set $X + \{\bot\}$ and let $(X_\bot, \sqsubseteq)$ be the \emph{lifted} partial order with $\bot$ as an adjoined bottom element, i.e. $\forall x \in X_\bot \ldotp \bot \sqsubseteq x$.

A \emph{table} on $X$ is a $|P| \times \omega$ matrix $T$ where each entry $T_{ij}$ is in $X_\bot$, and indices $(i,j)$ are in the set $\Set{Index} = \{(i,j) \mid {0 \leq i < |P|} \wedge {0 \leq j} \}$. The set of all tables on $X$ is denoted $\Set{Table}(X)$, and forms a partial order $(\Set{Table}(X), \sqsubseteq)$ by comparing entries pointwise: for $T, T' \in \Set{Table}(X)$, we write $T \sqsubseteq T'$ iff for all $(i,j) \in \Set{Index}$, we have $T_{ij} \sqsubseteq T_{ij}'$. Write $\bot \in \Set{Table}(X)$ for the table with all entries equal to $\bot \in X_\bot$. It is easy to verify that the partial order on $\Set{Table}(X)$ has the following structure:
\begin{description}
\item[complete partial order:] For all chains $T_0 \sqsubseteq T_1 \sqsubseteq ...$ where $T_i \in \Set{Table}(X)$, $i \in \{0,1, ...\}$, the least upper bound $\bigsqcup_i T_i$ exists.
\item[meet-semilattice:] For all non-empty subsets $S \subseteq \Set{Table}(X)$, the greatest lower bound $\bigsqcap S$ exists.
\end{description}

A function $F : \Set{Table}(X) \to \Set{Table}(X)$ is said to be \emph{continuous} if it preserves least upper bounds: For all $S \subseteq \Set{Table}(X)$, we have $F(\bigsqcup S) = \bigsqcup_{T \in S} F(T)$. A continous function is automatically \emph{monotone}, meaning that $T \sqsubseteq T'$ implies $F(T) \sqsubseteq F(T')$. A \emph{least fixed point} of $F$ is an element $T$ such that $F(T) = T$ ($T$ is a fixed point) and also $T \sqsubseteq T'$ for all fixed points $T'$. A general property of complete partial orders is that if $F$ is a continuous function then its least fixed point $\lfp F$ exists and is given by \[ \lfp F = \bigsqcup_n F^n(\bot) \] where $F^n$ is the $n$-fold composition of $F$ with itself. We will also rely on the following generalization:
\begin{restatable}[Lower bound iteration]{lemma}{lemrandomstart}
\label{lem:randomstart}
If $T \sqsubseteq \lfp F$, then $\lfp F = \bigsqcup_{n} F^n(T)$.
\end{restatable}

\subsection{Parse Tables}
We now recall the parse table used in the dynamic programming algorithm for linear time recognition~\cite{aho1972}, but presented here as a least fixed point.
The table will have entries in the set $\Set{Res} = \omega + \{\gfail\}$, i.e. either a natural number or $\gfail$ indicating failure.
Given a finite (respectively, infinite) string $w = a_0 a_1 ... a_{n-1}$ ($w=a_0 a_1 ...$), and an offset $0 \leq j < n$ ($0 \leq j$), write $u_j$ for the suffix $a_j a_{j+1} ... a_{n-1}$ ($a_j a_{j+1} ...$) obtained by skipping the first $j$ symbols.

\begin{definition}[Parse table]
Let $u \in \Sigma^*$.
Define a table operator $F^u$ on $\Set{Table}(\Set{Res})$ as follows.
Let $w = u \eof^\omega$, the infinite string starting with $u$ followed by an infinite number of repetitions of the end marker $\eof \not\in\Sigma$.
For any table $T \in \Set{Table}(\Set{Res})$ define $F^u(T) = T'$ such that for all $(i,j) \in \Set{Index}$:
\[
  T'_{ij} =
\begin{cases}
  \gfail & \text{$A_i \gto \gfail$ or $A_i \gto a$ and $a$ not a prefix of $w_j$} \\
  1 & \text{$A_i \gto a$ and $a$ is a prefix of $w_j$} \\
  0 & \text{$A_i \gto \geps$} \\
  m+m' & \text{$A_i \gto A_x[A_y,A_z]$; $T_{xj} = m$; $T_{y(j+m)} = m'$} \\
  \gfail & \text{$A_i \gto A_x[A_y,A_z]$; $T_{xj} = m$; $T_{y(j+m)} = \gfail$} \\
  T_{zj} & \text{$A_i \gto A_x[A_y,A_z]$; $T_{xj} = \gfail$} \\
  \bot & \text{otherwise}
\end{cases}
\]

The operator $F^u$ is easily seen to be continuous, and we define the \emph{parse table for $u$} by $T(u) = \lfp F^u$.
\end{definition}

For any $u \in \Sigma^*$, the table $T(u)$ is a tabulation of all parsing results on all suffixes of $u$:
\begin{restatable}[Fundamental theorem]{theorem}{thmfundamental}
  \label{thm:fundamental}
  Let $u \in \Sigma^*$ and consider $T(u)$ as defined above. For all $(i,j) \in \Set{Index}$:
\begin{enumerate}
\item \label{fund:cond:fail} $j \leq |u|$ and $T(u)_{ij} = \gfail$ iff $(A_i, u_j) \gparse_P \gfail$; and
\item \label{fund:cond:success} $j \leq |u|$ and $T(u)_{ij} = m \in \omega$ iff $(A_i, u_j) \gparse_P u_{j+m}$; and
\item \label{fund:cond:bot} $j \leq |u|$ and $T(u)_{ij} = \bot$ iff $(A_i, u_j) \not\gparse_P$;
\item \label{fund:cond:tail} if $j > |u|$ then $T_{ij} = T_{i{|u|}}$
\end{enumerate}
The converse also holds: for any $T$ satisfying the above, we have $T = T(u)$.
\end{restatable}
Property \ref{fund:cond:tail} is sufficient to ensure that all parse tables have a finitary representation of size $|P| \times |u|$. It is straightforward to extract a parse code from $T(u)$ by applying Definition~\ref{def:parse-tree} and the theorem.

\begin{example}
\label{ex:tables}
Consider the program $P$ from Example~\ref{ex:gtdpl1}. The tables $T = T(aa)$ and $T' = T(aaba)$ are shown below:
\begin{center}
\begin{tabular}{l|lll@{~}l@{}}
&$0$&$1$&$2$&$\cdots$\\
&$\Lit{a}$&$\Lit{a}$&$\eof$&$\cdots$\\
\hline
$A$&$1$&$1$&$\gfail$&\multirow{8}{*}{$\cdots$}\\
$B$&$\gfail$&$\gfail$&$\gfail$\\
$E$&$0$&$0$&$0$\\
$F$&$\gfail$&$\gfail$&$\gfail$\\
$L$&$0$&$0$&$0$\\
$P$&$\gfail$&$\gfail$&$\gfail$\\
$R$&$2$&$1$&$0$\\
$S$&$2$&$1$&$0$
\end{tabular} \qquad
\begin{tabular}{l|lllll@{~}l@{}}
&$0$&$1$&$2$&$3$&$4$&$\cdots$\\
&$\Lit{a}$&$\Lit{a}$&$\Lit{b}$&$\Lit{a}$&$\eof$&$\cdots$\\
\hline
$A$&$1$&$1$&$\gfail$&$1$&$\gfail$&\multirow{8}{*}{$\cdots$}\\
$B$&$\gfail$&$\gfail$&$1$&$\gfail$&$\gfail$\\
$E$&$0$&$0$&$0$&$0$&$0$\\
$F$&$\gfail$&$\gfail$&$\gfail$&$\gfail$&$\gfail$\\
$L$&$3$&$2$&$1$&$0$&$0$\\
$P$&$3$&$2$&$1$&$\gfail$&$\gfail$\\
$R$&$2$&$1$&$0$&$1$&$0$\\
$S$&$4$&$3$&$2$&$1$&$0$
\end{tabular}
\end{center}
Note that columns 1,2 in the left table equals columns 3,4 in the right table. In general, columns depend on the corresponding input suffix but are independent of the previous columns. This is a simple consequence of Theorem~\ref{thm:fundamental}.
\end{example}

For a table $T$ and $m \in \omega$, let $T[m]$ be the table obtained by removing the first $m$ columns from $T$, i.e. $T[m]_{ij} = T_{i(j+m)}$.

\begin{corollary}[Independence]
  \label{cor:independence}
  Let $u \in \Sigma^*$. For all $0 \leq m$, we have $T(u)[m] = T(u_m)$.
\end{corollary}
\begin{proof}
  By Theorem~\ref{thm:fundamental}. For example, if $T(u)_{i(j+m)} = m'$ for some $m'$ then $(A_i, u_{j+m}) \gparse u_{j+m+m'}$. Have $(u_m)_j = u_{m+j}$, so $(A_i, (u_m)_j) \gparse (u_m)_{j+m'}$, and therefore $T(u_m)_{ij} = m'$.
\end{proof}

Independence leads to the linear-time parsing algorithm of Aho and Ullman. For input $u$ with $|u| = n$, compute $T(u)$ column by column, starting from the right. For each $m \leq n$, we compute column $m$ by fixed point iteration of $F^{u_m}$ on the current table state. Since $T(u)[m+1] = T(u_{m+1})$ has already been computed, only $|P|$ entries need to be processed in each step, which takes time $O(|P|^2)$.

\section{Streaming Parsing with Tables}
\label{sec:streaming-parsing}

The linear-time parsing algorithm has asymptotically optimal time complexity. However, it always uses space linear in the length of the input string, since all columns of the parse table has to be computed before the final result can be obtained. For large grammars and inputs, this can be prohibitively expensive. In the following we describe a method for computing only an initial part of the table. The initial columns will in some cases provide enough information to construct a prefix of the parse code and allow us to continue parsing with a smaller table, saving space.

Let us illustrate the idea by an example. Let $w = uv$ be an input string, and let $A_i \gto A_x[A_y,A_z]$ be a rule in the program. Suppose that by analyzing only the prefix $u$, we can conclude that there is a constant $m$ such that $T(uv')_{x0} = m$ for all $v'$. In particular, this holds for $v'=v$, so $T(w)_{i0} \in \omega$ if and only if $T(w)_{i0} = m + m'$ where $m' = T(w)_{ym} = T(w)[m]_{y0} = T(w_m)_{y0}$ (the last equation follows by independence). By examining only the prefix $u$, we have thus determined that the result only depends on $T(w_m)$, freeing up $m$ columns of table space. The process can be repeated for the remaining input $w_m$.

We will need an analysis that can predict results as described. The theoretically optimal analysis is defined as follows:
\begin{definition}[Optimal prefix table]
\label{def:optimal-prefix}
Let $u \in \Sigma^*$, and define the \emph{optimal prefix table} $T^\sqcap(u) \in \Set{Table}(\Set{Res})$ as the largest approximation of all the complete tables for all extensions of $u$:
\[
  T^\sqcap(u) = \bigsqcap_{v \in \Sigma^*} T(uv)
\]
\end{definition}
\begin{theorem}
  \label{thm:prefix}
  For all $u,i,j$:
  \begin{enumerate}
  \item if $T^\sqcap(u)_{ij} \not= \bot$ then $\forall v\ldotp T(uv)_{ij} = T^\sqcap(u)_{ij}$;
  \item if $(\forall v\ldotp T(uv)_{ij} = r \not= \bot)$, then $T^\sqcap(u)_{ij} = r$.
  \end{enumerate}
\end{theorem}

Unfortunately, we cannot use this for parsing, as the optimal prefix table is too precise to be computable:
\begin{theorem}
  There is no procedure which computes $T^\sqcap(u)$ for all GTDPLs $P$ and input prefixes $u$.
\end{theorem}
\begin{proof}
  Assume otherwise that $T^\sqcap(u)$ is computable for any $u$ and GTDPL $P$. Then $L(P) = \emptyset$ iff $T^\sqcap(\varepsilon)_{i_S, 0} = \gfail$. Hence emptiness is decidable, a contradiction by Proposition~\ref{prop:undec:emptiness}.
\end{proof}

A conservative and computable approximation of $T^\sqcap$ can easily be defined as a least fixed point. Given a table operator $F$ and a subset $J \subseteq \Set{Index}$ define a restricted operator $F_J$ by
\[F_J(T)_{ij} = \begin{cases} F(T)_{ij} & \text{if $(i,j) \in J$} \\ T_{ij} & \text{otherwise} \end{cases} \]

If $J = \{(p,q)\}$ is a singleton, write $F_{pq}$ for $F_J$. Clearly, if $F$ is continuous then so is $F_J$.

For any $u \in \Sigma^*$, define an operator $F^{(u)}$ by $F^{(u)} = F^u_{J_u}$ where $J_u = \{ (i,j) \in \Set{Index} \mid j < |u| \}$. The \emph{prefix table} for $u$ is the least fixed point of this operator:
\[
  T^<(u) = \lfp F^{(u)}
\]

Intuitively, a prefix table contains as much information as can be determined without depending on column $|u|$.
Prefix tables are clearly computable by virtue of being least fixed points, and properly approximate the optimal analysis:

\begin{restatable}[Approximation]{theorem}{thmapproximation}
  \label{thm:approximation}
  For all $u \in \Sigma^*$, we have $T^<(u) \sqsubseteq T^\sqcap(u)$. In particular, if $T^<(u)_{ij} = m$ or $T^<(u)_{ij} = \gfail$, then $\forall v \ldotp T(uv)_{ij} = m$ or $\forall v \ldotp T(uv)_{ij} = \gfail$, respectively.
\end{restatable}

Perhaps not surprisingly, prefix tables become better approximations as the input prefix is extended. We will make use of this property and Lemma~\ref{lem:randomstart} to efficiently compute prefix tables in an incremental fashion:
\begin{proposition}[Prefix monotonicity]
  \label{prop:prefix-monotonicity}
  For all $u, v \in \Sigma^*$, we have $T^<(u) \sqsubseteq T^<(uv)$.
\end{proposition}

The full parse table can be recovered as a prefix table if we just append an explicit end marker to the input string:
\begin{proposition}[End marker]
  \label{prop:endmarker}
  For all $u \in \Sigma^*$ and $(i,j) \in \Set{Index}$, if $j \leq |u|$ then $T^<(u \eof)_{ij} = T(u)_{ij}$.
\end{proposition}

Independence carries over to prefix tables. For all $u \in \Sigma^*$ and $m \geq 0$, we thus have $T^<(u)[m] = T^<(u_m)$.

\subsection{Streaming Code Construction}
The resolved entries of a prefix table can be used to guide a partial leftmost expansion of a parse tree. We model this expansion process by a labeled transition system which generates the corresponding parse code. By constructing the expansion such that it is a prefix of all viable expansions, the parse code can be computed in a streaming fashion. In order to determine as much of the parse code as possible, we speculatively guess that choices succeed when a dynamic analysis can determine that the alternative must fail.

\newcommand{\fails}[2]{{#1}\textbf{ fails}_{#2}}
\begin{definition}[Leftmost parse tree expansion]
\label{def:leftmost-expansion}
Let $T \in \Set{Table}(\Set{Res})$ be a table and $d \in \omega$ a \emph{speculation constant}.
Define a labeled transition system $\C{E}_T = (Q,E)$ with states $Q = V^* \times \omega$ and transitions $E \subseteq \{ q \stackrel{c}{\to} q' \mid c \in \Two^*; q,q' \in Q \}$. Let $E$ be the smallest set such that for all $A_i \in V$, $\vec{K} \in V^*$ and $j \in \omega$:

\begin{enumerate}
\item If $A_i \gto A_x[A_y,A_z]$; and either $T_{xj} \in \omega$ or \highlight{$\fails{(A_z \vec{K}, j)}{d}$}, then:
\vspace{-1em}
\[ (A_i \vec{K}, j) \stackrel{0}{\to} (A_x A_y \vec{K}, j) \in E \]
\item If $A_i \gto A_x[A_y,A_z]$; and $T_{xj} = \gfail$, then:
\vspace{-1em}
\[ (A_i \vec{K}, j) \stackrel{1}{\to} (A_z \vec{K}, j) \in E \]
\item If $A_i \gto \geps$ or $A_i \gto a$; and $T_{ij} = m$, then:
\vspace{-1em}
\[ (A_i \vec{K}, j) \stackrel{\varepsilon}{\to} (\vec{K}, j) \in E \]
\item If $q \stackrel{c}{\to} q' \in E$ and $q' \stackrel{c'}{\to} q''$, then: $q \stackrel{cc'}{\to} q'' \in E$.
\end{enumerate}
where for all $\vec{K}, j, n$, write $\fails{(\vec{K}, j)}{n}$ if $\vec{K} = A_i \vec{K}'$ and either
\begin{enumerate}
\item $T_{ij} = \gfail$; or
\item $T_{ij} = m$, $n = n' + 1$ and $\fails{(\vec{K}',j+m)}{n'}$.
\end{enumerate}
\end{definition}
A state encodes the input offset and the stack of leaves that remain unexpanded. The node on the top of the stack is expanded upon a transition to the next state, with the expansion choice indicated in the label of the transition. The system is deterministic in the sense that every state can step to at most one other state in a single step (the label is determined by the source state).

The highlighted disjunct allows us to speculatively resolve a choice as succeeding when the failure branch is guaranteed to fail. This is determined by examining the table entries for at most $d$ nonterminals on the current stack $\vec{K}$.
\begin{example}
The partial parse tree of Example~\ref{ex:gtdpl1} corresponds to the following steps in $\C{E}_{T'}$ where $T'$ is the table from Example~\ref{ex:tables}:
\begin{align*}
  (S, 0) \stackrel{0}{\to} (LR, 0) \stackrel{0}{\to} (PER, 0) \stackrel{0}{\to} (APER,0) \stackrel{\varepsilon}{\to} (PER, 1)
\end{align*}
\end{example}

A state $q$ is \emph{quiescent} if there is no transition from it. Say that $q$ is \emph{convergent} and write $q \stackrel{c}{\to} q' \downarrow$ if either there is a path $q \stackrel{c}{\to} q'$ such that $q'$ quiescent; or, $q$ is already quiescent and $q' = q$ and $c = \varepsilon$. Clearly, if such $c$ and $q'$ exists, then they are unique and can be effectively determined. Otherwise, we say that $q$ is \emph{divergent}.

Expansions compute coded (matching) derivations in full parse tables:
\begin{proposition}
  \label{prop:leftmost-expansion}
  Let $u \in \Sigma^*$ and consider the system $\C{E}_{T(u)}$.
  \begin{enumerate}
  \item There is a derivation $\C{D} :: (A, u) \gparse_P u_m$ with $c = \C{C}_{\C{D}}$ if and only if $(A, 0) \stackrel{c}{\to} (\varepsilon, m) \downarrow$.
  \item We have $(A,u) \gparse_p \gfail$ if and only if $\fails{(A, 0)}{n}$ for some $n$.
  \end{enumerate}
\end{proposition}
It follows that a state $(A,0)$ is only divergent if the input is unhandled:
\begin{proposition}
  \label{prop:expansion-totality}
  Let $u \in \Sigma^*$ and consider the system $\C{E}_{T(u)}$. Then $(A,u) \not\gparse_P$ if and only if $(A,0)$ is divergent.
\end{proposition}
Hence, if $P$ is complete, then every state is convergent in $\C{E}_{T(u)}$, and the relation $q \stackrel{c}{\to} q' \downarrow$ becomes a total function $q \mapsto (c, q')$.

The function associating every input prefix $u$ with the code $c$ given by $(S,0) \stackrel{c}{\to} q' \downarrow$ in the system $\C{E}_{T^<(u)}$ is a streaming parse function as per Definition~\ref{def:streaming-parsing}. This is ensured by the following sufficient condition, which states that expansions never ``change direction'' as the underlying table is refined:
\begin{proposition}
  \label{prop:path-monotonicity}
  If $T \sqsubseteq T'$ and $(\vec{K}, j) \stackrel{c}{\to} (\vec{K}', j')$ in $\C{E}_T$, then either $(\vec{K}, j) \stackrel{c}{\to} (\vec{K}', j')$ in $\C{E}_{T'}$ or $(\vec{K},j)$ fails in $\C{E}_{T'}$.
\end{proposition}

Expansions also never backtrack in the input, that is, if $(\vec{K}, j) \stackrel{c}{\to} (\vec{K}', j')$ then $j \leq j'$. This allows us to discard the initial columns of a table as we derive a leftmost expansion:
\begin{proposition}
  \label{prop:path-shift}
  Let $T$ be a table. Then $(\vec{K}, m) \stackrel{c}{\to} (\vec{K}', n)$ in $\C{E}_T$ if and only if $(\vec{K}, 0) \stackrel{c}{\to} (\vec{K}', n-m)$ in $\C{E}_{T[m]}$.
\end{proposition}

\subsection{Progressive Tabular Parsing}
\label{sec:streaming-parsing-procedure}

\begin{figure*}
{\setlength{\tabcolsep}{5pt}
\def\arraystretch{0.6}
\begin{center}

\begin{tabular}[t]{l:llllll}
$(1)$          & $a$         \\
\hline\\
$A$          & $\cellcolor{orange!15} 1$\\
$B$          & $\cellcolor{orange!15} \gfail$\\
$E$          & $\cellcolor{orange!15} \mathbf{0}$\\
$F$          & $\cellcolor{orange!15} \mathbf{\gfail}$\\
$L$          & $\mathbf{\bot}$\\
$P$          & $\mathbf{\bot}$\\
$R$          & $\bot$      \\
$S$          & $\bot$      
\end{tabular}
\qquad
\begin{tabular}[t]{l:llllll}
$(2)$          & $a$          & $a$         \\
\hline\\
$A$          & $1$          & $\cellcolor{orange!15} 1$\\
$B$          & $\gfail$     & $\cellcolor{orange!15} \gfail$\\
$E$          & $\mathbf{0}$ & $\cellcolor{orange!15} 0$\\
$F$          & $\mathbf{\gfail}$ & $\cellcolor{orange!15} \gfail$\\
$L$          & $\mathbf{\bot}$ & $\bot$      \\
$P$          & $\mathbf{\bot}$ & $\bot$      \\
$R$          & $\bot$       & $\bot$      \\
$S$          & $\bot$       & $\bot$      
\end{tabular}
\qquad
\begin{tabular}[t]{l|lll:lll}
$(3)$          & $a$          & $a$          & $b$         \\
\hline\\
$A$          & $\mathbf{1}$ & $\mathbf{1}$ & $\cellcolor{orange!15} \mathbf{\gfail}$ & $\bot$\\
$B$          & $\gfail$     & $\gfail$     & $\cellcolor{orange!15} \mathbf{1}$ & $\bot$\\
$E$          & $\mathbf{0}$ & $0$          & $\cellcolor{orange!15} 0$ & $\bot$\\
$F$          & $\mathbf{\gfail}$ & $\gfail$     & $\cellcolor{orange!15} \gfail$ & $\bot$\\
$L$          & $\mathbf{\bot}$ & $\bot$       & $\bot$       & $\bot$\\
$P$          & $\cellcolor{orange!15} \mathbf{3}$ & $\cellcolor{orange!15} 2$ & $\cellcolor{orange!15} 1$ & $\bot$\\
$R$          & $\cellcolor{orange!15} 2$ & $\cellcolor{orange!15} 1$ & $\cellcolor{orange!15} 0$ & $\bot$\\
$S$          & $\bot$       & $\bot$       & $\bot$       & $\bot$
\end{tabular}

\begin{tabular}[t]{l|llll:ll}
$(4)$          & $a$          & $a$          & $b$          & $a$         \\
\hline\\
$A$          & $\mathbf{1}$ & $\mathbf{1}$ & $\mathbf{\gfail}$ & $\cellcolor{orange!15} \mathbf{1}$ & $\bot$\\
$B$          & $\gfail$     & $\gfail$     & $\mathbf{1}$ & $\cellcolor{orange!15} \gfail$ & $\bot$\\
$E$          & $\mathbf{0}$ & $0$          & $0$          & $\cellcolor{orange!15} \mathbf{0}$ & $\bot$\\
$F$          & $\mathbf{\gfail}$ & $\gfail$     & $\gfail$     & $\cellcolor{orange!15} \gfail$ & $\bot$\\
$L$          & $\mathbf{\bot}$ & $\bot$       & $\bot$       & $\bot$       & $\bot$\\
$P$          & $\mathbf{3}$ & $2$          & $1$          & $\bot$       & $\bot$\\
$R$          & $2$          & $1$          & $0$          & $\bot$       & $\bot$\\
$S$          & $\bot$       & $\bot$       & $\bot$       & $\bot$       & $\bot$
\end{tabular}
\qquad
\begin{tabular}[t]{l|llll:ll}
$(5)$          & $a$          & $a$          & $b$          & $a$          & $\eof$      \\
\hline\\
$A$          & $\mathbf{1}$ & $\mathbf{1}$ & $\mathbf{\gfail}$ & $\mathbf{1}$ & $\cellcolor{orange!15} \mathbf{\gfail}$\\
$B$          & $\gfail$     & $\gfail$     & $\mathbf{1}$ & $\gfail$     & $\cellcolor{orange!15} \gfail$\\
$E$          & $\mathbf{0}$ & $0$          & $0$          & $\mathbf{0}$ & $\cellcolor{orange!15} \mathbf{0}$\\
$F$          & $\mathbf{\gfail}$ & $\gfail$     & $\gfail$     & $\gfail$     & $\cellcolor{orange!15} \gfail$\\
$L$          & $\mathbf{\bot}$ & $\bot$       & $\bot$       & $\bot$       & $\cellcolor{orange!15} 0$\\
$P$          & $\mathbf{3}$ & $2$          & $1$          & $\bot$       & $\cellcolor{orange!15} \gfail$\\
$R$          & $2$          & $1$          & $0$          & $\bot$       & $\cellcolor{orange!15} 0$\\
$S$          & $\bot$       & $\bot$       & $\bot$       & $\bot$       & $\cellcolor{orange!15} 0$
\end{tabular}

\end{center}
$\begin{array}{@{}l}
(1): (S, a\textcolor{gray}{aba\eof})\stackrel{0}{\to} (LR, a\textcolor{gray}{aba\eof})
\\
(2): (LR, aa\textcolor{gray}{ba\eof})
\\
(3): (LR, aab\textcolor{gray}{a\eof})
\stackrel{0}{\to} (PER, aab\textcolor{gray}{a\eof})
\stackrel{0}{\to} (APER, aab\textcolor{gray}{a\eof})
\stackrel{}{\to} (PER, ab\textcolor{gray}{a\eof})
\\\qquad
\stackrel{0}{\to} (APER, ab\textcolor{gray}{a\eof})
\stackrel{}{\to} (PER, b\textcolor{gray}{a\eof})
\stackrel{1}{\to} (BER, b\textcolor{gray}{a\eof})
\stackrel{}{\to} (ER, \textcolor{gray}{a\eof})
\\
(4): (ER, a\textcolor{gray}{\eof})
\stackrel{}{\to} (R, a\textcolor{gray}{\eof})
\stackrel{0}{\to} (AR, a\textcolor{gray}{\eof})
\stackrel{}{\to} (R, \textcolor{gray}{\eof})
\\
(5): (R, \eof)
\stackrel{1}{\to} (E, \eof)
\stackrel{}{\to} (\varepsilon, \eof)
\end{array}$
}

%%% Local Variables:
%%% mode: latex
%%% TeX-master: "../thesis"
%%% End:

\caption[Example of prefix tables and online expansion.]{In the top is five consecutive tables during the parse of input $w = aaba$, using the program from Example~\ref{ex:gtdpl1}. Only the columns to the right of the dashed line has to be stored for the next iteration. Newly computed entries are colored; entries considered by the expansion process are written in bold face. The progression of the leftmost expansion is shown below. \label{fig:prefixes}}
\end{figure*}
Assume that $P$ is a complete program. We use the constructions of this section to define our \emph{progressive tabular parsing} procedure. The algorithmic issues of space and time complexity will not be of our concern yet, but will we be adressed in the following section.

Given an input string with end marker $w \eof = a_0 a_1 ... a_{n-1}$ ($a_{n-1} = \eof$), the procedure decides whether there exists a matching derivation $\C{D} :: (S, w) \gparse_P w_k$, and in that case produces $\C{C}_{\C{D}}$ in a streaming fashion. In each step $0 \leq k \leq n$, we compute a table $T^k \in \Set{Table}(\Set{Res})$, a stack $\vec{K}^k$, an offset $m^k \leq k$ and a code chunk $c^k \in \Two^*$. Upon termination, we will have $\C{C}_{\C{D}} = c^0 c^1 ... c^n$.

Initially $T^0 = T^<(\varepsilon)$, $q^0 = (S, 0)$ and $c^0 = \varepsilon$. For each $1 \leq k \leq n$, the values $T^k, \vec{K}^k, m^k$ and $c^k$ are obtained by
\begin{align*}
  T^k ={}& T^<(a_{m^{k-1}} ... a_{k-1}) \\
  m^k ={}& m^{k-1} + m' \\
  \text{where }& (\vec{K}^{k-1}, 0) \stackrel{c^k}{\to} (\vec{K}^k, m') \downarrow
\end{align*}
Since $P$ is complete, we have by Proposition~\ref{prop:expansion-totality} that the last line above can be resolved.

If $\vec{K}^n = \varepsilon$, accept the input; otherwise reject.

\begin{theorem}
  The procedure computes $\C{C}_{\C{D}}$ iff there is a derivation $\C{D} :: (S, w) \gparse_P w_k$.
\end{theorem}
\begin{proof}
We claim that after each step $k$, we have $(S,0) \stackrel{c^0...c^k}{\to} (\vec{K}^k, m^k) \downarrow$ in $\C{E}_{T(w)}$. This holds for $k=0$, as $(S,0)$ is quiescent. For $k>0$, we assume that it holds for $k-1$ and must show $(\vec{K}^{k-1},m^{k-1}) \stackrel{c^k}{\to} (\vec{K}^k, m^k) \downarrow$ in $\C{E}_{T(w)}$. By construction, we have a path $(\vec{K}^{k-1}, 0) \stackrel{c^k}{\to} (\vec{K}^k, m^k - m^{k-1})$ in $\C{E}_{T^k}$. By Proposition~\ref{prop:prefix-monotonicity} and Theorem~\ref{thm:approximation}, we have $T^k = T^<(a_{m^{k-1}} ... a_{k-1}) \sqsubseteq T^<(w_{m^{k-1}}) \sqsubseteq T(w_{m^{k-1}}) = T(w)[m^{k}-1]$, so by Proposition~\ref{prop:path-monotonicity} the path is in $\C{E}_{T(w)[m^{k-1}]}$, and by Proposition~\ref{prop:path-shift}, we obtain our subgoal.

If the procedure accepts the input, then we are done by Proposition~\ref{prop:leftmost-expansion}. If it rejects, it suffices to show that $(\vec{K}, m^k)$ is quiescent in $\C{E}_{T(w)}$ which by Proposition~\ref{prop:leftmost-expansion} implies that there is no matching derivation. Since $T^m = T^<(w \eof)$, we can apply Proposition~\ref{prop:endmarker} to easily show $\C{E}_{T^m} = \C{E}_{T(w)}$, and we are done.
\end{proof}

Figure \ref{fig:prefixes} shows an example of a few iterations of the procedure applied to the program in Example~\ref{ex:gtdpl1}.

In the next section we show that the above procedure can be performed using at most linear time and space. Linear space is easily seen to be obtained by observing that the table $T^{k-1}$ is no longer needed once $T^k$ has been computed. On the other hand, obtaining a linear time guarantee requires careful design: Computing each table $T^k$ using the classical right-to-left algorithm would take linear time in each step, and hence quadratic time in total. In the following section, we show how to obtain the desired time complexity by computing each table incrementally from the previous one.

\section{Algorithm}
\label{sec:algorithm2}

The streaming parsing procedure of Section~\ref{sec:streaming-parsing-procedure} can be performed in amortized time $O(|w|)$ (treating the program size as a constant). We assume that the program $P$ is complete.

Our algorithm computes each prefix table $T^k$ using a work set algorithm for computing fixed points. We save work by starting the computation from $T^{k-1}[m^{k-1}]$ instead of the empty table $\bot$. In order to avoid unnecessary processing, an auxiliary data structure is used to determine exactly those entries which have enough information available to be resolved. This structure itself can be maintained in constant time per step. Since at most $O(|w|)$ unique entries need to be resolved over the course of parsing $w$, this is also the time complexity of the algorithm.

The algorithm is presented in two parts in Figure~\ref{fig:alg:parse}. Algorithm~\ref{alg:parse} (\textsc{Parse}) takes as input a $\eof$-terminated input stream and maintains two structures: A table structure $T$ which incrementally gets updated to represent $T^<(u)$ for a varying substring $u = a_{m^{k-1}} ... a_{k-1}$; and a structure $R$ which keeps track of reverse data dependencies between the entries in $T$. In each iteration, any resolved code prefix is returned and the corresponding table columns freed. The main work is done in Algorithm~\ref{alg:fix} (\textsc{Fix}) which updates $T$ and $R$ to represent the next prefix table and its reverse dependencies, respectively.

We will sketch the correctness proof and highlight important lemmas during the presentation. Detailed proofs can be found in the appendix.

\begin{figure*}
  \input{p4-unpublished/fig-alg-parse}
  \caption{Parsing algorithm.}
  \label{fig:alg:parse}
\end{figure*}

\subsection{Work Sets}
Let $T$ be a table such that $T \sqsubseteq T^<(u)$ for some prefix $u$. The \emph{work set} $\Delta_u(T) \subseteq \Set{Index}$ consists of all indices of entries that can be updated to bring $T$ closer to $T^<(u)$ by applying $F^{(u)}$:
\[ \Delta_u(T) = \{ (i,j) \mid T_{ij} \sqsubset F^{(u)}(T)_{ij} \}. \]

It should be clear that $T = T^<(u)$ iff $\Delta_u(T) = \emptyset$, and that for all $(p,q) \in \Delta_u(T)$, we still have $F^{(u)}_{pq}(T) \sqsubseteq T^<(u)$ for the updated table.
In the following we show how $\Delta_u(F^{(u)}_{pq}(T))$ can be obtained from $\Delta_u(T)$ instead of recomputing it from scratch.

\subsection{Dependencies}
In order to determine the effect of table updates on the work set, we need to make some observations about the dependencies between table entries.

Consider an index $(i,j)$ such that  $A_i \gto A_x[A_y, A_z]$ and $T_{ij} = \bot$. The index $(i,j)$ cannot be in the work set for $T$ unless either $T_{xj} = m$ and $T_{y(j+m)} \not= \bot$; or $T_{xj} = \gfail$ and $T_{zj} \not=\bot$. We say that $(i,j)$ \emph{conditions} on $(x,j)$. The reverse condition map $C^{-1}$ in Figure~\ref{fig:alg:parse} associates every row index $x$ with the set of row indices $i \in C^{-1}_x$ such that $(i,j)$ conditions on $(x,j)$ for all $j$.

If $T_{xj} = m$ or $T_{xj} = \gfail$ then $(i,j)$ is in the work set iff $T_{y(j+m)} \not= \bot$ or $T_{zj} \not= \bot$, respectively. In either case we say that $(i,j)$ has a \emph{dynamic dependency} on $(y, j+m)$ or $(z,j)$, respectively. The dependency is \emph{dynamic} since it varies based on the value of $T_{xj}$. The partial map $D : \Set{Table}(\Set{Res}) \times \Set{Index} \to \Set{Index}_\bot$ defined in Figure~\ref{fig:alg:parse} associates every index $(i,j)$ with its unique dynamic dependency $D^T_{ij}$ in table $T$. The dynamic dependency is undefined ($\bot$) if the condition is unresolved or if the corresponding expression $g_i$ is simple.

By the observations above, we can reformulate the work set using dependencies:
\begin{restatable}[Work set characterization]{lemma}{lemworksetcharacterization}
\label{lem:workset-characterization}
For all $T$ we have
\[ \Delta_u(T) = \{ (i,j) \in J_u \begin{array}[t]{@{}l@{}l@{}} {}\mid{} & T_{ij} = \bot \\
                                               {}\wedge{} & (\text{$g_i$ complex} \Rightarrow D^T_{ij} \not= \bot \not= T_{D^T_{ij}}) \}
                          \end{array} \]
\end{restatable}

\subsection{Incremental Work Set Computation}
When a table $S$ is updated by computing $T = F^{(u)}_{pq}(S)$ for $(p,q) \in \Delta_u(S)$, Lemma~\ref{lem:workset-characterization} tells us that the changes to the work set can be characterized by considering the entries $(i,j)$ for which one or more of the values $D^T_{ij}$ and $T_{D^T_{ij}}$ differ from $D^S_{ij}$ and $S_{D^S_{ij}}$, respectively.

An important observation is that the dependency map only gets more defined as we go from $S$ to $T$:
\begin{restatable}[Dependency monotonicity]{lemma}{lemdependencymonotonicity}
  \label{lem:dependency-monotonicity}
  If $T \sqsubseteq T'$, then for all $(i,j) \in \Set{Index}$, we have $D^T_{ij} \sqsubseteq D^{T'}_{ij}$.
\end{restatable}
Using this and the fact that $S \sqsubseteq T$, it is easy to show that we must have $\Delta_u(T) \supseteq \Delta_u(S) \setminus \{(p,q)\}$.
Furthermore, we observe that $(i,j) \in \Delta_u(T) \setminus (\Delta_u(S) \setminus \{(p,q)\})$ iff
\begin{enumerate}
\item $D^S_{ij} \sqsubset D^T_{ij}$ and $T_{D^T_{ij}} \not=\bot$; or 
\item $D^S_{ij} = D^T_{ij} \not= \bot$ and $S_{D^S_{ij}} \sqsubset T_{D^T_{ij}}$.
\end{enumerate}
Since the second case can only be satisfied when $D^T_{ij} = (p,q)$, it is completely characterized by the reverse dependency set $(D^T)^{-1}_{pq}$, defined in Figure~\ref{fig:alg:parse}. The first case is when $(i,j)$ conditions on $(p,q)$ (equivalent to $D^S_{ij} \sqsubset D^T_{ij}$) and $T_{D^T_{ij}} \not= \bot$. The entries satisfying the former are completely characterized by the reverse condition map:

\begin{restatable}[Dependency difference]{lemma}{lemdependencydifference}
  \label{lem:dependency-difference}
  Let $S \in \Set{Table}(\Set{Res})$ such that $S \sqsubseteq T^<(u)$ and $(p,q) \in \Delta_u(S)$, and define $T = F^{(u)}_{pq}(S)$. Then $\{ (i,j) \mid D^S_{ij} \sqsubset D^{T}_{ij} \} = C^{-1}_p \times \{q\}$.
\end{restatable}

By Lemmas~\ref{lem:workset-characterization}, \ref{lem:dependency-monotonicity} and \ref{lem:dependency-difference}, we obtain the following incremental characterization of the work set:
\begin{restatable}[Work set update]{lemma}{lemworksetupdate}
  \label{lem:workset-update}
  Let $S \sqsubseteq F^{(u)}(S) \sqsubseteq T^<(u)$, $(p,q) \in \Delta_u(S)$ and $T=F^{(u)}_{pq}(S)$. Then
\begin{align*} \Delta_u(T) ={}& \Delta_u(S) \setminus \{(p,q)\} \\
  {}\cup{}& (D^S)^{-1}_{pq} \\
  {}\cup{}& \{ (i',q) \mid i' \in C^{-1}_p \wedge \bot \not= T_{D^T_{i'q}} \}
\end{align*}
\end{restatable}
The extra premise $S \sqsubseteq F^{(u)}(S)$ says that every entry in $S$ must be a consequence of the rules encoded by $F^{(u)}$, and can easily be shown to be an invariant of our algorithm.

Reverse dependency map lookups $(D^T)^{-1}_{pq}$ cannot easily be computed efficiently. To accomodate efficient evaluation of these lookups, the algorithm maintains a data structure $R$ to represent $(D^T)^{-1}$.
The following Lemma shows that the loop~\ref{alg:fix:for:begin}-\ref{alg:fix:for:end} will reestablish the invariant that $R = (D^T)^{-1}$:
\begin{restatable}[Dependency update]{lemma}{lemdependencyupdate}
  \label{lem:dependency-update}
  Let $S \sqsubseteq T^<(u)$, $(p,q) \in \Delta_u(S)$ and $T=F^{(u)}_{pq}(S)$. Then for all $(k,\ell) \in \Set{Index}$, we have
  $(D^T)^{-1}_{k\ell} = (D^S)^{-1}_{k\ell} \cup \{ (i',q) \mid i' \in C^{-1}_p \wedge (k, \ell) = D^T_{i'q} \}$.
\end{restatable}

\subsection{Correctness}

\begin{theorem}[Correctness of \textsc{Fix}]
If the precondition of \textsc{Fix} holds, then the postcondition holds upon termination.
\end{theorem}
\begin{proof}[Proof sketch]
We first remark that the algorithm never attempts to perform an undefined action. It suffices to check that line~\ref{alg:fix:pq} is always well-defined, and that Lemma~\ref{lem:dependency-difference} implies that the right of the equation in line~\ref{alg:fix:let} is always resolved.

The outer loop maintains that $R = (D^T)^{-1}$ and $W = \Delta_u(T)$. Initially, only the entries in the last column which are associated with simple expressions can be updated. If $S$ is the state of $T$ at the beginning of an iteration of loop \ref{alg:fix:while:begin}-\ref{alg:fix:while:end}, then at the end of the iteration $T$ will have the form of the right hand side of Lemma~\ref{lem:workset-update}. When the loop terminates we have $W = \Delta_u(T) = \emptyset$, so $T = T^<(u)$.
\end{proof}

\begin{theorem}[Correctness of \textsc{Parse}]
  The algorithm \textsc{Parse} performs the streaming parsing procedure of Section~\ref{sec:streaming-parsing-procedure}.
\end{theorem}
\begin{proof}[Proof sketch]
After executing lines \ref{alg:parse:uinit}-\ref{alg:parse:Rinit}, we verify that $R = (D^T)^{-1}$, and that for $k=0$:
\begin{align*}
T ={}& T^<(a_{m^k} ... a_{k-1}), &
\vec{K} ={}& \vec{K}^k, &
u ={}& a_{m^k} ... a_{k-1}
\end{align*}
The loop maintains the invariant: When entering the loop, we increment $k$ and thus have $R = (D^T)^{-1}$ and
\begin{align*}
T ={}& T^<(a_{m^{k-1}} ... a_{k-2}), &
\vec{K} ={}& \vec{K}^{k-1}, &
u ={}& a_{m^{k-1} ... a_{k-2}}
\end{align*}
After the assignment to $u$, we have $u = a_{m^{k-1}} ... a_{k-1}$. By running \textsc{Fix}, we then obtain $T = T^<(a_{m^{k-1}} ... a_{k-1}) = T^k$. By assumption that $P$ is complete, line~\ref{alg:parse:path} is computable, and we obtain
\begin{align*}
  \vec{K}' ={}& \vec{K}^k & c ={}& c^k & m' ={}& m^k - m^{k-1}
\end{align*}
The last updates in the loop thus reestablishes the invariant.
\end{proof}

\subsection{Complexity}
We give an informal argument for the linear time complexity. Let $d \in \omega$ be the constant from Definition~\ref{def:leftmost-expansion} limiting the number of stack symbols considered when resolving choices.

It can be shown that the three sets on the right hand side of the equation in Lemma~\ref{lem:workset-update} are pairwise disjoint; likewise for Lemma~\ref{lem:dependency-update}. We thus never add the same element twice to $W$ and $R$, meaning that they can be represented using list data structures, ensuring that all single-element operations are constant time.

The complexity argument is a simple aggregate analysis. To see that \textsc{Parse} runs in linear time, we observe that the work set invariant ensures that we execute at most $O(|u|)$ iterations of the loop \ref{alg:fix:while:begin}-\ref{alg:fix:while:end} in \textsc{Fix}. Since we only add unprocessed elements to the work list, and no element is added twice, the total number of append operations performed in lines \ref{alg:fix:wsextend} and \ref{alg:fix:wsappend} is also $O(|u|)$. The same reasoning applies for the total number of append operations in line \ref{alg:fix:rappend}. The remaining operations in \textsc{Fix} are constant time.

Line~\ref{alg:parse:path} in \textsc{Parse} computes an expansion of aggregate length $O(mn)$. For each expansion transition, we use at most $d$ steps to resolve choices, and we thus obtain a bound of $O(dmn)$.

The restriction operator $T[m]$ can be performed in constant time by moving a pointer. The restriction of the reverse dependency map $R[m]$ can be implemented in constant time by storing the offset and lazily performing the offset calculation $j-m$ and filtering by $j \leq m$ on lookup.

\section{Evaluation}
\label{sec:evaluation}

We have developed a simple prototype implementation for the purpose of measuring how the number of columns grow and shrink as the parser proceeds, which gives an indication of both its memory usage and its ability to resolve choices. The evaluation also reveals parts of the design which will require further engineering in order to obtain an efficient implementation. We have not yet developed an implementation optimized for speed, so a comparative performance comparison with other tools is reserved for future work.

We consider three programs:
\begin{enumerate*}[label={\alph*)}]
  \item a simplified JSON parser,
  \item a simplified parser for the fragment of statements and arithmetic expressions of a toy programming language,
  \item a tail-recursive program demonstrating a pathological worst-case.
\end{enumerate*}

All programs are presented as PEGs for readability. Nonterminals are underlined, terminals are written in \texttt{typewriter} and a character class $[\Pin{a} ... \Pin{z}]$ is short for $\Pin{a} / \Pin{b} / ... / \Pin{z}$.

\paragraph{JSON Parser}
We have written a simple JSON parser based on a simplification of the ECMA 404 specification\footnote{\url{http://www.ecma-international.org/publications/files/ECMA-ST/ECMA-404.pdf}} and taking advantage of the repetition operator of PEG. To keep the presentation uncluttered, we have left out handling of whitespace.
\[\begin{array}{rcl}
 \nonterminal{object} & \gto & \Pin{\{} \nonterminal{members} \Pin{\}}\\
 \nonterminal{members} & \gto & \nonterminal{pair} (\Pin{,} \nonterminal{pair})^* / \geps\\
 \nonterminal{pair} & \gto & \nonterminal{string} \Pin{:} \nonterminal{value}\\
 \nonterminal{array} & \gto & \Pin{[} \nonterminal{elements} \Pin{]}\\
 \nonterminal{elements} & \gto & \nonterminal{value} (\Pin{,} \nonterminal{value})^* / \geps\\
 \nonterminal{value} & \gto & \nonterminal{string} / \nonterminal{object} / \nonterminal{number} / \nonterminal{array} \\
    && {} / \Pin{t} \Pin{r} \Pin{u} \Pin{e} / \Pin{f} \Pin{a} \Pin{l} \Pin{s} \Pin{e} / \Pin{n} \Pin{u} \Pin{l} \Pin{l}\\
 \nonterminal{string} & \gto & \Pin{"} \Prange{a}{z}^* \Pin{"}\\
 \nonterminal{number} & \gto & \nonterminal{int} (\nonterminal{frac} / \geps) (\nonterminal{exp} / \geps)\\
 \nonterminal{int} & \gto & \Prange{1}{9} \nonterminal{digits} / \Pin{-} \Prange{1}{9} \nonterminal{digits} / \Pin{-} \Prange{0}{9} / \Prange{0}{9}\\
 \nonterminal{frac} & \gto & \Pin{.} \nonterminal{digits}\\
 \nonterminal{exp} & \gto & \nonterminal{e} \nonterminal{digits}\\
 \nonterminal{digits} & \gto & \Prange{0}{9} \Prange{0}{9}^*\\
 \nonterminal{e} & \gto & \Pin{e} \Pin{+} / \Pin{e} \Pin{-} / \Pin{e} / \Pin{E} \Pin{+} / \Pin{E} \Pin{-} / \Pin{E}
\end{array}\]
The desugared program contains 158 rules. We ran the program on a 364 byte JSON input with several nesting levels and syntactic constructs exercising all rules of the grammar. The resulting parse code is computed in $3530$ expansion steps based on the computed table information.

We would like to get an idea of how varying values of the speculation constant $d$ affects the amount of memory consumed and also the amount of work performed. Recall that $d$ specifies the number of stack symbols considered when determining whether a branch must succeed on all viable expansions. The results for the range $0$ to $12$ are summarized in the following table:
\begin{center}
\begin{tabular}{lllll}
$d$ & max cols & non-imm. entries & spec. steps & visited \\
 & (max $365$) & (max $23360$) & (rel. to $3530$) \\
\hline
$0$ & $362$ & $23348$ (99.95\%) & $0$ (0.00\%) & $2866$\\
$1$ & $229$ & $23248$ (99.52\%) & $6$ (0.17\%) & $2876$\\
$2$ & $229$ & $23248$ (99.52\%) & $9$ (0.25\%) & $2876$\\
$3$ & $10$ & $19321$ (82.71\%) & $271$ (7.68\%) & $3116$\\
$4$ & $10$ & $19321$ (82.71\%) & $283$ (8.02\%) & $3117$\\
$5$ & $10$ & $19284$ (82.55\%) & $295$ (8.36\%) & $3121$\\
$6$ & $10$ & $19200$ (82.19\%) & $312$ (8.84\%) & $3134$\\
$7$ & $10$ & $19200$ (82.19\%) & $321$ (9.09\%) & $3134$\\
$8$ & $2$ & $18936$ (81.06\%) & $419$ (11.87\%) & $3162$\\
$9$ & $2$ & $18921$ (81.00\%) & $431$ (12.21\%) & $3173$\\
$10$ & $2$ & $18921$ (81.00\%) & $442$ (12.52\%) & $3173$\\
$11$ & $2$ & $18789$ (80.43\%) & $453$ (12.83\%) & $3173$\\
$12$ & $2$ & $18789$ (80.43\%) & $453$ (12.83\%) & $3173$
\end{tabular}
\end{center}
The second column shows the maximum number of columns stored at any point. The worst case is $364+1=365$. We observe that $d=8$ results in just two columns needing to be stored in memory.

The third column measures the potential work saved as $d$ is increased. To explain it, we introduce the notion of an \emph{immediate rule}, which is either simple, or of the form $A \gto A[B,C]$ where $A$ and $B$ are immediate and either $C \gto \geps$ or $C \gto \gfail$. An entry $T_{ij}$ where $A_i$ is immediate is always resolved upon reading symbol $j$, and can thus be precomputed and looked up based on the symbol. The real run-time cost is therefore the number of computed non-immediate entries, which is shown in the third column together with the percentage compared to the worst case. The benchmark shows that for $d \geq 8$, an average of 52 complex entries must be resolved for each input symbol. This may turn out to be an issue for scalability, as the number of non-immediate entries can be expected to be proportional to the program size.

The fourth column is the number of steps spent evaluating the $\fails{(\vec{K},j)}{n}$ predicate, and the relative number compared to the number of expansion steps. For this particular program, the overhead is seen to be very small compared to the reduction in computed entries and the fact that parsing proceeds in practically constant memory.

The last column shows the total number of unique table entries visited by the expansion. This is much smaller than the number of entries actually computed, so there is ample room for optimization, e.g. by integration between the expansion process and the table computation in order to compute only the entries that are needed.

\paragraph{Statement/Expression Parser}
The following is inspired by an example from a paper on ALL(*)~\cite{parr2014}. The program parses a sequence of statements, each terminated by semicolon, with the whole sequence terminated by a single dot representing an end-of-program token. Each statement is either a single arithmetic expression or an assignment.

\[\begin{array}{rcl}
 \nonterminal{prog} & \gto & \nonterminal{stat} \nonterminal{stat}^* \Pin{.}\\
 \nonterminal{stat} & \gto & \nonterminal{sum} \Pin{=} \nonterminal{sum} \Pin{;} / \nonterminal{sum} \Pin{;}\\
 \nonterminal{sum} & \gto & \nonterminal{product} \Pin{+} \nonterminal{sum} / \nonterminal{product}\\
 \nonterminal{product} & \gto & \nonterminal{factor} \Pin{*} \nonterminal{product} / \nonterminal{factor}\\
 \nonterminal{factor} & \gto & \nonterminal{id} \Pin{(} \nonterminal{sum} \Pin{)} / \Pin{(} \nonterminal{sum} \Pin{)} / \nonterminal{id}\\
 \nonterminal{id} & \gto & \Prange{a}{z} \Prange{a}{z}^*
\end{array}\]

Top-down parsing of infix expressions may require unbounded buffering of the left operand, as the operator itself arrives later in the input stream. The following shows an input string, and below each symbol is the size of the parse table right after its consumption:
\begin{center}
{\setlength{\tabcolsep}{2.25pt}
\begin{tabular}{l|llllllllllllllllllllllll}
$a_j$ & \texttt{z} & \texttt{=} & \texttt{f} & \texttt{(} & \texttt{z} & \texttt{)} & \texttt{;} & \texttt{x} & \texttt{=} & \texttt{x} & \texttt{+} & \texttt{y} & \texttt{*} & \texttt{y} & \texttt{*} & \texttt{y} & \texttt{;} & \texttt{g} & \texttt{(} & \texttt{x} & \texttt{)} & \texttt{;} & \texttt{.} & \texttt{\eof}\\
size & 1 & 0 & 1 & 2 & 3 & 4 & 0 & 1 & 0 & 1 & 0 & 1 & 2 & 3 & 4 & 5 & 0 & 1 & 2 & 3 & 4 & 0 & 0 & 1
\end{tabular}
}
\end{center}
We are not concerned with the speculation constant; assume that it is unbounded. The example demonstrates how the method adapts the table size as input is consumed. Note that \texttt{;} and \texttt{=} resolves the sum expression currently being parsed, truncating the table, and also that the left operand of the \texttt{+} symbol is correctly resolved, while the \texttt{*} expression must be buffered.

\paragraph{Ambiguous Tail-Recursive Programs}
Any non-deterministic finite automaton (NFA) can be interpreted as a PEG program by assigning a nonterminal to each state, and for each state $q$ with transitions $q \stackrel{a_1}{\to} q_1, ..., q \stackrel{a_n}{\to} q_n$ creating a rule $\nonterminal{q} \gto a_1 \nonterminal{q_1} / ... / a_n \nonterminal{q_n}$. The ordering of transitions is significant and defines a disambiguation priority. The final state $q^f$ is assumed to have no transitions, and is given the rule $\nonterminal{q^f} \gto \geps$.

If the NFA contains no $\varepsilon$-loops then its language will coincide with that of its PEG encoding, which is a complete program implementing a backtracking depth-first search for an accepting path. The following shows a simple example of an NFA and its prioritized interpretation as a PEG:
\begin{center}
  \begin{minipage}{0.35\linewidth}
\begin{tikzpicture}[scale=0.7,every node/.style={scale=0.9},>=stealth',x=0.9cm]
  \node[state,initial left,initial text=] at (0,0) (S) {$S$};
  \node[state] at (2,0) (T) {$T$};
  \node[state,accepting] at (0,-2) (E) {$E$};
  \draw[loop above,->] (S) to node {$\texttt{a}$} (S);
  \draw[->,bend left] (S) to node[anchor=south] {$\texttt{a}$} (T);
  \draw[->,bend left] (T) to node[anchor=north] {$\texttt{a}$} (S);
  \draw[->] (S) to node[anchor=west] {$\texttt{b}$} (E);
\end{tikzpicture}
  \end{minipage}
\begin{minipage}{0.64\linewidth}
\[\begin{array}{r@{}c@{}lr@{}c@{}l}
    \nonterminal{S} & \gto & \multicolumn{4}{@{}l}{\Pin{a} \nonterminal{S} / \Pin{a} \nonterminal{T} / \Pin{b} \nonterminal{U}} \\
    \nonterminal{T} & \gto & \Pin{a} \nonterminal{S} \\
    \nonterminal{E} & \gto & \geps \\
    \hline
    S & \gto & P[E,Q] &  P & \gto & A[S,F] \\
    T & \gto & A[S,F] &     Q & \gto & V[E,W] \\
    E & \gto & \geps &     V & \gto & A[T,F] \\
    F & \gto & \gfail &     W & \gto & B[E,F] \\
    B & \gto & \Pin{b} \\
    A & \gto & \Pin{a}
\end{array}\]    
\end{minipage}
\end{center}
The NFA is ambiguous, as any string of the form $a^{n+2}b$, $n \geq 0$, can be matched by more than one path from $S$ to $E$. The priority enforced by the program dictates that $T$ is never invoked, as the production $\texttt{a} S$ \emph{covers} the production $\texttt{a} T$, meaning that every string accepted by the latter is also accepted by the former, which has higher priority in the choice.

The example triggers worst-case behavior for our method, which fails to detect coverage regardless of the speculation bound, resulting in a table size proportional to the input length. This is obviously suboptimal, as any regular language can be recognized in constant space.

The problem is in the tail recursion; the desugared program has every recursive call occur as a condition which remains unresolved until the end-of-marker input has been seen. The analysis is oblivious to coverage, and thus fails to detect that $T$ can never be on a viable expansion until the very end.

\section{Discussion}
\label{sec:discussion2}

We discuss our method in the context of the work of others, and point out directions for future work.

The workset algorithm is an instance of the scheme of \emph{chaotic iteration}~\cite{cousot77} for computing limits of finite iterations of monotone functions.
Our parsing formalism goes back to the TS/TDPL formalism introduced by Birman and Ullman~\cite{birman1970} and later generalized to GTDPL by Aho and Ullman~\cite{aho1972}. They also present the linear-time tabular parsing technique and show that GTDPL can express recognizers for all deterministic context-free languages, including all deterministic LR-class languages. On the other hand, there are context-free languages that cannot be recognised by GTDPL, as general context-free parsing is super-linear~\cite{lee2002}. Ford's Parsing Expression Grammars (PEG)~\cite{ford2004} have the same recognition power as GTDPL, albeit using a larger set of operators which arguably are better suited for practical use.

Packrat parsing~\cite{ford2002}, is a direct implementation of the PEG operational semantics with memoization. It can be viewed as ``sparse'' tabular parsing where only the entries encountered on a depth-first search for an expansion are computed. Our evaluation shows that \TT{} computes a very large portion of the table. Some of this overhead is unavoidable, as the dynamic analysis relies on the exploration of both branches of the choice currently being resolved, but most of the computed entries are never considered by the expansion process. A closer integration of expansion and table computation inspired by Packrat may turn out to be a rewarding implementation strategy.

Heuristic approaches include Kuramitsu's Elastic Packrat algorithm~\cite{kuramitsu2015} and Redziejowski's parser generator \emph{Mouse}~\cite{redziejowski2009}, both of which are Packrat implementations using memory bounded by a configurable constant. The former uses a sliding window to limit the number of stored table columns, and the latter limits the number of memoized calls per nonterminal. Both approaches risk triggering exponential behavior when backtracking exceeds the bounds of their configured constants, which however seems rare in practice. A disadvantage of heuristic memory reductions is that they have to store the full input string until the full parse is resolved, because they cannot guarantee that the parser will not backtrack.

\subsection{Packrat With Static Cut Annotations}
Mizushima, Maeda and Yamaguchi observes that when Packrat has no failure continuations on the stack, all table columns whose indices are less than the index of the current symbol can be removed from memory. To increase the likelihood of this, they extend PEG with cut operators à la Prolog to ``cut away'' failure continuations, and also devise a technique for sound automatic cut insertion, i.e. without changing the recognized language~\cite{mizushima2010}. Manually inserted cuts yield significant reductions in heap usage and increases in throughput, but automatic cut insertion seems to miss several opportunities for optimization. Redziejowski further develops the theory of cut insertion and identifies sufficient conditions for soundness, but notes that automation is difficult: \emph{``It appears that finding cut points in non-LL(1) grammars must to a large extent be done manually''}~\cite{redziejowski2016}.

The method of Mizushima et al. is subsumed by \TT{}. An empty stack of failure continuations corresponds to the case where the condition $A$ in a top-level choice $A[B,C]$ is resolved. Insertion of cuts is the same as refactoring the grammar using the GTDPL operator $A[B,C]$, which is the cut operator $A \uparrow B / C$ of Mizushima et al. in disguise. Increasing the speculation bound can achieve constant memory use without requiring any refactoring of the program.

\subsection{Cost vs Benefit of Memoization}
Several authors argue that the cost of saving parse results outweighs its benefits in practice~\cite{becket2008,ierusalimschy2009}. The PEG implementation for the Lua language~\cite{ierusalimschy2009} uses a backtracking parsing machine instead of Packrat in order to avoid paying the memory cost~\cite{medeiros2008}. Becket and Somogyi compares the performances of Packrat parsers with and without memoization using a parser for the Java language as benchmark~\cite{becket2008}. Their results show that full memoization is always much slower than plain recursive descent parsing, which never triggered the exponential worst case in any of their tests. On the other hand, memoizing only a few selected nonterminals \emph{may} yield in a speedup, suggesting that memoization does not serve as a performance optimization, but as a safeguard against pathological worst-case scenarios which are rare in practice.
However, another experiment by Redziejowki on PEG parsers for the C language show a significant overhead due to backtracking. This could not be completely eliminated by memoizing a limited number of nonterminals, but required manual rewriting of the grammar based on knowledge from the benchmark results~\cite{redziejowski2008}.

Our technique uses full tabulation rather than memoization, but the results still apply to suggest that a direct implementation will likely be slower than plain recursive descent parsers on common inputs and carefully constructed grammars. However, ad-hoc parsers cannot be expected to be constructed in such an optimal way, and thus may need memoization to prevent triggering worst-case behavior. Furthermore, our best-case memory usage---which is bounded---outperforms recursive descent parsers which must store the complete input string in case of backtracking. This is crucial in the case of huge or infinite input strings which cannot fit in memory, e.g. logging data, streaming protocols or very large data files.

\subsection{Parsing Using Regular Expressions}
Medeiros, Mascarenhas and Ierusalimschy embed backtracking regular expression matching in PEG~\cite{medeiros2014}. In fact, every regular expression corresponds to a right-regular context-free grammar\footnote{Contains only productions of the form $A \to \varepsilon$ and $A \to aB$, corresponding 1-1 to the transitions of an NFA.}, and one can easily check that interpreting this grammar as a PEG yields its backtracking matching semantics. Interestingly, the PEG encoding of ambiguous regular expressions make our method exhibit worst-case behavior with regards to streaming and memory usage, as the dynamic analysis is oblivous to detection of \emph{coverage}. Coverage is undecidable for PEG in general, but is decidable for right-regular grammars~\cite{grathwohl2014}.

Grathwohl, Henglein and Rasmussen give a streaming regular expression parsing technique which supports both approximate and optimal coverage analysis~\cite{grathwohl2014}. With Søholm and Tørholm they develop \emph{Kleenex}, which compiles grammars for regular languages into high-performance streaming parsers with backtracking semantics~\cite{grathwohl2016}. Since PEGs combine lexical and syntactic analysis, they can be expected to contain many regular fragments. Perhaps the technique of Kleenex can be combined with PTP to obtain better streaming behavior for these.

% \subsection{General CFG parsing}
% 
% Natural languages rarely have uniqueness of parses, and so require generalized CFG parsing algorithms which produc a \emph{forest} of all the possible parse trees for the input. Due to the lower bound of CFG parsing, they all have non-linear worst-case performance.
% 
% Popular techniques include the CYK algorithm~\cite{younger1967}, Earley parsers~\cite{earley1970}, GLR parsing due to Tomita~\cite{tomita1987} and GLL parsing due to Scott and Johnstone~\cite{scott2010}.
% 
% 
% Left recursion, but no hidden left recursion. Disambiguation at run-time: ALL(*) by \cite{parr2014}

% Classic parsing algorithms include LR parsing~\cite{knuth1965}, limited to LR grammars (generally only computer languages). Generalizations exist, e.g. \cite{tomita1987}, whose graph-structured stack might be related to path trees. General CFG parsing includes e.g. Earley parsing~\cite{earley1970}, CYK (TODO: Citation), parsing with derivatives~\cite{might2011}, etc.
% 
% TODO: Other top-down approaches: Parser combinators~\cite{swierstra1998,swierstra2009}.
% 
% TODO: Tabular LR parsing~\cite{nederhof1996}.

% GLL Parsing~\cite{scott2010}
\section{Conclusion}
\label{sec:conclusion}

We have presented \TT{}, a new streaming execution model for the TDPL family of recursive descent parsers with limited backtracking, together with a linear-time algorithm for computing progressive tables and a dynamic analysis for improving the streaming behavior of the resulting parsers. We have also demonstrated that parsers for both LL and non-LL languages automatically adapt their memory usage based on the amount of lookahead necessary to resolve choices.

A practical performance-oriented implementation will be crucial in order to get a better idea of the applicability of our method. Our prototype evaluation shows that a substantial amount of the computed table entries are never used, so future work should focus on minimizing this overhead.

We believe that our method will be useful in scenarios where a streaming parse is desired, either because all of the input is not yet available, or because it is too large to be stored in memory at once. Possible applications include read-eval-print-loops, implementation of streaming protocols and processing of huge structured data files.

\section{Proofs}

\proplinearparsetrees*
\begin{proof}
Observe that $\C{D}$ cannot contain a strict subderivation for the subject $(A,u)$, as determinism would imply that $\C{D}$ would be infinite.

We show by induction on $|u| - |v|$ that $|\C{T}(\C{D})| \leq 2^{|P|} (|u| - |v|)$.
\end{proof}

\subsection{Tabulation of Operational Semantics}

\lemrandomstart*
\begin{proof}
  We prove both directions of the equality.

  Claim: $\lfp F \sqsubseteq \bigsqcup_n F^n(T)$. We first remark that by definition, $\lfp F = \bigsqcup_n F^n(\bot)$ is the least upper bound of $\{F^n(\bot)\}$. Observe that for all $n$ we have $F^n(\bot) \sqsubseteq F^n(T) \sqsubseteq \bigsqcup_n F^n(T)$. Indeed, the last inequality follows by definition of least upper bounds. The former holds by induction, since we have $\bot \sqsubseteq T$ and by monotonicity of $F$, $F^m(\bot) \sqsubseteq F^m(T)$ implies $F^{m+1}(\bot) \sqsubseteq F^{m+1}(T)$. Since we have shown that $\bigsqcup_n F^n(T)$ is an upper bound of $\{ F^n(\bot) \}$, we have $\lfp F \sqsubseteq \bigsqcup_n F^n(T)$.

  Claim: $\bigsqcup_n F^n(T) \sqsubseteq \lfp F$. Observe that for all $n$ we have $F^n(T) \sqsubseteq \lfp F$. Indeed we have $T \sqsubseteq \lfp F$ by assumption, and by monotonicity $F^m(T) \sqsubseteq \lfp F$ implies $F^{m+1}(T) \sqsubseteq F (\lfp F) = \lfp F$. Since we have shown that $\lfp F$ is an upper bound of $\{F^n(T)\}$ it follows that $\bigsqcup_n F^n(T) \sqsubseteq \lfp F$.
\end{proof}

\thmfundamental*
\begin{proof}
We start by proving Property~\ref{fund:cond:tail}. Let $w = u \eof^\omega$ and observe that for all $(i,j)$ where $j > |u|$ we have $w_j = w_{|u|}$, and it follows that $F^u(\bot)_{ij} = F^u(\bot)_{i|u|} \in \{\bot, \gfail\}$. By a simple induction we obtain that $j > |u|$ implies $(F^u)^k(\bot)_{ij} = (F^u)^k(\bot)_{i|u|}$ for all $k \geq 0$. Therefore $j \geq |u|$ implies that $T(u)_{ij} = \bigsqcup_k (F^u)^k(\bot)_{ij} = \bigsqcup_k (F^u)^k(\bot)_{i|u|} = T(u)_{i|u|}$.

Before proving the remaining, we make the claim that for all $(i,j)$ such that $j \leq |u|$ we have
\begin{enumerate}
\item If $(A_i, u_j) \gparse u_{j+m}$ then $\exists k\ldotp (F^u)^k(\bot)_{ij} = m$.
\item If $(A_i, u_j) \gparse \gfail$ then $\exists k\ldotp (F^u)^k(\bot)_{ij} = \gfail$.
\item If $(A_i, u_j) \not\gparse$ then $\forall k\ldotp (F^u)^k(\bot)_{ij} = \bot$.
\end{enumerate}
If the claim holds, then one direction of Properties \ref{fund:cond:fail},\ref{fund:cond:success},\ref{fund:cond:bot} follow. For example, if $(A_i, u_j) \gparse u_{j+m}$ then there is a $k$ such that $m = (F^u)^k(\bot)_{ij} \sqsubseteq \bigsqcup_k (F^u)^k(\bot)_{ij} = T(u)_{ij}$. For the converse directions we use the fact that for all $(i,j)$ we have $(\exists m \ldotp (A_i, u_j) \gparse u_{j+m}) \vee ((A_i, u_j) \gparse \gfail) \vee ((A_i, u_j) \not\gparse)$. Using the previous claims, the value of $T_{ij}$ will be in contradiction with all but one of the three disjuncts.

The first two claims follow by induction on derivations. In the inductive cases we use monotonicity of $F^u$ to pick a large enough $k$. For the third claim we prove that $\exists k\ldotp (F^u)^k(\bot)_{ij} = \gfail/m$ then $(A_i, u_j) \gparse \gfail/u_{j+m}$ by induction on $k$. The contrapositive of this matches the third claim.

Using determinacy of the parsing relation, it is easily seen that the properties of the Theorem uniquely determines $T(u)$.
\end{proof}

\subsection{Prefix Tables}

\thmapproximation*
\begin{proof}
  Let $u \in \Sigma^*$. It suffices to show that for any $v \in \Sigma^*$, we have ${T^<(u) \sqsubseteq T(uv)}$.

  We first remark that for all $J \subseteq \Set{Index}$ and $T \in \Set{Table}$ we have $F^u_J(T) \sqsubseteq F^u(T)$.
  Furthermore, for all $v \in \Sigma^*$ we have $F^u_{J_u}(T) = F^{uv}_{J_u}(T)$. By these two remarks, we obtain via induction that for all $n \geq 0$, we have $(F^u_{J_u})^n(\bot) = (F^{uv}_{J_u})^n(\bot) \sqsubseteq (F^{uv})^n(\bot) \sqsubseteq \bigsqcup_n (F^{uv})^n(\bot)$. Hence $T(uv)$ is an upper bound of $\{ (F^u_{J_u})^n(\bot) \mid n \geq 0 \}$, but since $T^<(u)$ is the \emph{least} upper bound of this set, we obtain $T^<(u) \sqsubseteq T(uv)$.
\end{proof}

\subsection{Correctness of algorithm}

\lemworksetcharacterization*
\begin{proof}
   Let $(i,j) \in \Set{Index}$. For the forward direction, assume $T_{ij} \sqsubset F^{(u)}(T)_{ij}$. Then $T_{ij} = \bot$, and since $T_{ij} \not= F^{(u)}(T)_{ij} = F_{J_u}(T)_{ij}$, we must have $(i,j) \in J_u$. It remains to prove the implication. Assume $A_i \gto A_x[A_y,A_z]$. By cases on the definition of $F^u$ and the fact $F^u(T)_{ij} \not= \bot$, we have three possible cases:
 $T_{xj} = m$ and $T_{y(j+m)} = m'$; or
 $T_{xj} = m$ and $T_{y(j+m)} = \gfail$; or
 $T_{xj} = \gfail$ and $T_{zj} \not= \bot$.
 In the first two cases we have $D^T_{ij} = (y,j+m)$ and $T_{y(j+m)} \not= \bot$. In the last case we have $D^T_{ij} = (z,j)$ and $T_{zj} = \gfail \not= \bot$, and we are done.
 
 For the converse direction, assume $(i,j) \in J_u$, $T_{ij} = \bot$ and $(\text{$g_i$ complex} \Rightarrow D^T_{ij} \not= \bot \not= T_{D^T_{ij}})$. Since $(i,j) \in J_u$, we have $F^{(u)}(T)_{ij} = F^u(T)_{ij}$ and we need to show $F^u(T)_{ij} \not= \bot$. If $g_i$ is simple it is easy to check that $F^u(T)_{ij} \not= \bot$ in all cases. If $A_i \gto A_x[A_y,A_z]$, then by assumption we have $D^T_{ij} \not= \bot \not= T_{D^T_{ij}}$. We have three possible cases which are handled analogously. For the first case $T_{xj} = m$, $D^T_{ij} = (y,j+m)$ and $T_{y(j+m)} = m'$ for some $m, m'$. By definition $F^u(T)_{ij} = m + m' \not= \bot$, and we are done.
\end{proof}

Dependency monotonicity says that the dependency map seen as a table operator is monotone.
\lemdependencymonotonicity*
\begin{proof}
  Let $(i,j) \in \Set{Index}$ and assume $D^T_{ij} \not= \bot$ (the case $D^T_{ij} = \bot$ is trivial). Then $A_i \gto A_x[A_y,A_z]$ and either $D^T_{ij} = (y,j+m)$ and $T_{xj} = m$; or $D^T_{ij} = (z,j)$ and $t_{xj} = \gfail$. In the first case we get $T_{xj} \sqsubseteq T_{xj}' = m$ by assumption, so $D^{T'}_{ij} = (y,j+m) = D^T_{ij}$. The latter case is analogous.
\end{proof}

The following shows that upon updating a single entry in a table, the set of dependencies that will go from being undefined to being defined can be determined statically.
\lemdependencydifference*
\begin{proof}
For the converse direction, let $i \in C^{-1}_p$, which implies $A_i \gto A_p[A_y,A_z]$. Since $S_{pq} = \bot \not= T_{pq}$, we must have $\bot = D^S_{iq} \sqsubset D^T_{iq} \not= \bot$, and we are done.

  For the forward direction, assume $\bot = D^S_{ij} \sqsubset D^T_{ij} \not= \bot$. By the latter equality it follows that $A_i \gto A_x[A_y,A_z]$ where $C_i = x$, so $i \in C^{-1}_x$. By the first equality and definition, we have $S_{xj} = \bot$; by the latter equality we have $T_{xj} \not= \bot$. But then $S_{xj} \sqsubset T_{xj}$, which implies $(x,j) = (p,q)$. Since $i \in C^{-1}_p$ and $j = q$, we are done.
\end{proof}

We can now prove the main lemma of the correctness proof:

\lemworksetupdate*
\begin{proof}
  We initially remark that $S \sqsubseteq T$ by definition of $T$, and hence that $D^S \sqsubseteq D^T$ by Lemma~\ref{lem:dependency-monotonicity}. Since $(p,q) \in \Delta_u(S)$ we have $S_{pq} \sqsubset F^{(u)}(S)_{pq} = T_{pq}$, so in particular $T_{pq} \not= \bot$.

\textbf{Forward direction}. Assume $(i,j) \in \Delta_u(T)$. By Lemma~\ref{lem:workset-characterization} we have $T_{ij} = \bot$ and $(\text{$g_i$ complex} \Rightarrow D^T_{ij} \not= \bot \not= T_{D^T_{ij}})$. So, $(i,j) \not= (p,q)$ and $S_{ij} = \bot$.

Assume $g_i$ complex. Then $D^T_{ij} \not= \bot \not= T_{D^T_{ij}}$. Since $D^S_{ij} \sqsubseteq D^T_{ij}$, we have either (a) $D^S_{ij} = \bot \sqsubset D^T_{ij}$; or (b) $D^S_{ij} = D^T_{ij} \not= \bot$.

In case (a), we apply Lemma~\ref{lem:dependency-difference} to obtain $(i,j) \in C^{-1}_p \times \{q\}$, which implies $(i,j) \in \{ (i',q) \mid i' \in C^{-1}_p \wedge \bot \not= T_{D^T_{i'q}} \}$, and we are done.

In case (b) we consider the subcases ($\alpha$) $S_{D^S_{ij}} = \bot \sqsubset T_{D^T_{ij}}$; and ($\beta$) $S_{D^T_{ij}} = T_{D^T_{ij}} \not= \bot$. In subcase ($\alpha$),  we must have $D^S_{ij} = (p,q)$ so $(i,j) \in (D^S)^{-1}_{pq}$ and we are done. In subcase ($\beta$), observe that we have $D^S_{ij} \not= \bot \not= S_{D^S_{ij}}$ which by Lemma~\ref{lem:workset-characterization} implies $(i,j) \in \Delta_u(S) \setminus \{(p,q)\}$.

\textbf{Converse direction}. Assume that $(i,j)$ is in the set on the right hand side. By Lemma~\ref{lem:workset-characterization} it suffices to show $T_{ij} = \bot$ and $(\text{$g_i$ complex} \Rightarrow D^T_{ij} \not= \bot \not= T_{D^T_{ij}})$. We have three possible cases:

Case $(i,j) \in \Delta_u(S) \setminus \{(p,q)\}$. Since $(i,j) \not= (p,q)$ we have $S_{ij} = T_{ij}$ by definition of $T$. By Lemma~\ref{lem:workset-characterization} we obtain $S_{ij} = \bot = T_{ij}$ and $(\text{$g_i$ complex} \Rightarrow D^S_{ij} \not= \bot \not= S_{D^S_{ij}})$. Assuming $g_i$ complex, we thus have $D^S_{ij} \not= \bot \not= S_{D^S_{ij}}$, and since $S \sqsubseteq T$ and $D^S \sqsubseteq D^T$, this implies $D^T \not= \bot \not= T_{D^T_{ij}}$, and we are done.

Case $(i,j) \in (D^S)^{-1}_{pq}$. Then $(p,q) = D^S_{ij} = D^T_{ij}$. By Lemma~\ref{lem:dependency-strictness} and $S_{pq} = \bot$ we obtain $T_{ij} = \bot$. Since $D^T_{ij} \not= \bot \not= T_{pq} = T_{D^T_{ij}}$, we are done.

Case $i \in C^{-1}_p$, $j=q$ and $\bot \not= T_{D^T_{iq}}$. We have $D^T_{iq} \not= \bot \not= T_{D^T_{iq}}$, so it suffices to show $T_{iq} = \bot$. Since $C_i = p$, have $A_i \gto A_p[A_y,A_z]$. By $S_{pq} = \bot$, we therefore have $F^{(u)}(S)_{iq} = \bot$. By $S \sqsubseteq F^{(u)}(S)$, this implies $S_{iq} = \bot$. It suffices to show $i \not= p$, as this implies $T_{iq} = S_{iq} = \bot$.

Assume $i = p$. Then by $S_{pq} = \bot$ we have $T_{pq} = F^{(u)}(S)_{pq} = \bot \not= T_{pq}$, so $T_{pq} \not= T_{pq}$ a contradiction. Thus $i \not= p$, and we are done.
\end{proof}

The previous proof uses the following, which shows that the entry for a complex expression cannot be resolved if its dynamic dependency is undetermined.
\begin{lemma}[Dependency strictness]
  \label{lem:dependency-strictness}
  Let $T \in \Set{Table}$, $(i,j) \in \Set{Index}$ and $u \in \Sigma^*$. If $T \sqsubseteq T^<(u)$; $g_i$ complex and $D^T_{ij} = \bot$, then $F^{(u)}(T)_{ij} = \bot$.
\end{lemma}
\begin{proof}
If $(i,j) \not\in J_u$ then $F^{(u)}(T)_{ij} = T_{ij}$. Since $T_{ij} \sqsubseteq T^<(u)_{ij}$, the result follows by showing $T^<(u)_{ij} = \bot$. By $(i,j) \not\in J_u$ we have $F^{(u)}(T)_{ij} = T_{ij}$ for all $T$, and by induction we obtain $(F^{(u)})^n(\bot)_{ij} = \bot$ for all $n \geq 0$. We must therefore have $T^<(u)_{ij} = \bot$, since $T^<(u)$ is the least upper bound of all $(F^{(u)})^n(\bot)$.

In the other case, assume $(i,j) \in J_u$, so $F^{(u)}(T) = F^u(T)$.
We must have $A_i \gto A_x[A_y,A_z]$. By $D^T_{ij} = \bot$ and definition, we have $T_{xj} = \bot$ and hence $F^u(T)_{ij} = \bot$.
\end{proof}

Upon updating a single entry in a table, each entry in the updated reverse dependency map is obtained by appending a predetermined set of indices to the corresponding entry in the old reverse dependency map:
\lemdependencyupdate*
\begin{proof}
  Let $(i,j) \in \Set{Index}$ such that $D^T_{ij} = (k, \ell)$. Since $S \sqsubseteq T$ we have $D^S_{ij} \sqsubseteq D^T_{ij}$. We have $D_{ij}^S = D_{ij}^T = (k,\ell)$ if and only if $(i,j) \in (D^S)^{-1}_{k\ell}$. The other case, $D^S_{ij} \sqsubset D^T_{ij}$, holds if and only if $(i,j) \in C_p^{-1} \times \{q\}$ by Lemma~\ref{lem:dependency-difference}.
\end{proof}

\subsubsection{Correctness of \textsc{Fix}}

\begin{invariant}[Work loop]\label{inv:workloop}
  Assuming variables
  $u \in \Sigma^* (\eof + \varepsilon)$;
  $T \in \Set{Table}$;
  $R : \Set{Index} \to \Two^{\Set{Index}}$; and
  $W \subseteq \Set{Index}$:
\begin{enumerate}
\item\label{inv:workloop:T} $T \sqsubseteq F^{(u)}(T) \sqsubseteq T^<(u)$
\item\label{inv:workloop:R} $R = (D^T)^{-1}$
\item\label{inv:workloop:W} $W = \Delta_u(T)$
\end{enumerate}
\end{invariant}

\begin{lemma}[Initialization]
  When entering line \ref{alg:fix:while:begin} in \textsc{Fix}, Invariant~\ref{inv:workloop} holds.
\end{lemma}
\begin{proof}
Let $u' = a_m a_1 ... a_{n-1}$ and $u = u' a_n$. By the precondition, $T = T^<(u')$ and $R = (D^T)^{-1}$.

\paragraph{Property \ref{inv:workloop:T}.}
We first show $T \sqsubseteq F^{(u)}(T)$. Let $(i,j) \in \Set{Index}$. We either have $(i,j) \in J_{u'}$ or $(i,j) \not\in J_{u'}$.

In the first case we also have $(i,j) \in J_u$, so $F^{(u)}(T)_{ij} = F^{(u')}(T)_{ij} = T_{ij}$, where the last equality follows from the fact that $T$ is a fixed point of $F^{(u')}$.

In the second case $(i,j) \not\in J_{u'}$ we have $\forall T'\ldotp F^{(u')}(T')_{ij} = T'_{ij}$. Hence $\forall n \geq 0\ldotp (F^{(u')})^n(\bot)_{ij} = \bot$, and since $T = T^<(u')$ is the least upper bound of all $(F^{(u')})^n(\bot)$, we have $T_{ij} = \bot \sqsubseteq F^{(u)}(T)_{ij}$.

From the above we conclude $T \sqsubseteq F^{(u)}(T)$, and it remains to show $F^{u}(T) \sqsubseteq T^<(u)$.

Since $F^{(u)}(T^<(u)) = T^<(u)$, this follows by monotonicity of $F$ and $T = T^<(u') \sqsubseteq T^<(u)$, which in turn follows from Proposition~\ref{prop:prefix-monotonicity}.

\paragraph{Property \ref{inv:workloop:R}.} Follows by assumption.

\paragraph{Property~\ref{inv:workloop:W}.} Since $T=T^<(u')$, we have $\Delta_{u'}(T) = \emptyset$. Thus $\Delta_u(T) = \Delta_u(T) \setminus \Delta_{u'}(T)$, and by Lemma~\ref{lem:workset-characterization} $(i,j) \in T^<(u)$ if and only if $j = |u| - 1$, $T_{i(|u|-1)} = \bot$ and either $D^T_{i(|u|-1)} \not= \bot \not= T_{D^T_{i(|u|-1)}}$ or $g_i$ simple. But $D^T_{i(|u|-1)} = \bot$ for all $i < |P|$, so $\Delta_u(T) = \{ (i,|u|-1) \mid \text{$g_i$ simple} \}$.
\end{proof}

\begin{lemma}[Preservation of consistency]
  \label{lem:preservation-consistency}
  If $F:\Set{Table} \to \Set{Table}$ is monotone and $T \sqsubseteq F(T)$, then for all $(p,q) \in \Set{Index}$, we have $F_{pq}(T) \sqsubseteq F(F_{pq}(T))$.
\end{lemma}
\begin{proof}
  Since $T \sqsubseteq F(T)$ then in particular $T_{pq} \sqsubseteq F(T)_{pq}$, so $T \sqsubseteq F_{pq}(T)$. By monotonicity, we have $F(T) \sqsubseteq F(F_{pq}(T))$. But then
\[ \forall(i,j) \in \Set{Index}\ldotp T_{ij} \sqsubseteq F(T)_{ij} \sqsubseteq F(F_{pq}(T))_{ij} \]
We now prove $\forall (i,j) \in \Set{Index}\ldotp F_{pq}(T)_{ij} \sqsubseteq F(F_{pq}(T))_{ij}$. If $(i,j) = (p,q)$, then $F_{pq}(T)_{ij} = F(T)_{ij} \sqsubseteq F(F_{pq}(T))_{ij}$; and if $(p,q)\not\in(i,j)$, then $F_{pq}(T)_{ij} = T_{ij} \sqsubseteq F(F_{pq}(T))_{ij}$.
\end{proof}

\begin{lemma}[Maintenance]
  Invariant~\ref{inv:workloop} is maintained for each iteration of lines~\ref{alg:fix:while:begin}-\ref{alg:fix:while:end} in \textsc{Fix}.
\end{lemma}
\begin{proof}
Assume that Invariant~\ref{inv:workloop} holds, and let $S$ refer to the configuration of $T$ at the beginning of the iteration. When the iteration has finished, some $(p,q) \in \Delta_u(S)$ has been picked such that

\begin{enumerate}
\item[(a)]
$T = F^{(u)}_{pq}(S)_{ij}$

\item[(b)]
$S \sqsubseteq F^{(u)}(S) \sqsubseteq T^<(u)$

\item[(c)]
$\forall k,\ell \ldotp R_{k\ell} = (D^S)^{-1}_{k\ell} \cup \{ (i', q) \mid i' \in C^{-1}_p \} \cap (D^T)^{-1}_{k\ell}$

\item[(d)]
$W
\begin{array}[t]{@{}c@{}l@{}}
{}={} & \Delta_{u}(S) \setminus \{(p,q)\} \\
{}\cup{} & (D^S)^{-1}_{pq} \\
{}\cup{} & \{ (i',q) \mid i' \in C^{-1}_p \wedge D^T_{i'q} \not= \bot \not= T_{D^T_{i'q}} \}
\end{array}$
\end{enumerate}

By Lemma~\ref{lem:preservation-consistency} on (a), (b), Property~\ref{inv:workloop:T} is reestablished.

By Lemma~\ref{lem:dependency-update} on (a), (b), (c), Property~\ref{inv:workloop:R} is reestablished.

By Lemma~\ref{lem:workset-update} on (a), (b) and (d), Property~\ref{inv:workloop:W} is reestablished. \qedhere
\end{proof}

\begin{lemma}[Termination]
  Invariant~\ref{inv:workloop} entails the postcondition of \textsc{Fix} when the loop in lines~\ref{alg:fix:while:begin}-\ref{alg:fix:while:end} terminates.
\end{lemma}
\begin{proof}
  When the loop terminates we have $W = \emptyset$. By the invariant we have both $W = \Delta_u(T) = \emptyset$ and $T \sqsubseteq T^<(u)$, so $T = T^<(u)$.
\end{proof}

%%% Local Variables:
%%% mode: latex
%%% TeX-master: "popl17"
%%% End:

}

\putbib[bibliography]

\end{bibunit}

%%% Local Variables:
%%% mode: latex
%%% TeX-master: "thesis"
%%% End:

\backmatter

\clearpage

\thispagestyle{empty}
~
\vspace{24em}
\begin{center}
  \textcolor{gray}{
  (The following pages have intentionally been left blank)}
\end{center}

\end{document}